\documentclass[a4paper,twoside,11pt]{book}

\usepackage{amsmath}
\usepackage{amsfonts} 
\usepackage{amssymb} 
\usepackage{amsthm} 
\usepackage{bbm} 
\usepackage{dsfont}
\usepackage{graphicx}
\usepackage{psfrag}
\usepackage{appendix}
\usepackage{hyperref}
 \usepackage{a4wide}
 \usepackage{subfigure}
 \usepackage{fancyhdr}

\newcommand{\ch}{\mbox{(Ch)}}
\newcommand{\cc}{\,\mbox{c.c.}}
\newcommand{\eq}[1]{Eq. (\ref{#1})}

\newcommand{\acomm}[2]{\left\{#1,#2\right\}}

\newcommand{\ket}[1]{|{#1}\rangle}
\newcommand{\dket}[1]{|{#1}\rangle\!\rangle}

\newcommand{\ii}{\mathrm{i}}
\newcommand{\ee}{\mathrm{e}}
\newcommand{\one}{{\rm 1\kern -.9mm l}}

\newcommand{\Tr}{\mathrm{Tr}\,}
\newcommand{\tr}{\mathrm{tr}\,}

\newcommand{\re}{\mbox{Re}\,}
\newcommand{\im}{\mbox{Im}\,}

\newcommand{\taup}{\theta^\alpha}
\newcommand{\ldaup}{\lambda^{\dot\alpha}}
\newcommand{\lda}{\lambda_{\dot\alpha}}
\newcommand{\wda}{\bar w_{\dot\alpha}}
\newcommand{\wdb}{\bar w_{\dot\beta}}
\newcommand{\ta}{\theta_\alpha} 
\newcommand{\beq}{\begin{equation}}
\newcommand{\eeq}{\end{equation}}
\newcommand{\beqa}{\begin{eqnarray}}
\newcommand{\eeqa}{\end{eqnarray}}
\newcommand{\MM}{\mathcal{M}}
\newcommand{\DD}{\mathcal{D}}

\newcommand{\NN}{\mathcal{N}}
\newcommand{\FF}{\mathcal{F}}

\newcommand{\atopnew}[2]{\genfrac{}{}{0pt}{3}{#1}{#2}}
\newcommand{\cale}{\mathcal{E}}

\newcommand{\nn}{\nonumber}
\newcommand{\dslash}[1]{\displaystyle{\not} #1}

\begin{document}
\thispagestyle{empty}

\begin{flushright}
LAPTH-1366/09\\
\end{flushright}
\vskip 0.25in

\vskip 0.1in
\centerline{{\huge Applications of String Theory: }}
\centerline{{\huge Non-perturbative Effects in Flux Compactifications}}
\centerline{{\huge and Effective Description of Statistical Systems}} 
~\\
\vskip 0.25in
\begin{center}
{\bf
Livia Ferro
}
\end{center}
{\hspace{-0.5in}
  \begin{center}
{\it  LAPTH, Universit\'e de Savoie, CNRS\\
9, chemin de Bellevue, BP 110, 74941 Annecy le Vieux Cedex, France}
\end{center}
}
\begin{center}
{livia.ferro@lapp.in2p3.fr
}
\end{center}
\vskip 0.5in
\begin{center}
\bf{Abstract}
\end{center}
In this paper, which is a revised version of the author's PhD thesis, we analyze two different applications of string theory.  In the first part, we focus on four dimensional compactifications of
Type II string theories preserving $\mathcal N=1$ supersymmetry, in presence of intersecting or magnetized D-branes.  
We show, through world-sheet methods, how the insertion of closed string background fluxes may modify the effective interactions on Dirichlet and Euclidean branes. In particular, we compute flux-induced fermionic masses. The
generality of our results is exploited to determine the soft terms of the action on the instanton moduli space.
 Finally, we investigate how fluxes create new non-perturbative superpotential terms in presence of gauge and stringy instantons in SQCD-like models. 
The second part is devoted to the description of statistical systems through effective string models. In particular,  we focus our attention on $(d-1)$-dimensional interfaces, present in particular statistical systems defined on compact $d$-dimensional spaces.  We compute their exact partition function by resorting to standard covariant quantization of the Nambu-Goto theory, 
and we compare it with Monte Carlo data. 
Then, we propose an effective model to describe interfaces in 2$d$ space and test it against the dimensional reduction of the Nambu-Goto description of the 2$d$ interface.

\pagebreak

\normalsize

\begin{center}
\Large{UNIVERSIT\`A DEGLI STUDI DI TORINO} \\
\smallskip
\Large{Dipartimento di Fisica Teorica} \\
\vspace{0.6cm}
\Large{DOTTORATO DI RICERCA IN FISICA FONDAMENTALE, APPLICATA E
ASTROFISICA}\\
\vspace{0.2cm}
\Large{Ciclo XXI} \\

\vspace{0.3cm}
\begin{center}
\includegraphics[width=2.5cm]{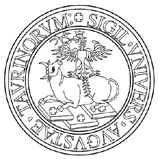}
\end{center}

\vspace{0.7cm}
\Large{\bfseries{Applications of String Theory: \\ Non-perturbative Effects \\in Flux Compactifications \\and\\
Effective Description of Statistical Systems}} \\
\end{center}

\vspace{1.0cm}
\noindent
\hspace{0.9cm} \large{Candidato:} \hspace{4.5cm} \large{Relatore:}  \\
\medskip
\noindent
\hspace{0.85cm} \large{Livia Ferro} \hspace{4.6cm} \large{Prof. Marco Bill\`{o}} \\

\noindent
\hspace{0.9cm} \large{Coordinatore:} \hspace{4.0cm} \large{Controrelatore:} \\
\medskip
\noindent
\hspace{0.85cm} \large{Prof. Stefano Sciuto} \hspace{2.95cm} \large{Prof. Angel Uranga} \\

\vspace{1.3cm}
\begin{center}
\noindent
\large{Anni accademici 2005/06 - 2006/07 - 2007/08} \\
\medskip
\noindent
\large{Settore scientifico disciplinare di afferenza: FIS/02} \\
\end{center}

\tableofcontents

\chapter{Introduction}
\label{intro}

String theory presents many different interesting facets; in this thesis we want to focus our attention to two of its various applications. We will first analyse how the presence of closed string background fluxes may modify the perturbative and non-perturbative sectors of the gauge theories realized by means of particular D-brane configurations; then we will explain how the stringy formalism can be applied to the effective description of certain aspects of Lattice Gauge Theories and more general statistical systems.

As many other theoretical
discoveries, 
string theory 
has a fascinating history, which goes back several decades.
String theory arose indeed in the late sixties, in a different form with respect to the modern one.  In that period experiments were providing an enormous proliferation of strongly interacting particles of higher spins. 

\paragraph{Regge trajectories}
Tullio Regge in 1957 introduced the \emph{complex angular momentum} method \cite{Regge:1959mz}.
In its relativistic formulation this helped to study the properties of scatterings as functions of angular momentum, after having analytically continued the scattering amplitude to the whole complex plane. The main characteristic was that the amplitude had an explicit exponential dependence on a \emph{Regge trajectory} function $J(s)$, which enclosed the information of the angular momentum $J$ with respect to the state energy $s$. 
Meanwhile regularities in the spectrum of strongly interacting particles were observed.
In 1960 G. Chew and S. Frautschi \cite{Chew:1961ev} conjectured for them
a simple dependence between their angular momentum and squared mass
\beq
J = \alpha(s) = \alpha(0) + \alpha' s ~;
\eeq
in other words the particles were aligned on Regge trajectories which were straight lines.
 The constant $\alpha'$ was called \emph{Regge slope} and $\alpha(0)$ is an additive shift. This description predicted the existence of infinitely many particle families, in function of  $\alpha(0)$.

\vspace{1cm}

The observation that the amplitudes for mesons scattering in the $s$-channel had a perfect match with amplitudes for the $t$-channel scattering (\emph{i.e.} there was a duality between the description in terms of Regge poles or of resonances) lead to \emph{Dual Resonance Models}. In 1968 Veneziano proposed his famous formula \cite{Veneziano:1968yb} which describes the scattering of four particles lying on Regge trajectories by means of the Euler Beta function
\beq
A(s,t) = \frac{\Gamma(-\alpha(s))\Gamma(-\alpha(t))}{\Gamma(-\alpha(s)-\alpha(t))}~,
\eeq
where $\alpha(s)$ is the Regge trajectory. Let us remark that this expression is explicitly $s$-$t$ crossing symmetric.
In 1970 it was argued independently by Nambu, Nielsen and Susskind \cite{Nambu:1997wf,Nielsen,Susskind:1970xm} that the Veneziano dual formula could be derived from the quantum mechanics of relativistic oscillating one-dimensional objects, \emph{strings}, \emph{i.e.} of an infinite tower of simple harmonic oscillators. In this description the $s$- and $t$-channel were naturally identified with the same process; indeed a tree level open string diagram at fixed external legs is unique while in the QFT limit it can be viewed as the $s$-channel or the $t$-channel and the straight-line Regge trajectories were then understood as arising from a rotating relativistic string of tension proportional to $\alpha(s)^{-1}$.
The idea was that the vibrational modes  of these one-dimensional objects coincide with hadronic particles but,
while particles are zero-dimensional objects, so that their classical motion is a one-dimensional line of minimal length, the string, which is a one-dimensional object, will classically describe a two-dimensional surface, the \emph{worldsheet}. The natural classical action is just the area of the worldsheet.

\paragraph{Nambu-Goto action}
Such an action was first introduced by Y. Nambu and T. Goto \cite{nambu-goto}:
\beq
\label{ng}
S_{NG} = T \int d\sigma d\tau \sqrt{-det\left(\eta^{\mu\nu}\partial_{\alpha}{X}_{\mu}\partial_{\beta}{X}_{\nu}\right)} 
\eeq
where $\alpha, \beta = (\sigma, \tau)$ are the coordinates of the worldsheet and $T = \frac{1}{2\pi\alpha'}$ is the tension of the string, which is therefore proportional to the Regge slope. The fields $X^{\mu} = X^{\mu}(\sigma, \tau)$ give the embedding of the world-sheet in space-time ($\mu = 1,...,d$). 
In 1976, by means of the definition of an independent metric on the worldsheet $h_{\alpha\beta}(\sigma,\tau)$, a first order version of Nambu-Goto action was proposed \cite{Polyakov_action}
 \beq
\label{SPol}
 S_{Pol} = -\frac{T}{2} \int d\sigma d\tau \sqrt{-h}~  h^{\alpha\beta}\left(\eta^{\mu\nu}\partial_{\alpha}{X}_{\mu}\partial_{\beta}{X}_{\nu}\right)
 \eeq
from which the NG action (\ref{ng}) is retrieved by integrating out $h$. 
In this theory both \emph{open} strings, with two distinct endpoints, and \emph{closed} strings, where the endpoints make a complete loop, can be naturally considered. 
For the closed string, where $ X^{\mu}(\sigma, \tau) =  X^{\mu}(\sigma + 2\pi, \tau)$, we can write the following mode expansion
\beq
X^{\mu}(\sigma, \tau) = x^\mu + \alpha' p^\mu \tau + i \sqrt{\frac{\alpha'}{2}} \sum_{n \neq 0} (\alpha_n^\mu e^{-in(\tau-\sigma)}+\tilde{\alpha}_n^\mu e^{-in(\tau+\sigma)}),
\eeq
where $x^{\mu}$ is the center of mass of the string and $p^\mu$ the momentum associated to it.
In the case of open strings, the equations of motion and the boundary conditions for the fields $X^\mu$ derived from (\ref{SPol})
\beqa
\label{bc}
\partial_\alpha \partial^\alpha X^\mu &=& 0 \\
\partial_\sigma X^\mu  \delta X^\mu |_{\sigma=0, \pi} &=& 0 
\eeqa
can be satisfied in different ways, 
depending on the chosen boundary conditions for each endpoint. Indeed, to solve (\ref{bc}) one can choose \emph{Neumann} boundary conditions 
\beq
\partial_\sigma X^\mu~= 0
\eeq
or \emph{Dirichlet} ones:
\beq
\delta X^\mu = 0,
\eeq
so that there can be Neumann-Neumann, Dirichlet-Dirichlet or Neumann-Dirichlet mode expansions.\\
The process of quantization promotes the oscillators $\alpha_n^\mu$ and $\tilde{\alpha}_n^\mu$ to annihilation and creation operators acting on a Fock space. The absence of non-physical states (or, equivalently, the cancellation of conformal anomalies) fixes the dimensions of the target space to $26$. 
\vspace{1cm}

The description of strong interactions based on bosonic strings was not satisfying because its spectrum contained only bosons, among which a tachyon responsible of instability.
Moreover, it made many predictions that directly contradicted experimental results and could not explain all the kinematical regimes. 
Indeed, dual models did not incorporate the parton-like behaviour; a different theory of strong interactions was required and since 1974 Quantum Chromodynamics was recognized to give a more accurate description of experimental data in the perturbative regime. It was indeed discovered that hadrons and mesons are made by quarks and well described by an $SU(3)$ gauge theory, QCD. However, QCD is very useful to describe the behaviour of strong interactions at high energies but, since at low energies it becomes strongly coupled,
calculations on items like confinement and chiral symmetry breaking are not easily performed.

The interaction between a quark and an antiquark can be instead well described  with a string-like colour flux tube (\emph{i.e.} an \emph{effective QCD string}) stretching between them, as we will see from Chapter \ref{introeff} on.  For large distances $R$, the potential goes like
where $T$ is the string \emph{tension} related to $\alpha'$. 
The asymptotic expansion of Nambu-Goto bosonic string gives just a confining term $V(R) = T R +...$.
For this reason the presence of a string-like behaviour in some regimes of strong interactions is still believed to be correct.
Strong support to this idea came from  G. 't Hooft suggestion \cite{'tHooft:1973jz} of studying gauge theories with $N$ colours in the large-$N$ limit.  
The diagrammatic expansion in the parameter $\frac{1}{N}$ turned out to be organized according to the genus $g$ of the diagram surface, just as an expansion of a perturbative theory with closed oriented strings.
This suggested that gauge theories admit a \emph{dual} representation by means of string models. 
The large-$N$ limit turned out to give a good qualitative description on confinement,  U(1) anomalies and other dynamical items.
For two-dimensional gauge theories much progress was made in the early nineties and dual string theories were found, starting with \cite{Gross:1992tu}.
The four-dimensional case is much more complicated but in 1997 J. Maldacena succeeded to find a dual of a specific four-dimensional gauge theory \cite{Maldacena:1997re}. 
He indeed  found a correspondence between $\NN = 4$ super Yang-Mills field theory in four space-time dimension and Type IIB string theory in a background of five-dimensional anti-de Sitter space times a five-sphere. 
His \emph{AdS/CFT correspondence}  brought back to the fore the idea of the effective QCD string. Nowadays, the comparison between results of effective QCD string and numerical lattice simulations can
give some important hints and insights for consistent models.

\vspace{0.5cm}

While string theory was being supplanted by QCD, a new discovery promoted it as a good candidate for a theory of quantum gravity.
Indeed, it was always in 1974 that a massless spin two excitation from the closed string sector was discovered \cite{Scherk:1974ca}, which could be interpreted as the \emph{graviton}. 
String theory therefore evolved to a more general theory of interactions and was examined as a possible ultimate theory of nature in the quest for a unified description of Fundamental Interactions, the so-called \emph{Theory of Everything} (TOE).
The Nambu-Goto theory had already been extended to a supersymmetric version including fermions, the \emph{superstring theory} \cite{Ramond:1971gb,Neveu:1971rx}.

\paragraph{Superstrings}
The fermionic worldsheet action (to be definite we will focus on Type II superstring theories) is based on the supersymmetry on the world-sheet
\beq
S = - \frac{T}{2} \int d\sigma d\tau  \left(\sqrt{-h}~  h^{\alpha\beta}\eta^{\mu\nu}\partial_{\alpha}{X}_{\mu}\partial_{\beta}{X}_{\nu}-i
\eta^{\mu\nu}\bar{\psi}_{\mu} \rho_{\alpha} \partial^{\alpha}\psi_{\nu} \right) 
\eeq
where $\psi^\mu$ are worldsheet Majorana spinors and the $\rho_\alpha$ matrices provide a representation of the Clifford algebra. This action is invariant under supersymmetric tranformations. As for the pure bosonic string, one can proceed by looking for the mode expansions given by the equations of motion and boundary conditions and then promoting the respective oscillators to operators. From the quantization of world-sheet spinors two sectors arise, the \emph{Ramond} (R) and the \emph{Neveu-Schwarz} (NS) one.

\begin{table}[h]
\begin{center}
\begin{tabular}{c||cc}

 &  NS-NS  & R-R\\
\hline\hline
Type IIA &  $g_{\mu\nu},B_{\mu\nu}$ & $C_1, C_3$ \\
\hline
Type IIB &  $g_{\mu\nu},B_{\mu\nu}$ & $C_0, C_2, C_4$ \\
\hline\hline
\end{tabular}

\caption{The closed string sector of Type IIA and Type IIB theories.}
  \label{listcharged}
\end{center}
\end{table}
In the Ramond sector of the open string the oscillators satisfy the anti-commutation relation
\beq
\{d_n^i,d_m^j\} = 2 \delta^{ij} \delta_{n+m}
\eeq
which for $n=m=0$ reduces to the Clifford algebra
\beq
\{d_0^i,d_0^j\} = 2 \delta^{ij}.
\eeq
The ground state of the Ramond sector is therefore a spacetime spinor. Enforcing the GSO projection, a supersymmetric spectrum is left.
For the superstring it can be shown that the absence of non-physical states requires a 10 dimensional spacetime $M_4 \otimes X_6$. As the space-time we can "feel" is only four-dimensional, this needs a way to \emph{compactify} the six spatial dimensions which are exceeding.
To compare this construction with particle physics, one needs \emph{low-energy effective} actions, which describe the dynamics of the massless states of the string in the field theory limit $\alpha'\rightarrow 0$, where the string reduces to a point particle.
To do this one has to find massless (or light) states and construct the effective interaction terms. (Also massive states can have impact on them, for instance if one computes a loop amplitude.) This can be performed for instance by computing string amplitudes 
 with these states and then going to the limit $\alpha'\rightarrow 0$.

\vspace{1cm}
Different kinds of consistent string theories were constructed; totally, they were five: Type I, Type IIA and Type IIB (on which we will mainly concentrate our attention), and two heterotic, E8 X E8 and SO(32). At that time it was believed that only one of these five candidates, the theory whose low energy limit after compactification would be able to match the physics observed, was the actual correct TOE.
They indeed present many different characteristics. For instance Type IIA is non-chiral, whereas the other four are chiral; Type I, Type IIA and IIB contain open and closed strings, while the heterotic theories only closed strings.

In 1984 the \emph{first superstring revolution} started by the discovery of M. Green and J. H. Schwarz  of anomaly cancellation in type I string theory (Green-Schwarz mechanism). String theory became to be accepted as an actual candidate for the unification theory.

\begin{figure}
\begin{center}
 \includegraphics[scale=0.8]{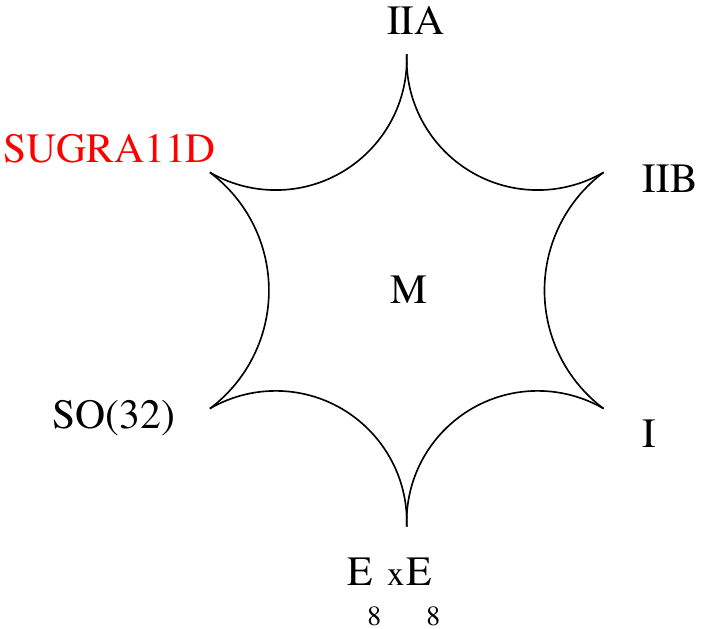}%
 \end{center}
\caption{The five superstring theories and the eleven dimensional supergravity are nothing else than different limits of M-theory.}
\label{five}
\end{figure}

Approximately between 1994 and 1997 the \emph{second superstring revolution} took place. It was realized that the five 10-dimensional string theories  were related through a web of duality transformations, which for instance connect large and small distance scales ($T$-duality), or strong and weak coupling constants ($S$-duality) of different theories. A particular combination of $T$- and $S$-duality is called $U$-duality. When dimensions are compactified other dualities arise.
In 1995 E. Witten discovered \cite{Witten:1995ex} that the five 10-dimensional superstring theories were not only related between them but actually were different limits of a new 11-dimensional theory called \emph{M-theory}, see Fig. (\ref{five}). Its fundamental objects should be membranes which appear as solitons of a 11-dimensional supergravity, but its understanding is not yet precise.

\noindent
This duality web required in some cases the matching of the non-perturbative spectrum of a theory with the perturbative one of the dual theory. Non-perturbative states were represented by higher-dimensional objects, \emph{branes}, which play a key r\^ole in this respect. In particular, Dirichlet branes, or \emph{D-branes}, which were been studied since 1990 and developed by J. Polchinski \cite{Polchinski:1995mt}, correspond to extended objects where open strings could end  (\emph{microscopic} decription) but can also be viewed as soliton solutions  of low energy superstring theory (\emph{macroscopic} description). 

\paragraph{D-branes}
D-branes were discussed during the quest for classical solutions of the low-energy string effective action and then they became an essential element to better understand the links between the five superstring theories, as their existence is required by various duality transformations. 
Later, it was understood that they can be efficiently used  in the construction of four-dimensional phenomenological models.
A Dp-brane is an extended object with $p$ spatial dimensions, where $D$ indicates that the endpoints of the strings attached to them have Dirichlet boundary conditions.
Their worldvolume action is the action of the massless open string modes embedded in a closed string background living in the bulk. It is divided into two pieces which involve respectively the NS-NS and the R-R sector. At leading order in the string coupling it reads
\beq
S = S_{DBI} [g,B,\phi] + S_{WZ} [C_p]
\eeq
where $S_{DBI}$ is the Dirac-Born-Infeld (DBI) action and is the generalization of the Maxwell theory with higher derivative couplings
\beq
S_{DBI} = - \mu_p \int_{\partial} d^{p+1}\xi e^{-\phi(X)} \sqrt{-det\left(g_{ab}(X) + \FF_{ab}(X) \right)}
\eeq
with $a,b = 0,...,p$. The Wess-Zumino (WZ) action $S_{WZ}$ measures the Ramond-Ramond charges of a Dp-brane and does not include the metric (so it is \emph{topologic}). It involves the R-R sector of the theory
\beq
S_{WZ} = \mu_p \int C \wedge e^{\FF}.
\eeq
At low energy, \emph{i.e.} at leading order in $\alpha'$, one can retrieve the Super Yang-Mills (SYM) theory. In particular, the DBI action leads to the gauge fields and scalar kinetic terms of SYM while the WZ part  to the $\theta$-term.\\
$N$ coincident D-branes support on their worldvolume the interactions of $U(N)$ gauge group, while gravity propagates on the whole ten-dimensional target space, the \emph{bulk}. Moreover, intersecting D-branes (or branes with world-volume fluxes) permit the existence of chiral matter localized at their intersection points. 
The presence of \emph{tadpoles} in Type II compactifications with D-branes led to the introduction of \emph{orientifold projections}, \emph{i.e.} transformations involving the world-sheet parity operator.

The enormous variety of possible constructions opened the way to the engineering of more and more models with semi-realistic properties. 

\vspace{1cm}

Let us mention that D-branes entered essentially in the AdS/CFT correspondence developed by Maldacena.
In these recent years many other developments have been made and many models constructed, in the search for a brane construction which could mimic the properties of Standard Model or of its minimal supersymmetric extension (MSSM). 
One of the main requirements is a way to break supersymmetry.

\paragraph{Engineering supersymmetry breaking}
In four-dimensional compactifications, the supersymmetry content depends on the choice of the compactification manifold and the embedding of D-branes.
There are several methods to reduce the supersymmetries in the bulk or the ones preserved by the theories living on D-branes. For instance a Calabi-Yau \emph{compactification manifold} preserves $\NN = 2$ supersymmetries in the bulk. D-branes can then be included in such a way to preserve $\NN =1$.
$\NN = 1$ supersymmetry is desirable from the phenomenological point of view, most due to hierarchy reasons, but it should be broken at some level to retrieve the physics we observe.
The search for precise supersymmetry breaking setups in string models is therefore very important.
If one does not want to spoil the good soft UV behaviour of the theory, supersymmetry has to be \emph{softly} broken, by adding explicit \emph{soft supersymmetry breaking} terms which respect the renormalization behaviour of supersymmetric gauge theories. One of the possible terms is the introduction of \emph{gaugino masses} 
\beq
-M^a \lambda_a \lambda_a~,
\eeq
where $a$ is the gauge group index. 
Other ones are scalar masses $(m^2)^{ab} \phi_a \phi_b^*$, Yukawa couplings $Y^{ijk} \phi_i \phi_j \phi_k$, quadratic terms in the potential for the scalar $B^{ij} \phi_i \phi_j$ (these terms, which are allowed by the symmetries of MSSM, give rise to the $\mu$-problem when the scalar is the Higgs field).
Usually these possibilities arise within a so-called \emph{mediated} supersymmetry breaking scheme. The supersymmetry is spontaneously broken at very high mass scales in some \emph{hidden} sector; then, through the \emph{messenger} sector, it is communicated to the \emph{visible} one, which can be for instance the MSSM, where soft terms are produced.

In the MSSM there is no microscopic description of these soft supersymmetry breaking terms.

To reproduce such terms in string theory, one can turn on background values for field strengths 

coming from the closed sector of the theory. Type IIB closed sector contains the antisymmetric tensor $B_{MN}$, coming from the NS-NS sector, and the $C_{p+1}$ forms, with $p=1,3,5$, coming from RR one. 

In the low-energy effective action the matter visible sector of MSSM is coupled to 4d $\NN =1$ supergravity and the gravitational interactions act as the messenger sector. The supersymmetry breaking terms can also be directly retrieved by computing the couplings between three-form fluxes and open string matter fields on Dp-branes through $S_{DBI}$ coupled to them \cite{Camara:2003ku} or from scattering amplitudes in closed string background.
They can lead to supersymmetry breaking in the bulk by giving mass to \emph{gravitinos} and/or in the open string sector via their coupling to D-branes, by generating soft supersymmetry breaking terms on the worldvolume of branes, such as \emph{gaugino} masses.

\vspace{1cm}

During the development of the theory, it became clear that general string compactifications have hundreds of parameters, called \emph{moduli}, which encode the data of the string model under consideration, such as the D-brane positions, size and shape of the manifold and so on. Each of them appears in the four-dimensional theory as a massless scalar field, giving rise to long-range interactions which are not observed and affecting the four dimensional effective action via its vacuum expectation value. Moreover they have a flat potential to all orders in perturbation theory. 
To stabilize them there are many possibilities. One is based on the introduction of background fluxes \cite{Taylor:1999ii,Dasgupta:1999ss,Giddings:2001yu} in the internal dimensions, to preserve Poincar\'e invariance in the Minkowskian space-time.

Background flux compactifications play therefore many non-trivial r\^oles in phenomenological models. As already pointed out, they can create an effective potential for the moduli and break supersymmetry by generating soft supersymmetry breaking terms on D-branes.

From their start in the mid eighties with the study of heterotic string compactiﬁcations in presence of three-form H-flux \cite{Hull:1986iu,Strominger:1986uh,de Wit:1986xg}, flux compactifications have enormously developed.

\vspace{0.5cm}
Other deep developments involved the \emph{non-perturbative} sector of gauge theories, starting from the discovery of Yang-Mills instantons \cite{Belavin:1975fg}. 
It was pointed out in 1995 \cite{Witten:1995gx,Douglas:1995bn} that gauge instantons could have a realization in the frame of string theory. Systems of suitably chosen D-branes, D-instantons and Euclidean branes can indeed support the stringy description of gauge instantons.
It was argued that non-perturbative effects, such as superpotentials arising from instantons and gaugino condensation, could for instance solve the problem of moduli stabilization. Moreover it was found that string theory could provide new kinds of instantons, called \emph{exotic}, which still do not have a complete field theory explanation.

As we will see, the interplay between fluxes and instantons is very deep. Indeed, in presence of fluxes, non-perturbative superpotentials can be generated by instantons giving rise to new low-energy effects. Moreover, fluxes can contribute to get non-vanishing results in presence of exotic instantons by lifting fermionic zero-modes which would make vanish instanton-generated interactions. 

We will discuss  these topics in detail in next Chapters, where we will give general informations about the models we will consider in our computations.

In particular, we will focus on four dimensional compactifications of
Type II string theories preserving $\mathcal N=1$ supersymmetry in the presence
of intersecting or magnetized D-branes, 
which constitute a promising scenario for phenomenological applications of string theory and realistic
model building. 
Indeed, in these compactifications, gauge interactions
similar to those of the supersymmetric extensions of the Standard Model of particle physics
can be engineered using space-filling D-branes that partially or totally wrap
the internal six-dimensional space. By introducing several
stacks of such D-branes, one can realize adjoint gauge fields for various
groups by means of the massless excitations of open strings that start and end
on the same stack, while open strings stretching between different
stacks provide  bi-fundamental matter fields. On the other hand, from the closed
string point of view, (wrapped) D-branes are sources for various fields
of Type II supergravity, which acquire a non-trivial profile in the bulk.
Thus the effective actions of these brane-world models describe interactions
of both open string (boundary) and closed string (bulk) degrees of freedom and have the
generic structure of $\mathcal N=1$ supergravity in four dimensions
coupled to vector and chiral multiplets.
Several important aspects of such effective actions have been intensively investigated over the years
from various points of view \cite{Blumenhagen:2005mu,Blumenhagen:2006ci,Marchesano:2007de}.
We will study, through world-sheet methods, how the insertion of background fluxes may modify effective interactions on Dirichlet and Euclidean branes and create new non-perturbative superpotential terms in presence of instantons.

 \section{Scheme of the thesis}
The thesis is divided into two parts, related to very different aspects of string theory. 
The first one is related to string theory viewed as the candidate for the theory of everything. In particular we will drive our attention to \emph{flux compactifications} and \emph{non-perturbative terms}, analyzing the interplay, given by fluxes, among soft
supersymmetry breaking, moduli stabilization and non-perturbative effects in the low-energy theory.

The second part of the thesis is devoted to the description of statistical systems, in particular \emph{interfaces}, via the effective string, coming back to the purpose string theory was born for. After the discovery of the AdS/CFT correspondence, the interest on QCD string has been renewed. We will show how the bosonic string of Nambu-goto model in the first order formulation can mimic very well the behaviour of interfaces. To support it we will present not only the theoretical evaluation but also the comparison with precise data provided by Monte Carlo simulations. 

For the detailed partial schemes see the corresponding introductions.

\chapter{Three-form Fluxes in $\NN = 1$ compactifications} 
\label{introflux}

As we already stressed in Chapter \ref{intro}, an important ingredient of Type II string theories compactifications preserving $\mathcal N=1$ supersymmetry in the presence
of intersecting or magnetized D-branes is
the possibility of adding internal (to preserve 4d Poincar\'{e} invariance) antisymmetric fluxes both in the Neveu-Schwarz-Neveu-Schwarz
 and in the Ramond-Ramond sector of the bulk theory
\cite{Grana:2005jc,Douglas:2006es,Denef:2007pq}. These fluxes bear important consequences
on the low-energy effective action of the brane-worlds,
such as moduli stabilization, supersymmetry breaking and also the generation of non-perturbative superpotentials. 

Indeed, as is well-known \cite{Cremmer:1982en}, four-dimensional ${\mathcal N}=1$ supergravity theories
are specified by the choice of a gauge group ${\mathcal G}$,
with the corresponding adjoint fields and gauge kinetic functions, by a K\"ahler potential $K$ 
and a superpotential $W$, which are, respectively, a real and a holomorphic function 
of some chiral superfields $\Phi^i$. 
The supergravity vacuum is parametrized by the expectation values of these chiral
multiplets that minimize the scalar potential
\begin{equation}
V\,=\,\ee^{K}\left( D_i \bar W D^i W - 3 \,|W|^2\right) +D^a D_a
\end{equation}
where $D_i W\equiv\partial_{\Phi^i} W + \big(\partial_{\Phi^i} K \big)\,W$ is the K\"ahler
covariant derivative of the superpotential and the $D^a$ ($a=1,\ldots,{\rm dim}(\mathcal G)$)
are the D-terms. Supersymmetric vacua, in particular, correspond to those
solutions of the equations 
$\partial_{\Phi^i} V=0$  satisfying the D- and F-flatness conditions $D^a=D_i W=0$.

The chiral superfields $\Phi^i$ of the theory
 comprise the fields $U^r$ and $T^m$ that
parameterize the deformations of the complex and K\"ahler structures of the three-fold,
the axion-dilaton field
\begin{equation}
\tau=C_0+\ii \,\ee^{-\varphi}~,
\end{equation} 
where $C_0$ is the R-R scalar and $\varphi$ the dilaton, and also some multiplets
$\Phi_{\mathrm{open}}$ coming from the open strings attached to the D-branes.
The resulting low energy ${\mathcal N}=1$ supergravity model has a highly degenerate vacuum.

One way to lift (at least partially) this degeneracy is provided by the addition
of internal 3-form fluxes of the bulk theory \cite{Grana:2005jc,Douglas:2006es,Denef:2007pq}
via the generation of a superpotential \cite{Gukov:1999ya,Taylor:1999ii}
\begin{equation}
W_{\mathrm{flux}}=\int G_3\wedge \Omega~,
\label{wflux3}
\end{equation}
where $\Omega$ is the holomorphic $(3,0)$-form of the Calabi-Yau three-fold and 
\beq
G_3={F}-\tau H
\eeq
is the complex 3-form flux given in terms of the R-R and NS-NS fluxes  
${F}$ and $H$. The flux superpotential (\ref{wflux3}) depends explicitly on $\tau$ through $G_3$ 
and implicitly on the complex structure parameters $U^r$ which specify $\Omega$, while it does not depend on Kahler structure moduli $T^m$.

Using standard
supergravity methods, F-terms for the various compactification moduli can be obtained from (\ref{wflux3}). 
Insisting on unbroken $\mathcal N=1$ supersymmetry requires the flux $G_3$ to be an Imaginary Self Dual 3-form of type $(2,1)$ \cite{Giddings:2001yu}, since 
the F-terms ,
$D_\tau W_{\mathrm{flux}}$, $D_{T^m} W_{\mathrm{flux}}$ and $D_{U^r} W_{\mathrm{flux}}$
are proportional to the $(3,0)$, $(0,3)$ and $(1,2)$
components of the $G$-flux respectively:
\beqa
D_\tau W_{\mathrm{flux}} = 0  &\rightarrow&  G_3^{3,0} = 0 \\
D_{T^m} W_{\mathrm{flux}} = 0 &\rightarrow&  G_3^{0,3} = 0 \\
D_{U^r} W_{\mathrm{flux}} = 0 &\rightarrow&  G_3^{1,2} = 0 
\eeqa
and only $G_3^{2,1}$ survives. So to preserve $\mathcal N=1$ supersymmetry the flux has to be Imaginary Self Dual and with vanishing $(0,3)$ part:
\beq
*G_3 = i G_3, \, ~G_3^{0,3} = 0.
\eeq

\begin{table}[ht]
\centering
\begin{tabular}{|ccc|}
\hline $\phantom{\vdots}G^{\mathrm{ISD}}$ & $\to$ & $G_{(0,3)}
\oplus G_{(1,2)_{\mathrm{NP}}}\oplus G_{(2,1)_{\mathrm{P}}}$\\
[1ex]
 \hline
$\phantom{\vdots}G^{\mathrm{IASD}}$ & $\to$  & $G_{(3,0)} \oplus
G_{(2,1)_{\mathrm{NP}}}\oplus G_{(1,2)_{\mathrm{P}}} $
\\[1ex]
\hline
\end{tabular}
 \caption{ Decomposition of the ISD and IASD parts of the 3-form $G$.}
 \end{table}

The requirement of existence of solutions to the supergravity equations of motions with fluxes imposes only \cite{Giddings:2001yu,Maldacena:2000mw}
\beq
\label{sugra_eq}
*G_3 = i G_3~ ,
\eeq
therefore for instance $G_3^{0,3}$ can break supersymmetry without destroying the solution.
A consistent model including gauge \emph{and} gravity would require fluxes which satisfy eq.(\ref{sugra_eq}). However, if in the setup under consideration the regime is such that the dynamical effects of gravity can be neglected (as in our model), gauge theories with "soft" couplings with all kinds of fluxes coming from closed strings can be considered.
The F-terms can also be interpreted as
the $\theta^2$ ``auxiliary'' components of the kinetic functions for the gauge theory
defined on the space-filling branes,
and thus are soft supersymmetry breaking terms for the brane-world
effective action. These soft terms have been computed in various scenarios of flux compactifications  \cite{Grana:2002nq}
\nocite{Camara:2003ku,Grana:2003ek,Camara:2004jj,Lust:2004fi,Conlon:2005ki,Conlon:2006wz}-
\cite{Berg:2007wt}
and their effects, such as flux-induced masses for the gauginos and the gravitino,
have been analyzed in various scenarios of flux compactifications
relying on the structure of the bulk supergravity Lagrangian and on
$\kappa$-symmetry considerations (see for instance the reviews \cite{Grana:2005jc,Douglas:2006es,Denef:2007pq} and references therein); here we derive them by a direct world-sheet analyisis.

So far the consequences of the presence of
internal NS-NS or R-R flux backgrounds onto the world-volume
theory of space-filling or instantonic branes
have been investigated relying entirely on space-time supergravity methods \cite{Grana:2002tu}\nocite{Marolf:2003vf,Marolf:2003ye,Tripathy:2005hv,Martucci:2005rb}
-\cite{Bergshoeff:2005yp},
rather than through a string world-sheet approach\footnote{For some recent developments using world-sheet methods see Ref.
\cite{Linch:2008rw}.}.
A paper recently appeared with an alternative approach which does not require a microscopic description, see \cite{Uranga:2008nh}.

In this thesis we fill this gap and derive the flux induced fermionic
terms of the D-brane effective actions with an explicit conformal
field theory calculation of scattering amplitudes among two open
string vertex operators describing the fermionic excitations at a
generic brane intersection and one closed string vertex operator
describing the background flux. Our world-sheet approach is quite
generic and allows to obtain the flux induced couplings in a unified
way for a large variety of different cases: space-filling or
instantonic branes, with or without magnetization, with twisted or
untwisted boundary conditions. Indeed, the scattering amplitudes we
compute are generic mixed disk amplitudes, {\it{i.e.}} mixed
open/closed string amplitudes on disks with mixed boundary
conditions, similar to the ones considered in Refs.
\cite{Billo:2004zq,Lust:2004cx,Billo:2005jw,Bertolini:2005qh,Billo:2006jm}.

Our approach
not only reproduces correctly all known results but can be applied also
to cases where the supergravity methods are less obvious, like
for example to study how NS-NS or R-R fluxes couple to fields with twisted
boundary conditions or how they modify the action which gives the measure of integration
on the moduli space of instantons. Finding the flux-induced soft terms on instantonic
branes of both ordinary and exotic type is a necessary step towards the investigations
of the non-perturbative aspects of flux compactifications we have mentioned above.

Indeed, in addition to fluxes, another important issue to study is the
non-perturbative sector of the effective actions coming from string
theory compactifications \cite{Witten:1995gx,Douglas:1995bn}. Only
in the last few years, concrete computational techniques have been
developed to analyze non-perturbative effects using systems of
branes with different boundary conditions
\cite{Green:2000ke,Billo:2002hm}.  Non-perturbative effects were also recently connected to topological strings \cite{Collinucci:2009nv}.
These non-perturbative contributions 
to the effective actions may play an important r\^ole in the moduli stabilization process
\cite{Denef:2005mm,Lust:2005dy} and bear phenomenologically relevant implications for
string theory compactifications.
In the framework we are considering, non-perturbative sectors are described by
configurations of D-instantons or, more generally, by wrapped Euclidean branes
which may lead to the generation of a non-perturbative superpotential of the form 
\begin{equation}
W_{\mathrm{n.p.}}= \sum_{ \{k_A\} } c_{\{k_A\}}(\Phi^i) \,\ee^{2\pi \ii\sum_A 
\!k_A \tau_A}~.
\label{wfluxnp}
\end{equation}
Here we have labeled the gauge group components (corresponding to different
stacks of D-branes) by an index $A$ and
denoted by $\tau_A$ their complexified gauge couplings. In general, the $\tau_A$'s depend
on the axion-dilaton modulus $\tau$ and the K\"ahler parameters $T^m$
that describe the volumes of the cycles which are wrapped by the D-branes%
\footnote{The explicit  dependence of $\tau_A$ on $\tau$ and $T^m$ 
can be derived from the Dirac-Born-Infeld action.}. Furthermore, in (\ref{wfluxnp}) the exponent represents the total classical
action for an instanton configuration with second Chern class $k_A$ with
respect to the gauge component A, and $c_{\{k_A\}}(\Phi^i)$ are (holomorphic) functions
of the chiral superfields whose particular form depends on the details of the
model.

The interplay of fluxes and non-perturbative contributions, leading to a combined
superpotential 
\begin{equation}
 \label{Wtot}
W = W_{\rm flux} + W_{\rm n.p.}~,
\end{equation}
offers new possibilities for finding supersymmetric vacua. 

\noindent
Indeed,
the derivatives $D_{U^r} W_{\mathrm{flux}}$,
$D_\tau W_{\mathrm{flux}}$ and $D_{T^m} W_{\mathrm{flux}}$   
might now be compensated by $D_{U^r} W_{\mathrm{n.p.}}$,
$D_\tau W_{\mathrm{n.p.}}$ and $D_{T^m} W_{\mathrm{n.p.}}$ \cite{Lust:2005dy}
so that also the $(1,2)$, $(3,0)$ and $(0,3)$
components of $G_3$ may become compatible with supersymmetry and help in removing the vacuum
degeneracy \cite{Lust:2006zg}.

Another option could be to arrange things in such a way to have a Minkowski vacuum with $V=0$
and broken supersymmetry. If the superpotential is divided into an observable and a hidden sector,
with the flux-induced supersymmetry breaking happening in the latter, 
this could be a viable model for supersymmetry breaking mediation.
If all moduli are present in $W$, the number of equations necessary to satisfy the
extremality condition for $V$ seems sufficient to obtain a complete moduli stabilization.
To fully explore these, or other, possibilities, it is crucial however 
to develop reliable techniques to compute non-perturbative corrections to the effective action
and determine the detailed structure of the non-perturbative superpotentials that can be generated,
also in presence of background fluxes.

These methods not only allow to
reproduce
\cite{Billo:2002hm}\nocite{Billo:2006jm,Akerblom:2006hx,Billo:2007sw}-\cite{Billo:2007py}
the known instanton calculus of (supersymmetric) field theories
\cite{Dorey:2002ik}, but can also be generalized to more exotic
configurations where a field theory explanation became avalaible only recently, but it is still far from being complete
 \cite{Blumenhagen:2006xt}
\nocite{Ibanez:2006da,Florea:2006si,Bianchi:2007fx,Argurio:2007vqa,
Bianchi:2007wy,Ibanez:2007rs,Antusch:2007jd,Blumenhagen:2007zk,Aharony:2007pr,Blumenhagen:2007bn,Camara:2007dy,Ibanez:2007tu,GarciaEtxebarria:2007zv,Petersson:2007sc,Bianchi:2007rb,Blumenhagen:2008ji,Argurio:2008jm,Cvetic:2008ws,Kachru:2008wt,GarciaEtxebarria:2008pi}-\cite{Fucito:2009rs}.
The study of these exotic instanton configurations has led to
interesting results in relation to moduli stabilization, (partial)
supersymmetry breaking and even fermion masses and Yukawa couplings
\cite{Blumenhagen:2006xt,Ibanez:2006da,Blumenhagen:2007zk,Ibanez:2008my} (for a recent systematic analysis see \cite{Cvetic:2009yh}). A
delicate point about these stringy instantons concerns the presence
of neutral anti-chiral fermionic zero-modes which completely
decouple from all other instanton moduli, contrarily to what happens
for the usual gauge theory instantons where they act as Lagrange
multipliers for the fermionic ADHM constraints \cite{Billo:2002hm}.
In order to get non-vanishing contributions to the effective action
from such exotic instantons, it is therefore necessary to remove
these anti-chiral zero modes \cite{Argurio:2007vqa,Bianchi:2007wy}
or lift them by some mechanism \cite{Blumenhagen:2007bn,Petersson:2007sc}. The
presence of internal background fluxes may allow for such a lifting
and points to the existence of an intriguing interplay among soft
supersymmetry breaking, moduli stabilization, instantons and
more-generally non-perturbative effects in the low-energy theory
which may lead to interesting developments and applications.

If really generated, such exotic interactions
could also become part of a scheme in which the supersymmetry breaking 
is mediated by non-perturbative soft-terms arising in the hidden sector of the theory, as recently 
advocated also in \cite{Buican:2008qe}. Nonetheless, the stringent conditions required for the 
non-perturbative terms to be different from zero, severely limit the freedom to
engineer models which are phenomenologically viable. 

To make this program more realistic, in this thesis we address the study of the generation of non-perturbative terms in presence of fluxes. In the following we will consider the 
interactions generated by gauge and stringy instantons in a specific 
setup consisting of fractional D3-branes 
at a $\mathbb C^3/(\mathbb Z_2\times \mathbb Z_2)$ singularity which engineer 
a ${\mathcal N}=1$ $\mathrm{U}(N_0)\times \mathrm{U}(N_1)$ quiver gauge theory
with bi-fundamental matter fields.
In order to simplify the treatment, still keeping the desired supergravity interpretation,
this quiver theory can thought of as a local description of a Type IIB
Calabi-Yau compactification on the toroidal orbifold 
$T^6/(\mathbb Z_2\times \mathbb Z_2)$. From this local standpoint, it is not necessary
to consider global restrictions on the number $N_0$ and $N_1$ of D3-branes, which can therefore be 
arbitrary, nor add orientifold planes for tadpole cancelation. In such a setup we then 
introduce background fluxes of type $G_{(3,0)}$ and $G_{(0,3)}$, and study the
induced non-perturbative interactions in the presence of gauge and stringy
instantons which we realize by means of fractional D-instantons. 
In this way we are able to obtain a very rich class of non-perturbative effects which range from
``exotic'' superpotentials terms in the effective gauge theory to non-supersymmetric multi-fermion couplings. We also show that stringy instantons in presence of $G$-fluxes can generate non-perturbative
interactions even for $\mathrm{U}(N)$ gauge theories. This has to be compared with the case without
fluxes where an orientifold projection \cite{Argurio:2007vqa,Bianchi:2007wy} (leading to orthogonal
or symplectic gauge groups) is required in order to solve the problem of the neutral fermionic zero-modes.
Notice also that since the $G_{(3,0)}$ and $G_{(0,3)}$ components of the $G_3$ are related 
to the gaugino and gravitino masses (see for instance \cite{Camara:2003ku,Camara:2004jj}), 
the non-perturbative flux-induced
interactions can be regarded as the analog of the Affleck-Dine-Seiberg (ADS) superpotentials 
\cite{Affleck:1983mk} for gauge/gravity theories with soft supersymmetry breaking terms.
In particular the presence of a $G_{(0,3)}$ flux has no effect on the gauge theory 
at a perturbative level but it generates new instanton-mediated effective interactions  \cite{GarciaEtxebarria:2007zv}.
 
For the sake of simplicity most of our computations will be carried out
for instantons with winding number $k=1$; however we also briefly discuss some multi-instanton effects.
In particular, from a simple counting of zero-modes we find that in our quiver gauge theory 
an infinite tower of D-instanton corrections can contribute to 
the low-energy superpotential, even in the field theory limit with no fluxes, in constrast to what
happens in theories with simple gauge groups where the ADS-like superpotentials are generated only by instanton with winding number $k=1$. These multi-instanton effects in the quiver theories certainly 
deserve further analysis and investigations.
For an interesting connection between matrix models and D-brane instanton calculus (and a perturbative way of computing stringy 
multi-instanton effects) see \cite{GarciaEtxebarria:2008iw}.  Results about multi-instanton processes have also appeared in Ref.
\cite{GarciaEtxebarria:2008pi}.

More specifically, this part of the thesis is organized as follows: in next Chapter we will briefly review the notion of instantons in gauge theories and how it can be derived in the stringy side.

Chapter \ref{flux}, based on the publication \cite{Billo':2008sp}, is devoted to the computation of interaction of massless fermions in presence of closed string background fluxes.
In Section \ref{sec:CFT} we describe in detail the
world-sheet derivation of the flux induced fermionic terms of the D-brane effective action
from mixed open/closed string scattering amplitudes. The explicit results for
various unmagnetized or magnetized branes as well as for instantonic branes are spelled
out in Section \ref{sec:effects} in the case of untwisted open strings and in Section \ref{sec:twisted}
in some case of twisted open strings. The flux-induced fermionic couplings are further analyzed
for the $\mathbb Z_2 \times \mathbb Z_2$ orbifold compactification which
we briefly review in Section \ref{sec:N1}. Later in Section \ref{sec:n1int}
we compare our world-sheet results for the flux couplings on fractional D3-branes
with the effective supergravity approach to the soft supersymmetry breaking terms,
finding perfect agreement. In Section \ref{sec:fD-1} we exploit the
generality of our world-sheet based results to determine the soft terms
of the action on the instanton moduli space.

Then in Chapter \ref{nonp} we present the results of \cite{Billo':2008pg}, where we analyse the non-perturbative side of flux compactifications.
In Section \ref{sec:effint} we discuss a quick method to infer the structure of the non-perturbative
contributions to the effective action based on dimensional analysis and symmetry considerations.
In Section \ref{secn:1inst} we analyze the ADHM instanton action and discuss in detail the one-instanton induced interactions in SQCD-like models without introducing $G$-fluxes. 
Finally in Sections \ref{secn:fluxeffects} and \ref{secn:strinst} we consider gauge and stringy instantons in presence of $G$-fluxes and compute the non-perturbative interactions they produce.

Chapter \ref{conclflux} is devoted to summary of results, conclusions and future perspectives.

Some more technical details, such as our conventions on spinors, on the $\mathbb Z_2 \times \mathbb Z_2$ orbifold
and on the flux couplings for wrapped fractional D9-branes are contained in the Appendix.

\chapter{Space-time Instantons in Gauge and String Theories}
\label{inst}

In this Chapter we want to briefly recall some basic facts about instantons in gauge theories and how they can be realized in string theory (many good reviews exist; see for instance \cite{Dorey:2002ik,Tong:2005un,Vandoren:2008xg}).
Setups which reproduce the usual Yang-Mills instantons (\emph{i.e.} gauge instantons) can be performed by means of D-brane models. As we will discuss, systems of Dp- and D(p-4)-branes in a suitably compactified target space give rise to instanton configurations of the gauge theory on the Dp's.
An important aspect of string theory realization is that new kinds of instantons can arise, which do not have an explanation on the gauge theory side yet. They are called \emph{exotic} instantons and, under appropriate conditions, can actually contribute to the low-energy effective actions. Moreover, other non-perturbative effects may arise when string corrections are taken into consideration.

\section{In gauge theory}
\label{ginst}

Instantons in gauge theories, defined in Minkowski spacetime, describe tunneling processes from one vacuum to another. The simplest models which exhibit this phenomenon are the quantum mechanical point particle with a double-well potential having two vacua, or a periodic potential with infinitely many vacua. There is no classical allowed trajectory for a particle to travel from one vacuum to the other, but quantum mechanically tunneling occurs. The tunneling amplitude can be computed in the WKB approximation and is exponentially suppressed. \\
Sometimes it is useful to perform a Wick rotation since path integrals are more conveniently computed in Euclidean spacetime. In the Euclidean regime instantons are defined as finite action solutions to the fields equations of motion.  \\
When a theory admits different topological sectors, in each of them a configuration of lowest finite Euclidean action can be identified.
Euclidean path integral requires to keep in consideration all these configurations, where fields assume a non-trivial profile, by summing over them.
The contribution of instantons to the path integral is very tiny as it turns out to be exponentially suppressed. Moreover, as we will see, when fermions are present strong selection rules appear and may eventually lead to a vanishing instanton contribution.

\subsection{Instantons in pure Yang-Mills}

Let's take the 4-dimensional SU(N) pure Yang-Mills:
\beq
\label{euclYM}
S = -\frac{1}{2g^2}\int d^4 x ~ tr_N F_{\mu\nu}F^{\mu\nu} ~.
\eeq
As we said instantons are Euclidean solutions of motion equations with finite action.
The requirement of finite action implies that the field strength $F$ goes to zero faster than $|x|^{-2}$  at infinity. This requires that the gauge field approaches a \emph{pure gauge}
\beq
A_{\mu} =^{|x|^{-2}\rightarrow\infty} U^{-1} \partial_\mu U
\eeq
for some $U(x) \in SU(N)$. Actually, there is a way to classify such fields into sectors characterized by an integer number
\beq
k = -\frac{1}{16\pi^2} \int d^4x tr F_{\mu\nu} {}^*F^{\mu\nu},
\eeq
where
\beq
{}^*F_{\mu\nu}=\frac{1}{2} \epsilon_{\mu\nu\rho\sigma}F^{\rho\sigma}~.
\eeq
$k$ is called instanton number and corresponds to the second Chern class of the theory.
By means of the Bogomoln'yi trick one can write the following bound for the action
\beq
S \geq \frac{8\pi}{g^2} |k|
\eeq
which is saturated by (anti)self-dual configurations
\beq
\label{eqinst}
F = \pm {}^*F~.
\eeq
The self-dual configuration is called instanton and corresponds to $k>0$ while $k<0$ yields the antiself-dual one, called anti-instanton. They satisfy the equations of motion
\beq
\mathcal{D}_\mu F_{\mu\nu} = 0
\eeq
by means of the Bianchi identity.
The action for an instanton, as well as for an anti-instanton, is simply:
\beq
S_{cl}  = \frac{8\pi^2}{g^2} |k| ~.
\eeq
If we have a $\theta$-angle term 
\beq
-i \frac{\theta}{16\pi^2} \int d^4 x ~ tr F_{\mu\nu} *F^{\mu\nu} =  i \theta k~,
\eeq
the classical action for an instanton number $k$ 
becomes
\beqa
\label{sinst0}
S = \frac{8\pi^2}{g^2} k + i \theta k = - 2 \pi \ii k \tau
\eeqa
where $\tau$ is the complex gauge constant
\beq
\label{tau}
\tau = \frac{4\pi \ii}{g^2} + \frac{\theta}{2\pi}~.
\eeq
The goal of the so-called instanton calculus is to evaluate correlation functions in the instanton sectors. 
Correlators are expressed as 
\beq
<O_1(x_1)...O_n(x_n)> = \sum_k \int \DD A^{(k)} e^{-S^{(k)}} ~O_1(x_1)...O_n(x_n)
\eeq
where the field insertions can be replaced at first order by their values in the instanton background.

\subsubsection{Moduli space and partition function}
The partition function is obtained by integrating over all the possible inequivalent histories, \emph{i.e.} over the inequivalent configurations of the fields. 
This can be traded for an integral over the so-called moduli space $\MM_{k,N}$, which is the space of inequivalent solutions of self-dual SU(N) Yang-Mills equations. The moduli correspond to the parameters on which the gauge profile depends. 
For instance, in a $k=1$ SU(2) theory, this field assumes the following profile:
\beq
A_{cl,\mu}^a(x;x_0,\rho) = 2 \frac{\eta_{\mu\nu}^a (x-x_0)^\nu}{(x-x_0)^2+\rho^2}
\eeq
where $x_0$ is the position and $\rho$ the size of the instanton. These, together with the moduli associated to the gauge orientation of the instanton\footnote{Remember that the solutions of self-dual e.o.m. are equivalent for \emph{local} gauge transformations but inequivalent for \emph{global} ones.}, form the \emph{collective coordinates}.
We outline here a simple example, which can clarify this notion. Suppose we have only one massless field $A$ depending on a unique collective coordinate $X$ and we want to perform the path integral
\beq
\int \DD A ~e^{-S[A]}
\eeq
in a one-instanton background with vanishing $\theta$-term. With the saddle-point approximation, the field can be expanded around the instanton solution
\beq
A  
= A_{cl}(x,X) + A_{qu}(x,X), 
\eeq
where $A_{cl}$ satisfies the self-dual equation. Therefore at first order the action reads
\beq
S \sim S_{cl} + \frac{1}{2} A_{qu} ~ M(A_{cl}) ~A_{qu} + ...~.
\eeq
The quantum fluctuation $A_{qu}$ can be written as a linear combination of the eigenfunctions $F_n$ of $M$
\beq
A_{qu}(x,X) = \sum_{n=0}^\infty \xi_n F^n = \xi_0 F^0 + \sum_{n=1}^\infty \xi_n F^n ~,
\eeq
where the coefficient $\xi_0$ is called zero mode and corresponds to an eigenfunction of $M$ with zero eigenvalue. It indeed represents the fluctuations which do not change the action.
The path integral measure can be rewritten as
\beq
 \DD A~ \propto~  d\xi_0 \prod_n [d\xi_n]~.
\eeq
To perform the computation, the integral over $\xi_0$ has to be converted to an integral over the corresponding collective coordinate with a Fadeev-Popov-like method. This procedure is necessary to get a finite result from the integral, as $\xi_0$ corresponds to an eigenfunction with zero eigenvalue and does not appear in the expansion of the action.
If instead a mass term is present into the action, the zero modes are said to be \emph{lifted} and behave like the other $\xi_n$'s.

We have just used the fact that zero modes are associated to collective coordinates.
When the gauge theory is not pure but the gauge fields
couple to other fields, it can happen that not every zero mode is connected to a collective
coordinate. Nevertheless, one continues to call moduli space the space constructed by zero modes.
In general, the dimension of moduli space (\emph{i.e.} the number of zero modes) can be evaluated through index theorem techniques and turns out to be
\beq
dim(\MM_{k,N}) = 4 |k| N 
\eeq
in the case of pure gauge theory, where the zero modes are only \emph{bosonic}.

The most powerful method to solve the (anti)self-dual equations (\ref{eqinst}) is the ADHM construction \cite{adhm}. 
It realizes the instanton moduli space as a hyper-K\"{a}hler quotient of a flat space by an auxiliary U(k) gauge theory.
The Higgs branch of this U(k) theory, which is related to $\MM_{k,N}$, is defined through a triplet of algebraic equations, the \emph{ADHM constraints}, for each solution of which a solution to the set of equations $F = \pm {}*F$ can be built.
We do not want to enter into technical details here, but we will see that the ADHM construction can be naturally embedded in a stringy setup of D-branes.

We finally remark that usually the entire moduli space can be rewritten as 
\beq
\MM_{k,N} = \mathbb{R}^4 \times \widehat{\MM}_{k,N}~,
\eeq
where $\widehat{\MM}_{k,N}$ is the \emph{centered} moduli space which defines the \emph{centered} partition function.

\subsection{Adding fermionic and scalar fields}
We now want to consider the Dirac equation for a massless fermion $\psi$ in an (anti-)instanton background 
\beq
\dslash{\DD}^{cl}\psi = 0~ ,
\eeq
where the covariant derivatives are evaluated in the (anti-)instanton background. Decomposing $\psi$ into its chiral and antichiral parts ($\lambda$ and $\bar{\lambda}$ respectively)
\beq
\dslash{\bar{\DD}}^{cl}\lambda = 0~, ~~~\dslash{\DD}^{cl}\bar{\lambda} = 0~,
\eeq
where $\dslash{\DD} = \sigma_{\mu} \DD^{\mu}$. 

One can demonstrate that $\dslash{\bar{\DD}}^{cl}\lambda = 0$ has non-trivial solutions only in an instanton background, while $\dslash{\DD}^{cl}\bar{\lambda} = 0$ only in an anti-instanton one. Therefore in the background of an instanton only $\lambda$ picks up zero modes
(and the reverse is true for the anti-instanton). 
Since the current which modifies the pure gauge equations of motion $\DD_{\mu} F^{\mu\nu} = 0$ is bilinear in $\lambda$ and $\bar{\lambda}$,
the (anti-)instanton configurations described before in the case of a pure gauge theory still remain exact solutions of this background.
The non-trivial solutions of Dirac equations lead to further zero modes, \emph{fermionic} ones, which obviously are Grassmann variables. 
The Atiyah-Singer index theorem tells us that, if the massless fermions are in the adjoint representation, the number of fermionic zero-modes is $2 |k| N$.
The presence of fermionic zero-modes is a very delicate point because Grassmann variables must be in some way saturated to give a non-vanishing result in the path integral. This provides strong \emph{selection rules} determining which correlation functions admit instanton corrections. In particular, for each zero mode there must be one external fermion leg, which may be given for instance by introducing fermionic mass terms or external interactions; zero modes are therefore \emph{lifted}. The fermionic zero-modes can then be saturated by bringing down enough powers of the action; the correlator
\beq
<\psi(x_1)...\psi(x_m)>_{inst}
\eeq
is therefore non-vanishing only if $ m = 2|k|N $.

A more delicate and subtle procedure should be used if scalar fields are introduced. One can indeed demonstrate that bosonic fields other than gauge ones do not lead to new zero-modes but can drastically modify the equations of motion, with an important impact on solutions.
(Anti-)Instantons turn out to no longer be exact solutions of the coupled equations of motion. Nevertheless, different methods have been developed which lead to approximate solutions.

In the case of absent scalar vev's, the equations of motions can be solved perturbatively in the gauge coupling constant $g^2$ and the resulting non-exact configuration is called $quasi$-$instanton$.
The action turns out to explicitly depend on Grassmann collective coordinates (besides possible bosonic ones), meaning that the corresponding zero modes are not \emph{exact} 
but rather \emph{quasi}-zero modes.

If scalars acquire a non-vanishing vacuum expectation values, the classical action gets modified by the instanton scale size. For instance in the SU(2) case with $k=1$ it becomes
\beq
S^{cl} = \frac{8\pi^2}{g^2} - i\theta + 4\pi^2 \rho^2 (\phi^o)^2~ , 
\eeq
where $\rho$ is the scale size and $\phi^0$ the scalar vev. The term proportional to $\rho$ is essential to make converge the path integral on the scale size. To leading order in $(\rho^2 \phi^o)^2$ they can be well approximated by an ordinary instanton.
These were called \emph{constrained} instantons by Affleck and are examples of more general quasi-instantons as collective coordinates appear in the action and therefore some zero modes are lifted. 
\vspace{0.5cm}

\noindent
These considerations can be generalized to supersymmetric theories.

\subsection{Instanton-generated superpotentials and F-terms}

Instantons play a leading r\^{o}le in the understanding of non-perturbative regime of four-dimensional supersymmetric gauge theories. As shown by Affleck, Dine and Seiberg \cite{Affleck:1983mk}, instantons in SQCD with gauge group $SU(N_c)$ and $N_f$ massless flavors generate a superpotential in the case $N_f = N_c -1$.
This is not the end of the story because, even in cases where instantons do not generate such superpotential, they can deform the complex structure of the moduli space of supersymmetric vacua, which we will call $\NN$, via the creation of an F-term \cite{Beasley:2004ys}, which cannot be integrated to retrieve a corresponding superpotential but is nevertheless a genuine F-term.
The properties of SQCD are often listed in function of the number of flavours $N_f$ with respect to the number of colours $N_c$:
\begin{itemize}
\item
$N_f < N_c -1$: $W_{ADS}$ is generated but not by instantons 
\item
$N_f = N_c -1$: instantons generate $W_{ADS}$  which lifts all flat directions on the moduli space $\NN$
\item
$N_f = N_c$: instantons do not generate a superpotential but deform the complex structure of $\NN$. It is described by an F-term which is a 4-fermion interaction on $\NN$
\item
$N_f > N_c$: a superpotential is not generated and the moduli space $\NN$ is undeformed. However, there are F-terms which generate, for instance, $2(N_f - N_c) + 4$ fermions interactions, called \emph{multi-fermion F-terms}. Indeed, far from the origin of the moduli space the theory has gauge instantons which generate them. 
\end{itemize}

\section{In string theory}
\label{stringinst}
Instanton configuration are realized in string theory by systems of D(p-4)- and Dp-branes suitably wrapped. As we already mentioned, besides the ordinary gauge instantons, other non-perturbative effects may appear in a stringy construction.
There is indeed the possibility of \emph{exotic} instantons, which have not a gauge field realization yet. These instantons present unbalanced fermionic zero-modes which may combine to make vanish some contributions. To \emph{lift} them, one needs to drastically change the background by means of orientifold planes, deformations of the Calabi-Yau geometry or introduction of fluxes. They are very attractive because they seem to stabilize the gauge theory and can give a possible explanation for neutrino Majorana masses in the context of string phenomenology (see for instance the models constructed in \cite{Ibanez:2006da,Ibanez:2007rs}).

\subsection{Gauge instantons from string theory}

Yang-Mills instantons have a simple realization in string theory, by systems involving D(p-4)- and Dp-branes. Witten first showed this in the maximal case $p=5$ \cite{Witten:1995gx} in Type I string theory.  
Let us consider, in Type IIB theory, $N$ D9-branes, which we know supporting on their world-volume a ten-dimensional U(N) supersymmetric gauge theory. Their world-volume theory contains also the couplings to the various R-R fields of the bulk. In particular it includes the term 
\beq
\int_{\MM_{10}} C_{(6)}\wedge F\wedge F
\eeq
which comes from the expansion of the Wess-Zumino part of the world-volume action. The 6-form field $C_{(6)}$ is also the same R-R form which minimally couples to the D5-branes. Therefore an instanton configuration of the gauge theory living on the D9-branes with non-zero second Chern class $k$ corresponds to  $k$ units of the D5-brane charge \cite{Douglas:1995bn}. In more detail, one can demonstrates that the mass and the charge of the D5-brane are the same of the instanton.
This can obviously be extended to generic $p$ and so 
\beq
\mathrm{k ~ D(p-4) ~ branes~ on~ top ~of~ Dp-branes~ \equiv~ instantons ~of~ Dp-branes}.
\eeq
We now list some examples. In the uncompactified case we for instance have
\begin{itemize}
\item
\emph{D3/D(-1) system}. This model exactly gives the 4-dimensional Yang-Mills instanton. 
\item
\emph{D5/D1 system}. The 6- plus the 2-dimensional gauge theories which are supported by this setup do not describe the 4-dimensional Yang-Mills instanton; however the D1 represents, with respect to the D5, a configuration with non-trivial $F \wedge F$ in 4 of the 6 directions. The spectrum of mixed strings corresponds to the ADHM construction. The D9/D5 setup shows the same characteristics.
\end{itemize}
In the compactified case, $\mathbb R^{1,3} \times \mathcal{Y}$ we can have for instance
\begin{itemize}
\item
\emph{D9/E5 system}. The D9 is completely wrapped in the 6d $\mathcal Y$ so that on its non-compact 4d world-volume lives a Yang-Mills theory. Its istantons are represented by 6d branes completely wrapped on $\mathcal Y$ (with the sam magnetization of D9) which are points in 4d. These are called \emph{euclidean} branes because their world-volume directions are euclidean, indipendently of a possible Wick rotation of the non-compact part.
\item
More generally, let us consider a D$p$-brane wrapping a $(p-3)$-cycle $\mathcal C$ on $\mathcal Y$, with a field strength $F$ only in the 4dimensional spacetime,  and denote by
$\tau$ the complexified gauge coupling of the resulting four-dimensional (super) Yang-Mills theory as defined in (\ref{tau}).
A gauge instanton in this theory can be described in terms of a Euclidean $(p-4)$-brane 
wrapping the same $(p-3)$-cycle $\mathcal C$. The instanton induces non-perturbative interactions weighted by $\ee^{-k S^{\mathrm E(p-4)}}$ with $k$ being
the number of instantonic branes and $S$ the action for a single instanton. 
We want to demonstrate that
\begin{equation}
S^{\mathrm E(p-4)}=-2\pi \ii \tau~.
\label{sinst}
\end{equation}
Eq. (\ref{sinst}) follows from a comparison of the world-volume action of the Euclidean
E$(p-4)$-brane with that of the wrapped D$p$-brane \cite{Billo:2007sw}.
To get consistency with previous sections, we move to Euclidean signature; the action of the Dp-brane is%
\footnote{Here we assume $\big[F_{\mu\nu},F_{\sigma \rho}\big]=0$ and take
$F=F_i T^j$ with   ${\mathrm{Tr}}\big(T^i T^j\big)=\frac{1}{2} \delta^{ij}$ and $i,j$
running in the adjoint of the gauge.}
\beqa
S^{\mathrm{D}p} &=& \mu_{p} \,{\mathrm{Tr}}\left[\int_{\mathbb R^4\times \mathcal C}
\!\!\ee^{-\varphi}\,\sqrt{\,\det \,\big( g+2\pi \alpha' F\big)} \right. \nn \\ 
  && \left. -
\ii \int_{\mathbb R^4\times \mathcal C} \,\sum_{n}\, C_{2n} \,\ee^{2\pi \alpha' F}\right]~,
\label{cal}
\eeqa
where $\mu_p=(2\pi)^{-p} (\alpha')^{-(p+1)/2}$ is the D$p$-brane tension, $\varphi$ the
dilaton, $g$ the string frame metric and $C_{2n}$ the R-R $2n$-form potentials.
Expanding (\ref{cal}) to quadratic order in $F$ and comparing with the standard
form of the Yang-Mills action in Euclidean signature, we find
that the complexified four-dimensional gauge coupling is 
\begin{equation}
\tau=  2\pi (2\pi \alpha')^2 \, \mu_{p}\,\int_{\mathcal C}
\left[ C_{p-3}+\ii \, \ee^{-\varphi}\,
\sqrt{\,\det \,g }  \right]~. 
\label{taua}
\end{equation}
On the other hand the action for a Euclidean $(p-4)$-brane
wrapping $\mathcal C$ is given by
\begin{equation}
S^{\mathrm{E}(p-4)} =\mu_{p-4} \left[\int_{\mathcal C} \ee^{-\varphi}\,
\sqrt{\,\det \,g } ~-~\ii \int_{\mathcal C}\, C_{p-3}\right] =-2\pi\ii\, \tau.
   \label{cal2}
  \end{equation}

\end{itemize}

\subsection{Exotic instantons}

As we mentioned at the beginning, exotic instantons are instantons which arise from string constructions and do not have a field theory interpretation yet (for recent developments see \cite{Amariti:2008xu}). They have been intensively studied in the last years as they may give contribution to neutrino Majorana masses and to moduli stabilizing terms. 
They can arise from many different brane setups; for instance
\begin{itemize}
\item
in quiver gauge theories where instanton branes are in an unoccupied node of the theory,
\item
in D9/E5 systems with different magnetization,
\item
in brane systems with more than 4 mixed ND directions.
\end{itemize}
Their characteristic is to present an unbalanced fermionic zero mode integration which, if not suitably handled, leads to a vanishing contribution to the instanton calculus. Unlike in gauge string instantons, these Grassmann variables no more appear in the action and therefore they must be removed, for instance with an orientifold projection \cite{Argurio:2007vqa}, or lifted, for instance with bulk fluxes, as we will show later.

\subsection{The D3/D(-1) model}
\label{d3d01}

We now focus our attention on a particular model, which we will study in detail in next chapters. We consider a configuration of $N$ parallel D3-branes and $k$ D(-1)-branes (or \emph{D-instantons.}). For reviews see \cite{Billo:2002hm,Argurio:2007vqa}.

A stack of $N$ D3-branes in flat space gives rise to a four-dimensional $\mathrm U(N)$ gauge
theory with $\mathcal N=4$ supersymmetry. 
Its field content, corresponding to the massless excitations
of the open strings attached to the D3-branes, can be organized into
a $\mathcal N=1$ vector multiplet $V$ and three $\mathcal N=1$ chiral multiplets $\Phi^I$ ($I=1,2,3$).
These are $N\times N$ matrices:
\begin{equation}
\big\{ V,\Phi^I\big\}^u_{~v}
\label{gauge0}
\end{equation}
with $u,v,\ldots=1,\ldots,N$. In $\mathcal N=1$ superspace notation,
the action of the $\mathcal N=4$ theory is
\beqa
S&=& \frac{1}{4\pi}\,\mathrm{Im}\left[\tau\!\!\int\!\!d^4x\, d^2\theta\, d^2 \bar \theta ~
\mathrm{Tr}\big(\bar\Phi_I\,\ee^{2V}\Phi^I\big) \right. \nn \\
&& \left. + ~\tau
\!\!\int \!\!d^4x\, d^2\theta ~\mathrm{Tr}\Big(\frac{1}{2}W^\alpha\, W_\alpha +
\frac{1}{3!}\epsilon_{IJK}\Phi^I \Phi^J\Phi^K\Big) \right]
\label{actgauge0}
\eeqa
where $\tau$ is the axion-dilaton field (\ref{tau}) and $W_\alpha=-\frac{1}{4}\bar D_{\dot\alpha}
\bar D^{\dot\alpha} D_\alpha V$ is the chiral superfield whose lowest component is the
gaugino.
Strings of D(-1)/D(-1) type give rise to the \emph{neutral} sector. They have no longitudinal Neumann directions, so the fields arising from these strings do not have momentum and they are moduli rather than dynamical fields. They do not transform under the gauge group (and so do not couple to gauge sector) but couple to the charged sector.
They are $k \times k$ matrices, where $k$ is the instanton number, \emph{i.e.} the number of D(-1)'s. The bosonic massless fields are called $a_\mu,\, \chi^a$, where the index $\mu = 1,...4$ identifies is the space-time directions and $a = 5,...,10$ the internal ones. There are then fermionic zero modes $M^{\alpha A},\, \lambda_{\dot{\alpha}A}$, (let's assume negative chirality) which are treated independently in Euclidean space. Moreover one can introduce a triplet of auxiliary fields $D^c$.
The \emph{charged} sector is made by strings stretching between D(-1)/D3. They have mixed boundary conditions, so the fields have no momentum. The Neveu-Schwarz sector is made up by $\omega_{\dot{\alpha}}$ (where negative chirality is imposed by GSO projection) and $\bar{\omega}_{\dot{\alpha}}$ for the conjugate sector. The Ramond sector, after the GSO, leaves us with a pair of fermions $\mu^A,\, \bar{\mu}^A$, which respectively are $N \times k$ and $k \times N$ matrices.\\
If there are $N$ D3's and $k$ D(-1), we have to introduce Chan-Paton factors $\xi_{ui} [\bar{\xi}^{ui}]$ in the bifundamental of $(N,k) [(\bar{N},\bar{k})]$.
The interactions between D3/D3 fields give the usual 4d gauge theory, whereas instantons corrections are obtained by constructing the interaction terms between gauge and charged sectors and then integrating out all zero-modes (charged and neutral).
\vspace{0.3cm}
\begin{table}
\begin{center}
\begin{tabular}{c||cccc}
 & ADHM  & Meaning & Vertex & Chan-Paton \\
\hline\hline

NS & $a_\mu$ & \emph{centers} &
$\psi^{\mu}\,\ee^{-\varphi}$ & \phantom{$\vdots$}adj. $\mbox{U}(k)$ \\
   & $\chi$ & \emph{aux.} &
${\overline\Psi}\,\ee^{-\varphi(z)}$ &
   $\vdots$ \\
&  $D_c$ & \emph{Lagrange mult.} &
$\bar\eta_{\mu\nu}^c\psi^\nu\psi^\mu$ & $\vdots$ \\
&  & & \\
\hline
R & $M^{\alpha A}$ &  \emph{partners} &
$S_{\alpha} S_- S_{A}\,\ee^{-\frac{1}{2}\varphi}$ & $\vdots$ \\
   & $\lambda_{\dot\alpha A}$ & \emph{Lagrange mult.} &
$S^{\dot\alpha}S^+S^{A}\,\ee^{-\frac{1}{2}\varphi}$ & $\vdots$\\
\hline\hline
\end{tabular}

\caption{ADHM moduli in the neutral sector.}
  \label{listneutral}
\end{center}
\end{table}
\vspace{0.5cm}
\begin{table}
\begin{center}
\begin{tabular}{c||cccc}

 & ADHM  & Meaning & Vertex & Chan-Paton \\
\hline\hline
NS &  $w_{\dot\alpha}$ & \emph{sizes} &
$\Delta S^{\dot\alpha}\,\ee^{-\varphi}$ &\phantom{$\vdots$}
$k\times N$\\
  & ${\bar w}_{\dot\alpha}$ & \emph{sizes} &
$\overline\Delta S^{\dot\alpha}\,\ee^{-\varphi}$ & \phantom{$\vdots$}
$N \times k$\\
&  & & \\
\hline
R & ${\mu}^A$ & \emph{partners}  &
$\Delta S_{-}S_A\ee^{-{\frac12}\varphi}$ & \phantom{$\vdots$}
$k\times N$\\
  & ${\bar \mu}^A$ & $\vdots$ &
$\overline\Delta S_-S_{A}\,\ee^{-{\frac12}\varphi}$ & \phantom{$\vdots$}$N \times k$\\
\hline\hline

\end{tabular}

\caption{ADHM moduli in the charged sector.}
\end{center}
\end{table}

\pagebreak

The moduli space is then
\begin{equation}
\mathfrak M= \big\{ a_{\mu}, \chi_m, D_c, M^{\alpha A},
\lambda_{\dot\alpha A}\big\}^i_{~j} ~\cup~
\big\{ w_{\dot \alpha}, \mu^A \big\}^u_{~i} ~\cup~ \big\{
\bar w_{\dot \alpha}, \bar\mu^A \big\}^i_{~u}
\label{mod40}
\end{equation}
with $i,j=1,\ldots,k $ and $u,v=1,\ldots,N$ labeling the $k$ D$(-1)$ and the $N$ D3 boundaries respectively.
The other indices run over the following domains:
$\mu,\nu=0,\ldots,3$; $\alpha,\dot\alpha=1,2$; $m,n=1,\ldots,6$; $A,B=0,\ldots,3$, labeling,
respectively, the vector and spinor representations of the $\mathrm{SO}(4)$ Lorentz group and of
the $\mathrm{SO}(6) \sim \mathrm{SU}(4)$ internal-symmetry group%
\footnote{See footnote \ref{foot:1}.}, while $c=1,2,3$.
This system is described in terms of a $\mathrm{U}(N) \times \mathrm{U}(k)$ matrix theory whose action is \cite{Dorey:2002ik}
\begin{equation}
S_{\mathrm{D3/D(-1)}}={\mathrm{Tr}}_{k}\,\left[\frac{1}{2g_0^2}\,
S_G+S_{K}+S_{D}+S_\phi\right]
\label{cometipare}
\end{equation}
with
\begin{equation}
\begin{aligned}
S_{G} =& ~ D_{c}D^{c}-\frac{1}{2}\,[\chi_m,\chi_n]^2 - {\ii}\,
{\lambda}_{\dot{\alpha}A}\,[{\chi}^{AB},{\lambda}^{\dot{\alpha}}_{~B}] ~,\\
S_{K} =& ~ \chi_m \bar{w}_{\dot\alpha} w^{\dot\alpha}\chi^m -[\chi_m,a_{\mu}]^2
-\ii\,M^{\alpha A}\,[\overline\chi_{AB},M_{\alpha}^{~B}]+ \frac{\ii}{2}\,
\overline\chi_{AB} \,\bar{\mu}^{A} \mu^{B} ~,\\
S_{D} =& ~ \ii\,D_{c}\big(\bar{w}_{\dot\alpha}
(\tau^c)^{\dot\alpha}_{~\dot\beta} w^{\dot\beta} -\ii\,
\bar{\eta}_{\mu\nu}^c [a^\mu,a^\nu] \big)
+\ii \,\lambda_{\dot\alpha A}
\big( \bar{\mu}^{A} w^{\dot{\alpha}}
+\bar{w}^{\dot{\alpha}}\mu^{A} -[a^{\alpha\dot{\alpha}},M_{\alpha}^{~A}]\big)
 ,\\
S_\phi = &\frac{1}{8}\,\epsilon^{ABCD}\bar w_{\dot\alpha}\overline\phi_{AB}
\overline\phi_{CD}w^{\dot\alpha}+\frac{1}{2}\,\bar w_{\dot\alpha}\phi^{AB}
w^{\dot\alpha}\overline\chi_{AB}
+\frac{\ii}{2}\,\bar\mu^A\overline\phi_{AB}\mu^B~.
\end{aligned}
\label{Sd1d3}
\end{equation}
where we have also defined
\begin{equation}
\begin{aligned}
\chi^{AB} &= \chi_m \,(\Sigma^{m})^{AB}\quad,\quad
\overline{\chi}_{AB}= \chi_m\,(\overline\Sigma^{\,m})_{AB}
= \frac{1}{2}\epsilon_{ABCD}\,\chi^{CD}~,\\
\phi^{AB} &= \phi_m \,(\Sigma^{m})^{AB}\quad,\quad
\overline{\phi}_{AB}= \phi_m\,(\overline\Sigma^{\,m})_{AB}
= \frac{1}{2}\epsilon_{ABCD}\,\phi^{CD}
\end{aligned}
\label{chi}
\end{equation}
where $\Sigma^m$ and $\overline{\Sigma}^{\,m}$ are the chiral and anti-chiral blocks of the
Dirac matrices in the six-dimensional internal space, and $\phi_m$ are the six 
vacuum expectation values of the scalar fields $\Phi^I$ in the real basis.
Finally,
\begin{equation}
\frac{1}{g^2_0}=\frac{\pi}{g_s}\,(2\pi \alpha')^{2}
\label{g0}
\end{equation}
is the coupling constant of the gauge theory on the D$(-1)$ branes.
The scaling dimensions of
the various moduli appearing in (\ref{Sd1d3}) are listed in Tab. \ref{dimensions}.
\begin{table}[ht]
\begin{small}
\centering
\begin{tabular}{c|ccccccccc}
\hline\hline
  moduli &$\phantom{\vdots}a_\mu$ & $\chi_m$ & $D_c$  & $M^{\alpha A}$ & $\lambda_{\dot\alpha A}$
& $w_{\dot\alpha}$ & $\bar w_{\dot\alpha}$ & $\mu^A$ & $\bar \mu^A$ \\
    \hline
  dimension &$\phantom{\vdots} M_s^{-1}$ & $M_s$ & $M_s^2$ & $M_s^{-\frac{1}{2}}$ & $M_s^{\frac{3}{2}}$
& $M_s^{-1}$ & $M_s^{-1}$ & $M_s^{-\frac{1}{2}}$ & $M_s^{-\frac{1}{2}}$ \\
   \hline\hline
 \end{tabular}
\end{small}
 \caption{Scaling dimensions of the ADHM moduli in terms of the string
scale $M_s=(2\pi\alpha')^{-1/2}$.}
 \label{dimensions}
 \end{table}

\noindent
The action (\ref{cometipare}) follows from dimensional reduction of the six-dimensional
action of the D5/D9 brane system down to zero dimensions and, as discussed in detail in Ref. \cite{Billo:2002hm}, it can be explicitly derived from scattering amplitudes of open strings with
D$(-1)$/D$(-1)$ or D$(-1)$/D3 boundary conditions on (mixed) disks.
In the field theory limit $\alpha'\to 0$ ({\it {i.e.}} $g_0\to \infty$), the term $S_G$ 
in (\ref{cometipare}) can be discarded and the fields $D^c$ and $\lambda_{\dot \alpha A}$
become Lagrange multipliers for the super ADHM constraints that realize 
the D- and F-flatness conditions in the matrix theory.

The instanton partition function is defined to be
\beq
\label{Sinst}
S_{eff} = \mathcal{C} \int~ d\MM e^{-S_{\mathrm{D3/D(-1)}}}
\eeq
where $\MM$ are all the moduli. As we will see in next Section, if in a $\NN = 1$ supersymmetric theory a term $\int d^4x d^2\theta$ can be factored out, whatever is left will define a superpotential or more generally an F-term. As $a^{\mu}$ and $M^{\alpha A}$ come from the supertranlations broken by the instanton, they can be identified with the spacetime coordinates and their superpartners. (\ref{Sinst}) becomes
\beqa
S_{eff} &=& \mathcal{C} \int d^4x d^2\theta \int~ d\widehat{\MM} e^{-S_{\mathrm{D3/D(-1)}}} \\
&=& \mathcal{C} \int d^4x d^2\theta ~W_{np}
\eeqa
where $\widehat{\MM}$ is the \emph{centered} moduli space.

\subsection{D3/D(--1)-branes on $\mathbb{C}^3/\big(\mathbb{Z}_2\times \mathbb{Z}_2\big)$ }
\label{secn:d3d-1}

In this section we want to discuss the dynamics of the D3/D$(-1)$ brane system on the orbifold $\mathbb{C}^3/\big(\mathbb{Z}_2\times \mathbb{Z}_2\big)$ where the elements of
$\mathbb Z_2 \times \mathbb Z_2$ act on the orthonormal three complex coordinates $z^I$ ($I=1,2,3$) of $\mathbb C^3$
as follows
\begin{table}[ht]
\centering
\begin{tabular}{c|cccc}
\hline \hline
  coordinates & $h_0$ & $h_1$ & $h_2$ & $h_3$  \\
   \hline
$z^1$   &  $+$ & $+$ & $-$  & $-$\\
$z^2$ &  $+$ & $-$ & $+$ & $-$  \\
$z^3$ & $+$ & $-$ & $-$  &$+$\\
  \hline \hline
\end{tabular}
\caption{Orbifold group action on the complex
coordinates $z^i$ of $\mathcal{T}^6$.} 
\label{tablez2z2Z}
\end{table}

The fundamental types of D-branes which can be placed transversely
to an orbifold space are called fractional branes
\cite{Douglas:1996sw}. Such branes must be localized at one of the
fixed points of the orbifold group (which in our case are 64). For
simplicity we focus on fractional D3 branes sitting at a specific
fixed point (say, the origin) and work around this configuration.

The fractional branes are in correspondence with the irreducible
representations $R_A$ of the orbifold group: in fact the
Chan-Paton indices of an open string connecting two fractional
branes of type $A$ and $B$ transform in the representation $R_A\otimes
R_B$. In addition the orbifold group acts on the $\mathrm{SO}(6)$ internal
indices of the open string fields as indicated in Tab. \ref{tablez2z2lp}.
\begin{table}[ht]
\centering
\begin{tabular}{c|ccc}
\hline\hline
irrep $R_A$& & ~~~~~~~~~fields &\\
\hline
$R_0\phantom{\vdots}$ & $A_\mu$ & $\Lambda^0\equiv\Lambda^{---}$ &$ \bar\Lambda_{0}\equiv\bar\Lambda_{+++}$
\\
$R_1\phantom{\vdots}$ & $\phi^1$ & $\Lambda^1\equiv\Lambda^{-++}$ &$ \bar\Lambda_{1}\equiv\bar\Lambda_{+--}$
\\
$R_2\phantom{\vdots}$ & $\phi^2$ & $\Lambda^2\equiv\Lambda^{+-+}$ &$ \bar\Lambda_{2}\equiv\bar\Lambda_{-+-}$
\\
$R_3\phantom{\vdots}$ & $\phi^3$ & $\Lambda^3\equiv\Lambda^{++-}$ &$ \bar\Lambda_{3}\equiv\bar\Lambda_{--+}$
\\
\hline \hline
\end{tabular}
\caption{Representation of the orbifold group on the ${\mathcal N}=4$ open string
fields}
\label{tablez2z2lp}
\end{table}
This action should be compensated by that on the
Chan-Paton indices in such a way that the whole field is
$\mathbb{Z}_2\times\mathbb{Z}_2$ invariant. For example the
vector $A_\mu$ and the gaugino $\Lambda^0$ are invariant under the
orbifold group and therefore they should carry indices $(N_A,\overline N_A)$ 
in the adjoint of the $\prod_{A=0}^3 \mathrm{U}(N_A)$. The remaining
fields fall into chiral multiplets transforming in the
bifundamental representations $(N_A,\overline N_B)$.

When the D3-branes are placed in the $\mathbb{C}^3/\big(\mathbb{Z}_2\times \mathbb{Z}_2\big)$ 
orbifold, the supersymmetry of the gauge theory is reduced to $\mathcal N=1$ and only
the $\big(\mathbb{Z}_2 \times \mathbb{Z}_2\big)$-invariant components of $V$ and $\Phi^I$
are retained. Since $V$ is a scalar under the internal $\mathrm{SO}(6)$ group, while the chiral multiplets $\Phi^I$ form a vector, it is immediate
to find the transformation properties of these fields under the orbifold group elements $h_I$. 
These are collected in Tab. \ref{tablez2z2}, where in the first columns 
we have displayed the eigenvalues of
$h_I$ and in the last column we have indicated
the $\mathbb{Z}_2\times \mathbb{Z}_2$
irreducible representation $R_A$ ($A=0,1,2,3$) under which each field transforms%
\footnote{Quantities carrying an index $A$ of the chiral or anti-chiral spinor representation of $\mathrm{SO}(6)$, like for example the gauginos $\Lambda^{\alpha A}$ or
$\bar\Lambda_{\dot\alpha A}$ of the $\mathcal N=4$ theory,
transform in the representation $R_A$ of the orbifold group; thus there is a one-to-one
correspondence between the spinor indices of $\mathrm{SO}(6)$ and those labeling the irreducible
representations of $\mathbb Z_2 \times \mathbb Z_2$; for this reason
we can use the same letters $A,B,\ldots$ in the two cases.\label{foot:1}}.
\begin{table}[ht]
\centering
\begin{tabular}{cccccc}
\hline\hline
  fields & $h_0$ & $h_1$ & $h_2$  & $h_3$ & Rep's \\
    \hline
   V & + & + & + & + & $R_0$\\
   $\Phi^1$ & +& $+$ & $-$ & $-$ & $R_1$ \\
   $\Phi^2$ & +&$-$ & + & $-$ & $R_2$ \\
   $\Phi^3$ & +&$-$ &$-$ & + & $R_3$ \\
   \hline\hline
 \end{tabular}
 \caption{$\mathbb{Z}_2\times \mathbb{Z}_2$ eigenvalues of D3/D3 fields}
 \label{tablez2z2}
 \end{table}

\begin{figure}[hbt]
\begin{center}
\begin{picture}(0,0)%
\includegraphics{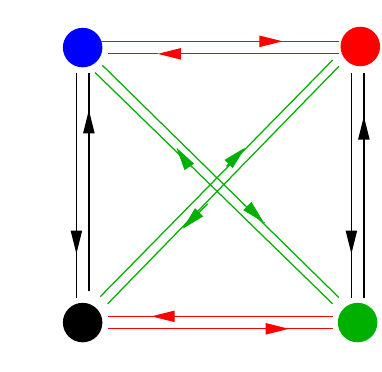}%
\end{picture}%
\setlength{\unitlength}{1381sp}%
\begingroup\makeatletter\ifx\SetFigFontNFSS\undefined%
\gdef\SetFigFontNFSS#1#2#3#4#5{%
  \reset@font\fontsize{#1}{#2pt}%
  \fontfamily{#3}\fontseries{#4}\fontshape{#5}%
  \selectfont}%
\fi\endgroup%
\begin{picture}(4559,4450)(211,-4247)
\put(4726,-136){\makebox(0,0)[lb]{\smash{{\SetFigFontNFSS{8}{9.6}{\familydefault}{\mddefault}{\updefault}$N_1$}}}}
\put(4726,-4111){\makebox(0,0)[lb]{\smash{{\SetFigFontNFSS{8}{9.6}{\familydefault}{\mddefault}{\updefault}$N_2$}}}}
\put(301,-136){\makebox(0,0)[lb]{\smash{{\SetFigFontNFSS{8}{9.6}{\familydefault}{\mddefault}{\updefault}$N_0$}}}}
\put(226,-3961){\makebox(0,0)[lb]{\smash{{\SetFigFontNFSS{8}{9.6}{\familydefault}{\mddefault}{\updefault}$N_3$}}}}
\end{picture}%
\end{center}
\caption{The quiver diagram encoding the field content and the charges for
fractional D-branes of the orbifold
$\mathbb{C}^3/(\mathbb{Z}_2\times\mathbb{Z}_2)$. The dots represent the branes associated with the irrep $R_A$
of the orbifold group. A stack of $N_A$ such branes supports a $\mathrm{U}(N_A)$ gauge theory.
An oriented link from the $A$-th to the $B$-th dot
corresponds to a chiral multiplet transforming in the
$\big({N_A},{\overline N_B}\big)$ representation of the gauge group and in the $R_A \otimes R_B$ representation of the orbifold group.}
\label{fig:quiver}
\end{figure}
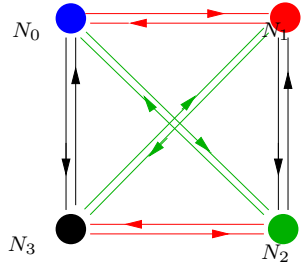

To each representation $R_A$ of $\mathbb Z_2 \times \mathbb Z_2$
one associates a fractional D3 brane type. Let $N_A$ be the number
of D3-branes of type $A$ with
\begin{equation}
 \sum_{A=0}^3 N_A = N
\label{N}
\end{equation}
so that the $N\times N$ adjoint fields $V$, $\Phi^I$ of the parent theory break into
$N_A \times N_B$ blocks transforming in the representation $R_A \otimes R_B$.
Explicitly, writing $A=(0,I)$ with $I=1,2,3$, we have
\begin{equation}
R_0\otimes R_A = R_A \quad\mbox{and}\quad R_I\otimes R_J = \delta_{IJ}  R_0 + |\epsilon_{IJK}|
\,R_K~.
\label{RR}
\end{equation}
The invariant components of $V$ and $\Phi^I$ surviving the orbifold projection are
given by those blocks 
where the non-trivial transformation properties of the fields are compensated by those of
their Chan-Paton indices and are
\begin{equation}
\big\{ V\big\}^{u_A}_{~v_A} ~\cup ~\big\{\Phi^I\big\}^{u_A}_{~v_{A\otimes I}}~.
\label{gauge}
\end{equation}
Here the symbol $\{\,\}^{u_A}_{~ v_B}$ denotes the components of the $N_A \times N_B$ block, and 
the subindex $A\otimes I$ is a shorthand for the representation product $R_A \otimes R_I$, namely
\begin{equation}
 0\otimes I = I\quad\mbox{and}\quad J\otimes I = |\epsilon_{JIK}| \,K
\label{otimes}
\end{equation}
as follows from (\ref{RR}). 
Eq. (\ref{gauge}) represents the field content of a ${\mathcal N}=1$
gauge theory with gauge group $\prod_A \mathrm U(N_A)$
and matter in the bifundamentals
$\big({N_A},{\overline N_B}\big)$, which is encoded in
the quiver diagram displayed in Fig. \ref{fig:quiver}.

The fractional D3-branes can also be thought of as D5-branes
wrapping exceptional ({\emph i.e.} vanishing) 2-cycles $\mathcal C_A$ of the orbifold.
Note that there are
only three independent such cycles on $\mathbb{C}^3/\big(\mathbb{Z}_2\times \mathbb{Z}_2\big)$
which are associated to the three exceptional $\mathbb{P}^1$'s corresponding to the non-trivial elements $h_I$ of $\mathbb{Z}_2\times \mathbb{Z}_2$.
This implies that only three linear combinations of the $\mathcal C_A$'s are really independent.
Indeed, the linear combination $\sum_{A=0}^3\mathcal C_A$ is trivial in the homological sense since
a D5 brane wrapping this cycle transforms in the regular representation and can
move freely away from the singularity because it is made of a D5-brane plus
its three images under the orbifold group. The gauge kinetic functions
$\tau_A$ of the four $\mathrm{U}(N_A)$ factors can be expressed in terms of the three
K\"ahler parameters describing the complexified string volumes of the three non-trivial 
independent 2-cycles and the axion-dilaton field $\tau$.
In the unresolved (singular) orbifold limit, which from the string point of view corresponds
to switching off the fluctuations of all twisted closed string fields, we simply have
\begin{equation}
\tau_A = \frac{\theta_A}{2\pi} + \ii\,\frac{4\pi^2}{g_A^2} = \frac{1}{4}\,\tau
\label{tauA}
\end{equation}
for all $A$'s. However, by turning on twisted closed string moduli, one can introduce
differences among the $\tau_A$'s and thus distinguish the gauge couplings of
the various group factors.

Now let us consider the $\mathbb Z_2 \times \mathbb Z_2$ orbifold projection of fields arising from strings with at least one endpoint on D(-1). The
group $\mathrm{U}(N)\times \mathrm{U}(k)$ breaks down
to $\prod_A \mathrm{U}(N_A)\times \mathrm{U}(k_A)$ with $N_A$ and $k_A$ being the numbers
of fractional D3 and D$(-1)$ branes of type $A$ such that
\begin{equation}
N=\sum_{A=0}^3 N_A \quad\mbox{and}\quad k=\sum_{A=0}^3 k_A~.
\label{nk}
\end{equation}
Consequently, the indices $u$ and
$i$ break into $u_A=1,\ldots,N_A$ and $i_A=1,\ldots,k_A$,
while the spinor index $A$ splits into $A=(0,I)$ with $A=0$ denoting the ${\mathcal N}=1$
unbroken symmetry. For the sake of simplicity from now on we always omit the index 0
and write
\begin{equation}
M^{\alpha 0}\equiv M^{\alpha}\quad,\quad
\lambda_{\dot\alpha 0}\equiv \lambda_{\dot\alpha} \quad,\quad
\mu^0\equiv \mu \quad,\quad\bar\mu^0\equiv\bar\mu~.
\label{00}
\end{equation}
Furthermore we set
\begin{equation}
\begin{aligned}
\overline{\chi}_{0I} &\equiv \overline{\chi}_{I}\quad,&\quad  &\chi^{0I} =\frac{1}{2}\,\epsilon^{IJK}\overline{\chi}_{JK}\equiv\chi^I~,\\
\overline{\phi}_{0I} &\equiv \overline{\phi}_{I}=\langle \,\bar{\Phi}_I\rangle
\quad,&\quad  &\phi^{0I}=\frac{1}{2}\,\epsilon^{IJK}\overline{\phi}_{JK}\equiv \phi^I
=\langle \,{\Phi}^I\rangle~.
\label{varchi}
\end{aligned}
\end{equation}
This notation makes more manifest which zero-modes couple to the holomorphic superfields
and which others couple to the anti-holomorphic ones. Indeed, 
the action $S_\phi$ in (\ref{Sd1d3}) becomes
\begin{equation}
\begin{aligned}
S_\phi = &~\frac{1}{2}\,\bar{w}_{\dot\alpha}\big(\phi^I\bar\phi_I+
\bar\phi_I\phi^I\big) w^{\dot\alpha} + \bar w_{\dot\alpha} \phi^I 
 w^{\dot\alpha} \overline\chi_{I} + \chi^{I} \bar w_{\dot\alpha}
 \bar\phi_I w^{\dot\alpha}\\
 &-\frac{\ii}{2}\,\bar\mu \,\bar\phi_I \mu^I
+\frac{\ii}{2}\,\bar\mu^I \bar\phi_I \mu  -\frac{\ii}{2}\,
 \epsilon_{IJK}\bar\mu^I\phi^J\mu^K~.
\end{aligned}
\label{Sphi1}
\end{equation}

Taking into account the $\mathbb Z_2 \times \mathbb Z_2$ transformation properties 
of the various fields and of their Chan-Paton labels, one finds that the moduli that survive
the orbifold projection are
\begin{equation}
\begin{aligned}
{\mathfrak M} ~=~& \big\{ a_{\mu}, D_c, M^{\alpha}, \lambda_{\dot \alpha}\big\}^{i_A}_{~ j_A}
     ~~\cup ~~ \big\{ w_{\dot \alpha},   \mu \big\}^{u_A}_{~ j_A}
~~ \cup ~~\big\{\bar w_{\dot \alpha},  \bar\mu \big\}^{i_A}_{~ u_A}
\\
&\cup~~\big\{\chi^I, \overline{\chi}_{I}, M^{\alpha I},
\lambda_{\dot\alpha I}\big\}^{i_A}_{~ j_{A\otimes I}}
~~\cup~~\big\{\mu^I\big\}^{u_A}_{~ i_{A\otimes I}}
~~\cup~~\big\{\bar\mu^I\big\}^{i_A}_{~ u_{A\otimes I}}~.
\end{aligned}
\label{mod4}
\end{equation}
Like for the chiral multiplets in (\ref{gauge}), the non-trivial transformation
of the instanton moduli carrying an index $I$ is compensated by a similar transformation 
of the Chan-Paton labels making the whole expression invariant under the orbifold
group, as indicated in the second line of (\ref{mod4}).

Among the moduli in ${\mathfrak M}$, the bosonic combinations
\begin{equation}
x^\mu \equiv \frac{1}{k}\,\sum_{A=0}^3\,\sum_{{i_A}=1}^{k_A} \big\{a^\mu\big\}^{i_A}_{~i_A}
\label{xmu}
\end{equation}
represent the center of mass coordinates of the D$(-1)$-branes and can be interpreted as the
Goldstone modes associated to the translational symmetry of the D3-branes that is broken by the
D-instantons. Thus they can be identified with the space-time coordinates, and indeed have dimensions
of a length. Similarly, the fermionic combinations
\begin{equation}
\theta^\alpha \equiv \frac{1}{k}\,\sum_{A=0}^3\,\sum_{{i_A}=1}^{k_A} \big\{M^\alpha\big\}^{i_A}_{~i_A}
\label{theta}
\end{equation}
are the Goldstinos for the two supersymmetries of the D$(-1)$-branes that are broken
by the D3-branes, and thus they can be identified with the chiral fermionic superspace coordinates.
Indeed they have dimensions of (length)$^{1/2}$.
Notice that neither $x^\mu$ nor $\theta^\alpha$ appear in the moduli action obtained by
projecting (\ref{cometipare}).

The moduli (\ref{mod4}) account for both gauge and stringy instantons. 
Gauge instantons correspond to D$(-1)$-branes that sit on non-empty
nodes of the quiver diagram, so that their number $k_A$ can be interpreted as the second Chern class
of the Yang-Mills bundle of the $\mathrm{U}(N_A)$ component of the gauge group.
Stringy instantons correspond instead to D$(-1)$-branes
occupying empty nodes of the quiver, meaning in this case $k_A N_A=0$. {F}rom (\ref{mod4}) 
one sees that in the stringy instanton case
the modes $w_{\dot \alpha}$, $\bar w_{\dot \alpha}$, $\mu$ and $\bar\mu$ 
are missing since there are no $k_A \times N_A$ invariant blocks and thus
the only moduli are the charged fermionic fields $\mu^I$ and $\bar\mu^I$.
Recalling that the bosonic $w$-moduli describe the instanton sizes and gauge orientations, one can
say that exotic instantons are entirely specified by their spacetime positions.
The presence (absence) of bosonic modes in the D3/D$(-1)$ sector for gauge (stringy)
instantons can be understood in geometric terms by blowing up 2-cycles at the singularity
and reinterpreting the fractional D3/D$(-1)$ system in terms of a D5/E1 bound state wrapping
an exceptional 2-cycle of $\mathbb{C}^3/\big(\mathbb{Z}_2\times \mathbb{Z}_2\big)$.
Gauge (stringy) instantons correspond to the cases when the D5 brane and the E1 are (are not)
parallel in the internal space. In the first case the number of Neumann-Dirichlet
directions is 4 and therefore the NS ground states is massless. In the stringy case, the number
of Neumann-Dirichlet directions exceeds 4 and
therefore the NS ground state is massive and all charged moduli come only from the fermionic R sector.

We will see how to retrieve the superpotentials or F-terms of four-dimensional SQCD from a direct stringy calculation.

\chapter{Flux Interactions on Branes}
\label{flux}

In this Chapter we present the general evaluation, using world-sheet methods, of the interactions between
closed string background fluxes and massless open string excitations living on a generic D-brane intersection.
We focus on fermionic terms (like for example mass terms for gauginos),
but our conformal field theory techniques could be applied to other terms of the brane
effective action.

\section{Flux interactions on D-branes from string diagrams}
\label{sec:CFT}

In order to keep the discussion as general as possible, we adopt here a ten-dimensional notation.
Later, in Sections \ref{sec:effects} and \ref{sec:twisted} we will rephrase our findings using
a four-dimensional language suitable to discuss compactifications of Type IIB string theory to $d=4$.

\begin{figure}[t]
\begin{center}
\begin{picture}(0,0)%
\includegraphics{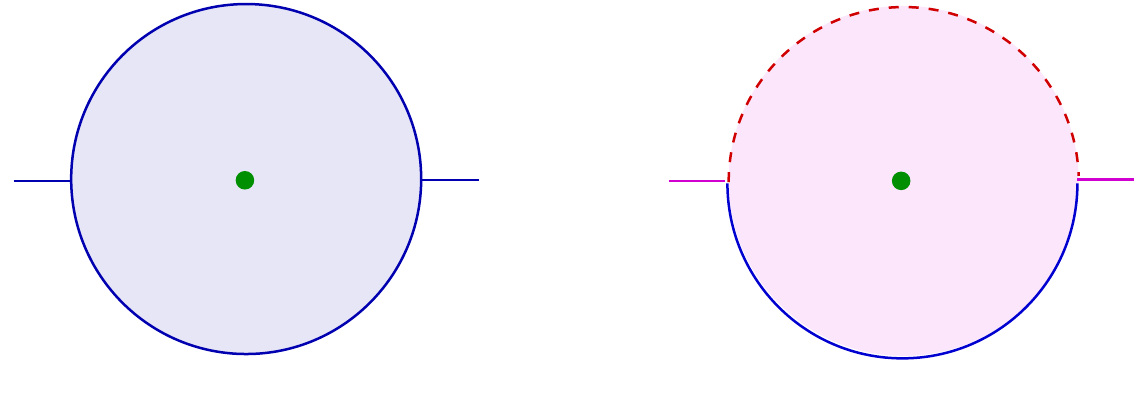}%
\end{picture}%
\setlength{\unitlength}{1579sp}%
\begingroup\makeatletter\ifx\SetFigFontNFSS\undefined%
\gdef\SetFigFontNFSS#1#2#3#4#5{%
  \reset@font\fontsize{#1}{#2pt}%
  \fontfamily{#3}\fontseries{#4}\fontshape{#5}%
  \selectfont}%
\fi\endgroup%
\begin{picture}(13638,4765)(271,-3920)
\put(286,-1825){\makebox(0,0)[lb]{\smash{{\SetFigFontNFSS{10}{12.0}{\familydefault}{\mddefault}{\updefault}$V_\Theta$}}}}
\put(5596,-1810){\makebox(0,0)[lb]{\smash{{\SetFigFontNFSS{10}{12.0}{\familydefault}{\mddefault}{\updefault}$V_{\Theta^\prime}$}}}}
\put(8146,-1823){\makebox(0,0)[lb]{\smash{{\SetFigFontNFSS{10}{12}{\familydefault}{\mddefault}{\updefault}$V_\Theta$}}}}
\put(13456,-1808){\makebox(0,0)[lb]{\smash{{\SetFigFontNFSS{10}{12.0}{\familydefault}{\mddefault}{\updefault}$V_{\Theta^\prime}$}}}}
\put(451,434){\makebox(0,0)[lb]{\smash{{\SetFigFontNFSS{10}{12.0}{\familydefault}{\mddefault}{\updefault}\emph{a)}}}}}
\put(8221,404){\makebox(0,0)[lb]{\smash{{\SetFigFontNFSS{10}{12.0}{\familydefault}{\mddefault}{\updefault}\emph{b)}}}}}
\put(1546,-3811){\makebox(0,0)[lb]{\smash{{\SetFigFontNFSS{10}{12.0}{\familydefault}{\mddefault}{\updefault}$\vec\vartheta = 0$}}}}
\put(9331,-3811){\makebox(0,0)[lb]{\smash{{\SetFigFontNFSS{10}{12.0}{\familydefault}{\mddefault}{\updefault}$\vec\vartheta \not= 0$}}}}
\put(10606,-983){\makebox(0,0)[lb]{\smash{{\SetFigFontNFSS{10}{12.0}{\familydefault}{\mddefault}{\updefault}$V_F, V_H$}}}}
\put(2731,-970){\makebox(0,0)[lb]{\smash{{\SetFigFontNFSS{10}{12.0}{\familydefault}{\mddefault}{\updefault}$V_F, V_H$}}}}
\end{picture}%
\end{center}
\caption{Quadratic coupling of untwisted, $a)$, and twisted, $b)$, open string states to closed
string fluxes. }
\label{fig:flux}
\end{figure}
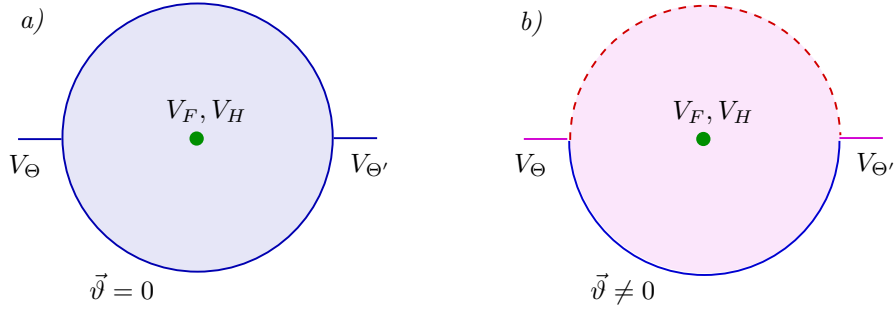

At the lowest order, the fermionic interaction terms can be derived from disk 3-point
correlators involving two vertices describing massless open-string fermions and one closed string
vertex describing the background flux, as represented in Fig. \ref{fig:flux}.
At a brane intersection, massless open string modes can arise either from open strings
starting and ending on the same stack of D-branes, or from open strings connecting
two different sets of branes. In the former case the open string fields satisfy
the standard untwisted boundary conditions and the corresponding vertex operators transform in the
adjoint representation of the gauge group. In the latter case the string coordinates
satisfy twisted boundary conditions characterized by twist parameters $\vartheta$ and the
associated vertices carry Chan-Paton factors in the bi-fundamental representation of the gauge
group; by inserting twisted open string vertices, one splits the disk boundary into different portions
distinguished by their boundary conditions and Chan-Paton labels, see Fig. \ref{fig:flux}\emph{b)}. We now give some details on
these boundary conditions and later describe the physical vertex operators and their interactions with
R-R and NS-NS background fluxes.

\subsection{Boundary conditions and reflection matrices}
\label{sec:bc}
The boundary conditions for the bosonic coordinates $x^M$ ($M=0,\ldots, 9$) of the open string
are given by
\begin{equation}
\Big(\delta_{MN}\,\partial_\sigma x^N+{\rm i}({\cal F}_\sigma)_{MN}\,\partial_\tau
x^N\Big)\Big|_{\sigma=0,\pi}=0~,
\label{bc2}
\end{equation}
where $\delta_{MN}$ is the flat background metric%
\footnote{Here, for convenience, we assume the space-time to have an Euclidean signature. Later, in Section \ref{sec:effects} we revert to a Minkowskian signature when appropriate.} and
\begin{equation}
({\mathcal F}_\sigma)_{MN}
=B_{MN} + 2\pi\alpha' \,({F}_\sigma)_{MN}
\label{calF}
\end{equation}
with $B_{MN}$ the anti-symmetric tensor of the NS-NS sector and $({F}_\sigma)_{MN}$
the background gauge field strength at the string end points $\sigma=0,\pi$.
Introducing the complex variable $z={\rm e}^{\tau+{\rm i}\sigma}$ and the reflection matrices
\begin{equation}
{R}_\sigma=\big(1-{\mathcal F}_\sigma\big)^{-1}\,\big(1+{\mathcal F}_\sigma\big)~,
\label{Rot}
\end{equation}
the conditions (\ref{bc2}) become
\begin{equation}
\overline\partial x^M\Big|_{\sigma=0,\pi}=({R}_\sigma)^M_{~N}\,\partial
x^N\Big|_{\sigma=0,\pi}~.
\label{bc1}
\end{equation}
The standard Neumann boundary conditions ({\it i.e.} ${R}_\sigma=1$)
are obtained by setting ${\mathcal F}_\sigma=0$, whereas the Dirichlet case
({\it i.e.} ${R}_\sigma=-1$) is recovered in the limit ${\mathcal F}_\sigma\to \infty$.
A convenient way to solve (\ref{bc1}) is to
define multi-valued holomorphic fields $X^M(z)$ such that
\begin{equation}
X^M({\rm e}^{2\pi{\rm i}}z) =
\big({R}_\pi^{-1}\,{R}_0\big)^{M}_{~N}\,X^N(z) \equiv R^{M}_{~N}\,X^N(z)
\label{chiraly}
\end{equation}
where $R \equiv {R}_\pi^{-1} {R}_0$ is the monodromy matrix.
Then, putting the branch cut just below the negative real axis of the $z$-plane,
the conditions (\ref{bc1}) are solved by
\begin{equation}
x^M(z,\overline z) = q^M+ \frac{1}{2}\Big[X^M(z) +\big({R}_0\big)^{M}_{~N}\,X^N(\overline
z)\Big]~,
\label{sol}
\end{equation}
where $z$ is restricted to the upper half-complex plane, and $q^M$
are constant zero-modes.

For simplicity, we will take the reflection matrices ${R}_0$ and ${R}_\pi$
to be commuting. Then, with a suitable $\mathrm{SO}(10)$ transformation
we can simultaneously diagonalize both matrices and write
\begin{subequations}
\begin{align}
{R}_\sigma &= \mathrm{diag}\left(\ee^{2\pi\ii\theta_\sigma^1},\ee^{-2\pi\ii\theta_\sigma^1}\,
\ldots,\ee^{2\pi\ii\theta_\sigma^5},\ee^{-2\pi\ii\theta_\sigma^5}\right)~,
\label{diagonalRs}\\
{R}   &= \mathrm{diag}\left(\ee^{2\pi\ii\vartheta^1},\ee^{-2\pi\ii\vartheta^1},\ldots,
\ee^{2\pi\ii\vartheta^5},\ee^{-2\pi\ii\vartheta^5}\right)~,
\label{diagonalR1}
\end{align}
\end{subequations}
with $\vartheta^I=\theta_0^I-\theta_\pi^I$.
In this basis the resulting (complex) coordinates, denoted by $Z^I$ and ${\overline Z}^I$
with%
\footnote{In the subsequent sections we will take the space-time to be the
product of a four-dimensional part and an internal six-dimensional part.
For notational convenience, we label the complex coordinates of the four-dimensional part by $I=4,5$ and those of the internal six-dimensional part by $I\equiv i=1,2,3$.} $I=1,\ldots,5$, satisfy
\begin{equation}
\partial{Z}^{I}({\ee}^{2\pi{\ii}}z) = {\ee}^{2\pi\ii\vartheta^I}\partial{Z}^{I}(z)
\quad\mbox{and}\quad
\partial{\overline Z}^{I}({\ee}^{2\pi\ii}z) = {\ee}^{-2\pi\ii\vartheta^I}
\partial{\overline Z}^{I}(z)~,
\label{calY1}
\end{equation}
and hence have an expansion in powers of $z^{n+\vartheta^I}$ and $z^{n-\vartheta^I}$, respectively,
with $n\in \mathbb{Z}$. The corresponding oscillators act on a twisted vacuum $|\vec{\vartheta}\rangle$
created by the twist operator $\sigma_{\vec{\vartheta}}(z)$, which is a conformal field of dimension
$h_{\sigma_{\vec{\vartheta}}}=\frac{1}{2}\sum_I\vartheta^I(1-\vartheta^I)$ satisfying the following OPE
\begin{equation}
\sigma_{\vec{\vartheta}}(z) \, \sigma_{-\vec{\vartheta}}(w)
\sim (z-w)^{\sum_I\vartheta^I(1-\vartheta^I)}~.
\label{OPE}
\end{equation}

For our purposes it is necessary to consider also the boundary conditions on the
fermionic fields $\psi^M$ of the open superstring in the RNS formalism, which are
\begin{equation}
\psi^M({\rm e}^{2\pi{\rm i}}z) = \eta\, R^{M}_{~N}\,\psi^N(z)
\label{chiralpsi}
\end{equation}
where $\eta=1$ in the NS sector and $\eta=-1$ in the R sector.
In the complex basis these boundary conditions become
\begin{equation}
\Psi^I({\ee}^{2\pi{\ii}}z) = \eta\,{\ee}^{2\pi\ii\vartheta^I}{\Psi}^{I}(z)\quad\mbox{and}\quad
{\overline \Psi}^{I}({\ee}^{2\pi\ii}z) = \eta\,{\ee}^{-2\pi\ii\vartheta^I}{\overline \Psi}^{I}(z)
~.
\label{psi}
\end{equation}
Thus, in the NS sector $\Psi^I$ and ${\overline\Psi}^{I}$ admit an
expansion in powers of $z^{n+\vartheta^I}$ and
$z^{n-\vartheta^I}$, respectively, with $n\in \mathbb{Z}$, so that
their oscillators are of the type
$\psi^I_{n+\vartheta^I + \frac 12}$. In the R sector they have a mode
expansion in powers of $z^{n+\vartheta^I}$ and
$z^{n-\vartheta^I}$, respectively, with $n\in
\mathbb{Z}+\frac{1}{2}$. Note that if $\vartheta^I\not=0$ neither
the NS nor the R sector possesses zero-modes and the corresponding
fermionic vacuum is non-degenerate. On the other hand if
$\vartheta^I=0$ there are zero-modes in the R sector, while if
$\vartheta^I=\frac{1}{2}$ there are zero-modes in the NS sector.
In these cases the corresponding fermionic vacuum is degenerate
and carries the spinor representation of the rotation group acting
on the directions in which the $\vartheta$'s vanish. For this
reason it is in general necessary to determine the boundary
reflection matrices also in the spinor representation, which we
will denote by $\mathcal{R}_\sigma$.

To find these matrices, we first note that in the vector representation $R_\sigma$
simply describes the product of five rotations with angles $2\pi\theta_\sigma^I$
in the five complex planes defined by the complex coordinates $Z^I$ and $\overline{Z}^I$, as is clear from (\ref{diagonalRs}).
Then, recalling that the infinitesimal generator of such rotations in the spinor representation is
$\frac{\ii}{2}\Gamma^{I\bar I} =\frac{\ii}{4}\left[\Gamma^I,\Gamma^{\bar I}\right]$
with $\Gamma^I$ being the $\mathrm{SO}(10)$ $\Gamma$-matrices in the
complex basis (see Appendix \ref{app:conventions} for our conventions),
we easily conclude that
\begin{equation}
\mathcal{R}_\sigma = \pm \prod_{I=1}^5\ee^{\ii\pi\theta_\sigma^I\Gamma^{I\bar I}}
= \pm \prod_{I=1}^5\frac{\left(1+\ii f_\sigma^I\Gamma^{I\bar I}\right)}{\sqrt{1+(f_\sigma^I)^2}}
\label{rspinor}
\end{equation}
where $f_\sigma^I=\tan\pi\theta_\sigma^I$ and the overall sign
depends on whether we have a D-brane or an anti D-brane. This
general formula is particularly useful to derive the explicit
expression for $\mathcal{R}_\sigma$ in the limits $f_\sigma^I\to 0$
or $f_\sigma^I\to\infty$ corresponding, respectively, to Neumann or
Dirichlet boundary conditions in the $I$-plane. For example, for an
open string starting from a D$p$-brane extending in the directions
$(01\ldots p)$ we have
\begin{equation}
\mathcal{R}_0 = \prod_{I=1}^{\frac{9-p}{2}} \big(\ii\Gamma^{I\bar I}\big)
= \Gamma^{(p+1)}\cdots\Gamma^9~.
\label{R0D3}
\end{equation}
Being particular instances of rotations, the reflection matrices in the vector and spinor representations satisfy the
following relation
\begin{equation}
\mathcal{R}_\sigma^{-1} \Gamma^M\mathcal{R}_\sigma= ({R}_\sigma)^M_{~N}\Gamma^N~.
\label{vectspin}
\end{equation}

\subsection{Open and closed string vertices}
\label{subsec:vertices}
A generic brane intersection can describe different physical situations depending
on the values of the five twists $\vartheta^I$.

When $\vartheta^I=0$ for all $I$'s, all fields are untwisted:
this is the case of the open strings
starting and ending on the same stack of D-branes which account for dynamical gauge excitations
in the adjoint representation when the branes are space-filling, or for neutral instanton moduli
when the branes are instantonic.

When $\vartheta^4=\vartheta^5=0$ but the $\vartheta^i$'s with
$i=1,2,3$ are non vanishing, only the string coordinates in the
space-time directions are untwisted and describe open strings
stretching between different stacks of D-branes. The corresponding
excitations organize in multiplets that transform in the
bi-fundamental representation of the gauge group and always contain
massless chiral fermions. When suitable relations among the
non-vanishing twists are satisfied ({\it e.g.}
$\vartheta^1+\vartheta^2+\vartheta^3=2$) also massless scalars
appear in the spectrum and they can be combined with the fermions to
form $\mathcal{N}=1$ chiral multiplets suitable to describe the
matter content of brane-world models.

Finally, when $\vartheta^4=\vartheta^5=\frac{1}{2}$, the string
coordinates have mixed Neumann-Dirichlet boundary conditions in the
last four directions and correspond to open strings connecting a
space-filling D-brane  with an instantonic brane. In this situation,
if the $\vartheta^i$'s ($i=1,2,3$) are vanishing, the instantonic
brane describes an ordinary gauge instanton configuration and the
twisted open strings account for the charged instanton moduli of the
ADHM construction
\cite{Witten:1995gx,Douglas:1995bn,Green:2000ke,Billo:2002hm}; if
instead also the $\vartheta^i$'s are non vanishing the instantonic
branes represent exotic instantons of truly stringy nature whose
r\^ole in the effective low-energy field theory has been recently the
subject of intense investigation
\cite{Blumenhagen:2006xt}\nocite{Ibanez:2006da,Bianchi:2007fx,Argurio:2007vqa,
Bianchi:2007wy,Ibanez:2007rs,Antusch:2007jd,Blumenhagen:2007zk,Aharony:2007pr,Blumenhagen:2007bn,Camara:2007dy,Ibanez:2007tu,GarciaEtxebarria:2007zv,Petersson:2007sc,Bianchi:2007rb,Blumenhagen:2008ji,Argurio:2008jm,Cvetic:2008ws,Kachru:2008wt,GarciaEtxebarria:2008pi}-\cite{Buican:2008qe}. {F}rom these
considerations it is clear that by considering open strings that are
generically twisted we can simultaneously treat all configurations
that are relevant  for the applications mentioned in the
Introduction.

\paragraph{Open String Vertices} Let us now focus on the  R sector
of the open strings at a generic brane intersection. Here the vertex operator for the
lowest fermionic excitation $\Theta_{\mathcal{A}}$ is
\begin{equation}
V_{\Theta}(z) = \mathcal{N}_{\Theta}\,\Theta_{\!\mathcal{A}}
\big[\sigma_{\vec{\vartheta}}\,s_{ \vec{\epsilon}_{\mathcal{A}}+\vec{\vartheta}}\,
\ee^{- {\frac{1}{2}\phi}} \,\ee^{\ii \,k \cdot X}\big](z)
\label{vertexferm}
\end{equation}
where we understand that the momentum $k$ is defined only in untwisted directions.
In this expression the
index $\mathcal{A}=1,\ldots,16$ labels a spinor representation of $\mathrm{SO}(10)$ with definite
chirality and runs over all possible choices of signs in the weight vector
\begin{equation}
\vec{\epsilon}_{\mathcal{A}} = \frac{1}{2}\Big(\pm,\pm,\pm,\pm,\pm\Big)
\label{epsilonA}
\end{equation}
with, say, an odd number of $+$'s, and the symbol $s_{\vec{q}}(z)$ stands for the fermionic
spin field
\begin{equation}
s_{\vec{q}}(z)=\ee^{\ii\sum_Iq^I\varphi^I(z)}
\label{spinfield}
\end{equation}
where $\varphi^I(z)$ are the fields that bosonize the world-sheet fermions according to
$\Psi^I=\ee^{\ii\varphi^I}$ (up to cocycle factors). Finally, $\phi(z)$ is the boson entering the superghost
fermionization formulas, $\sigma_{\vec{\vartheta}}(z)$ is the bosonic twist field introduced
above and $\mathcal{N}_\Theta$ is a normalization factor which will be discussed in the following
sections.

The conformal weight of the vertex operator (\ref{vertexferm}) is
\beqa
h &=& \frac{k^2}{2}+\frac{1}{2}\sum_I\left[|\vartheta^I|(1-|\vartheta^I|)
+(\epsilon_{\mathcal{A}}^I+\vartheta^I)^2\right]+\frac{3}{8} \nn \\
&=&
\frac{k^2}{2}+1+\frac{1}{2}\sum_I\left(|\vartheta^I|+2\vartheta^I\epsilon_{\mathcal{A}}^I\right)
\label{confweight}
\eeqa
and hence $V_{\Theta}$ describes a physical massless fermion $h=1$,
$k^2=0$, when the last term vanishes. This condition restricts the
number of the allowed polarization components of $\Theta$ as follows
\begin{equation}
\Theta_{\!\mathcal{A}}\not = 0 \quad\mbox{only if}
\quad \epsilon_{\mathcal{A}}^I = \left\{  \begin{array}{ll}
 \pm \frac{1}{2}   &  ~~\mbox{for}~~ \vartheta^I=0  \\
  -\frac{1}{2}   &  ~~\mbox{for}~~\vartheta^I > 0\\
  ~~\frac{1}{2}   &  ~~\mbox{for}~~\vartheta^I < 0\\
\end{array}
\right.
\label{theta0}
\end{equation}
For example, when all $\vartheta^I$'s are vanishing we have a chiral
spinor in ten dimensions, but if only $\vartheta^4=\vartheta^5=0$
  we have a chiral spinor in the four untwisted directions
along the $(Z^4,Z^5)$ complex plane. On the other hand, in the
instantonic brane constructions mentioned above, for which
$\vartheta^4=\vartheta^5=\frac{1}{2}$, we see from the second line
in (\ref{theta0}) that the R sector describes fermions that do not
carry a spinor index under Lorentz rotations along the ND
four-dimensional plane, in perfect agreement with the ADHM
realization of the charged fermionic instanton moduli.

\paragraph{Closed String Vertices}
We now describe the closed string vertex operators corresponding to background fluxes.
In the closed string sector all fields (both bosonic and fermionic) are untwisted
due to the periodic boundary conditions%
\footnote{Even if in later sections we will consider an orbifold compactification, we will
include background fluxes from the untwisted closed string sector only.}. However,
in the presence of D-branes a suitable identification
between the left and the right moving components of the closed string has to be
enforced at the boundary and a non-trivial dependence on the angles $\theta_\sigma^I$ appears through
the matrices $R_\sigma$ or $\mathcal{R}_\sigma$.

Let us first consider the R-R sector of the Type IIB theory,
where the physical vertex operators for the field strengths of the
anti-symmetric tensor fields are, in the $(-\frac{1}{2},-\frac{1}{2})$ superghost picture,
\begin{equation}
V_{F}(z,\overline z)=\mathcal{N}_{F}\,F_{\mathcal{AB}}~
\ee^{-\ii\pi\alpha'k_{\mathrm L}\cdot k_{\mathrm R}} \big[s_{
\vec{\epsilon}_{\mathcal{A}}} \,\ee^{-\frac{1}{2}\phi} \,\ee^{\ii
\,k_{\mathrm L} \cdot X}\big](z)
 \times
\big[\widetilde s_{ \vec{\epsilon}_{\mathcal{B}}}
 \,\ee^{-\frac{1}{2}\widetilde\phi} \,\ee^{\ii \,k_{\mathrm R} \cdot \widetilde X}\big](\overline z)~.
\label{vertexRR}
\end{equation}
In this expression $\mathcal{N}_F$ is a normalization factor that will be discussed later,
$k_{\mathrm L}$ and $k_{\mathrm R}$ are the left and right momenta, and the
tilde sign denotes the right-moving components.
Furthermore, the factor $\ee^{-\ii\pi\alpha'k_{\mathrm L}\cdot k_{\mathrm R}}$ is a cocycle that allows
for an off-shell extension of the closed string vertex with $k_{\mathrm L}\not =k_{\mathrm R}$, as discussed in Ref. \cite{Bertolini:2005qh}. The bi-spinor polarization $F_{\mathcal{AB}}$
comprises all R-R field strengths
of the Type IIB theory according to
\begin{equation}
F_{\mathcal{AB}}=\sum_{n=1,3,5}\frac{1}{n!}\,F_{M_1\ldots M_n}
\left(\Gamma^{M_1\ldots M_n}\right)_{\mathcal{AB}}~,
\label{F135}
\end{equation}
even if in our applications only the 3-form part will play a r\^ole.
In the presence of D-branes the left and right
moving components of the vertex operator $V_F$ must be identified using the reflection rules
discussed above. In practice (see for example Ref. \cite{Bertolini:2005qh} for more details)
this amounts to set
\begin{equation}
\widetilde X^M(\overline{z}) = (R_0)^M_{~N}\,X^N(\overline{z})
\quad,\quad \widetilde s_{ \vec{\epsilon}_{\mathcal{A}}}(\overline{z})=
(\mathcal{R}_0)^{\mathcal{A}}_{~\,\mathcal B}\,
s_{ \vec{\epsilon}_{\mathcal{B}}}(\overline{z})\quad,\quad\widetilde\phi(\overline{z})=
\phi(\overline{z})
\label{leftright}
\end{equation}
and modify the cocycle factor in the vertex operator (\ref{vertexRR}) to
$\ee^{-\ii\pi\alpha'k_{\mathrm L}\cdot (k_{\mathrm R}R_0)}$.
As a consequence of the identifications (\ref{leftright}),
the R-R field-strength $F_{\mathcal{AB}}$
gets replaced by the bi-spinor polarization $(F\,\mathcal{R}_0)_{\mathcal{AB}}$ that
incorporates also the information on the type of boundary conditions enforced by the D-branes.

Let us now turn to the NS-NS sector of the closed string. Here it is possible to write
an effective BRST invariant vertex operator for the derivatives of the anti-symmetric tensor $B$
that are related to the 3-form flux $H$.
In the $(0,-1)$ superghost picture%
\footnote{This particular asymmetric picture
is chosen in view of the calculations of the disk amplitudes described in
Section \ref{subsec:RR}.}, this effective vertex is
\begin{equation}
V_H(z,\overline{z}) = {\mathcal{N}_H}\,\big(\partial_MB_{NP}\big)\,
\ee^{-\ii\pi\alpha'k_{\mathrm L}\cdot k_{\mathrm R}}
\big[\psi^M\psi^N\ee^{\ii \,k_{\mathrm L} \cdot X}\big](z) \times
\big[\widetilde \psi^P\,\ee^{-\widetilde\phi} \,\ee^{\ii
\,k_{\mathrm R} \cdot \widetilde X}\big] (\overline z)
\label{vertexNS}
\end{equation}
where again we have introduced a normalization factor and a cocycle. When we insert this vertex in
a disk diagram, we must identify the left and right moving sectors using the reflection rules
(\ref{leftright}) supplemented by
\begin{equation}
\widetilde \psi^M(\overline{z}) = (R_0)^M_{~N}\psi^N(\overline{z})~.
\label{leftright1}
\end{equation}
Consequently, in (\ref{vertexNS}) the polarization $(\partial B)$ is effectively replaced by
$(\partial BR_0)$. Notice that the NS-NS polarization combines with the boundary reflection matrix in the
vector representation $R_0$, in contrast to the R-R case
where one finds instead the reflection matrix in the spinor representation $\mathcal{R}_0$.

\subsection{The string correlator with R-R and NS-NS fluxes}
\label{subsec:RR}
We now evaluate the string correlation functions among two massless open string fermions and the
background closed string flux, as represented in Fig. \ref{fig:flux}.
It is a mixed open/closed string amplitude on a disk which, generically, has mixed boundary conditions. From the conformal field theory point of view such fermionic correlation functions are similar to the mixed amplitudes considered in
Ref.s \cite{Billo:2004zq,Lust:2004cx,Billo:2005jw,Bertolini:2005qh}.
Let us analyze first the interaction with the R-R flux.

\paragraph{R-R flux} We take two fermionic open string vertices
(\ref{vertexferm}) and one closed string R-R vertex (\ref{vertexRR}), and compute the amplitude
\begin{equation}
\mathcal{A}_{F}= \Big\langle
V_{\Theta}(x)\,V_F(z,\overline{z})\,V_{\Theta'}(y) \Big\rangle
= c_F~ {\Theta}_{\!\mathcal {A}_1}(F\mathcal{R}_0)_{\mathcal{A}_2\mathcal{A}_3}\,
{\Theta'}_{\!\!\mathcal{A}_4}\,\times\,
A^{\mathcal{A}_1\mathcal{A}_2\mathcal{A}_3\mathcal{A}_4}
\label{amplF}
\end{equation}
where the prefactor
\begin{equation}
c_F =
\mathcal{C}_{(p+1)}\,\mathcal{N}_{\Theta}\,\mathcal{N}_{\Theta'}\,\mathcal{N}_F~,
\label{cF}
\end{equation}
accounts for the normalizations of the vertex operators and the topological
normalization $\mathcal{C}_{(p+1)}$
of any disk amplitude with the boundary conditions of a D$p$-brane \cite{Billo:2002hm,DiVecchia:1996uq}, whose explicit
expression will be given in Section \ref{subsec:masses} for D3-branes and D-instantons,
see Eqs. (\ref{c4}) and (\ref{c0}).
The last factor in (\ref{amplF}) is the 4-point correlator
\begin{equation}
A^{\mathcal{A}_1\mathcal{A}_2\mathcal{A}_3\mathcal{A}_4}
=\int\frac{\prod_{i=1}^4dz_i}{dV_{\mathrm{CKG}}}~\ee^{-\ii\pi\alpha'k_2\cdot k_3}\,
\Big\langle\prod_{i=1}^4
\big[\sigma_{\vec{\vartheta}_i}\,s_{ \vec{\epsilon}_{i}+\vec{\vartheta}_i}\,
\ee^{- {\frac{1}{2}\phi}} \,\ee^{\ii \,k_i \cdot X}\big](z_i)\Big\rangle
\label{amplF1}
\end{equation}
where we have used the convenient notation
\begin{equation}
\begin{aligned}
z_1&=x~\,,~\,z_2=z~\,,~\,z_3=\overline{z}~\,,~\, z_4=y~,\\
k_1&=k~\,,~k_2=k_{\mathrm{L}}~\,,~\,k_3=k_{\mathrm{R}}R_0~\,,~k_4=k'~,\\
\vec{\vartheta}_1&=\vec{\vartheta}~\,,~\,
\vec{\vartheta}_2=0~\,,~\,\vec{\vartheta}_3=0~\,,~\,
\vec{\vartheta}_4=-\vec{\vartheta}~,
\end{aligned}
\label{zi}
\end{equation}
and we have set
$\vec{\epsilon}_i\equiv\vec{\epsilon}_{\mathcal{A}_i}$. Since the
closed string vertex is untwisted, the two open string vertices
must have opposite twists in order to have a non-vanishing
amplitude. This explains the third line above which, according to
Eq. (\ref{theta0}), implies that when $\vartheta^I\not=0$
the polarizations $\Theta_{\!\mathcal{A}_1}$ and $\Theta'_{\!\mathcal{A}_4}$ are 
not vanishing only if $\epsilon^I_{1}=-\epsilon^I_{4}$. Therefore, if
$\vartheta^I\not=0$ for all $I$'s, the spinor weights
$\vec{\epsilon}_1$ and $\vec{\epsilon}_4$ have different GSO
parity and the amplitude (\ref{amplF}) ceases to exist. To avoid
this, from now on we will assume that
at least one of the $\vartheta^I$'s be vanishing%
\footnote{As pointed out in Ref. \cite{Billo:2007py}
when all five $\vartheta^I$'s are non vanishing, the simplest tree-level diagram involving
massless fermions of the twisted R sector requires at least three different types of boundary conditions
and thus it is not of the type of amplitudes we are discussing here, which involve only two boundary
changing operators.}. The evaluation of the correlator in (\ref{amplF1}) can be simplified by
assuming that the closed string vertex does not carry momentum in the twisted directions ({\it i.e.}
$k_2^I=k_3^I=0$ if $\vartheta^I\not=0$). This is not a restrictive choice for our purposes,
since we will be interested in the effects induced by \emph{constant} background fluxes.

In the correlator (\ref{amplF1}) the open string positions $z_1$ and $z_4$ are integrated
on the real axis while the closed string variables $z_2$ and $z_3$ are integrated in the
upper half complex plane, modulo the $\mathrm{Sl}(2;\mathbb{R})$ projective invariance
that is fixed by the conformal Killing group volume $dV_{\mathrm{CKG}}$. Using this fact we have
\begin{equation}
\frac{\prod_{i=1}^4dz_i}{dV_{\mathrm{CKG}}}= d\omega \,(1-\omega)^{-2} \,\big(z_{14}z_{23}\big)^2
\label{dz}
\end{equation}
where $\omega$ is the anharmonic ratio
\begin{equation}
\omega= \frac{z_{12}z_{34}}{z_{13}z_{24}} \quad\quad(\,|\omega|=1\,)
\label{omega}
\end{equation}
with $z_{ij}=z_i-z_j$. Due to our kinematical configuration, the contribution
of the twist fields and the bosonic exponentials to the correlator (\ref{amplF1}) can be
factorized and becomes
\begin{equation}
\Big\langle
\sigma_{\vec{\vartheta}}(z_1)\,\sigma_{-\vec{\vartheta}}(z_4)\Big\rangle
\,\Big\langle\prod_{i=1}^4 \ee^{\ii \,k_i \cdot X(z_i)}\Big\rangle
= z_{14}^{\,\sum_I\vartheta^I(1-\vartheta^I)}~\omega^{\alpha't}\,(1-\omega)^{\alpha' s}
\label{x}
\end{equation}
where we have used (\ref{OPE}), introduced the two kinematic invariants
\begin{equation}
s=(k_1+k_4)^2=(k_2+k_3)^2\quad\mbox{and}\quad t=(k_1+k_3)^2=(k_2+k_4)^2~,
\label{stu}
\end{equation}
and understood the momentum conservation.

Also the contribution of the spin fields and the
superghosts can be easily evaluated using the bosonization formulas,
that allow to write
\begin{equation}
\Big\langle \prod_{i=1}^4 s_{\vec{\epsilon}_{i}
+\vec{\vartheta}_i}(z_i)\,\ee^{-\frac{1}{2}\phi(z_i)}
\Big\rangle =
\Big\langle \prod_{i=1}^4 s_{\vec{\epsilon}_i}(z_i)
\,\ee^{-\frac{1}{2}\phi(z_i)}\Big\rangle\,\times\,
\prod_{i<j}\,z_{ij}^{\vec{\epsilon}_i
\cdot \vec{\vartheta}_j + \vec{\epsilon}_j
\cdot \vec{\vartheta}_i +\vec{\vartheta}_i\cdot \vec{\vartheta}_j}~.
\label{spinc}
\end{equation}
The first factor in the right hand side is the
four fermion correlator of the Type IIB superstring in ten dimensions which has been computed
for example in Ref. \cite{Friedan:1985ge}, namely
\beqa
\Big\langle \prod_{i=1}^4 s_{\vec{\epsilon}_i}(z_i)
\,\ee^{-\frac{1}{2}\phi(z_i)}\Big\rangle
&=& \frac{1}{2}\prod_{i<j}z_{ij}^{-1}
\left[z_{13}z_{24}\big(\Gamma_M\big)^{\!\mathcal{A}_1\mathcal{A}_4}
\!\big(\Gamma^M\big)^{\!\mathcal{A}_2\mathcal{A}_3}
\! \right. \nn \\
&& \left. + ~z_{14}z_{23}\big(\Gamma_M\big)^{\!\mathcal{A}_1\mathcal{A}_3}
\!\big(\Gamma^M\big)^{\!\mathcal{A}_2\mathcal{A}_4}
\right]
\label{4spin}
\eeqa
where we have understood the ``charge'' conservation $\sum_i \vec{\epsilon}_i=0$.
Furthermore, the $\vartheta$-dependent factor in (\ref{spinc}) can be simplified using
the relations
\begin{equation}
\vec{\epsilon}_2\cdot\vec{\vartheta}=-\vec{\epsilon}_3\cdot\vec{\vartheta}
\quad,\quad\vec{\epsilon}_1\cdot\vec{\vartheta}=
-\vec{\epsilon}_4\cdot\vec{\vartheta}=\frac{1}{2}\sum_I\vartheta^I~,
\label{rel}
\end{equation}
that follow from (\ref{theta0}) and the ``charge'' conservation of the spinor weights. Indeed, using
(\ref{rel}) we have
\begin{equation}
\prod_{i<j}\,z_{ij}^{\vec{\epsilon}_i
\cdot \vec{\vartheta}_j + \vec{\epsilon}_j
\cdot \vec{\vartheta}_i +\vec{\vartheta}_i\cdot \vec{\vartheta}_j}
=z_{14}^{\,\sum_I\vartheta^I(\vartheta^I-1)}~\omega^{-\vec{\epsilon}_3
\cdot\vec{\vartheta}}~.
\label{spin41}
\end{equation}
Collecting everything we find that the amplitude (\ref{amplF1}) can be written as
\begin{equation}
A^{\mathcal{A}_1\mathcal{A}_2\mathcal{A}_3\mathcal{A}_4}
=\big(\Gamma_M\big)^{\!\mathcal{A}_1\mathcal{A}_4}
\!\big(\Gamma^MI_1\big)^{\!\mathcal{A}_2\mathcal{A}_3}
+\big(\Gamma_MI_2\big)^{\!\mathcal{A}_1\mathcal{A}_3}
\!\big(\Gamma^M\big)^{\!\!\mathcal{A}_2\mathcal{A}_4}
\label{ampl2}
\end{equation}
where we have introduced the two $\vec\vartheta$-dependent diagonal matrices with entries
\begin{equation}
\begin{aligned}
\big(I_1\big)_{\!\mathcal{A}_3}^{~\mathcal{A}_3} &=\frac{1}{2}~\ee^{-\frac{\ii\pi\alpha's}{2}}
\int_\gamma d\omega\,(1-\omega)^{\alpha's-1}\,\omega^{\alpha't
-\vec{\vartheta}\cdot \vec{\epsilon}_3-1}~,\\
\big(I_2\big)_{\!\mathcal{A}_3}^{~\mathcal{A}_3} &=\frac{1}{2}~\ee^{-\frac{\ii\pi\alpha's}{2}}
\int_\gamma d\omega\,(1-\omega)^{\alpha's}\,\omega^{\alpha't
-\vec{\vartheta}\cdot \vec{\epsilon}_3-1}~,
\end{aligned}
\label{integrals}
\end{equation}
where $\mathcal{A}_3$ is the spinor index corresponding to the spinor weight $\vec{\epsilon}_3$. 
Here the integrals run around the clockwise oriented unit circle $\gamma: |\omega|=1$,
and can be evaluated to be \cite{Bertolini:2005qh}
\begin{equation}
\begin{aligned}
\big(I_1\big)_{\!\mathcal{A}_3}^{~\mathcal{A}_3}
&=\frac{1}{2}~\ee^{-\frac{\ii\pi\alpha's}{2}}
\Big(\ee^{-2\pi\ii\big(\alpha't-\vec{\vartheta}\cdot
\vec{\epsilon}_3\big)}-1\Big)
\,B\big(\alpha's;\alpha't-\vec{\vartheta}\cdot \vec{\epsilon}_3\big)~,\\
\big(I_2\big)_{\!\mathcal{A}_3}^{~\mathcal{A}_3}
&=\frac{1}{2}~\ee^{-\frac{\ii\pi\alpha's}{2}}
\Big(\ee^{-2\pi\ii\big(\alpha't-\vec{\vartheta}\cdot
\vec{\epsilon}_3\big)}-1\Big)
\,B\big(\alpha's+1;\alpha't-\vec{\vartheta}\cdot
\vec{\epsilon}_3\big)~,
\end{aligned}
\label{integrals1}
\end{equation}
where $B(a;b)$ is the Euler $\beta$-function.
Plugging (\ref{ampl2}) into (\ref{amplF}), with some simple manipulations we find
\begin{equation}
\mathcal{A}_{F}= -c_F\Big[\Theta'\Gamma^M\Theta\,\,\mathrm{tr}\big(
F\mathcal{R}_0I_1\Gamma_M\big)+
\Theta'\Gamma^MF\mathcal{R}_0I_2\Gamma_M\Theta\Big]
\end{equation}
where the trace is understood in the $16\times16$ block spanned by the spinor indices $\mathcal{A}_i$'s.
This expression can be further simplified by expanding the matrices $F\mathcal{R}_0I_{1}$
and $F\mathcal{R}_0I_{2}$ as
\begin{equation}
\big(F\mathcal{R}_0I_{a}\big)_{\mathcal{AB}}= \sum_{n=1,3,5}\frac{1}{n!}\,\big(F\mathcal{R}_0I_{a}\big)_{N_1\ldots N_n}
\left(\Gamma^{N_1\ldots N_n}\right)_{\mathcal{AB}}\quad\quad(a=1,2)~,
\label{FRI}
\end{equation}
and by using the $\Gamma$-matrix identities
\begin{equation}
\begin{aligned}
\tr\big(\Gamma^M\Gamma^N\big)=&16\,\delta^{MN}\quad,\quad
\tr\big(\Gamma^M\Gamma^{N_1N_2N_3}\big)= \tr\big(\Gamma^M\Gamma^{N_1\ldots N_5}\big)=0~,\\
&\Gamma_M\Gamma^{N_1\ldots N_n}\Gamma^M = (-1)^n (10-2n)\,\Gamma^{N_1\ldots N_n}~.
\end{aligned}
\label{identities}
\end{equation}
After some straightforward algebra we find
\begin{equation}
\mathcal{A}_{F}= -8c_F\Theta'\Gamma^M\Theta\,\big[F\mathcal{R}_0(2I_{1}-I_2)\big]_M
+\frac{4c_F}{3!}\Theta'\Gamma^{MNP}\Theta\,\big[F\mathcal{R}_0I_2\big]_{MNP}~.
\label{amplFfinal}
\end{equation}
This formula is one of the main results of this section. It describes the tree-level bilinear
fermionic couplings induced by R-R fluxes on a general brane intersection.

\paragraph{NS-NS flux}
Let us now turn to the fermionic couplings induced by the NS-NS 3-form flux effectively
described by the vertex operator (\ref{vertexNS}). Such couplings arise from
the following mixed disk amplitude
\begin{equation}
\mathcal{A}_{H}= \Big\langle
V_{\Theta}(x)\,V_H(z,\overline{z})\,V_{\Theta'}(y) \Big\rangle
= {c_H}~ {\Theta}_{\!\mathcal A}(\partial B{R}_0)_{{MNP}}\,
{\Theta'}_{\!\mathcal B}\,\times\,
A^{\mathcal{AB};MNP}
\label{amplH}
\end{equation}
where the normalization factor is
\begin{equation}
c_H =
\mathcal{C}_{(p+1)}\,\mathcal{N}_{\Theta}\,\mathcal{N}_{\Theta'}\,\mathcal{N}_H
\label{cH}
\end{equation}
and the 4-point correlator is
\begin{eqnarray}
A^{\mathcal{AB};MNP} &=&\int\frac{\prod_{i=1}^4dz_i}{dV_{\mathrm{CKG}}}
~\ee^{-\ii\pi\alpha'k_2\cdot k_3}\,
\Big\langle
\sigma_{\vec{\vartheta}}(z_1)\,\sigma_{-\vec{\vartheta}}(z_4)\Big\rangle
\,\Big\langle\prod_{i=1}^4 \ee^{\ii \,k_i \cdot X(z_i)}\Big\rangle
\nn\\
&&\times
\Big\langle
s_{\vec{\epsilon}_{\mathcal A}+\vec{\vartheta}}(z_1)
\,\psi^M\!\psi^N(z_2)\,\psi^P(z_3)\,
s_{\vec{\epsilon}_{\mathcal B}-\vec{\vartheta}}(z_4)
\Big\rangle \nn \\
&& \times
\Big\langle
\ee^{- {\frac{1}{2}\phi(z_1)}}
\ee^{- {\phi(z_3)}}
\ee^{- {\frac{1}{2}\phi(z_4)}}\Big\rangle~.
\label{amplH1}
\end{eqnarray}
Here we have used a notation similar to that of Eq. (\ref{zi}) for the
bosonic and twist fields, whose contribution is the
same as in Eq. (\ref{x}) because of our kinematical configuration.
Due to the Lorentz structure of the fermionic correlator, the
second and third lines of (\ref{amplH1}) can be written as
\begin{equation}
\begin{aligned}
\Big\langle
s_{\vec{\epsilon}_{\mathcal A}+\vec{\vartheta}}(z_1)
\,\psi^M\!&\psi^N(z_2)\,\psi^P(z_3)\,
s_{\vec{\epsilon}_{\mathcal B}-\vec{\vartheta}}(z_4)
\Big\rangle
\Big\langle
\ee^{- {\frac{1}{2}\phi(z_1)}}
\ee^{- {\phi(z_3)}}
\ee^{- {\frac{1}{2}\phi(z_4)}}\Big\rangle\\
&= f(z_{ij}) \big(\Gamma^{MNP}\big)^{\!\mathcal{A}\mathcal{B}}
+g(z_{ij}) \Big[\delta^{MP}\big(\Gamma^{N}\big)^{\!\mathcal{A}\mathcal{B}}
-\delta^{NP}\big(\Gamma^{M}\big)^{\!\mathcal{A}\mathcal{B}}\Big]
\end{aligned}
\end{equation}
where the two functions $f$ and $g$ can be determined, for example,
by using the bosonization technique.
If we pick a configuration such that the field $\psi^M\!\psi^N(z_2)$
can be bosonized as $\ee^{\ii\vec{\epsilon}_2\cdot\vec{\varphi}}$
with weight vectors of the form
\begin{equation}
\vec{\epsilon}_2=\big(0,\ldots,\pm1,\ldots,\pm1,\ldots,0)~,
\label{epsilon2}
\end{equation}
corresponding to roots of $\mathrm{SO}(10)$,
we can use the same strategy we have described before in the R-R case to find
\begin{equation}
\begin{aligned}
f(z_{ij}) &= \prod_{i<j}z_{ij}^{-1}\times \big(z_{14}z_{23}\big)\times
z_{14}^{\,\sum_I\vartheta^I(\vartheta^I-1)}~\omega^{-\vec{\epsilon}_3
\cdot\vec{\vartheta}}~,\\
g(z_{ij})  &= \prod_{i<j}z_{ij}^{-1}\times\big(z_{12}z_{34}+z_{13}z_{24}\big)\times
z_{14}^{\,\sum_I\vartheta^I(\vartheta^I-1)}~\omega^{-\vec{\epsilon}_3
\cdot\vec{\vartheta}}~,
\end{aligned}
\end{equation}
where the last factors are the same as in Eq. (\ref{rel}) and
$\vec{\epsilon}_3$ is the weight vector in the vector representation associated to $\psi^P(z_3)$, of the form
\begin{equation}
\vec{\epsilon}_3=\big(0,\ldots,\pm1,\ldots,0)~.
\label{epsilon3}
\end{equation}
Collecting everything, and introducing the diagonal matrices (with vector indices) $(I_1)^P_{~P}$ and $(I_2)^P_{~P}$
defined analogously to Eq. (\ref{integrals}), after some simple manipulations we obtain
\beqa
\mathcal{A}_{H} &=& -{4c_H}\Theta'\Gamma^N\Theta \,\delta^{MP}\,\big[\partial BR_0(2I_{1}-I_2)\big]_{[MN]P}
\nn \\
&&
+{2c_H}\Theta'\Gamma^{MNP}\Theta\,\big[\partial B R_0I_2\big]_{MNP}
\label{amplHfinal}
\eeqa
which is the NS-NS counterpart of the R-R amplitude (\ref{amplFfinal}) on a generic D brane intersection
and shares with it the same type of fermionic structures.

\section{Flux couplings with untwisted open strings ($\vec{\vartheta}=0$)}
\label{sec:effects}
We now exploit the results obtained in the previous section to analyze how constant background fluxes
couple to untwisted open strings, {\it i.e.} strings starting and ending on a single stack of D-branes.
This corresponds to set $\vec{\vartheta}=0$ in all previous formulas which drastically simplify.
Note that the condition $\vec{\vartheta}=0$ implies that $\vec{\theta}_0=\vec{\theta}_\pi$, so that the
reflection rules are the same at the two string end-points. We can
distinguish two cases, namely when these reflection rules are just signs
({\it i.e.} $\theta_\sigma^I=0$ or $1$) and when they instead depend on generic angles $\theta_\sigma^I$.
In the first case the branes are unmagnetized,  while the second corresponds to
magnetized branes.

Since we are interested in constant background fluxes, we can set the momentum of the closed string
vertices to zero; this corresponds to take the limit $s=-2t \to 0$ in the integrals (\ref{integrals1})
which yields
\begin{equation}
 2I_1=I_2= -\ii\pi~.
\label{intr0}
\end{equation}
Using this result in the R-R and NS-NS amplitudes (\ref{amplFfinal}) and (\ref{amplHfinal}),
we see that the fermionic couplings with a single $\Gamma$ matrix vanish and only the terms with three
$\Gamma$'s survive, so that the total flux amplitude is
\begin{equation}
\mathcal A \equiv \mathcal{A}_F + \mathcal{A}_H =-2\pi\ii\,
\Theta\Gamma^{MNP}\Theta\,\Big[\frac{c_F}{3}\big(F\mathcal{R}_0\big)_{MNP}
+c_H\big(\partial B R_0\big)_{MNP}\Big]~. \label{ampltot}
\end{equation}
Here we used the fact that the untwisted fermions $\Theta$ and
$\Theta'$ in (\ref{amplFfinal}) and (\ref{amplHfinal}) actually
describe the same field and only differ because they carry opposite
momentum. For this reason we multiplied the above amplitudes by a
symmetry factor of 1/2 and dropped the $'$ without introducing
ambiguities.

It is clear from Eq. (\ref{ampltot}) that once the flux configuration is given, the structure of the
fermionic couplings for different types of D-branes depends crucially
on the boundary reflection matrices $R_0$ and $\mathcal R_0$.
Notice that the R-R piece of the amplitude (\ref{ampltot}) is generically non zero for 1-form, 3-form
and 5-form fluxes. However, from now on we will restrict our analysis only to the 3-form  and hence the
bi-spinor to be used is simply
\begin{equation}
F_{\mathcal{AB}} = \frac{1}{3!} F_{MNP}
\left(\Gamma^{MNP}\right)_{\mathcal{AB}}~.
\label{F3}
\end{equation}
We can now specify better
the normalization factors $c_F$ and $c_H$. In fact the vertex (\ref{vertexRR}) for a R-R 3-form and the NS-NS vertex (\ref{vertexNS})
should account for the following quadratic terms of the bulk theory
in the ten-dimensional Einstein frame:
\begin{equation}
\frac{1}{2\kappa_{10}^2}\int d^{10}x\,\sqrt{g_{(E)}}\,\Big(\frac{1}{3!}\,\ee^{\varphi}F^2
+ \frac{1}{3!}\,\ee^{-\varphi}dB^2\Big)~,
\label{bulk10}
\end{equation}
where $\varphi$ is the dilaton and $\kappa_{10}$ is the gravitational Newton constant in ten dimensions.
In order to reproduce the above dilaton dependence, the normalization factors $\mathcal{N}_F$
and $\mathcal{N}_H$ of the R-R and NS-NS vertex operators must scale with the string coupling $g_s
=\ee^\varphi$ as
\begin{equation}
\mathcal{N}_F \sim g_s^{1/2}\quad\mbox{and}\quad\mathcal{N}_H\sim g_s^{-1/2}~,
\label{NfNh}
\end{equation}
so that from Eqs. (\ref{cF}) and (\ref{cH}) we obtain
\begin{equation}
c_H=c_F/g_s~.
\label{cfch}
\end{equation}
Taking all this into account, we can rewrite the total amplitude (\ref{ampltot}) as
\begin{equation}
\mathcal A \equiv \mathcal{A}_F + \mathcal{A}_H =-\frac{2\pi\ii}{3!}\,c_F\,
\Theta\Gamma^{MNP}\Theta\,
T_{MNP}
\label{ampltot1}
\end{equation}
where
\begin{equation}
T_{MNP} = \big(F\mathcal{R}_0\big)_{MNP}+\frac{3}{g_s}\,
\big( \partial B R_0\big)_{[MNP]} ~.
\label{tmnl}
\end{equation}

Up to now we have used a ten-dimensional notation. However, since we are interested
in studying the flux induced couplings for gauge theories and instantons in four dimensions, it becomes
necessary to split the indices $M,N,\ldots=0,1,\ldots,9$ appearing in the above equations into four-dimensional space-time indices $\mu,\nu,\ldots=0,1,2,3$, and six-dimensional indices
$m,n,\ldots=4,5,\ldots,9$ labeling the directions of the internal space
(which we will later take to be compact, for example a 6-torus $\mathcal T_6$ or an orbifold thereof).
Clearly background fluxes carrying indices along the space-time break the four-dimensional
Lorentz invariance and generically give rise to deformed gauge theories. Effects of this
kind have already been studied using world-sheet techniques in
Refs. \cite{Billo:2004zq,Billo:2005jw} where
a non vanishing R-R 5-form background of the type $F_{\mu\nu mnp}$ was shown to originate the
${\mathcal N}\!=\!1/2$ gauge theory, and in Ref. \cite{Billo:2006jm} where the so-called $\Omega$
deformation of the $\mathcal N =2$ gauge theory was shown to derive from a R-R 3-form flux of
the type ${F}_{\mu\nu m}$. In the following, however, we will consider only internal fluxes,
like $F_{mnp}$ or $(\partial B)_{mnp}$, which preserve the four-dimensional Lorentz invariance, and
the fermionic amplitudes we will compute are of the form
\begin{equation}
\mathcal A =-\frac{2\pi\ii}{3!}\,c_F\,\Theta\Gamma^{mnp}\Theta\,T_{mnp}~ .
\label{ampltot2}
\end{equation}
We now analyze the structure of these couplings beginning with the simplest case of space-filling
unmagnetized D-branes; later we examine
unmagnetized Euclidean branes and finally branes with a non-trivial world-volume magnetic field.

\subsection{Unmagnetized D-branes}
\label{subsec:unmagnetizedD}

Even if the fermionic couplings (\ref{ampltot1}) have been derived in Section \ref{sec:CFT} assuming a
Euclidean signature, when we discuss space-filling D-branes with $\vec{\vartheta}=0$, the
rotation to a Minkowskian signature poses no problems. In this case,
$\Theta$ becomes a Majorana-Weyl spinor in ten dimensions which in particular satisfies
\begin{equation}
\Theta\Gamma^{mnp}\Theta =-\,\big(
\Theta\Gamma^{mnp}\Theta\big)^*~.
\label{cc}
\end{equation}
Furthermore for an unmagnetized D$p$-brane that fills the four-dimensional Minkowski space
and possibly extends also in some internal directions, the reflection matrices $R_0$
and $\mathcal R_0$ are very simple: indeed in the vector representation
\begin{equation}
R_0 = \mathrm{diag}(\pm {1},\pm 1, \ldots)~,
\label{r0vec}
\end{equation}
where the entries specify whether a direction is longitudinal ($+$) or transverse ($-$), while in the spinor representation
\begin{equation}
\mathcal R_0 =  \Gamma^{p+1}\cdots\Gamma^{9}~.
\label{r0spin}
 \end{equation}
Using these matrices we easily see that $T_{mnp}$ is a {\it real} tensor, so that
in view of Eq. (\ref{cc}) also the total fermionic amplitude (\ref{ampltot2})
is real, as it should be.

The explicit expression of $T_{mnp}$ is particularly simple in the case of brane configurations
which respect the $4+6$ structure of the space-time, {\it i.e.} D3- and D9- branes. For space-filling
D3-branes all internal indices are transverse, so that $\left.R_0\right|_{\mathrm{int}}=-1$
and $\mathcal R_0=\Gamma^4\cdots\Gamma^9$. From Eq. (\ref{tmnl}) it follows then
\begin{equation}
T_{mnp}= (*_6 F)_{mnp} -\frac{1}{g_s}\,H_{mnp}
\label{td3}
\end{equation}
where $*_6$ denotes the Poincar\`{e} dual in the six-dimensional internal space and
$H=dB$
\footnote{In our conventions $(*_6 F)_{mnp}= \frac{1}{3!}\epsilon_{mnprst}\,F^{rst}$ and $H_{mnp}=
3\partial_{[m} B_{np]}= \big(\partial_m B_{np}+\partial_n B_{pm}+\partial_p B_{mn}\big)$.}.

For D9-branes, instead, all internal indices are longitudinal, and to emphasize this fact  we denote them
as $\hat m, \hat n, \ldots$ In this case we simply have $R_0=1$ and $\mathcal R_0=1$
so that
\begin{equation}
T_{\hat m\hat n \hat p}= F_{\hat m\hat n \hat p}+\frac{1}{g_s}\,H_{\hat m \hat n \hat p}~.
\label{td9}
\end{equation}
Note however that D9-branes must always be accompanied by orientifold 9-planes (O9)
for tadpole cancellation and that the corresponding orientifold projection kills the
NS-NS flux $H_{\hat m \hat n \hat p}$. If we take this fact into account,
the coupling tensor for D9-branes reduces to
\begin{equation}
T_{\hat m\hat n \hat p}= F_{\hat m\hat n \hat p}~.
\label{td91}
\end{equation}

The case of space-filling D7- and D5-branes is slightly more involved since for these branes
the internal directions are partially longitudinal and partially transverse. In particular,
for D7-branes the longitudinal internal indices $\hat m, \hat n\ldots$ take four values while the transverse indices $p,q,\ldots$ take two values. Eq.~(\ref{tmnl}) implies then that the only non vanishing
components of the $T$ tensor for D7-branes are
\begin{equation}
T_{\hat m \hat n \hat p}=\frac{1}{g_s}\,H_{\hat m \hat n \hat p}~,\quad
T_{\hat m \hat n p} = F_{\hat m \hat n}^{\phantom{\hat m \hat n}\!q}\,\epsilon_{qp}
+\frac{1}{g_s}\,H_{\hat m \hat n p}
\quad\mbox{and}\quad T_{\hat mnp} = -\frac{1}{g_s}\,H_{\hat mnp}~.
\label{tD7}
\end{equation}
If one introduces O7-planes to cancel the tadpoles produced by the D7-branes, one can see that
the corresponding orientifold projection\footnote{The extra $(-1)^{F_L}$ appearing in the case of
O3/O7-planes ensure that the corresponding orientifold actions square to one, i.e.
$(\Omega I_{4n+2} (-1)^{F_L})^2=I^2_{4n+2} (-1)^{F_L+F_R}=1$. } $\Omega I_2 (-1)^{F_L}$
removes all $F$ and $H$ components with an even number of
transverse indices so that the only surviving couplings are
\begin{equation}
T_{\hat m \hat n p} = F_{\hat m \hat n}^{\phantom{\hat m \hat n}\!q}\,\epsilon_{qp}
+\frac{1}{g_s}\,H_{\hat m \hat n p}~.
\label{tD7a}
\end{equation}

For D5-branes the situation is somehow complementary, since
the longitudinal internal indices take two values while the transverse ones run over four values.
In this case one can show that
the non vanishing components of the $T$ tensor are
\begin{equation}
T_{\hat m \hat n p}=\frac{1}{g_s}\,H_{\hat m \hat n p}~,\quad
T_{\hat m np}=-\frac{1}{2}\,F_{\hat m}^{\phantom{\hat m}\!qr}\epsilon_{qrnp}
-\frac{1}{g_s}\,H_{\hat m np}
~,\quad T_{mnp} = -\frac{1}{g_s}\,H_{mnp}~.
\label{tD5}
\end{equation}
Again the O5-planes required for tadpole cancellation enforce an orientifold projection $\Omega I_4$
which removes the components of $H$($F$) with an even(odd) number of transverse indices.
Thus, the coupling $T_{\hat m np}$ reduces to
\begin{equation}
T_{\hat m np}=-\frac{1}{2}\,F_{\hat m}^{\phantom{\hat m}\!qr}\epsilon_{qrnp}~.
\label{tD5a}
\end{equation}
The fermionic couplings for the various D-branes we have discussed,
taking into account the appropriate orientifold projections, are summarized in Tab.~\ref{Dbranes}.
\begin{table}[ht]
\begin{small}
\centering
\begin{tabular}{cccccccccccc}
\hline\hline
  & \phantom{\vdots}\!0\!&\!1\!&\!2\!&\!3\!&\!4\!&\!5\!&\!6\!&\!7\!&\!8\!&\!9\! & $T_{mnp}$ \\ [1ex]
 \hline
D3     & \phantom{\vdots}\!$-$\!&\!$-$\!&\!$-$\!&\!$-$\!
&\!$\times$\!&\!$\times$\!&\!$\times$\!&\!$\times$\!&\!$\times$\!&\!$\times$\! &
$(*_6F)_{mnp} - \frac{1}{g_s}H_{mnp} $
\\ [1ex]
D5    & \phantom{\vdots}\!$-$\!&\!$-$\!&\!$-$\!&\!$-$\!
&\!$-$\!&\!$-$\!&\!$\times$\!&\!$\times$\!&\!$\times$\!&\!$\times$\! &
$\frac{1}{g_s}H_{\hat m \hat n p};~
-\frac{1}{2}F_{\hat m}^{\phantom{\hat m}\!qr}\epsilon_{qrnp};~
-\frac{1}{g_s}H_{mnp}
$
\\[1ex]
D7    & \phantom{\vdots}\!$-$\!&\!$-$\!&\!$-$\!&\!$-$\!
&\!$-$\!&\!$-$\!&\!$-$\!&\!$-$\!&\!$\times$\!&\!$\times$\! &
$F_{\hat m \hat n}^{\phantom{\hat m \hat n}\!q}\,\epsilon_{qp}
+\frac{1}{g_s}\,H_{\hat m \hat n p}
$
\\[1ex]
D9    & \phantom{\vdots}\!$-$\!&\!$-$\!&\!$-$\!&\!$-$\!
&\!$-$\!&\!$-$\!&\!$-$\!&\!$-$\!&\!$-$\!&\!$-$\! &
$F_{\hat m\hat n \hat p}s
$
\\[1ex]
\hline\hline
\end{tabular}
\end{small}
\caption{Structure of the fermionic couplings $T$ induced by background fluxes on D3, D5, D7 and
D9-branes after taking into account the appropriate orientifold projections;
longitudinal internal directions are labeled by $\hat m, \hat n, \ldots$ and
internal transverse ones by $m, n, \ldots$~.}
 \label{Dbranes}
 \end{table}

These results clearly exhibit the fact that the R-R and NS-NS 3-form fluxes do
not appear on equal footing in the effective couplings $T$.  This is due to the
different ${\cal R}_0$ and $R_0$ reflection matrices
entering in the definition of the R-R and NS-NS vertex operators as discussed in Section \ref{sec:CFT}.
It is interesting to observe in Tab. \ref{Dbranes} that,
while for D9- and D5- branes the fermionic couplings depend either on $F$ or on $H$,
for D3- and D7-branes they depend on a combination of the R-R and NS-NS fluxes. This
follows from the fact that O3- and O7-planes act on the same way on R-R and NS-NS 3-forms.
By introducing the complex 3-form\footnote{Self-duality of type IIB can be used to
promote this expression to its $SL(2,\mathbb{Z})$-covariant version $G=F-\tau H$ with
$\tau=C_0-\ii \ee^{-\varphi}$. A direct evaluation of the $C_0$-dependent term however requires
a string amplitude involving two closed and two open string insertions in the disk. }
\begin{equation}
G=F-\frac{\ii}{g_s}\,H~,
\label{G}
\end{equation}
it is possible to rewrite the D3 brane coupling (\ref{td3}) as
\begin{equation}
T_{mnp}=(*_6 F)_{mnp} - \frac{1}{g_s}\,H_{mnp} =  \re\big( \!*_6 \!G-\ii G\big)_{mnp}~.
\label{tD31}
\end{equation}
Thus our explicit conformal field theory calculation confirms
that an imaginary self-dual (ISD) 3-form flux $G$ does not couple to
unmagnetized D3-branes, a well-known result that has been previously obtained using
purely supergravity methods
\cite{Grana:2002tu,Marolf:2003ye,Camara:2003ku,Camara:2004jj,Martucci:2005rb}.

Also the fermionic couplings (\ref{tD7a}) for the D7 branes can be written in terms of the
3-form flux $G$.
Indeed, introducing a complex notation and denoting as $i$ and $\overline i$
the complex directions of the plane transverse to the D7-branes
(sometimes in the literature also called D$7_i$-branes), we have
\begin{equation}
T_{\hat m \hat n i}= \ii \,G_{\hat m \hat n i}\quad\mbox{and}\quad
T_{\hat m \hat n \overline{i}}= -\ii \,G^*_{\hat m \hat n \overline{i}}~,
\label{tD71}
\end{equation}
in agreement with the structure of soft fermionic mass terms found
in Ref. \cite{Camara:2004jj}.

\subsection{Unmagnetized Euclidean branes}
\label{subsec:unmagnetizedE}

Euclidean branes that are transverse to the four-dimensional space-time and extend partially
or totally in the internal directions are relevant to discuss non-perturbative instanton effects
in the framework of branes models. In this case, to treat consistently the flux induced couplings
it is necessary to work in a space with Euclidean signature as we have done in
Section \ref{sec:CFT}. Then, the massless fermions $\Theta$ cannot satisfy a Majorana condition,
and relations like (\ref{cc}) do not hold any more. On the other hand, in Euclidean space
there is no issue about the reality of a fermionic amplitude and, as we will see, also the
coupling tensor $T$ is in general complex.

Let us begin by considering the D-instantons (or D$(-1)$-branes)
for which all ten directions are transverse. In this case we have
\begin{equation}
R_0=-1\quad\mbox{and}\quad {\mathcal R}_0=\Gamma^0\Gamma^1\cdots\Gamma^9 \equiv
\ii \,\Gamma_{(11)}^{\mathrm E}
\label{rd-1}
\end{equation}
where $\Gamma_{(11)}^{\mathrm E}$ is the chirality matrix in ten Euclidean dimensions. Thus,
recalling our chirality choice for the spinors $\Theta$, we easily see that for D-instantons
the $T$ tensor (\ref{tmnl}) is simply
\begin{equation}
T_{mnp}= -\ii\, F_{mnp} - \frac{1}{g_s}\,H_{mnp} = -\ii\,G_{mnp}~.
\label{tD-1}
\end{equation}

Let us now turn to Euclidean instantonic 5-branes (or E5-branes)
extending in the six internal directions.
In this case the reflection matrix  in the vector representation entering in
the fermionic coupling $T$ is $R_0|_{\rm int}=1$ along the internal
directions, while the matrix in the spinor representation is
\begin{equation}
\mathcal R_0 = \Gamma^0\Gamma^1\Gamma^2\Gamma^3
= -\ii \,\Gamma^4\cdots\Gamma^9\Gamma_{(11)}^{\mathrm E}~.
\label{re5}
\end{equation}
Therefore, for unmagnetized E5-branes we obtain from Eq. (\ref{tmnl})
\begin{equation}
T_{\hat m\hat n\hat p}= \ii\, (*_6 F)_{\hat m\hat n\hat p} + \frac{1}{g_s}\,H_{\hat m\hat n\hat p}
\label{tE5}
\end{equation}
where we have used the same index notation introduced in the previous subsection.
The above coupling simply reduces to
\begin{equation}
T_{\hat m\hat n\hat p}= \ii\, (*_6 F)_{\hat m\hat n\hat p}
\label{tE5a}
\end{equation}
in an orientifold model with O9-planes.

In the literature some attention has been devoted also to
Euclidean 3-branes (or E3-branes) extending along four of the six
internal directions \cite{Tripathy:2005hv,Bergshoeff:2005yp}.
These branes have some similarity with the D7-branes considered in
the previous subsection, and thus our discussion can follow the
same path. Using again the convention of splitting the internal
indices into longitudinal (hatted) and transverse (unhatted) ones,
we can show that the flux-induced fermionic couplings on E3-branes
are
\begin{equation}
T_{\hat m \hat n \hat p} = \frac{1}{g_s}\,H_{\hat m \hat n \hat p}~,\quad
T_{\hat m \hat n p} = -\frac{\ii}{2}\,\epsilon_{\hat m \hat n \hat r \hat s}\,
F^{\hat r \hat s}_{\phantom{\hat r \hat s}p}
+\frac{1}{g_s}\,H_{\hat m \hat n p}
~,\quad T_{\hat m np} = -\frac{1}{g_s}\,H_{\hat m np}~.
\label{tE3}
\end{equation}
If we consider the appropriate orientifold projections, which in this case remove both
$H_{\hat m \hat n \hat p}$ and $H_{\hat m np}$, we see that the only non-vanishing coupling is
\begin{equation}
T_{\hat m \hat n p} = -\frac{\ii}{2}\,\epsilon_{\hat m \hat n \hat r \hat s}\,
F^{\hat r \hat s}_{\phantom{\hat r \hat s}p}
+\frac{1}{g_s}\,H_{\hat m \hat n p}~.
\label{tE3a}
\end{equation}
This is in perfect agreement with the result of Refs. \cite{Tripathy:2005hv,Bergshoeff:2005yp}
that has been derived with pure supergravity methods.
To make the comparison easier, we observe that the E3-fermionic terms can be rewritten as
\begin{equation}
\Theta \Gamma^{\hat m \hat n p}\Theta\,T_{\hat m\hat n p} =
\Theta \Gamma^{\hat m \hat n p}\,{\widetilde G}_{\hat m\hat n p}\,\Theta
\label{aE3}
\end{equation}
where
\begin{equation}
{\widetilde G} = \frac{1}{g_s}\, H + \ii\,F \,\gamma_{(5)}
\label{e33}
\end{equation}
is the flux combination that is usually introduced in this case, with
$\gamma_{(5)}$ being the chirality matrix for the
four-dimensional brane world-volume. We further remark
that our general formula (\ref{ampltot1}) accounts for all flux-induced fermionic terms of the E3-brane
effective action discussed in Refs. \cite{Tripathy:2005hv,Bergshoeff:2005yp} including those
which break the Lorentz invariance in the first four directions.

For completeness we also mention that the fermionic couplings for the Euclidean 1-branes (or E1-branes)
are given by
\begin{equation}
T_{\hat m \hat n p}=\frac{1}{g_s}\,H_{\hat m \hat n p}~,\quad
T_{\hat m  n p} = -\ii\,\epsilon_{\hat m \hat q}\,F^{\hat q}_{\phantom{\hat q}np}
-\frac{1}{g_s}\,H_{\hat m n p}
~,\quad T_{mnp} = -\frac{1}{g_s}\,H_{mnp}~;
\label{tE1}
\end{equation}
note that $H_{\hat m n p}$ is removed by the orientifold projection
when the E1-branes are considered together with D5/D9-branes and the corresponding orientifold planes.
The structure of the various fermionic couplings for the instantonic branes discussed above is
summarized in Tab.~\ref{Ebranes}.
\begin{table}[ht]
\begin{small}
\centering
\begin{tabular}{cccccccccccc}
\hline\hline
  & \phantom{\vdots}\!0\!&\!1\!&\!2\!&\!3\!&\!4\!&\!5\!&\!6\!&\!7\!&\!8\!&\!9\! & $T_{mnp}$ \\ [1ex]
 \hline
D$(-1)$     & \phantom{\vdots}\!$\times$\!&\!$\times$\!&\!$\times$\!&\!$\times$\!
&\!$\times$\!&\!$\times$\!&\!$\times$\!&\!$\times$\!&\!$\times$\!&\!$\times$\! &
$-\ii F_{mnp} - \frac{1}{g_s}H_{mnp} $
\\ [1ex]
E1    & \phantom{\vdots}\!$\times$\!&\!$\times$\!&\!$\times$\!&\!$\times$\!
&\!$-$\!&\!$-$\!&\!$\times$\!&\!$\times$\!&\!$\times$\!&\!$\times$\! &
$\frac{1}{g_s}H_{\hat m \hat n p};
-\ii \epsilon_{\hat m \hat q}F^{\hat q}_{\phantom{\hat q}np};
-\frac{1}{g_s}H_{mnp}
$
\\ [1ex]
E3    & \phantom{\vdots}\!$\times$\!&\!$\times$\!&\!$\times$\!&\!$\times$\!
&\!$-$\!&\!$-$\!&\!$-$\!&\!$-$\!&\!$\times$\!&\!$\times$\! &
$-\frac{\ii}{2}\,\epsilon_{\hat m \hat n \hat r \hat s}\,
F^{\hat r \hat s}_{\phantom{\hat r \hat s}p}
+\frac{1}{g_s}\,H_{\hat m \hat n p}$
\\[1ex]
E5    & \phantom{\vdots}\!$\times$\!&\!$\times$\!&\!$\times$\!&\!$\times$\!
&\!$-$\!&\!$-$\!&\!$-$\!&\!$-$\!&\!$-$\!&\!$-$\! &
$\ii\,(*_6F)_{\hat m \hat n \hat p}$
\\[1ex]
\hline\hline
\end{tabular}
\end{small}
\caption{Structure of the fermionic couplings $T$ induced by background fluxes on
D$(-1)$, E1, E3 and E5 instantonic branes
after taking into account the appropriate orientifold projections;
the longitudinal internal directions are labeled by $\hat m, \hat n, \ldots$, the
internal transverse ones by $m, n, \ldots$ .}
\label{Ebranes}
\end{table}

We conclude our analysis by observing that in presence of E-branes,
the spacetime filling D$p$-branes live in the Euclidean
ten-dimensional space. Still, the couplings of such D$p$-branes are
again given by the same linear combinations of $F$ and $H$ like in
the Minkowskian case considered in last section, since $R_0$ and
${\cal R}_0$ are trivial along the would be time direction.

\subsection{Magnetized branes}
\label{subsec:magnetized}

The results of the previous subsections can be generalized in a rather straightforward way
to branes with a non-trivial magnetization on their world-volume for which
the longitudinal coordinates satisfy non-diagonal boundary conditions.
Indeed we can start from the same brane configurations we have analyzed before, introduce a
world-volume gauge field $A$ that couples to the open string end-points and obtain a
magnetization ${\mathcal F}_0={\mathcal F}_\pi= 2\pi\alpha'(dA)$. In this way we can use the
same R-R and NS-NS background fluxes of the previous subsections and simply study the new
couplings induced by the world-volume magnetization through the reflection matrices
$R_0$ and $\mathcal R_0$ given in Eqs. (\ref{Rot}) and (\ref{rspinor}).

As an example we briefly discuss the case of the magnetized E5 branes which play an important
r\^ole in the instanton calculus of the gauge theory engineered with wrapped D9-branes and O9-planes \cite{Billo:2007sw,Billo:2007py}. Adopting the same index notation as before, one can
easily realize that the spinor reflection matrix (\ref{rspinor}) for a magnetized E5 brane
can be written in the real basis as
\begin{equation}
\mathcal R_0= \Gamma^0\cdots\Gamma^3\,\mathcal U_0 = -\ii
\,\Gamma^4\cdots\Gamma^9\Gamma_{(11)}^{\mathrm E}\,\mathcal U_0
\label{rrm}
\end{equation}
where
\begin{equation}
\mathcal U_0 = \frac{1}{\sqrt{\det(1-{\mathcal F}_0)}}
~;\,\ee^{\frac{1}{2}({\mathcal F}_0)_{\hat m \hat n}
\Gamma^{\hat m\hat n}}\,;
\label{uu}
\end{equation}
in which the symbol $;\cdots ;$ means antisymmetrization on the vector indices of the $\Gamma$'s,
so that only a finite number of terms appear in the expansion of the exponential.
In our case we explicitly have
\begin{equation}
\begin{aligned}
;\,\ee^{\frac{1}{2}({\mathcal F}_0)_{\hat m \hat n}
\Gamma^{\hat m\hat n}}\,;~
= \,\,&1 \,+\,\frac{1}{2}\, ({\mathcal F}_0)_{\hat m \hat n}
\Gamma^{\hat m\hat n}
\,+\,\frac{{\ii}}{16} \,
({\mathcal F}_0)^{\hat m \hat n}({\mathcal F}_0)^{\hat p \hat q}
\epsilon_{\hat m \hat n \hat p \hat q \hat r \hat s}\,
\Gamma^{\hat r \hat s}\Gamma_{(7)}
\\
&
- \frac{{\rm i}}{3! \cdot 8} ({\mathcal F}_0)^{\hat m \hat n}
({\mathcal F}_0)^{\hat p \hat q}({\mathcal F}_0)^{\hat r \hat s}
\epsilon_{\hat m \hat n \hat p \hat q \hat r \hat s} \,\Gamma_{(7)}
\end{aligned}
\label{uu5}
\end{equation}
where $\Gamma_{(7)}=\ii\Gamma^4\ldots\Gamma^9$ is the chirality matrix of the E5-brane world volume.
Using this expression in Eq. (\ref{tmnl}) and focusing for
simplicity only on R-R fluxes since the NS-NS fluxes are anyhow removed by the orientifold
projection, after simple manipulations we find that the fermionic couplings of a magnetized E5-brane
are described by the tensor
\begin{equation}
 \begin{aligned}
T_{\hat m \hat n \hat p} =& \frac{1}{\sqrt{\det(1-{\mathcal F}_0)}}\,
\Big[\,\ii\,(*_6F)_{\hat m \hat n \hat p}
+ 3\ii\,(*_6F)_{\hat m \hat n}^{\phantom{mn} \hat q}\,({\mathcal F_0})_{\hat q \hat p}
\\
&+\frac{3\ii}{8}F_{\hat m \hat n}^{\phantom{mn}\hat q}
\,({\mathcal F_0})^{\hat r \hat s}({\mathcal F_0})^{\hat t \hat u}\,
\epsilon_{\hat q \hat r \hat s \hat t \hat u \hat p}
 - \frac{\ii}{3! \,8}\, F_{\hat m \hat n \hat p}
({\mathcal F_0})^{\hat q \hat r}({\mathcal F_0})^{\hat s \hat t}
({\mathcal F_0})^{\hat u \hat v}\,\epsilon_{\hat q \hat r \hat s \hat t \hat u \hat v}\Big].
\end{aligned}
\label{te5m}
\end{equation}
In the same way, and always starting from the general formula (\ref{tmnl}) one can discuss all other
types of magnetized branes.

\subsection{Flux-induced fermionic mass and lifting of instanton zero-modes}
\label{subsec:masses}

To complete the previous analysis we write the fermion bilinear $\Theta\Gamma^{mnp}\Theta$
using a four-dimensional spinor notation; in this way the structure of the flux-induced
fermionic masses will be more clearly exposed.  According to our $4+6$ splitting, the anti-chiral
ten dimensional spinor $\Theta_{\mathcal A}$ decomposes as
\begin{equation}
\Theta_{\mathcal A} ~\to~\big(\Theta^{\alpha A},\Theta_{\dot\alpha A}\big)
\label{Theta}
\end{equation}
where $\alpha$ ($\dot\alpha$) are chiral (anti-chiral) indices in four dimensions,
and the lower (upper) indices $A$ are chiral (anti-chiral) spinor indices of
the internal six dimensional space.
Furthermore, by decomposing the $\Gamma$ matrices according to
\begin{equation}
\Gamma^\mu= \gamma^\mu \otimes 1\quad,\quad\Gamma^m=\gamma_{(5)}\otimes\gamma^m~,
\label{gamma}
\end{equation}
one can show that
\begin{equation}
\Theta \Gamma^{mnp}\Theta = -\ii\,\Theta^{\alpha A} \Theta_{\alpha}^{{\phantom\alpha} B}
\big(\overline\Sigma^{mnp}\big)_{AB}
-\ii\,\Theta_{\dot\alpha A}
\Theta^{\dot\alpha}_{{\phantom\alpha} B}\big(\Sigma^{mnp}\big)^{AB}
\label{decomp}
\end{equation}
where $\Sigma^{mnp}$ and $\overline\Sigma^{mnp}$ are respectively the chiral and anti-chiral blocks
of $\gamma^{mnp}$ (see Appendix \ref{app:conventions} for details).
It is important to notice that
\begin{equation}
*_6\Sigma^{mnp} = -\ii\,\Sigma^{mnp}\quad,\quad
*_6\overline\Sigma^{mnp} = +\ii\,\overline\Sigma^{mnp}
\label{sigmaSD}
\end{equation}
so that $\Sigma^{mnp}$ only couples to an imaginary self-dual (ISD) tensor, while
$\overline{\Sigma}^{mnp}$ only couples to an imaginary anti-self dual (IASD)
tensor.
More explicitly, we have
\begin{equation}
\begin{aligned}
\Theta \Gamma^{mnp}\Theta\,T_{mnp} &= -\ii\,\Theta^{\alpha A}
\Theta_{\alpha}^{{\phantom\alpha} B}
\big(\overline\Sigma^{mnp}\big)_{AB}\,T_{mnp}^{\mathrm{IASD}}
-\ii\,\Theta_{\dot\alpha A} \Theta^{\dot\alpha}_{{\phantom\alpha}
B}\big(\Sigma^{mnp}\big)^{AB} \,T_{mnp}^{\mathrm{ISD}}\\
&= -\ii\,\Theta^{\alpha A} \Theta_{\alpha}^{{\phantom\alpha} B}
T_{AB} -\ii\,\Theta_{\dot\alpha A}
\Theta^{\dot\alpha}_{{\phantom\alpha} B} T^{AB}
\end{aligned}
\label{decomp1}
\end{equation}
where
\begin{equation}
T_{mnp}^{\mathrm{ISD}} =\frac{1}{2}\big(T-\ii*_6\!T\big)_{mnp}\quad,\quad
T_{mnp}^{\mathrm{IASD}} =\frac{1}{2}\big(T+\ii*_6\!T\big)_{mnp}~.
\label{isdiasd}
\end{equation}
In the second line of Eq. (\ref{decomp1}) we have adopted a $\mathrm{SU}(4)\sim \mathrm{SO}(6)$
notation and defined the IASD and ISD parts of the $T$-tensor as the following
$4\times 4$ symmetric matrices
\begin{equation}
T_{AB}=(\overline\Sigma^{mnp}\big)_{AB}\,T_{mnp}^{\mathrm{IASD}}\quad,\quad
T^{AB}=\big(\Sigma^{mnp}\big)^{AB} \,T_{mnp}^{\mathrm{ISD}}~,
\label{isdiasd1}
\end{equation}
with upper (lower) indices $A,B$ running
over the 
${\bf 4}$ (${\bf\bar 4}$) representations of $\mathrm{SU}(4)$.
Fixing a complex structure, the 3-form tensors
$T^{\mathrm{ISD}},T^{\mathrm{IASD}}$  can be decomposed into their
(3,0),(2,1),(1,2) and (0,3) parts as indicated in Tab. \ref{TIASD}.
The $(2,1)$ components are distinguished into six primitive ones (P),
satisfying $g^{j\bar k}T_{ij\bar k}=0$,
and three non-primitive ones (NP), satisfying $T_i=g^{j\bar k}T_{ij\bar k}$.
A similar decomposition holds for the $(1,2)$ part.
The various components transform in irreducible representations of
the $\mathrm{SU}(3)\in \mathrm{SU}(4)$ holonomy group under which the internal coordinates $Z^i,\bar
Z^i$ transform as ${\bf 3}$ and ${\bf \bar 3}$
respectively and spinors like ${\bf 4}={\bf 1}+{\bf 3}$ and
${ \bf \bar4}= { \bf \bar1}+{ \bf\bar 3}$.
The $\mathrm{SU}(3)$ content of the $T$-tensor is displayed in the last column in
Tab. \ref{TIASD}.
\begin{table}[ht]
\centering
\begin{tabular}{|ccc|}
\hline $\phantom{\vdots}T^{\mathrm{ISD}}$ & $\to$ & $T_{(0,3)}
\oplus T_{(1,2)_{\mathrm{NP}}}\oplus T_{(2,1)_{\mathrm{P}}}={\bf
\bar 1}
\oplus {\bf \bar 3} \oplus {\bf \bar 6}$\\
[1ex]
 \hline
$\phantom{\vdots}T^{\mathrm{IASD}}$ & $\to$  & $T_{(3,0)} \oplus
T_{(2,1)_{\mathrm{NP}}}\oplus T_{(1,2)_{\mathrm{P}}}={\bf 1} \oplus
{\bf 3} \oplus {\bf 6} $
\\[1ex]
\hline
\end{tabular}
 \caption{ Decomposition of the ISD and IASD parts of the 3-form $T$.
The $(2,1)$ and $(1,2)$ components are distinguished into primitive (P)
and non-primitive (NP) parts. The last column displays the $\mathrm{SU}(3)$
content of the various pieces.}
 \label{TIASD}
 \end{table}

Let us now use this information to rewrite the fermionic terms we have discussed in
the previous subsections, focusing in particular on D3-branes and D-instantons
on flat space.
In the case of D3-branes, we can use a Minkowski
signature and the Majorana-Weyl fermion $\Theta$ decomposes as in (\ref{decomp})
where the four-dimensional chiral and anti-chiral
components are related by charge conjugation and assembled into four Majorana spinors.
These are the four gauginos living on the world-volume of the D3-brane, and for future
notational convenience we will denote their chiral and anti-chiral parts
as $\Lambda^{\alpha A}$ and $\bar\Lambda_{\dot\alpha A}$ (instead of $\Theta^{\alpha A}$
and $\Theta_{\dot\alpha A}$).
Then, using Eqs. (\ref{tD31}) and (\ref{decomp}) in the general
expression (\ref{ampltot2}), we obtain
\begin{equation}
\mathcal A_{\mathrm D3} =
\frac{2\pi\ii}{3!}\,c_F\,{\mathrm{Tr}}\Big[\, \Lambda^{\alpha A}
\Lambda_{\alpha}^{{\phantom \alpha}B}
\big(\overline\Sigma^{mnp}\big)_{AB}\,G_{mnp}^{\mathrm{IASD}} -
\bar\Lambda_{\dot\alpha A}\bar\Lambda^{\dot\alpha}_{{\phantom
\alpha}B}
\big(\Sigma^{mnp}\big)^{AB}\,\big(G_{mnp}^{\mathrm{IASD}}\big)^*
\,\Big] \label{massD3}
\end{equation}
where we have made explicit the colour trace
generators%
\footnote{We use the following normalization:
${\mathrm{Tr}}\left(T^aT^b\right)=\frac{1}{2}\,\delta^{ab}$.}.

Recalling that the topological normalization of any disk
amplitude with D3-strings is \cite{Billo:2002hm}
\begin{equation}
\mathcal C_{(4)} = \frac{1}{\pi^2\,{\alpha'}^2\,g_{\mathrm{YM}}^2}~,
\label{c4}
\end{equation}
one can show that in order to obtain gauginos with canonical dimension of
(length)$^{-3/2}$ and standard kinetic term of the form
\begin{equation}
\frac{1}{g_{\mathrm{YM}}^2}\int d^4x \,{\mathrm{Tr}}\big(\!-2\ii\,
\bar\Lambda_{\dot\alpha A}\bar D\!\!\!\!/^{\,\dot\alpha \beta}
\Lambda_\beta^{\,A}\big)~,
\label{kinetic}
\end{equation}
one has to normalize the gaugino vertices with
\begin{equation}
\mathcal N_\Lambda= (2\pi\alpha')^{\frac{3}{4}}~.
\label{nlambda}
\end{equation}
Then, using these ingredients the prefactor appearing in \eq{massD3} becomes
\begin{equation}
c_F=\frac{4}{
g_{\mathrm{YM}}^2}\,(2\pi\alpha')^{-\frac{1}{2}}\,{\mathcal N}_F~.
\label{cfD3}
\end{equation}
{F}rom the explicit expression of the amplitude (\ref{massD3})
we see that an IASD $G$-flux configuration
induces a Majorana mass%
\footnote{Notice that this is the mass term for gauginos which are
not canonically normalized, as we do not rescale away the overall
factor of $1/g_{\mathrm{YM}}^2$ appearing in \eq{kinetic}.} for the
gauginos leading to supersymmetry breaking on the gauge theory
\cite{Grana:2002tu,Camara:2003ku,Grana:2003ek,Camara:2004jj}. Notice
that the mass terms for the two different chiralities are complex
conjugate of each other: $T^{\mathrm{IASD}}=-\ii\,
G^{\mathrm{IASD}}$ and $T^{\mathrm{ISD}}=\ii(G^{\mathrm{IASD}})^*$. This
is a consequence of the Majorana condition that the four-dimensional
spinors inherit from the Majorana-Weyl condition of the fermions in
the original ten-dimensional theory.

If we decompose $G^{\mathrm{IASD}}$ as indicated in Tab.
\ref{TIASD}, we see  that a $G$-flux of type $(1,2)_{\mathrm P}$
gives mass to the three gauginos transforming non-trivially under
$\mathrm{SU}(3)$ but keeps the $\mathrm{SU}(3)$-singlet gaugino massless, thus
preserving ${\mathcal N}=1$ supersymmetry. On the other hand, a $G$-flux of type $(3,0)$,
or $(2,1)_{\mathrm{NP}}$ gives mass also to the $\mathrm{SU}(3)$-singlet gaugino.

Things are rather different instead on D-instantons whose fermionic
coupling is given by Eq. (\ref{tD-1}). Indeed, by inserting such
coupling in Eq. (\ref{ampltot2}) and using again Eq. (\ref{decomp})
we obtain
\begin{equation}
\mathcal A_{{\mathrm D}(-1)} =
\frac{2\pi\ii}{3!}\,c_F(\Theta)\,\Big[\, \Theta^{\alpha A}
\Theta_{\alpha}^{{\phantom\alpha} B}
\big(\overline\Sigma^{mnp}\big)_{AB}\,G_{mnp}^{\mathrm{IASD}} +\bar
\Theta_{\dot\alpha A}\bar \Theta^{\dot\alpha}_{{\phantom \alpha}B}
\big(\Sigma^{mnp}\big)^{AB}\,G_{mnp}^{\mathrm{ISD}} \,\Big]
\label{massD-1}
\end{equation}
where now the prefactor $c_F(\Theta)$ contains the topological
normalization of the D$(-1)$ disks (the value of the gauge instanton
action ), namely \cite{Billo:2002hm}
\begin{equation}
\mathcal C_{(0)} = \frac{8\pi^2}{g_{\mathrm{YM}}^2} \quad
\Rightarrow \quad c_F(\Theta)=\frac{8\pi^2}{g_{\mathrm{YM}}^2}{\cal
N}_{\Theta}^2\,{\mathcal N}_F \label{c0}
\end{equation}
{}From the amplitude (\ref{massD-1}) we explicitly see that
both the IASD and the ISD components of the $G$-flux couple to the D-instanton fermions;
however the couplings are different and independent for the two chiralities since
they are not related by complex conjugation, as always in Euclidean spaces.
In particular, comparing Eqs. (\ref{massD3}) and
(\ref{massD-1}), we see that an ISD $G$-flux does not give a mass to any gauginos
but instead induces a ``mass'' term for the anti-chiral instanton zero-modes which are therefore
lifted. This effect plays a crucial r\^ole in discussing the non-perturbative contributions
of the so-called ``exotic'' D-instantons for which the neutral anti-chiral zero modes $\bar \Theta_{\dot\alpha A}$ must be removed \cite{Argurio:2007vqa,Bianchi:2007wy} or lifted by some mechanism \cite{Blumenhagen:2007bn,Petersson:2007sc}.
Introducing an ISD $G$-flux is one of such mechanisms as we
will discuss in more detail in Section \ref{sec:fD-1}.

\section{Flux couplings with twisted open strings ($\vec{\vartheta}\not =0$)}
\label{sec:twisted}

As we have emphasized, the general world-sheet calculation presented
in Section \ref{sec:CFT} allows to obtain the couplings between
closed string fluxes and open string fermions at a generic D-brane
intersection, even for non-vanishing twist parameters
$\vec{\vartheta}$. 
Here we just analyze a simple case
of such twisted amplitudes which will be relevant for the
applications discussed in Section \ref{sec:fD-1}.

The case we discuss is that of the 3-form flux couplings with the
twisted fermions stretching between a D3-brane and a D-instanton
which represent the charged (or flavored) fermionic moduli of the
${\cal N}=4$ ADHM construction of instantons (see for example Refs.
\cite{Dorey:2002ik,Billo:2002hm}) and are usually denoted as $\mu^A$
and $\bar\mu^A$ depending on the orientation. In the notation of
Section \ref{sec:CFT} the D3/D$(-1)$ and D$(-1)$/D3 strings are
characterized by twist vectors of the form
\begin{equation}
\vec{\vartheta} = \Big(0,0,0,+\frac{1}{2},+\frac{1}{2}\Big)
\quad\mbox{and}\quad
\vec{\vartheta}' = \Big(0,0,0,-\frac{1}{2},-\frac{1}{2}\Big)
\label{vartheta3-1}
\end{equation}
respectively,
and thus, according to \eq{theta0}, the open string fermions in these
sectors have weight vectors
\begin{equation}
\vec{\epsilon}_1 =\Big(\vec{\epsilon}_{A},-\frac{1}{2},-\frac{1}{2}\Big)
\quad\mbox{and}\quad
\vec{\epsilon}_4 =
\Big(\vec{\epsilon}_{A},+\frac{1}{2},+\frac{1}{2}\Big)~.
\label{epsilonmu}
\end{equation}
The notation $\vec{\epsilon}_{A}$ ($\vec{\epsilon}^{\,A}$) denotes
an anti-chiral (chiral) spinor weight of the internal
$\mathrm{SO}(6)$ rotation group. The vertex operators corresponding
to (\ref{epsilonmu}) (see \eq{vertexferm}) are then
\begin{equation}
V_\mu(z) =\mathcal{N}_{\mu}\,\mu^{A}
\big[\sigma_{\vec{\vartheta}}\,s_{ \vec{\epsilon}_A}\, \ee^{-
{\frac{1}{2}\phi}}\big](z) \quad\mbox{and}\quad
V_{\bar\mu}(z) = \mathcal{N}_{\bar\mu}\,{\bar\mu}^{A}
\big[\sigma_{\vec{\vartheta}'}\,s_{ \vec{\epsilon}_A}\, \ee^{-
{\frac{1}{2}\phi}}\big](z)
 \label{vertexmu}
\end{equation}
where $\mathcal{N}_{\mu},\mathcal{N}_{\bar\mu}$ are suitable
normalizations. Notice that since the last two components of
$(\vec{\epsilon}_1+\vec{\vartheta})$ and
$(\vec{\epsilon}_4+\vec{\vartheta}')$ are zero, only a spin-field
$s_{ \vec{\epsilon}_A}=S_A$ of the internal $\mathrm{SO}(6)$ appears
in the vertices (\ref{vertexmu}); furthermore there is no momentum
in any direction since $\mu^{A},\bar\mu^A$ are moduli rather than
dynamical fields. Note also that both $\mu^A$ and $\bar\mu^A$ carry the
same $\mathrm{SO}(6)$ chirality.

On the other hand, the vertex operator for a R-R field strength contains two
parts: one with left and right weights of the type
\begin{equation}
\vec{\epsilon}_2=\big(\vec{\epsilon}^{\,A},
\vec{\epsilon}^{\,\dot\alpha} \big) \quad\mbox{and}\quad
\vec{\epsilon}_3=\big(\vec{\epsilon}^{\,B},
\vec{\epsilon}^{\,\dot\beta} \big) ~,
\label{epsilon22}
\end{equation}
and one with weights of the type
\begin{equation}
\vec{\epsilon}_2=\big(\vec{\epsilon}_{A},
\vec{\epsilon}_\alpha \big) \quad\mbox{and}\quad
\vec{\epsilon}_3=\big(\vec{\epsilon}_{B}, \vec{\epsilon}_\beta
\big)~,
\label{epsilon21}
\end{equation}
where $\vec{\epsilon}_\alpha$ ($\vec{\epsilon}^{\,\dot\alpha}$) are
the chiral (anti-chiral) spinor weights%
\footnote{In our conventions, as explained in Appendix \ref{app:conventions},
$\alpha \in \{\frac12(++),\frac12(--)\}$
and  $\dot\alpha\in\{\frac12(-+),\frac12(+-)\}$.} 
of $\mathrm{SO}(4)$.
We now show that when the R-R field is an internal 3-form $F_{mnp}$ only the part in
(\ref{epsilon22}) couples to the $\mu$ and $\bar\mu$'s.

Let us consider the general R-R amplitude (\ref{amplFfinal}) 
and take, for example, the $\sigma=0$ boundary on the D$(-1)$-brane,
{\it i.e.} $R_0=-1$ and $\mathcal R_0=\ii\Gamma_{(11)}^{\mathrm E}$.
Let us then observe that the spinor reflection matrix can be effectively replaced by
$\mathcal R_0=-\ii$, since our GSO projection selects the anti-chiral sector, and that
the two $\vec{\vartheta}\cdot \vec{\epsilon}_3$-dependent integrals $I_1$ and $I_2$,
defined in (\ref{integrals}), are scalars along the six internal directions 
because the internal components of $\vec\vartheta$ are vanishing (see Eq. (\ref{vartheta3-1})).
All this implies that the
term with a single $\Gamma$ in (\ref{amplFfinal}) vanishes, so that the only
non-trivial contribution comes from the term with three $\Gamma$'s.

To proceed we need to evaluate the integral $I_2$. In the limit $s=-2t\to 0$, from
Eq. (\ref{integrals1}) we easily find
\begin{equation}
\big(I_2\big)_{\!\mathcal{A}_3}^{~\mathcal{A}_3} =\frac{1}{2
\vec{\vartheta}\cdot \vec{\epsilon}_3}~ \Big(1-\ee^{2\pi\ii
\vec{\vartheta}\cdot \vec{\epsilon}_3 }\Big)~;
\label{integralsi2}
\end{equation}
recall that $\mathcal{A}_3$ is the spinor index corresponding to the weight $\vec{\epsilon}_3$.
There are two distinct cases, corresponding to the two possibilities
(\ref{epsilon22}) and (\ref{epsilon21}) respectively. In the first case we have 
\begin{equation}
\vec\vartheta\cdot\vec\epsilon_3=0 
\quad\quad  \Rightarrow  \quad\quad \big(I_2\big)_{\!\mathcal{A}_3}^{~\mathcal{A}_3}=-\pi \ii~,
\label{case1}
\end{equation}
while in the second case we have
\begin{equation}
 \begin{aligned}
\vec\vartheta\cdot\vec\epsilon_3&=+\frac12\quad\mbox{if}\quad
 \vec{\epsilon}_\beta=\frac12\big(++\big)
\quad\quad  \Rightarrow  \quad\quad \big(I_2\big)_{\!\mathcal{A}_3}^{~\mathcal{A}_3}=+2~,\\
\vec\vartheta\cdot\vec\epsilon_3&=-\frac12\quad\mbox{if}\quad
 \vec{\epsilon}_\beta=\frac12\big(--\big)
\quad\quad  \Rightarrow  \quad\quad \big(I_2\big)_{\!\mathcal{A}_3}^{~\mathcal{A}_3}=-2~.
 \end{aligned}
\label{case2}
\end{equation}
Using the explicit expression of the $\Gamma$ matrices given in Appendix \ref{app:conventions},
it is not difficult to realize that the above results can be summarized by writing
\begin{equation}
 I_2=-\pi\ii\,\left(\frac{1+\Gamma_{(7)}}{2}\right)-2\ii\,\Gamma^{01}
\left(\frac{1-\Gamma_{(7)}}{2}\right)
\label{I2}
\end{equation}
where $\Gamma_{(7)}$ is the chirality matrix in the six-dimensional internal space.
Indeed, restricting to the anti-chiral block, one can check that the first
term in (\ref{I2}) accounts for the matrix elements (\ref{case1}), while the second
term for the matrix elements (\ref{case2}). At this point it is clear that 
with an internal R-R 3-form flux $F_{mnp}$, the coefficient $\big(F{\mathcal R}_0I_2\big)_{mnp}$
of the amplitude (\ref{amplFfinal}) can only receive contribution from the first
term in (\ref{I2}), which yields
\begin{equation}
 {\mathcal A}_F \sim 
{\bar\mu}^{A}\mu^{B}
\,\big(\overline\Sigma^{mnp}\big)_{AB}\,F_{mnp}^{\mathrm{IASD}}~.
\label{mumubF}
\end{equation}

The evaluation of coupling with an internal NS-NS 3-form flux $H_{mnp}$ is much simpler.
In fact the left and right weights appearing in the NS-NS vertex operator are
\begin{equation}
\vec{\epsilon}_2=\big(\pm \vec{e}_{m}\pm
\vec{e}_{n}, \vec 0 \big) \quad\mbox{and}\quad
\vec{\epsilon}_3=\big(\pm\vec{e}_{p}, \vec 0 \big)~,
\label{epsilon21nsns}
\end{equation}
with $\vec{e}_{m,n,p}$ unit vectors in the $SO(6)$ weight space
specifying the $H_{mnp}$-hyperplane. Thus, we always have $\vec\theta\cdot
\vec{\epsilon}_3=0$ which implies that the entries of the two diagonal
matrices $I_1$ and $I_2$ (with vector indices) are
\begin{equation}
 2\,\big(I_1\big)^P_{\,P}= \big(I_2\big)^P_{\,P} = -\pi\ii~.
\label{i1i2}
\end{equation}
Thus, from Eq. (\ref{amplHfinal}) we see that the term with a single $\Gamma$ vanishes, while the
term with three $\Gamma$'s yields
\begin{equation}
 {\mathcal A}_H \sim 
{\bar\mu}^{A}\mu^{B}
\,\big(\overline\Sigma^{mnp}\big)_{AB}\,H_{mnp}^{\mathrm{IASD}}~.
\label{mumubH}
\end{equation}

Collecting the two contributions (\ref{mumubF}) and (\ref{mumubH}) and reinstating the
appropriate normalizations, we finally obtain
\begin{equation}
{\mathcal A}_{\mathrm{D3/D(-1)}} \equiv {\mathcal A}_F +{\mathcal
A}_H =\frac{4\pi\ii}{3!}\,c_F(\mu)\,{\bar\mu}^{A}\mu^{B}
\,\big(\overline\Sigma^{mnp}\big)_{AB}\,G_{mnp}^{\mathrm{IASD}}
\label{mumubtot}
\end{equation}
where
$c_F(\mu)=\mathcal{C}_{(0)}\,\mathcal{N}_{\mu}\,\mathcal{N}_{\bar\mu}\,\mathcal{N}_F$
with $\mathcal C_{(0)}$ given in \eq{c0}. Notice that no symmetry factors has to
be included in this amplitude, since $\mu$ and $\bar\mu$ are really
distinct and independent quantities. This amplitude together with the one
in \eq{massD-1} accounts for the flux induced fermionic couplings on
the D-instanton effective action, and their meaning will be
discussed in Section \ref{sec:fD-1}.

\section{Flux couplings in an ${\mathcal N}=1$ orbifold set-up}
\label{sec:N1}
The results of the previous sections clearly show that internal
NS-NS and R-R fluxes bear important consequences on the brane
effective action and may be relevant in phenomenological
applications. Therefore it is particularly interesting to study
such flux interactions in models with $\mathcal{N}=1$
supersymmetry. To do so we adopt a toroidal orbifold
compactification scheme where string theory remains calculable and
the flux couplings are basically those described in Section
\ref{sec:effects}. We consider type IIB theory compactified on a
Calabi-Yau 3-fold with the $\mathcal{N}=2$ bulk supersymmetry
further broken down to $\mathcal{N}=1$ by the introduction of
D-branes and O-planes. To be specific we consider the orbifold
$\mathcal{T}_6/(\mathbb{Z}_2\times\mathbb{Z}_2)$ with $\mathcal{T}_6$ completely
factorized as a product of three 2-torii. 
Locally our system is undistinguishable from the theory living on
the non-compact orbifold $\mathbb{C}^3/(\mathbb{Z}_2\times\mathbb{Z}_2)$, described in \ref{secn:d3d-1}.

Let us briefly recall the structure
of the closed string multiplets  and the pattern of fractional
D-branes that can be introduced in this orbifold.

\paragraph{Closed and open string sectors in the $\mathbb Z_2\times\mathbb{Z}_2$ orbifold} Let us start by considering the oriented closed string states
before the introduction of O-planes. The massless closed string
states in the   orbifold organize into a gravity multiplet,
$h_{2,1}$ vector multiplets and $h_{1,1}+1$ hypermultiplets of the
${\cal N}=2$ supersymmetry. For strings defined on the quotient
space, the orbifold projection onto
$\mathbb{Z}_2\times\mathbb{Z}_2$ invariant states has to be
enforced and as usual we distinguish between untwisted and twisted
sectors.

The untwisted sector follows from that on $\mathcal{T}^6$ after
restricting to $\mathbb{Z}_2\times\mathbb{Z}_2$-invariant
components. It contains: the gravity multiplet; the universal
hypermultiplet having as bosonic components the dilaton $\varphi$,
the axion $C_0$ and the dualized NS-NS and R-R 2-forms $B_2$ and
$C_2$ with four dimensional indices;
$h^{\mathrm{untw}}_{21}=3$ vector multiplets with bosonic components $(V^i,u^i)$
where the scalars $u^i$ parametrize the complex structure
deformations; $h^{\mathrm{untw}}_{11}=3$ hypermultiplets containing
the scalars $(v_i,b_i,c_i,\tilde c^i)$ with $v_i$ representing the
K\"ahler parameter of the $i$-th torus, $b_i$ and $c_i$ the components of the
NS-NS and R-R 2-forms along the $i$-th torus, and $\tilde c^i$
the R-R 4-form components along the dual 4-cycle.

Closed strings on the $\mathbb{Z}_2\times\mathbb{Z}_2$ orbifold
have also twisted sectors associated to each of the three non
trivial elements $h_i$ and localized at the 16 possible fixed loci
of their action; the total number of twisted sectors is therefore
$3\times 16 = 48$. To fully specify the
orbifold model, we must declare the action of the group elements
$h_i$ also on the twist fields. There are two consistent
possibilities, which correspond to the singular limits of two
different CY manifolds with $(h_{11},h_{21})=(51,3)$ and
$(h_{11},h_{21})=(3,51)$. Here we restrict ourselves to the first
choice%
\footnote{The second choice corresponds to declare that the
twist field $\Delta_{ij}$ in the four-dimensional plane (ij)
transforms in the representation $|\epsilon_{ijk}|R_k$.}, 
corresponding to take all twisted fields invariant under $h_i$. With
this choice, the twisted sectors contribute to the massless
spectrum with $h^{\mathrm{tw}}_{11}=48$ hypermultiplets containing the
scalars $(v_{\hat i},b_{\hat i}, c_{\hat i},\tilde c^{\hat i})$,
$\hat i=1,...48$, which describe, respectively, the deformations of the blow-up modes
of the vanishing 2-cycles and the exceptional components of the NS-NS 2-form 
and of the R-R 2- and 4-forms. It is important to recall that
the orbifold limit is attained with a non-zero background value of
the NS-NS $B$-field on the vanishing cycles \cite{Aspinwall:1996mn},
so that the scalar fields $b_{\hat i}$ mentioned above represent fluctuations
around this value.

Finally the introduction of O-planes projects the spectrum onto the
subset of $\Omega I$-invariant states with $\Omega$ the worldsheet
parity and $I$ some involution of the CY threefold. The resulting
spectrum falls into vectors and chiral multiplets of the unbroken
${\cal N}=1$ supersymmetry. The details depend on the choice of
the O-planes. For example, for a vacuum built out of O3/O7-planes,
$B_2$ and $C_2$ are odd while for O5/O9 planes, $B_2$ and $C_4$
are odd. As a consequence, in the twisted sector either $b_{\hat i}$ and
$c_{\hat i}$, or $b_{\hat i}$ and $\tilde c^{\hat i}$ would be
projected out for the O3/O7 and O5/O9-choices respectively.

To discuss the couplings of fractional branes to closed strings,
it may be convenient to describe the branes by means of boundary
states \cite{DiVecchia:1999rh,DiVecchia:1999fx}, which we indicate
schematically as $\ket A$ for a brane of type $A$. It turns out 
(see for example Ref. \cite{Billo:2000yb}) that these boundary states 
$\ket A$ are suitable combinations of boundary
states $\dket{I}$ associated to the $h_I$-twisted sector, namely
\begin{equation}
 \label{ctoi}
\ket{A} = \frac{1}{4} \sum_I \,\ch^I_{A} \dket{I}~.
\end{equation}
with $\ch^I_{A}={\mathrm{tr}}_{R_A}(h_I)$. In our orbifold, using the
character table (\ref{frac3}) given in Appendix \ref{subapp:T6orb}, these sums explicitly read
\cite{Bertolini:2001gg}
\begin{equation}
 \label{fbbs}
\begin{aligned}
\ket{0} & = \frac 14\Big( \dket{0} + \dket{1} + \dket{2} +
\dket{3}\Big)\quad,
\quad \ket{1}  = \frac 14 \Big(\dket{0} + \dket{1} - \dket{2} - \dket{3}\Big)~,\\
\ket{2} & = \frac 14 \Big(\dket{0} - \dket{1} + \dket{2} -
\dket{3}\Big)\quad, \quad \ket{3} = \frac 14 \Big(\dket{0} -
\dket{1} - \dket{2} + \dket{3}\Big)~.
\end{aligned}
\end{equation}
These boundary states show that the fractional D3-branes
couple not only to twisted closed string fields but to
\emph{untwisted} ones as well, with a fractional tension and
a fractional charge given by $1/4$ of the ones of the regular branes (see \ref{tauA}).

As already discussed, the fractional branes corresponding to $R_A$ ($A\not= 0$) can also
be interpreted geometrically as D5-branes
suitably wrapped on exceptional%
\footnote{Since we focus on branes localized at the origin, for each
$A$ we consider only one of the 16 possible exceptional cycles
$e^{\hat A}$.} 2-cycles $e^{\hat A}$ in the blown-up space. To this
extent, the background value of the NS-NS 2-form $B_2$ in the orbifold limit plays a
crucial r\^ole in accounting for the untwisted couplings of the
branes. We will take advantage of this description in the remaining
of this section when we interpret the flux couplings computed in
Section \ref{sec:effects} in the effective low-energy supergravity
theory.

\subsection{Gauge kinetic functions and soft supersymmetry breaking on D3-branes}
\label{sec:n1int}

In presence of D-branes the $\mathcal{N}=2$ bulk supersymmetry of
the chosen compactification is reduced to a specific $\mathcal{N}=1$
slice depending on the boundary conditions imposed by the branes on
the spin fields, which are encoded  in the spinor reflection matrix
$\mathcal{R}_0$ of \eq{rspinor}. The supersymmetry left unbroken by
D-branes should be aligned to that preserved by O-planes and tadpole
conditions should be enforced.
As a consequence, the field content of the bulk theory is reorganized into $\mathcal N=1$
multiplets; in particular the compactification moduli, as well as
the dilaton and axion fields, are assembled into complex scalars
within suitable chiral superfields, which couple to the $\mathcal
N=1$ vector and chiral multiplets living on the D-branes.

The tree-level effective action on the D-branes can be obtained in the field theory
limit $\alpha'\to 0$ from disk diagrams and takes the standard
form of an $\mathcal{N}=1$ supersymmetric action
in which the couplings are actually functions of the moduli due to the
possible interactions with closed string fields. In particular, the
gauge Lagrangian depends on the bulk moduli $M$ via its ``gauge kinetic function''
$f(M)$ which encodes the information on the Yang-Mills coupling $g_{\mathrm{YM}}$ and
the $\theta$-angle $\theta_{\mathrm{YM}}$ according to
\begin{equation}
f(M)=\frac{\theta_{\mathrm{YM}}}{2\pi}\,+\,\ii\,\frac{4\pi}{g_{\mathrm{YM}}^2}~,
\label{gym}
\end{equation}
so that the quadratic part in the gauge field strengths reads
\begin{equation}
 \label{biwzf}
-\frac{1}{8\pi}\int d^4x\,\, \im f(M) \, \Tr \big(F_{\mu\nu} F^{\mu\nu}\big)
+ \frac{1}{8\pi}\int d^4x\,\,\re f(M)\,  \Tr \big(
F_{\mu\nu} {}^*\!F^{\mu\nu}\big)~.
\end{equation}
Actually, the residual supersymmetry implies that the gauge Lagrangian
takes the form
\begin{equation}
 \label{gkf}
-\frac{\ii}{8\pi}\int d^2\theta \,\, f\big(M(\theta)\big)\, \Tr \big(W^\alpha(\theta)
W_\alpha(\theta)\big) + \cc
\end{equation}
where $W^\alpha(\theta)$ is the $\mathcal{N}=1$ gauge superfield whose
lowest component is the gaugino $\Lambda^\alpha$, while
the moduli $M$ in the gauge kinetic function $f$ get promoted to chiral superfields $M(\theta)$.

This is very interesting in two respects. First, the determination
of the gauge kinetic functions for different types of branes
preserving the same $\mathcal{N}=1$ supersymmetry suggests a way
to assemble the bulk moduli and their superpartners into
$\mathcal{N}=1$ chiral multiplets. Second, the Lagrangian
(\ref{gkf}) contains a gaugino mass term, which arises whenever
the $\theta^2$ component of $f\big(M(\theta)\big)$ assumes a
non-zero vacuum expectation value. As we will see, such mass terms
can be related to the flux-induced fermionic couplings computed in
Section \ref{sec:effects} (see in particular \eq{massD3}). To
establish the precise relation we need to determine both the gauge
kinetic functions for the D-branes used to engineer the gauge
theory, and the appropriate complex combinations of the
compactification moduli $M$ that can be promoted to chiral
superfields. This is what we do in the following.

\paragraph{Gauge kinetic function}
Let us take a fractional D3-brane, say of type $A$. To deduce its gauge
kinetic function $f_A$ we have several possibilities. We can derive the
quadratic terms in the gauge fields of \eq{biwzf} from disk diagrams, with the
boundary attached to the brane and with two open string vertices for the gauge field
inserted on the boundary and closed string scalar vertices in the interior.
Alternatively, we can compute the coupling among closed strings and the boundary
state $\ket{A}_F$ representing the fractional D3-brane with a
constant magnetic field $F$ turned on in the world-volume and infer from it the gauge
kinetic function $f_A$ (see {\it e.g.} Ref. \cite{DiVecchia:2005vm}).
Finally, we can simply read off the coupling from the
Dirac-Born-Infeld (DBI) and Wess-Zumino (WZ) actions of the fractional D3-brane.

The last option is viable if one regards the fractional D3-branes of type $A$
as D5-branes wrapped%
\footnote{This extends to the case of the fractional brane of type
$0$ by interpreting it as a D5-wrapped on $\sum_{A=1}^3 {\hat
e}^A$ with negative orientation and with a suitable magnetic flux
turned on.} on the twisted 2-cycle $\hat e^A$, as recalled in
Section \ref{sec:N1}. In this case the DBI action with an
additional Wess-Zumino term for the D5-brane  (in the string
frame) is
\begin{equation}
 \label{bid5t}
S_{D3,A}=-T_5\int_{D3}\int_{{\hat e}^A} \ee^{-\varphi} \sqrt{-\det\left(G +
\mathcal{F}\right)}+T_5\int_{D3}\int_{{\hat e}^A} \sum_{n=0}^3
C_{2n}\, e^{ \mathcal F   } ~,
\end{equation}
where $\mathcal F= B_2+2\pi\alpha'F$, and $T_5 =
T_3/(4\pi^2\alpha')$ with $T_3 =(2\pi)^{-1} (2\pi\alpha')^{-2}$
being the D3-brane tension.
Expanding to quadratic order in F and
using the non-zero background value of $B_2$ along
the vanishing cycles \cite{Aspinwall:1996mn}
\begin{equation}
\int_{{\hat e}^A} B_2=\frac 14(4\pi^2\alpha')
\end{equation}
one finds
\begin{equation}
 \label{ymtermt}
S_{D3,A}=-\frac{1}{16\pi}\int_{D3} \,\ee^{-\varphi}
\, F_{\mu\nu} F^{\mu\nu}+ \frac{1}{16\pi}\int_{D3} \,C_0\,F_{\mu\nu} {}^*\!F^{\mu\nu}+\ldots  ~.
\end{equation}
Promoting these expressions to the non-abelian case, which results in an extra factor of
1/2 due to the normalization of the colour trace, and comparing
with \eq{biwzf}, we can read off that the gauge kinetic function for the fractional D3-brane
of type $A$ is
\begin{equation}
 \label{gkfA}
f_A(M) = \frac{\tau}4  ~,
\end{equation}
with
\begin{equation}
 \label{gkf3}
\tau \equiv C_0 + \ii\ee^{-\varphi}
\end{equation}
the axion-dilaton field.
Combining \eq{gkfA} with Eqs. (\ref{gym}) and (\ref{gkf3}) leads to
\begin{equation}
g_{\mathrm{YM}}^2=16\pi\ee^{\varphi}\quad\mbox{and}\quad\theta_{\mathrm{YM}}=\frac{\pi C_0}{2}~.
\label{gym1}
\end{equation}
The way to arrange the remaining untwisted and twisted scalars as
the complex bosons of suitable chiral multiplets is suggested, as
remarked above, by the gauge kinetic functions of other D-branes
maintaining the same $\mathcal{N}=1$ supersymmetry selected by the
fractional D3-branes. For the untwisted scalars, we can just
consider ``regular'' branes, such as D7-ones wrapped on one of the
untwisted 4-cycles. Starting from the wrapped D7-brane DBI-WZ
action, in the end one finds (see for instance Ref.
\cite{Blumenhagen:2006ci}) that the gauge kinetic function for
these branes is
\begin{equation}
 \label{gkf7}
f_i(M) = t^i\quad\mbox{with}\quad
t^i \equiv \tilde c^i + \frac{\ii}{2} |\epsilon^{ijk}| v_j v_k
\end{equation}
where $v_i$ and $\tilde c^i$ have been defined at the beginning of this section.
The complex fields $t^i$ represent
the correct (untwisted) K\"ahler coordinates to be used for the $\mathcal{N}=1$
supergravity associated to CY compactifications with D3/D7-branes and O3/O7-planes,
together with the $\tau$ variable defined in \eq{gkf3}.
Notice also that the imaginary parts
of the coordinates (\ref{gkf7}) are related to the volume $\mathcal{V}$ of
the $\mathcal{T}_6/\mathbb{Z}_2\times\mathbb{Z}_2$ orbifold, measured in the Einstein frame; in fact
\begin{equation}
 \label{vol}
\Big(\im t^1\, \im t^2\, \im t^3\Big)^\frac{1}{2} = v_1 v_2 v_3 =
\mathcal{V}~.
\end{equation}

Let us now return to the gauge theory defined on the fractional D3-branes and on its gauge kinetic function $f_A=\tau/4$. The modulus $\tau$ is connected by
the residual supercharges to other closed string states and it
can be promoted to a chiral superfield $\tau(\theta)$. 
The complete Lagrangian of the fractional D3-branes, given in \eq{gkf},
contains then also the coupling of the gaugino $\Lambda^\alpha$ to the
auxiliary component $F^\tau$ of $\tau(\theta)$, namely
\begin{equation}
\label{gaugaux}
-\,\frac{\ii}{8\pi}\,\frac{F^\tau}{4}\,\Tr \big(\Lambda^\alpha \Lambda_\alpha\big) + \cc ~,
\end{equation}
which corresponds to the following mass
\begin{equation}
\label{cangaugmass}
m_\Lambda = \frac{1}{2 \im f_A}\, \frac{F^\tau}4
= \frac{\ee^\varphi\,F^\tau}{2}
\end{equation}
for canonically normalized gaugino fields.

The bilinear term (\ref{gaugaux}) must coincide
with the flux-induced coupling we have computed in Section \ref{subsec:masses}. In fact,
in presence of a $G$-flux the gauginos acquire mass terms given
by \eq{massD3} which must be adapted to our $\mathcal N=1$ orbifold model. This is
easily done by taking only the invariant gaugino $\Lambda^0\equiv\Lambda$.
Using Appendix \ref{app:conventions} (and in particular \eq{sigma00}) we find that
the only component of the $G^{\mathrm{IASD}}_{mnp}$ tensor
which contributes to \eq{massD3} when $A=B=0$, is its $(3,0)$ part; thus, after combining
Eqs. (\ref{cfD3}) and (\ref{gym1}), we find that the flux-induced
gaugino mass term for fractional D3-branes reads
\begin{equation}
 \label{gmt}
-\frac{\ii}{2}\,\ee^{-\varphi}\,(2\pi\alpha')^{-\frac{1}{2}}\mathcal{N}_F\,
G_{(3,0)} \, \Tr \big(\Lambda^\alpha
\Lambda_\alpha\big) + \cc ~.
\end{equation}
Comparing with \eq{gaugaux}, we finally deduce that
\begin{equation}
 \label{Ftaubi}
F^\tau =  16\pi\, \ee^{-\varphi}\,(2\pi\alpha')^{-\frac{1}{2}}\mathcal{N}_F\,
G_{(3,0)} ~.
\end{equation}
Later in this section, we will fix the normalization $\mathcal{N}_F$ of the
flux vertex by requiring that this expression for $F^\tau$
matches the one obtained by constructing the bulk low energy Lagrangian.

\paragraph{Comparison with the bulk theory}
It is well known (see for example Refs.
\cite{Grana:2003ek,Grimm:2004uq}) that the bulk theory for our
toroidal compactification yields, after a dimensional reduction to four
dimensions and a Weyl rescaling to the $d=4$ Einstein frame, a
$\mathcal{N}=1$ supergravity theory coupled to vector and matter
multiplets in the standard form. This effective theory is
therefore specified, besides the gauge kinetic function for the
bulk vector multiplets, by the K\"ahler potential $K$ and by the
holomorphic superpotential $W$ for the chiral multiplets. To
simplify the treatment, in the following we consider as dynamical
only a subset of the compactification moduli; in particular we
keep the dependence on the universal chiral multiplet $\tau$ of
\eq{gkf3}, but restrict to a slice of the K\"ahler moduli space in
which an overall scale
\begin{equation}
 \label{kms}
t \equiv t^1 = t^2 = t^3
\end{equation}
is considered. Such a scale is related to the compactification
volume by $\mathcal{V} = (\im t)^{3/2}$ as it follows from \eq{vol}.
We also freeze the complex structure moduli $u^i$ to their
``trivial'' value corresponding to $\mathcal{T}_6$ being the product
of three upright tori, {\it {i.e.}} we set $u^1= u^2 = u^3 = \ii$;
furthermore we neglect the dependence on all the remaining twisted
and untwisted moduli.

With these assumptions, the K\"ahler potential for the bulk theory is
\begin{equation}
\label{KP}
 K =-\log(\im \tau) - 3 \log(\im t)~.
\end{equation}
When internal 3-form fluxes are turned on, a non-trivial bulk superpotential appears \cite{Gukov:1999ya,Taylor:1999ii} and its expression is
\begin{equation}
 \label{WGO}
W = \frac{1}{\kappa_{10}^2} \int G\wedge \Omega =
\frac{4}{\kappa_4^2} G_{(0,3)}
\end{equation}
where $\Omega$ is the holomorphic 3-form of the internal space, and $\kappa_{10}$ and $\kappa_4$
are, respectively, the gravitational constants in ten and four dimensions%
\footnote{In our case the holomorphic 3-form is simply given by
$\Omega= dZ^1\wedge dZ^2\wedge dZ^3$. In our conventions, we have
$\int \bar\Omega\wedge \Omega = (2\pi\sqrt{\alpha'})^6$, while $\kappa_{10}^2/\kappa_4^2 =
(2\pi\sqrt{\alpha'})^6/4$, where the factor of $1/4$ represents the order of the orbifold group.}.
In \eq{WGO} the 3-form flux is
\begin{equation}
G= F -\tau \,H
\label{G3}
\end{equation}
which is the natural extension of \eq{G} when $g_s$ is promoted to $\ee^\varphi$ and the
presence of a non-vanishing axion $C_0$ is taken into account.
Note that $W$ has the correct dimensions of (length)$^{-3}$, since $\kappa_4$ is a length and
the flux is a mass, and that only the ISD component $G_{(0,3)}$ of $G$ is responsible
for a non-vanishing $W$.

In presence of a superpotential $W$, the auxiliary fields in the chiral multiplets are given
by the standard supergravity expressions which in our case become
\begin{equation}
 \label{FW}
\begin{aligned}
{\overline F}^{\,\bar \tau} & = -\ii \,\kappa_4^2\,\ee^{K/2} \,K^{\bar\tau \tau}\, D_{\tau} W
\,=\, 8 \,\frac{\ee^{-\varphi/2}}{\mathcal{V}} \,\overline{G}_{(0,3)}\quad \Rightarrow \quad
{F}^{\tau} \,=\, 8 \,\frac{\ee^{-\varphi/2}}{\mathcal{V}} \,G_{(3,0)}~,\\
{\overline F}^{\,\bar t} & =- \ii\, \kappa_4^2\,\ee^{K/2}\, K^{\bar t t} \,D_{t} W
\,=\, 8\, \frac{\ee^{\varphi/2}}{\mathcal{V}^{\frac 13}}\, {G}_{(0,3)}
\quad \Rightarrow \quad
{F}^{t} \,=\,   8\, \frac{\ee^{\varphi/2}}{\mathcal{V}^{\frac 13}}\, \overline{G}_{(3,0)}    ~,
\end{aligned}
\end{equation}
where $\overline{G}$ is the complex conjugate of $G$,
$K^{\bar\tau \tau}$ and $K^{\bar t t}$ are the inverse K\"ahler metrics for $\tau$ and
$t$ respectively, and the K\"ahler covariant derivatives of the superpotential
are defined as $D_iW=\partial_iW+\big(\partial_iK\big)W$.
Thus, by comparing the expression of $F^\tau$ derived from the
flux-induced gaugino mass and given in \eq{Ftaubi} with \eq{FW}, we find perfect
agreement in the structure and can fix the normalization
of the closed string vertex for the flux to be
\begin{equation}
 \label{NFvalue}
\mathcal{N}_F = \frac{\ee^{\varphi/2}}{2\pi\mathcal{V}} \,(2\pi\alpha')^{\frac{1}{2}}~.
\end{equation}
{F}rom \eq{NfNh} we also infer that (promoting $g_s$ to $\ee^{\varphi}$)
\begin{equation}
 \label{NHvalue}
\mathcal{N}_H = \frac{\ee^{-\varphi/2}}{2\pi\mathcal{V}}\,(2\pi\alpha')^{\frac{1}{2}}~.
\end{equation}
With these normalizations, the closed string vertices (\ref{vertexRR})
and (\ref{vertexNS}) can be used to derive directly from string amplitudes  the
terms in the four dimensional effective Lagrangian in the Einstein frame, and the
resulting expressions do indeed have the correct normalization that follows from
the dimensional reduction of the original Type IIB action in ten dimensions.
In this perspective, we point out that
the scalar potential due to the chiral multiplets, which has the form
\begin{equation}
 \label{scalpot}
 \begin{aligned}
 V_F  &= \kappa_4^2\,\,\ee^K \left(K^{\tau\bar\tau} D_\tau W D_{\bar\tau}\bar
 W + K^{t\bar t} D_t W D_{\bar t}\overline W
 - 3\, |W|^2\right) \\
 &= \frac{16}{\kappa_4^2}\,\, \frac{\ee^\varphi}{\mathcal{V}^2} \,
 {G}_{(3,0)}\,\overline{G}_{(0,3)} = \frac{1}{\kappa_4^2}\,\,
 \Big|4\,\frac{\ee^{\varphi/2}}{\mathcal{V}} \,{G}_{(3,0)}\Big|^2 ~,
 \end{aligned}
\end{equation}
coincides with the kinetic terms for the R-R
and NS-NS 3-forms in the ten dimensional Einstein frame, given in  \eq{bulk10},
after dimensional reduction to $d=4$ and rescaling
to the four dimensional Einstein frame,
if only the (3,0) and (0,3) components of the fluxes
are turned on.

Let us also recall that the last term of $V_F$ in the first line of \eq{scalpot} is a purely ``gravitational'' contribution, related to the gravitino mass
\begin{equation}
 \label{gravmass}
m_{3/2} = \kappa_4^2\, \ee^{K/2} \,|W| =
\Big|4\,\frac{\ee^{\varphi/2}}{\mathcal{V}} \,G_{(0,3)}\Big|~.
\end{equation}
{F}rom these equations, we see clearly the very different r\^ole of the ISD flux $G_{(0,3)}$,
which induces a gravitino mass, with respect to the
IASD flux $G_{(3,0)}$, which is instead responsible for the gaugino mass term. The latter
is described by Eq.s (\ref{gaugaux}) and (\ref{FW}) which correspond, according to \eq{cangaugmass},
to a canonical gaugino mass
\begin{equation}
 \label{gmG}
m_\Lambda = \Big|4\,\frac{\ee^{\varphi/2}}{\mathcal{V}}\,G_{(3,0)}\Big|~.
\end{equation}
These very well-known results \cite{Camara:2003ku,Grana:2003ek,Camara:2004jj} will
be generalized and extended to instantonic branes in the following section, and the effects
on the instanton moduli space of a flux-induced mass term for the gaugino or the gravitino
will be determined. We conclude this section by mentioning that the same analysis we have
described for fractional D3 branes can be performed without any difficulty
in the case of fractional D9 branes.
Some details on this are provided in Appendix \ref{subsec:fD9}.

\section{The r\^ole of fluxes on fractional D-instantons}
\label{sec:fD-1} 
In Sections \ref{sec:effects} and \ref{sec:twisted}
we have computed the fermionic bilinear couplings of the NS-NS and
R-R bulk fluxes to open strings with at least one end-point on the
D-instanton. The results (\ref{massD-1}) and (\ref{mumubtot})
describe deformations of the instanton moduli space of the
${\cal N}=4$ gauge theory living on the D3-brane. We now discuss the
meaning of these interaction terms in a simple example within the
context of the $\mathcal N=1$ orbifold compactification introduced
in the last section.

Consider a specific node $A$ of the quiver diagram represented in
Fig. \ref{fig:quiver} and put on it $N$ fractional D3 branes and one
fractional D-instanton. The latter describes the $k=1$ gauge
instanton for the $\mathcal N=1$ $\mathrm U(N)$ Yang-Mills theory
defined on the world-volume of the space-filling D3-branes, as we already explained in Chapter \ref{inst}. 

The action of the $\mathcal N=1$ fractional D-instanton zero-modes turns out to be
(see {\it e.g.} Ref. \cite{Dorey:2002ik})
\begin{equation}
S_{\mathrm{inst}}=2\pi\ii\,f_A +\ii\,\lambda_{\dot\alpha}\big(
\bar\mu_u w^{\dot\alpha u} + \bar w^{\dot\alpha}_{\,\,u} \mu^u\big)
-\ii D_c\,\bar w_{\dot\alpha u}
(\tau^c)^{\dot\alpha}_{\,\,\dot\beta}w^{\dot\beta u} \label{sd-1}
\end{equation}
where $f_A=\tau/4$ is the gauge kinetic function (\ref{gkfA}) and
$\tau^c$ are the three Pauli matrices. Note that neither $x^\mu$ nor
$\theta^\alpha$ appear in $S_{\mathrm D(-1)}$; in fact they are the
Goldstone modes of the supertranslation symmetries broken by the
instanton and as such can be identified with the superspace
coordinates of the $\mathcal N=1$ theory. On the other hand
$\lambda_{\dot\alpha}$ and $D_c$ appear only linearly in
$S_{\mathrm D(-1)}$: they are Lagrange multipliers
enforcing the so-called super ADHM constraints. The action
(\ref{sd-1}) can be easily derived by computing (mixed) disk
amplitudes with insertions of vertex operators representing the
various zero-modes \cite{Billo:2002hm}.

Let us focus in particular on the fermionic moduli. The neutral zero-modes
$\theta^\alpha$ and $\lambda_{\dot\alpha}$ are clearly described by the chiral and anti-chiral
components of the D$(-1)$/D$(-1)$ fermion that is invariant under the orbifold
action ({\it
i.e.} with an internal spinor index 0), namely
\begin{equation}
 \Theta^{\alpha 0}\, \sim\, g_0\, \theta^{\alpha}\,\quad\mbox{and}\quad
\Theta_{\dot\alpha 0} \,\sim\, \lambda_{\dot\alpha}\,~.
\label{Thetalambda}
\end{equation}
The extra power of the D$(-1)$ gauge coupling
$g_0=\frac{1}{\sqrt\pi}(2\pi\alpha')^{-1}\ee^{\varphi/2}$
accounts for the correct scaling dimensions that allow to interpret $\theta^\alpha$ as
the fermionic superspace coordinate with dimensions of (length)$^{1/2}$. On the
other hand, as mentioned above, $\lambda_{\dot\alpha}$ is the Lagrange multiplier
for the fermionic ADHM constraint and carries dimensions of (length)$^{-{3}/{2}}$, so that
no rescaling is needed.

In the charged sector the fermionic moduli $\mu^u$ and $\bar\mu_u$
correspond to the $\mathbb{Z}_2\times\mathbb{Z}_2$ invariant fermions
of the strings stretching between the D3-branes and the D-instanton so that, using the
notation of Section \ref{sec:twisted}, for each colour we have
\begin{equation}
\mu^0\,\sim\,g_0\,\mu \quad\mbox{and}\quad\bar \mu^0\,\sim\,g_0\,\bar \mu~.
\label{mumubar}
\end{equation}
As before, an extra power of $g_0$ is included to account for the correct
(length)$^{{1}/{2}} $ dimensions of the charged moduli $\mu,\bar \mu$.
The normalizations%
\footnote{The extra factor of
$4\sqrt\pi\,\ee^{-\varphi/2}$ in the definition of $\mathcal{N}_\theta$ with
respect to \cite{Billo:2002hm} is needed, as we will see, in order
to identify $\theta$ with the superspace coordinates. } of the fermionic string vertices can then be
written as \cite{Billo:2002hm} 
\begin{equation}
{\mathcal N}_{\lambda} = {(2\pi\alpha')^{\frac{3}{4}}}~,\quad
{\mathcal N}_\theta =  4\sqrt\pi\,\ee^{-\varphi/2}\,\frac{g_0}{\sqrt{2}} \,(2\pi\alpha')^{\frac{3}{4}}~,\quad
 {\mathcal N}_\mu ={\mathcal N}_{\bar\mu} =
\frac{g_0}{\sqrt{2}}{(2\pi\alpha')^{\frac{3}{4}}} ~.
\label{nmu}
\end{equation}

We are now ready to study how the bulk R-R and NS-NS fluxes modify the moduli
action. Actually in Section \ref{sec:effects} we have already computed the flux interactions
with the untwisted fermions of a D$(-1)$-brane (see Eq. (\ref{massD-1})) while in
Section \ref{sec:twisted} we computed the flux couplings to the twisted fermions
of the D3/D$(-1)$ system (see Eq. \ref{mumubtot}). So what we have to do now is simply
to insert in these equations the appropriate normalizations discussed above and
take into account the identifications of the fluxes with the bulk chiral multiplets explained
in the previous section. The flux induced terms in the instanton moduli action are thus%
\footnote{Recall that in Euclidean spaces there is a minus sign in going from an amplitude to
an action.}
\begin{equation}
S_{\mathrm{inst}}^{\mathrm{flux}} = - \mathcal A^{\rm flux}_{{\mathrm D}(-1)} - \mathcal A^{\rm flux}_{\mathrm{D3/D(-1)}}
\label{sflux}
\end{equation}
where $\mathcal A^{\rm flux}_{{\mathrm D}(-1)}$ and $\mathcal A^{\rm flux}_{\mathrm{D3/D(-1)}} $
are the $A=B=0$ parts of the amplitudes (\ref{massD-1})
and (\ref{mumubtot}), {\it i.e.}
\begin{equation}
\begin{aligned}
\mathcal A^{\mathrm{flux}}_{{\mathrm D}(-1)}{\phantom{\vdots}} &= -2\pi\ii\,c_F(\theta)\,\theta^\alpha\theta_\alpha\,{G}_{(3,0)}
+ 2\pi\ii\,c_F(\lambda)\,\lambda_{\dot\alpha}\lambda^{\dot\alpha}\,{G}_{(0,3)}~,\\
\mathcal A^{\mathrm{flux}}_{\mathrm{D3/D(-1)}}{\phantom{\vdots}}
&=-4\pi\ii\,c_F(\mu)\,\bar\mu_u\,\mu^u\,{G}_{(3,0)}~.
\end{aligned}
\label{sflux01}
\end{equation}
In these expressions we have distinguished the $c_F$ coefficients
for the various terms to account for the appropriate normalizations
of the moduli as discussed before.
Recalling that the normalization
$\mathcal C_{(0)}$ of the disk amplitudes is given by Eq. (\ref{c0})
with $g_{\mathrm{YM}}^2$ defined in (\ref{gym1}), and that the $G$
fluxes are normalized as indicated in Eq. (\ref{NFvalue}), we find
\begin{equation}
\begin{aligned}
c_F(\theta)&= \mathcal C_{(0)}\,\mathcal N_F\,\mathcal N_\theta^2=
2\,\frac{\ee^{-\varphi/2}}{\mathcal{V}}~,\\
c_F(\lambda)&=  \mathcal C_{(0)}\,\mathcal
N_F\,\mathcal N_{\lambda}^2
= (2\pi\alpha')^{2}\,\frac{\ee^{-\varphi/2}}{4\mathcal{V}}~,\\
c_F(\mu)&= \mathcal C_{(0)}\,\mathcal N_F\,\mathcal N_\mu^2
= \frac{\ee^{\varphi/2}}{8\pi\mathcal{V}}~.
\end{aligned}
\label{cfs}
\end{equation}
Notice that all factors of $\alpha'$ cancel in $c_F(\theta)$ and
$c_F(\mu)$, but they survive in $c_F(\lambda)$ whose scaling
dimension of (length)$^4$ is the correct one for the $\lambda^2$
term of $\mathcal A^{\mathrm{flux}}_{{\mathrm D}(-1)}$ in (\ref{sflux01}) to be dimensionless.
Using these coefficients in (\ref{sflux01}) and exploiting the results of the previous section
(in particular Eqs. (\ref{gmG}) and (\ref{gravmass})), we can
rewrite the flux-induced moduli action as follows
\begin{equation}
\begin{aligned}
S_{\mathrm{inst}}^{\mathrm{flux}}\,& =\,4\pi\ii\frac{\ee^{-\varphi/2} G_{(3,0)}}{\mathcal V}\,\theta^\alpha\theta_\alpha
\,-\,\ii\pi\,(2\pi{\alpha'})^2\,\frac{\ee^{-\varphi/2} G_{(0,3)}}{2{\mathcal V}} \,\lambda_{\dot\alpha}\lambda^{\dot\alpha} \,+\,\ii\,
\frac{ G_{(3,0)}}{2{\mathcal V}} \,\bar\mu_u \mu^u\\
& =\,\frac{\ii\pi}{2}\,F^\tau\,\theta^\alpha\theta_\alpha
\,-\,\ii\pi\,(2\pi{\alpha'})^2\,\frac{\ee^{-\varphi/2} G_{(0,3)}}{2{\mathcal V}} \,\lambda_{\dot\alpha}\lambda^{\dot\alpha} \,+\,
\frac{\ii\,\ee^{\varphi}}{16} F^\tau \,\bar\mu_u \mu^u~.
\end{aligned}
\label{sflux02}
\end{equation}
The $\theta^2$ term represents the auxiliary component of the gauge
kinetic function $f_A=\tau/4$, which is therefore promoted to the
full chiral superfield $f_A(\theta)=\tau(\theta)/4$ in complete (and
expected) analogy with what happened on the D3-branes. The other two
terms are less obvious: they represent the explicit effects of a
background $G$ flux on the instanton moduli space, and are the
strict analogue for the instanton action of the soft supersymmetry
breaking terms of the gauge theory. In particular the $\bar\mu\mu$
term is related to the IASD flux component $G_{(3,0)}$ which is
responsible for the gaugino mass $m_\Lambda$, while the $\lambda^2$
term represents a truly stringy effect on the instanton moduli space
and is related to the ISD flux component $G_{(0,3)}$ which gives
rise to the gravitino mass $m_{3/2}$.

The study of these terms, of their consequences for the instanton
calculus and of the non-perturbative effects that they induce in
the gauge theory will be presented in the next Chapter. 

The above analysis can be easily generalized to
SQCD models with flavored matter in the fundamental
(anti-fundamental) representation and also to configurations in
which the fractional D-instanton occupies a node of the quiver
diagram which is {\em not} occupied by the colored or flavored
space-filling branes. For these ``exotic'' instanton configurations
there are no bosonic moduli $w_{\dot\alpha}$ and $\bar
w_{\dot\alpha}$, as we explained in Chapter \ref{inst}, and the action (\ref{sd-1}) simply reduces to the first
term involving the gauge kinetic function. Since the neutral
anti-chiral fermionic moduli $\lambda_{\dot\alpha}$ do not couple
to anything, to avoid a trivial vanishing result upon integration
over the moduli space, it is necessary to remove them
\cite{Argurio:2007vqa,Bianchi:2007wy} or to lift them
\cite{Blumenhagen:2007bn,Petersson:2007sc}. As we have explicitly seen, by coupling
the fractional D-instanton to an ISD $G$-flux of type (0,3) it is
possible to achieve this goal exploiting the $\lambda^2$ term
proportional to the gravitino mass.

Finally, we observe that using the explicit expression (\ref{tE3a}) of the
fermionic coupling, the flux-induced moduli amplitude on a E3 instanton
$\mathcal A^{\mathrm{flux}}_{{\mathrm E}3}$ contains terms of the form
\begin{equation}
\theta^\alpha\theta_\alpha\,{\bar G}_{(3,0)} \quad\mbox{and}\quad
\lambda_{\dot\alpha}\lambda^{\dot\alpha}\,{\bar G}_{(0,3)}
\label{AE3}
\end{equation}
where ${\bar G}_{(3,0)}$ and ${\bar G}_{(0,3)}$ are, respectively, the $(3,0)$ and
$(0,3)$ components of $\bar G= F +\frac{\ii}{g_s}\,H$. Thus, on a E3 instanton
a $G$-flux of type (2,1) or (0,3) cannot lift the $\lambda_{\dot\alpha}$'s since its
conjugate flux $\bar G$ does not contain a $(0,3)$-component, in full agreement with the
findings of Ref. \cite{Blumenhagen:2007bn}. However, as is clear from (\ref{AE3}),
a $G$-flux of type (3,0) can lift the anti-chiral zero-modes $\lambda_{\dot\alpha}$.

\chapter{Non-perturbative interactions}
\label{nonp}

In this Chapter we analyse the non-perturbative effects generated by fractional
D-instantons on the $\NN = 1$ SQCD theory realized with space-filling fractional D3 branes at $\mathcal{C}^3/(\mathbb{Z}_2\times\mathbb{Z}_2)$ singularity. In particular, we want to study the gauge instantons in presence or without background fluxes and how these fluxes affect the exotic instantons. 

\section{D-instanton partition function}
\label{sec:effint}

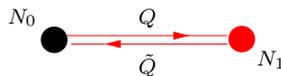
\begin{figure}[hbt]
\begin{center}
\begin{picture}(0,0)%
\includegraphics{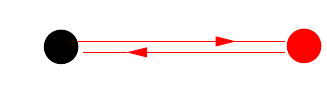}%
\end{picture}%
\setlength{\unitlength}{1381sp}%
\begingroup\makeatletter\ifx\SetFigFontNFSS\undefined%
\gdef\SetFigFontNFSS#1#2#3#4#5{%
  \reset@font\fontsize{#1}{#2pt}%
  \fontfamily{#3}\fontseries{#4}\fontshape{#5}%
  \selectfont}%
\fi\endgroup%
\begin{picture}(4455,1375)(361,-1097)
\put(2701,-61){\makebox(0,0)[lb]{\smash{{\SetFigFontNFSS{8}{9.6}{\familydefault}{\mddefault}{\updefault}$Q$}}}}
\put(2701,-961){\makebox(0,0)[lb]{\smash{{\SetFigFontNFSS{8}{9.6}{\familydefault}{\mddefault}{\updefault}${\tilde Q}$}}}}
\put(4801,-811){\makebox(0,0)[lb]{\smash{{\SetFigFontNFSS{8}{9.6}{\familydefault}{\mddefault}{\updefault}$N_1$}}}}
\put(376,-61){\makebox(0,0)[lb]{\smash{{\SetFigFontNFSS{8}{9.6}{\familydefault}{\mddefault}{\updefault}$N_0$}}}}
\end{picture}%
\end{center}
\caption{This simple quiver gauge theory is (from the point of view of one of the nodes) 
just $\mathcal{N}=1$ SQCD.}
\label{fig:quiver_SQCD}
\end{figure}

We study the simplest configuration that allows us to discuss both gauge and
stringy instanton effects (see Fig. \ref{fig:quiver_SQCD})
\begin{equation}
N_2=N_3=0\quad\mbox{with}\quad N_0~\mbox{and}~N_1~\mbox{arbitrary}~.
\label{NA}
\end{equation}

This brane system describes a $\mathcal N=1$ theory with gauge group 
$\mathrm U(N_0)\times \mathrm U(N_1)$ and a single bifundamental multiplet
\begin{equation}
 \label{Phimult}
\Phi^1(x,\theta)\equiv \Phi(x,\theta)=\phi(x)+\sqrt2 \theta \psi(x)+ \theta^2 F(x)~,
\end{equation}
which in block form is
\begin{equation}
\Phi=\left(
         \begin{array}{cc}
           0 & Q^{u}_{~ f} \\
           \widetilde Q^{f}_{~ u} & 0 \\
         \end{array}
       \right)
\label{Phi}
\end{equation}
with $u=1,\ldots N_0$ and $f=1,\ldots N_1$. The two off-diagonal blocks $Q$ and $\widetilde Q$ represent the quark and anti-quark superfields
which transform respectively in the fundamental and anti-fundamental of $\mathrm U(N_0)$, and in the anti-fundamental and fundamental of $\mathrm U(N_1)$. Both quarks and anti-quarks are neutral
under the diagonal $\mathrm U(1)$ factor of the gauge group, which decouples. 
On the other hand \cite{Intriligator:2005aw}
the relative $\mathrm U(1)$ group, under which both $Q$ and $\widetilde Q$ are charged, is IR free 
and thus at low energies the resulting effective gauge group
is $\mathrm {SU}(N_0)\times \mathrm {SU}(N_1)$. Therefore,
from the point of view of, say, the $\mathrm{SU}(N_0)$ factor
this theory is just $\mathcal N=1$ SQCD with $N_c=N_0$ colors
and $N_f=N_1$ flavors. In the following we will study
the non-perturbative properties of this theory in the
Higgs phase where the gauge invariance is completely broken by
giving (large) vacuum expectation values to the lowest components of the matter superfields.
This requires $N_f\geq N_c-1$. The moduli space of this SQCD is obtained by imposing
the D-flatness conditions. As remarked in \cite{Franco:2005zu}, even if the effective gauge
group is $\mathrm{SU}(N_c)$, we have to impose the D-term equations also for the (massive)
$\mathrm{U}(1)$ factors to obtain the correct moduli space of the quiver theory; 
in our case these D-term conditions lead to the constraint
\begin{equation}
Q\,\bar Q - \bar{\widetilde Q}\,\widetilde Q = \xi\,\one_{N_c\times N_c}
\label{D-flat1}
\end{equation}
where $\xi$ is a Fayet-Iliopoulos parameter related to twisted closed string fields which vanish
in the singular orbifold limit. For $N_f\geq N_c$ the D-term constraints allow for flat directions parameterized by meson fields
\begin{equation}
M^{f_1}_{~f_2} \equiv {\widetilde Q}^{f_1}_{~u} \,Q^{u}_{~f_2}
\label{meson}
\end{equation}
and baryon fields
\begin{equation}
B_{f_1\ldots f_{N_c}}=\epsilon_{u_1\ldots u_{N_c}}\, Q^{u_1}_{~f1} \ldots Q^{u_{N_c}}_{~\,f_{N_c}} 
\quad,\quad
{\widetilde B}^{f_1\ldots f_{N_c}}
=\epsilon^{u_1\ldots u_{N_c}}\, {\widetilde Q}^{f1}_{~u_1} \ldots 
{\widetilde Q}^{f_{N_c}}_{~\,u_{N_c}} 
\label{baryon}
\end{equation}
which are subject to constraints whose specific form depends on the difference $(N_f-N_c)$ (see for instance Ref. \cite{Intriligator:1995au}). These are the good observables 
of the low-energy theory in the Higgs phase.
For $N_f=N_c-1$, instead, the baryons cannot be formed and only the meson fields are present.

To have a quick understanding of the non-perturbative effects that
can be obtained in our stringy set-up, it is convenient to use dimensional analysis and exploit the
symmetries of the D3/D$(-1)$ brane system; we will see that besides the well-known
one-instanton effects like the ADS superpotential at $N_f=N_c-1$ \cite{Affleck:1983mk}, in the
quiver theory an infinite tower of multi-instanton corrections to the superpotential are 
in principle allowed. This is what we are going to show in the remainder of this section.
In Section \ref{secn:1inst} we specialize our discussion to the one-instanton sector and,
using again dimensional analysis and symmetry considerations, we analyze various types of non-perturbative effects in the low-energy theory, 
as a preparation for the study of the flux-induced terms presented in Section \ref{secn:fluxeffects}.

\subsection{The moduli space integral}
\label{integral}
The non-perturbative effects produced by a configuration of fractional D-instantons 
with numbers $k_A$ can be analyzed by studying the centered partition function
\begin{equation}
{W}_{\mathrm{n.p.}}  = \int d\,{\widehat{\mathfrak M}}\,\,\,\prod_{A=0}^3\!\Big(
M_s^{\,k_A \beta_A}\,\ee^{2\pi \ii k_A \tau_A }\,
\ee^{-\mathrm{Tr}_{k_A}[S_K+S_D+S_\phi]}\Big)~,
\label{Z}
\end{equation}
where the integration is over all moduli listed in (\ref{mod4}) except for the
center of mass supercoordinates $x^\mu$ and $\theta^\alpha$ defined in (\ref{xmu}) and
(\ref{theta}). These centered moduli are collectively denoted by $\widehat{\mathfrak M}$.
The action $\mathrm{Tr}_{k_A}[S_K+S_D+S_\phi]$ is obtained from (\ref{Sd1d3}) by restricting
the moduli to their $\mathbb Z_2 \times \mathbb Z_2$ invariant blocks for each $A$, while
the term $2\pi\ii k_A\tau_A$ represents the classical action of $k_A$ fractional
D-instantons of type $A$ (see Eqs. (\ref{wfluxnp}) and (\ref{tauA})).
Finally the power of the string scale $M_s$ compensates for the scaling dimensions
of the measure over the centered moduli space so that
the centered partition function ${W}_{\mathrm{n.p.}}$ has mass dimension 3,
as expected.
Indeed, using Tab. \ref{dimensions} one can easily show that the mass dimension $D$ of the instanton measure is
\begin{equation}
D\big[d\,{\widehat{\mathfrak M}}\big]= 3 -\sum_{A=0}^3 k_A\beta_A
\label{DdM}
\end{equation}
where $\beta_A$ 
is the one-loop $\beta$-function coefficient of the $\mathcal N=1$ $\mathrm{SU}(N_A)$ gauge theory
with $\sum_{I=1}^3 N_{A\otimes I}$ fundamentals and anti-fundamentals, namely
\begin{equation}
\beta_A = 3\, \ell(\mathrm{Adj}_A)- 2\sum_{I=1}^3 
N_{A \otimes I}\,\,\ell(N_A)
=3 N_A-\sum_{I=1}^3 N_{A\otimes I}
 \label{betagauge}
 \end{equation} 
where $\ell({r})$ denotes the index%
\footnote{The index $\ell(r)$ is defined by 
${\mathrm Tr}_{r}\big(T^a T^b\big)=\ell(r) \,\delta^{ab}$. 
For $\mathrm{SU}(N)$ gauge groups, the indices in the adjoint, fundamental, symmetric and 
antisymmetric represetations are, respectively, given by  $\ell(\mathrm{Adj})=N$, $\ell(N)=\frac12$,
$\ell\big(\frac12N(N+1)\big)=N+2$ and $\ell\big(\frac12N(N-1)\big)=N-2$.}
of the representation ${r}$.
It is interesting to remark that the explicit expression of $\beta_A$
is well-defined even in the case
$N_A=0$ where it cannot be interpreted as the $\beta$-function coefficient 
of any gauge theory. Keeping this in mind, all formulas in this section can be
applied to both gauge and stringy instanton configurations.

Coming back to the centered partition function (\ref{Z}) one can ask which dependence on
the scalar vacuum expectation values is generated by the integral over the instanton
moduli. A quick answer to this question follows by requiring that the form of
${W}_{\mathrm{n.p.}}$ be consistent with the symmetries of the D3/D$(-1)$ system.

The $(\mathbb Z_2 \times \mathbb Z_2)$ projected theory is  indeed invariant under the $\mathrm{U}(1)^3\subset \mathrm{SO}(6)$ global symmetries 
corresponding to the Cartan subgroup of the $\mathrm{SO}(6)$ ${\cal R}$-symmetry 
invariance of the ${\mathcal N}=4$ action (\ref{actgauge0}):
 \begin{equation}
\begin{aligned}
&\Phi^I\to \ee^{\ii\zeta_I}\, \Phi^I\quad,\quad V\to V\quad,\quad
W_\alpha \to \ee^{\frac{\ii}{2}\sum_I \zeta_I} \,W_\alpha\\
& d\theta\to  \ee^{-\frac{\ii}{2}\sum_I \zeta_I} \, d\theta\quad,
\quad d\bar \theta\to  \ee^{\frac{\ii}{2}\sum_I \zeta_I}\, d\bar \theta 
 ~;\end{aligned} 
\label{symgauge}
\end{equation}
these transformations encode the charges $q_I$  of the various fields w.r.t. to
the three $\mathrm{U}(1)$'s. The symmetry extends, 
as we will see, to the zero modes of the gauge fields in instantonic sectors, and can be exploited to constrain the form of the allowed non-perturbative interactions.
To this aim, it will prove useful to take linear combinations of the $\mathrm{U}(1)^3$ symmetries (\ref{symgauge}) corresponding to
introducing the charges
\begin{equation}
\label{new_charges}
q= q_1 + q_2 + q_3~, \quad\quad 
q'= q_1 - q_2 ~, \quad\quad 
q''= q_1 - q_3~. 
\end{equation}
The values of these charges for the various gauge fields are displayed in Table \ref{tableQG} while the charges of the moduli with respect to the same choice of $\mathrm{U}(1)^3$  made in Eq. (\ref{new_charges})
are given in Table \ref{tableu1}.

 \begin{table}[ht]
\centering
\begin{tabular}{c|cccc}
\hline\hline
     fields$\phantom{\vdots}$  & $A_\mu$ &$\Lambda^\alpha$ 
     & $\phi^I$& $\psi^{\alpha I}$ \\
  \hline
charge $\phantom{\vdots}q$  &$0$ & $+\frac32$   & $+1$ & $-\frac12$  \\
charge $\phantom{\vdots}q^\prime$  &$0$ &  $0$
 & $\delta_{I1}-\delta_{I2}$ & $\delta_{I1}-\delta_{I2}$\\
charge $\phantom{\vdots}q^{\prime\prime}$  &$0$ &$0$
 &$\delta_{I1}-\delta_{I3}$  & $\delta_{I1}-\delta_{I3}$   \\
\hline\hline
\end{tabular}
\caption{The charges $q$, $q^\prime$ and $q^{\prime\prime}$ of the various fields of the
${\mathcal N}=1$ quiver gauge theory.  Here $A_\mu$ is the gauge vector, $\Lambda^\alpha$ is the corresponding gaugino while $\psi^{\alpha I}$ is the fermion of the
bifundamental matter superfield $\Phi^I$ of which $\phi^I$ is the
lowest component. The complex conjugate fields $\bar \phi_I$,
$\bar \Lambda_{\dot\alpha}$ and $\bar\psi_{\dot\alpha I}$ transform oppositely to the ones displayed. Finally $\theta$ transforms as $\Lambda$ and oppositely to $\bar \theta$.}
\label{tableQG}
\end{table}

\begin{table}[ht]
\centering
\begin{tabular}{c|cccccc}
\hline\hline
     moduli$\phantom{\vdots}$  & $\chi^I$ & $\bar \chi_I$ &
     $\mu,\bar \mu,M^\alpha$ & $\mu^I,\bar \mu^I,M^{\alpha I}$  & $\lambda_{\dot\alpha}$
 &$\lambda_{\dot\alpha I}$ \\
  \hline
charge $\phantom{\vdots}q$  & $+1$ & $-1$ & $+\frac32$ & $-\frac12$ & $-\frac32$ & $+\frac12$ \\
charge $\phantom{\vdots}q^\prime$   &  $\delta_{I1}-\delta_{I2}$
&  $\delta_{I2}-\delta_{I1}$  & $0$ & $\delta_{I1}-\delta_{I2}$
&  $0$ & $\delta_{I2}-\delta_{I1}$\\
charge $\phantom{\vdots}q^{\prime\prime}$ &$\delta_{I1}-\delta_{I3}$
& $\delta_{I3}-\delta_{I1} $& $0$ & $\delta_{I1}-\delta_{I3}$ &
$0$ & $\delta_{I3}-\delta_{I1}$ \\
\hline\hline
\end{tabular}
\caption{The charges $q$, $q^\prime$ and $q^{\prime\prime}$ of the various fields of the
D3/D$(-1)$ brane system. The bosonic moduli  $a^\mu$, $w_{\dot\alpha}$,
$\bar w_{\dot\alpha}$ and $D_c$ are neutral under all three $\mathrm U(1)$'s.}
\label{tableu1}
\end{table}

These are symmetries of the D3/D$(-1)$ action but not of the instanton measure. Indeed, since
there are unpaired moduli, like $\mu^A$ and $\bar\mu^A$, which transform in the same way under $U(1)^3$,
the charges of the centered instanton measure, and hence of ${W}_{\mathrm{n.p.}}$, 
are non-trivial. In particular, the charge $q$ is
 \begin{equation}
q\big[d\,{\widehat{\mathfrak M}}\big]=-2n_\mu q(\mu)-2n_{\mu^I} q(\mu^I)-2 q(\lambda)
\label{condq}
\end{equation}
where $n_\mu$ and $n_{\mu^I}$ are the numbers of $\mu$ and $\mu^I$ moduli, and
the factors of $2$ come from the identical contributions of $\mu^A$ and $\bar \mu^A$ and 
from the two components of the anti-chiral fermion
\begin{equation}
 \lambda_{\dot\alpha}
 \equiv \frac{1}{k}\,\sum_{A=0}^3\,\sum_{{i_A}=1}^{k_A} \big\{\lambda_{\dot\alpha}\big\}^{i_A}_{~i_A}
 \label{lambda}
 \end{equation} 
which are unpaired since their partners, namely the fermionic superspace coordinates $\theta^\alpha$ defined in (\ref{theta}), have been taken out from the centered measure
$d\,{\widehat{\mathfrak M}}$. The minus signs in (\ref{condq}) come from the fact that a
fermionic differential transforms oppositely to the field itself.
Using the charges listed in Tab. \ref{tableu1}, it is easy to rewrite (\ref{condq}) as
\begin{equation}
q\big[d\,{\widehat{\mathfrak M}}\big]= 3-\sum_{A=0}^3k_A\Big(3
N_A-\sum_{I=1}^3 N_{A\otimes I}\Big)=3-\sum_{A=0}^3 k_A \beta_A~.
\label{qdm}
\end{equation}
In a similar way one finds
\begin{equation}
\begin{aligned}
q^\prime\big[d\,{\widehat{\mathfrak M}}\big] &=-2n_\mu q^\prime(\mu)
-2n_{\mu^I} q^\prime(\mu^I)-2 q^\prime(\lambda)=-2 \sum_{A=0}^3k_A \big(N_{A\otimes 1}-
N_{A\otimes 2}\big)
~,\\
q^{\prime\prime}\big[d\,{\widehat{\mathfrak M}}\big] &= -2n_\mu q^{\prime\prime}(\mu)-2n_{\mu^I} q^{\prime\prime}(\mu^I)-2 q^{\prime\prime}(\lambda)=-2 \sum_{A=0}^3k_A \big(N_{A\otimes 1}
-N_{A\otimes 3}\big)~.
\end{aligned}
\label{qdm1}
\end{equation}
One can check that the $\mathrm{U}(1)^3$ charges of the ADHM measure coincide
with the ones of the moduli space of instanton zero-modes. In fact, since in an instanton 
background the bosonic zero-modes always come together with their complex conjugates, 
the charges $q$, $q^\prime$ and $q^{\prime\prime}$ of the instanton measure depend
only on the number of fermionic zero-modes, namely on the number $n_{\Lambda}$ of gaugino zero-modes, 
and on the number $n_{\psi^I}$ of zero-modes of the fundamental matter fields%
\footnote{Remember that in an instanton background $\bar\Lambda=\bar\psi_I=0$}.
These numbers are given by the index of the Dirac operator evaluated,
respectively, in the adjoint and fundamental $\mathrm{SU}(N_A)$ representations 
under which the fields transform, {\it i.e.}
\begin{equation}
\begin{aligned}
n_{\Lambda} &=2 k_A \,\ell(\mathrm{Adj}_A)=2 k_A \,N_A~,\\
n_{\psi^I} &= 2 k_A\,\left(2 N_{A\otimes I}\right)\,\ell(N_A)=2 k_A \,N_{A\otimes I}~.
\label{indexgauge}
 \end{aligned}
\end{equation}
Taking into account the contribution of the two fermionic superspace coordinates
$\theta^\alpha$ to the charges of the instanton measure and using the values reported
in Tab. \ref{tableQG}, we have
\begin{equation}
\begin{aligned}
q&= -n_{\Lambda}\,q(\Lambda) - \sum_{I=1}^3 n_{\psi^I}\,q(\psi^I) - 2q(\theta) 
= -\frac32\,n_{\Lambda}
+\frac12\,\sum_{I=1}^3n_{\psi^I}+3~,\\
q^\prime&= -n_{\Lambda}\,q^\prime(\Lambda) - \sum_{I=1}^3 n_{\psi^I}\,q^\prime(\psi^I) - 2q^\prime(\theta)=
-n_{\psi^1}+n_{\psi^2}~,\\
q^{\prime\prime}&= - n_{\Lambda}\,q^{\prime\prime}(\Lambda) - \sum_{I=1}^3 n_{\psi^I}\,q(\psi^I) 
- 2q^{\prime\prime}(\theta)= -n_{\psi^1}+n_{\psi^3}~.
 \end{aligned}
\label{charges00}
\end{equation}
Exploiting (\ref{indexgauge}), it is immediate to see that these charges coincide with the
ones given in (\ref{qdm}) and (\ref{qdm1}) and computed using the ADHM construction.

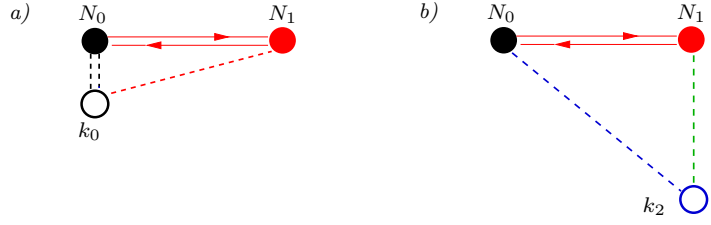
\begin{figure}[hbt]
\begin{center}
\begin{picture}(0,0)%
\includegraphics{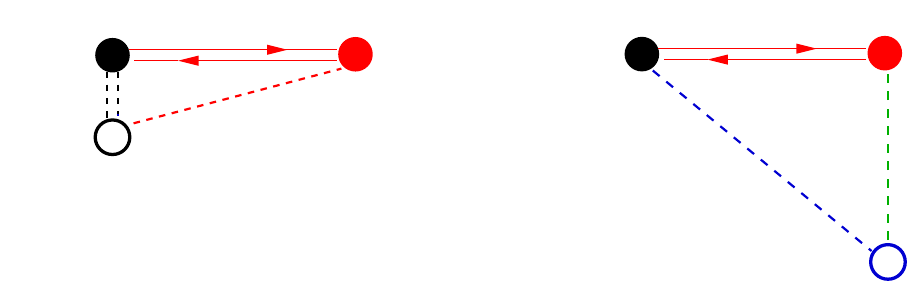}%
\end{picture}%
\setlength{\unitlength}{1381sp}%
\begingroup\makeatletter\ifx\SetFigFontNFSS\undefined%
\gdef\SetFigFontNFSS#1#2#3#4#5{%
  \reset@font\fontsize{#1}{#2pt}%
  \fontfamily{#3}\fontseries{#4}\fontshape{#5}%
  \selectfont}%
\fi\endgroup%
\begin{picture}(12447,3931)(-344,-3557)
\put(6946, 59){\makebox(0,0)[lb]{\smash{{\SetFigFontNFSS{8}{9.6}{\familydefault}{\mddefault}{\updefault}\emph{b)}}}}}
\put(901,-2086){\makebox(0,0)[lb]{\smash{{\SetFigFontNFSS{8}{9.6}{\familydefault}{\mddefault}{\updefault}$k_0$}}}}
\put(-329, 29){\makebox(0,0)[lb]{\smash{{\SetFigFontNFSS{8}{9.6}{\familydefault}{\mddefault}{\updefault}\emph{a)}}}}}
\put(901, 14){\makebox(0,0)[lb]{\smash{{\SetFigFontNFSS{8}{9.6}{\familydefault}{\mddefault}{\updefault}$N_0$}}}}
\put(4276, 14){\makebox(0,0)[lb]{\smash{{\SetFigFontNFSS{8}{9.6}{\familydefault}{\mddefault}{\updefault}$N_1$}}}}
\put(10936,-3421){\makebox(0,0)[lb]{\smash{{\SetFigFontNFSS{8}{9.6}{\familydefault}{\mddefault}{\updefault}$k_2$}}}}
\put(8161, 29){\makebox(0,0)[lb]{\smash{{\SetFigFontNFSS{8}{9.6}{\familydefault}{\mddefault}{\updefault}$N_0$}}}}
\put(11536, 29){\makebox(0,0)[lb]{\smash{{\SetFigFontNFSS{8}{9.6}{\familydefault}{\mddefault}{\updefault}$N_1$}}}}
\end{picture}%
\end{center}
\caption{D3/D$(-1)$-quiver for SQCD with \emph{a)} gauge instantons and \emph{b)} stringy instantons. 
Filled and empty circles represent stacks of D3 and D$(-1)$ branes,
solid lines stand for chiral bifundamental matter, dashed lines for charged instanton moduli.
A single dashed line represents the fermions $\mu$, while a double dashed line is a $(\mu,w)$ pair. }
\label{fk0k2}
\end{figure}

So far we have considered a generic D-instanton configuration. From now on we will focus on 
two cases, namely gauge and stringy instantons, which correspond to
the following choices of $k_A$'s
\begin{equation}
\begin{aligned}
{\mathrm{gauge}~:~~~} & \big(k_0,k_1,0,0\big)
~,\\
{\mathrm{stringy}~:~~~} & \big(0,0,k_2,k_3\big)~.
\end{aligned}
\label{gagstr}
\end{equation}
Some gauge and stringy instanton quiver diagrams are displayed in Fig. \ref{fk0k2}.

As noticed above, for stringy instantons we have $k_A N_A=0$ and therefore the only charged
ADHM moduli that survive are $\mu^I$ and $\bar\mu^I$, while $w$,
$\bar w$, $\mu$ and $\bar\mu$ are absent. As a consequence, the
fermion $\lambda$ of Eq. (\ref{lambda}) decouples from the moduli action and in the
centered partition function of stringy instantons there is an unbalanced fermionic zero-mode
integration. Therefore, unless such zero-modes are removed, for example with an orientifold projection
\cite{Argurio:2007vqa,Bianchi:2007wy}, or lifted with some mechanism \cite{Petersson:2007sc,GarciaEtxebarria:2008pi}, one gets a vanishing result.
We will return to the stringy instanton configurations in Section \ref{secn:fluxeffects}
where we discuss how bulk fluxes can cure this problem. In the remaining part of this section
we instead analyze in more detail the instanton partition function for a generic configuration
of gauge instantons.

\subsection{Zero-mode counting for gauge instantons}
\label{subsec:zeromode}

For gauge instantons the centered partition function (\ref{Z}) 
is a function of $\phi$ and $\bar\phi$, {\it{i.e.}} of the vacuum
expectation values of the matter superfield (\ref{Phi}) and its conjugate, 
with scaling dimension 3.
The most general ansatz for ${W}_{\mathrm{n.p.}}$ is therefore
\begin{equation}
{W}_{\mathrm{n.p.}}=  \mathcal C
\,M_s^{(k_0\beta_0+k_1\beta_1)}\,\ee^{2\pi\ii(k_0\tau_0+k_1\tau_1)}\,\,
\bar\phi^{n}\,{\phi}^{m}
\label{Wn}
\end{equation}
with $n+m+k_0\beta_0+k_1\beta_1=3$ and $\mathcal C$ a numerical constant.
Using Tab. \ref{tableQG}, it is easy to see that the $\mathrm U(1)^3$ charges of 
this expression are all equal and given by
\begin{equation}
q\big[{W}_{\mathrm{n.p.}}\big] =
q^\prime\big[{W}_{\mathrm{n.p.}}\big] =
q^{\prime\prime}\big[{W}_{\mathrm{n.p.}}\big] =3-2n -k_0\beta_0 -k_1\beta_1~.
\label{qw}
\end{equation}
We must require that these charges match those of the centered measure,
which, as follows from (\ref{qdm}) and (\ref{qdm1}), in this case are given by
\begin{equation}
q\big[d\,{\widehat{\mathfrak M}}\big] = 3-k_0\beta_0 -k_1\beta_1\quad,\quad
q^\prime\big[d\,{\widehat{\mathfrak M}}\big] = q^{\prime\prime}\big[d\,{\widehat{\mathfrak M}}\big]
=-2k_0N_1-2k_1N_0 ~.
\label{qqq}
\end{equation}
Then we immediately find that $n=0$ and
\begin{equation}
(k_0-k_1)(N_0-N_1)=1 ~.
\end{equation}
This equation is solved by
\begin{equation}
k_1=k_0-1 \quad,\quad N_1=N_0-1~,
\label{sol1}
\end{equation}
or by
\begin{equation}
k_0=k_1-1 \quad,\quad N_0=N_1-1~.
\label{sol2}
\end{equation}
Thus the partition function generated by gauge instantons is, as expected, an holomorphic function
of $\phi$ and is given by
\begin{equation}
{W}_{\mathrm{n.p.}}=  \mathcal C
\,M_s^{(k_0\beta_0+k_1\beta_1)}\,\ee^{2\pi\ii(k_0\tau_0+k_1\tau_1)}\,\,
\phi^{(3-k_0\beta_0-k_1\beta_1)}~.
\label{Wn1}
\end{equation}
Notice that this is a formal expression in which $\phi$ stands for the vacuum expectation
values of either the quark and anti-quark superfields $Q$ and $\widetilde Q$ defined in
(\ref{Phi}), and actually only the appropriate gauge invariant combinations of these should
appear in the final result. The solutions above with $(k_0=1,k_1=0)$ and $(k_0=0,k_1=1)$ 
reproduce the well-known ADS superpotential \cite{Affleck:1983mk} 
for $\mathrm{SU}(N_0)$ and $\mathrm{SU}(N_1)$ SQCD's
respectively. The multi-instanton corrections with $k_0,k_1 > 0$ are instead a distinct feature of
the quiver gauge theory we have engineered with the fractional D3 branes.
We conclude by observing that we could have arrived at the same results without
referring to the charges of the ADHM moduli but using instead those of the gauge field
zero-modes in the instanton background. 

It is possible to generalize
the previous analysis of the instanton partition function by including a dependence on the
entire matter fields and not only on their vacuum expectation values $\phi$ and $\bar\phi$.
This leads to a very rich structure of non-perturbative interactions that include the
holomorphic ADS superpotential when $N_f=N_c-1$ and the multi-fermion F-terms of the
Beasley-Witten (BW) type \cite{Beasley:2004ys} when $N_f\geq N_c$, plus their possible
multi-instanton extensions. In the following sections we will analyze in detail such
non-perturbative effective interactions in the one-instanton case.

\section{Effective interactions from gauge instantons}
\label{secn:1inst}
In this section we discuss the non-perturbative effective interactions induced by instantons using the explicit string construction of the ADHM moduli provided by fractional D3 and D$(-1)$ branes,
and show that in the field theory limit $\alpha'\to 0$ we recover the known 
non-perturbative F-terms, such as the ADS superpotential \cite{Affleck:1983mk} 
and the BW multi-fermion couplings \cite{Beasley:2004ys}.
{From} now on we will consider one-instanton effects in the quiver gauge theory
corresponding to a D3-brane system with $N_2=N_3=0$; this should not be regarded as a limitation of
our procedure but only a choice made for the sake of simplicity.

\subsection{The gauge instanton action}
\label{subsec:instact}
To discuss the D-instanton induced effective action on the D3 brane volume in the Higgs branch,
we first have to generalize the results of Section \ref{d3d01}
and introduce in the moduli action a
dependence on the entire matter superfields and not only on their vacuum expectation values.
As discussed in detail in \cite{Green:2000ke,Billo:2002hm,Billo:2006jm}, the couplings of the matter
fields with the ADHM instanton moduli can be obtained by computing mixed disk diagrams
with insertions of vertex operators for dynamical 3/3 strings on the portion of the boundary attached to the D3-branes.
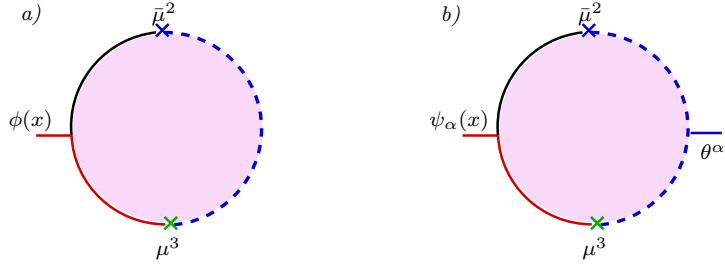
\begin{figure}[htb]
 \begin{center}
 \begin{picture}(0,0)%
\includegraphics{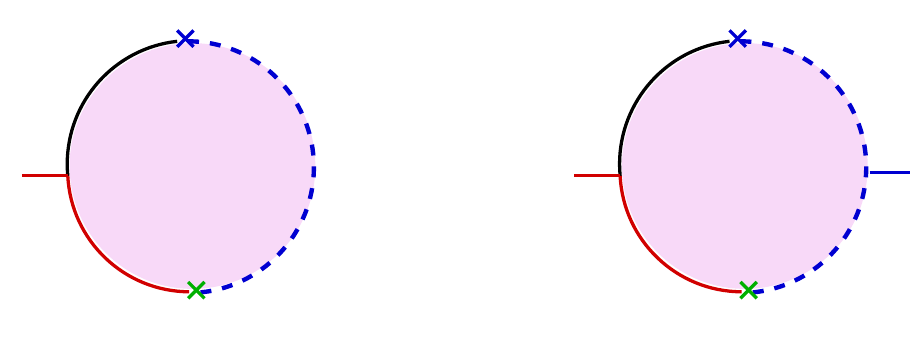}%
\end{picture}%
\setlength{\unitlength}{1381sp}%
\begingroup\makeatletter\ifx\SetFigFontNFSS\undefined%
\gdef\SetFigFontNFSS#1#2#3#4#5{%
  \reset@font\fontsize{#1}{#2pt}%
  \fontfamily{#3}\fontseries{#4}\fontshape{#5}%
  \selectfont}%
\fi\endgroup%
\begin{picture}(12513,4627)(496,-3845)
\put(311,-1426){\makebox(0,0)[lb]{\smash{{\SetFigFontNFSS{8}{9.6}{\familydefault}{\mddefault}{\updefault}$\phi(x)$}}}}
\put(2926,-3736){\makebox(0,0)[lb]{\smash{{\SetFigFontNFSS{8}{9.6}{\familydefault}{\mddefault}{\updefault}$\mu^3$}}}}
\put(2851,464){\makebox(0,0)[lb]{\smash{{\SetFigFontNFSS{8}{9.6}{\familydefault}{\mddefault}{\updefault}${\bar\mu}^2$}}}}
\put(511,479){\makebox(0,0)[lb]{\smash{{\SetFigFontNFSS{8}{9.6}{\familydefault}{\mddefault}{\updefault}\emph{a)}}}}}
\put(7981,464){\makebox(0,0)[lb]{\smash{{\SetFigFontNFSS{8}{9.6}{\familydefault}{\mddefault}{\updefault}\emph{b)}}}}}
\put(10426,464){\makebox(0,0)[lb]{\smash{{\SetFigFontNFSS{8}{9.6}{\familydefault}{\mddefault}{\updefault}${\bar\mu}^2$}}}}
\put(7786,-1426){\makebox(0,0)[lb]{\smash{{\SetFigFontNFSS{8}{9.6}{\familydefault}{\mddefault}{\updefault}$\psi_\alpha(x)$}}}}
\put(10501,-3736){\makebox(0,0)[lb]{\smash{{\SetFigFontNFSS{8}{9.6}{\familydefault}{\mddefault}{\updefault}$\mu^3$}}}}
\put(12601,-2011){\makebox(0,0)[lb]{\smash{{\SetFigFontNFSS{8}{9.6}{\familydefault}{\mddefault}{\updefault}$\theta^\alpha$}}}}
\end{picture}%
 \end{center}
\caption{Disk diagrams leading to the interaction between the scalar $\phi(x)$, or its
superpartner $\psi_\alpha(x)$, and the fermionic instanton moduli $\mu^3$ and $\bar\mu^2$.
}
\label{fig:phimu2mu3}
\end{figure}

An example of a coupling of $\phi(x)$ with the fermionic moduli $\mu$ and $\bar\mu$ is
provided by the diagram of Fig. \ref{fig:phimu2mu3}\emph{a)}, whose 
explicit evaluation leads to
\begin{equation}
 \frac{\ii}{2}\,\bar\mu^2\,\phi(x)\,\mu^3~.
\label{phimumu}
\end{equation}
If $\phi(x)$ is frozen to its vacuum expectation value, this coupling precisely accounts for one
of the last terms of $S_\phi$ given in (\ref{Sphi1}), once we specify our D3 brane configuration%
\footnote{As is clear from Fig. \ref{fig:phimu2mu3}, this contribution is present only when the
instanton is of stringy nature, {\it i.e.} $k_2=1$ or $k_3=1$, since the D$(-1)$ boundary must be of a different type with respect to the D3 boundaries. This type of contributions will play a crucial r\^ole in Section \ref{secn:strinst}, but
we discuss it here to illustrate in a simple example how the couplings with the holomorphic matter superfields
can be obtained.}.
Another possible diagram,
represented in Fig. \ref{fig:phimu2mu3}\emph{b)}, gives rise to the following coupling:
\begin{equation}
-\frac{\ii}{\sqrt2}\,\theta^\alpha\bar\mu^2\,\psi_\alpha(x)\mu^3~.
\label{psimumu}
\end{equation}
As discussed in \cite{Green:2000ke,Billo:2002hm,Billo:2006jm}, diagrams like those in Fig.
\ref{fig:phimu2mu3}\emph{a)} and \emph{b)} are related to each other by the action
of the two supersymmetries of the D-instanton broken by the D3 branes.
A further application of these supersymmetries leads to
\begin{equation}
 \frac{\ii}{2}\,\theta^2\,\bar\mu^2\,F(x)\,\mu^3~
\label{Fmumu}
\end{equation}
where $F$ is the auxiliary field of the matter multiplet. Also this coupling arises from a mixed disk diagram with two $\theta$-insertions on the D(--1) boundary and one insertion of $F$ on the
D3 boundary \footnote{For details on the calculations of disk amplitudes involving auxiliary fields
in this orbifold model see for example \cite{Billo:2005jw}.}. 
Adding the contributions (\ref{phimumu}), (\ref{psimumu}) and (\ref{Fmumu}), 
we reconstruct the combination
\begin{equation}
 \phi(x) + \sqrt2 \theta^\alpha\psi_\alpha(x) + \theta^2\,F(x)
\label{phipsi}
\end{equation}
which is the component expansion of the matter chiral 
superfield $\Phi(x,\theta)$ of our SQCD model.
Proceeding systematically in this way, one can show that the same pattern appears everywhere,
so that we can simply promote the vacuum expectation value $\phi$ to the complete superfield $\Phi(x,\theta)$, {\it i.e.} 
perform in the action (\ref{Sphi1}) the following replacement:
\begin{equation}
\phi~\to~\Phi(x,\theta)~
\label{promo0}
\end{equation}
in order to obtain all instanton couplings with the holomorphic scalar and its superpartners.
\begin{figure}[htb]
 \begin{center}
 \begin{picture}(0,0)%
\includegraphics{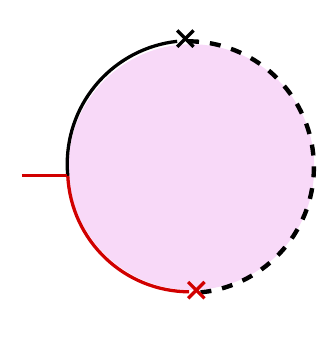}%
\end{picture}%
\setlength{\unitlength}{1381sp}%
\begingroup\makeatletter\ifx\SetFigFontNFSS\undefined%
\gdef\SetFigFontNFSS#1#2#3#4#5{%
  \reset@font\fontsize{#1}{#2pt}%
  \fontfamily{#3}\fontseries{#4}\fontshape{#5}%
  \selectfont}%
\fi\endgroup%
\begin{picture}(4344,4626)(496,-3859)
\put(2851,464){\makebox(0,0)[lb]{\smash{{\SetFigFontNFSS{8}{9.6}{\familydefault}{\mddefault}{\updefault}$\bar\mu$}}}}
\put(511,-1426){\makebox(0,0)[lb]{\smash{{\SetFigFontNFSS{8}{9.6}{\familydefault}{\mddefault}{\updefault}$\bar \phi$}}}}
\put(2926,-3736){\makebox(0,0)[lb]{\smash{{\SetFigFontNFSS{8}{9.6}{\familydefault}{\mddefault}{\updefault}$\mu^1$}}}}
\end{picture}%
 \end{center}
\caption{An example of a disk interaction between the (anti-holomorphic) scalar $\bar\phi(x)$ and the instanton moduli leading to the coupling (\ref{barphimumu}).}
\label{fig:psimumu1}
\end{figure}
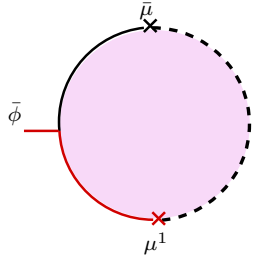 

Let us now turn to the anti-holomorphic variables. An example of a mixed disk
amplitude involving the scalar $\bar\phi(x)$ is represented in 
Fig. \ref{fig:psimumu1}.
It accounts for the coupling
\begin{equation}
-\frac{\ii}{2}\,\bar\mu\,\bar\phi(x)\,\mu^1~,
\label{barphimumu}
\end{equation}
which is the obvious generalization of the first term in the second line of
(\ref{Sphi1}) when the anti-holomorphic vacuum expectation value $\bar\phi$ is promoted
to a dynamical field.
The same pattern occurs in all
terms involving the anti-holomorphic vacuum expectation values $\bar\phi$, so that 
we can promote the latter with the replacement
\begin{equation}
\bar\phi~\to~\bar\phi(x) =\bar\Phi(x,\bar\theta)\Big|_{\bar\theta=0}~.
\label{promo00}
\end{equation}
Notice that no $\bar\theta$ dependence arises, due to the half-BPS nature of the D3/D$(-1)$ system.

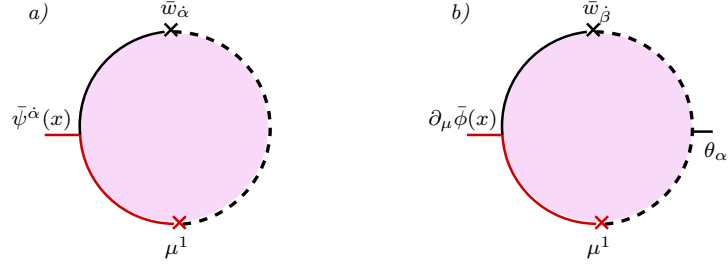
\begin{figure}[hbt]
 \begin{center}
  \begin{picture}(0,0)%
\includegraphics{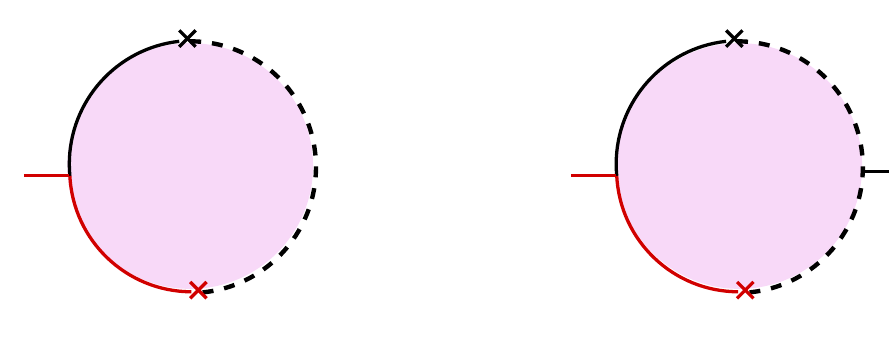}%
\end{picture}%
\setlength{\unitlength}{1381sp}%
\begingroup\makeatletter\ifx\SetFigFontNFSS\undefined%
\gdef\SetFigFontNFSS#1#2#3#4#5{%
  \reset@font\fontsize{#1}{#2pt}%
  \fontfamily{#3}\fontseries{#4}\fontshape{#5}%
  \selectfont}%
\fi\endgroup%
\begin{picture}(12228,4653)(466,-3859)
\put(10351,564){\makebox(0,0)[lb]{\smash{{\SetFigFontNFSS{8}{9.6}{\familydefault}{\mddefault}{\updefault}$\wdb$}}}}
\put(7611,-1426){\makebox(0,0)[lb]{\smash{{\SetFigFontNFSS{8}{9.6}{\familydefault}{\mddefault}{\updefault}$\partial_\mu\bar\phi(x)$}}}}
\put(10426,-3736){\makebox(0,0)[lb]{\smash{{\SetFigFontNFSS{8}{9.6}{\familydefault}{\mddefault}{\updefault}$\mu^1$}}}}
\put(481,479){\makebox(0,0)[lb]{\smash{{\SetFigFontNFSS{8}{9.6}{\familydefault}{\mddefault}{\updefault}\emph{a)}}}}}
\put(7996,464){\makebox(0,0)[lb]{\smash{{\SetFigFontNFSS{8}{9.6}{\familydefault}{\mddefault}{\updefault}\emph{b)}}}}}
\put(12511,-1996){\makebox(0,0)[lb]{\smash{{\SetFigFontNFSS{8}{9.6}{\familydefault}{\mddefault}{\updefault}$\ta$}}}}
\put(2851,564){\makebox(0,0)[lb]{\smash{{\SetFigFontNFSS{8}{9.6}{\familydefault}{\mddefault}{\updefault}$\wda$}}}}
\put(211,-1426){\makebox(0,0)[lb]{\smash{{\SetFigFontNFSS{8}{9.6}{\familydefault}{\mddefault}{\updefault}$\bar\psi^{\dot\alpha}(x)$}}}}
\put(2926,-3736){\makebox(0,0)[lb]{\smash{{\SetFigFontNFSS{8}{9.6}{\familydefault}{\mddefault}{\updefault}$\mu^1$}}}}
\end{picture}%
 \end{center}
\caption{Examples of disk diagrams responsible for the coupling of the superfield $D_{\dot\alpha}\bar\Phi(x,\bar\theta)\Big|_{\bar\theta=0}$ to the instanton moduli.
In \emph{b)} the vertex for the scalar $\bar\phi$ 
is in the 0-th superghost picture, which leads to a derivative coupling.}
\label{fig:psiwmu1}
\end{figure}

When one considers dynamical gauge fields, there
are new types of mixed disk amplitudes 
that correspond to couplings which do not depend on the vacuum expectation values of the scalars; 
since they are not present in the action (\ref{Sphi1}), they cannot be obtained with the
replacements (\ref{promo0}) and (\ref{promo00}).
These new types of interactions typically 
involve the D3/D3 anti-chiral fermions $\bar\psi_{\dot\alpha}(x)$ and
correspond to the following couplings:
\begin{equation}
\ii\,\bar w_{\dot \alpha}\bar\psi^{\,\dot\alpha}\!(x)\,\mu^1
-\ii\,\bar\mu^1\,\bar\psi_{\dot\alpha}(x)\,w^{\dot\alpha}~.
\label{wpsimu}
\end{equation}
Fig. \ref{fig:psiwmu1}\emph{a)} represents the disk diagram corresponding to the first term
of (\ref{wpsimu}). 
Furthermore, using the D3 supersymmetries that are broken by
the D$(-1)$ branes, we can produce the following terms:
\begin{equation}
-\theta_{\alpha}\bar w_{\dot \beta}(\bar\sigma^\mu)^{\alpha\dot\beta}
\partial_\mu\bar\phi(x)\,\mu^1
+\theta^{\alpha}\bar\mu^1\,(\bar\sigma^\mu)_{\alpha\dot\beta}
\partial_\mu\bar\phi(x)\,w^{\dot\beta}~.
\label{wpsimu2}
\end{equation}
The diagram responsible for the first term in (\ref{wpsimu2}) is represented in Fig. \ref{fig:psiwmu1}\emph{b)}.
The couplings (\ref{wpsimu2}) can be obtained from (\ref{wpsimu}) 
by means of the replacement
\begin{equation}
\bar\psi_{\dot\alpha}(x) ~\to~ \bar D_{\dot\alpha}\bar\Phi(x,\bar\theta)\Big|_{\bar\theta=0}~,
\label{promo2}
\end{equation}
where $\bar D_{\dot\alpha}$ is the standard spinor covariant derivative .  Note that
$\bar D_{\dot\alpha}\bar\Phi(x,\bar\theta)\Big|_{\bar\theta=0}$ is a chiral superfield \footnote{
Couplings
involving instanton moduli and chiral superfields of the form 
$\bar D_{\dot\alpha}\bar \Phi\big|_{\bar\theta=0}$ have been recently considered in Refs.
\cite{Matsuo:2008nu,GarciaEtxebarria:2008pi,Uranga:2008nh}.
}.

We are now  in the position of writing the action for the D3/D$(-1)$ system in 
presence of dynamical bi-fundamental matter fields, including string corrections.
In the case of a single gauge instanton configuration for the $\mathrm{SU}(N_0)$ factor, which corresponds to take $k_0=1$, this action is given by%
\footnote{Remember that in the one-instanton case for $\mathcal N=1$ models there are no
$\chi$-moduli.} 
\begin{equation}
\begin{aligned}
S_{\mathrm{D3/D(-1)}}(\Phi,\bar\Phi) = &~\frac{2\pi^3{\alpha'}^2}{g_s}\,D_cD^c+\ii\,D_{c}\big(\bar{w}_{\dot\alpha}
(\tau^c)^{\dot\alpha}_{~\dot\beta} w^{\dot\beta}\big)
+ \ii\,\lambda_{\dot\alpha}
\big( \bar{\mu}\,w^{\dot{\alpha}}
+\bar{w}^{\dot{\alpha}}\,\mu\big)
\\
&+\Big[\,\frac{1}{2}\,\bar{w}_{\dot\alpha}\big(\Phi\,\bar\Phi+
\bar\Phi\,\Phi\big) w^{\dot\alpha}+\frac{\ii}{2}\,\bar\mu^1\,
\bar\Phi\,\mu - \frac{\ii}{2}\,\bar\mu\, \bar\Phi\,\mu^1 \\
& ~~~~+\ii\,\bar w_{\dot \alpha}\big(\bar D^{\dot\alpha}\bar\Phi\big)\mu^1
-\ii\,\bar\mu^1\big(\bar D_{\dot\alpha}\bar\Phi\big) w^{\dot\alpha}
\Big]_{\bar\theta=0}~.
\end{aligned}
\label{sk2} 
\end{equation}
In the first line above the quadratic term in the auxiliary fields 
$D_c$ with an $\alpha'$-dependent coefficient comes from the gauge action $S_G$ in (\ref{Sd1d3})
written for a one-instanton configuration.
The second line in (\ref{sk2}) is the result of the replacements 
(\ref{promo0}) and (\ref{promo00}) in $S_\phi$, whereas
the third line arises from (\ref{wpsimu}) upon use of (\ref{promo2}).

In the following we will discuss the non-perturbative effective terms that are
induced on the D3 brane world volume by this instanton configuration.

\subsection{Field theory results: non-perturbative F-terms}
\label{subsecn:fieldtheory}

In the field theory limit $\alpha'\to 0$,
the instanton action (\ref{sk2}) simplifies to
\begin{equation}
\begin{aligned}
S^{(0)}_{\mathrm{D3/D(-1)}}(\Phi,\bar\Phi) & =
~\ii\,D_{c}\big(\bar{w}_{\dot\alpha}
(\tau^c)^{\dot\alpha}_{~\dot\beta} w^{\dot\beta}\big)
+\ii \,\lambda_{\dot\alpha} \big( \bar{\mu}\,w^{\dot{\alpha}}
+\bar{w}^{\dot{\alpha}}\,\mu\big)
\\
& +\Big[\,\frac{1}{2}\,\bar{w}_{\dot\alpha}\big(\Phi\,\bar\Phi+
\bar\Phi\,\Phi\big) w^{\dot\alpha}+\frac{\ii}{2}\,\bar\mu^1\,
\bar\Phi\,\mu - \frac{\ii}{2}\,\bar\mu\, \bar\Phi\,\mu^1 \\
&~~~~ +\ii\,\bar w_{\dot \alpha}\big(\bar D^{\dot\alpha}\bar\Phi\big)\mu^1
-\ii\,\bar\mu^1\big(\bar D_{\dot\alpha}\bar\Phi\big) w^{\dot\alpha}\Big]_{\bar\theta=0}~.
\end{aligned}
\label{sk1} 
\end{equation}
Note that in this action $D_c$ and $\lambda_{\dot\alpha}$ appear only linearly and act as Lagrange multipliers for the bosonic and fermionic ADHM constraints, and that,
as in (\ref{sk2}),
the dependence on the superspace coordinates 
$x^\mu$ and $\theta^\alpha$ is only through the matter superfields.

Integrating over all instanton moduli we obtain the following non-perturbative F-terms:
\begin{equation}
S_{\mathrm{n.p.}}= \int d^4x\,d^2\theta\,\,{W}_{\mathrm{n.p.}}
~,\quad
 {W}_{\mathrm{n.p.}}= \Lambda^{\beta_0} \int d\,{\widehat{\mathfrak M}}
\,~\ee^{-S^{(0)}_{\mathrm{D3/D(-1)}}(\Phi,\bar\Phi)}~,
\label{weff} 
\end{equation}
where
$\Lambda$ is the dynamically generated scale of the effective $\mathrm{SU}(N_0)$ SQCD theory we are considering, namely
\begin{equation}
 \Lambda^{\beta_0}=M_s^{\beta_0}\,\ee^{2\pi\ii\tau_0}
\quad\mbox{with}\quad \beta_0=3N_0-N_1~.
\end{equation}
Despite the notation we have adopted, one should not immediately
conclude that ${W}_{\mathrm{n.p.}}$ defined in (\ref{weff})
be a superpotential since, as we will see momentarily, gauge instantons
can induce also other types of non-perturbative F-terms.

In view of the explicit form of the field dependent moduli action (\ref{sk1}),
we can make the following general Ansatz:
\begin{equation}
 {W}_{\mathrm{n.p.}} = {\mathcal C}\, \Lambda^{\beta_0}\,
{\bar\Phi}^{n}\,\Phi^{m}\,\big({\bar D_{\dot\alpha}
\bar\Phi\,\bar D^{\dot\alpha}\bar\Phi}
\big)^{p}\Big|_{\bar\theta=0}~,
\label{ansatz1}
\end{equation}
where $p$ is restricted to positive values to avoid the appearance of fermionic fields in the 
denominator. We now proceed as in Section \ref{subsec:zeromode} and require
${W}_{\mathrm{n.p.}}$ to 
be a quantity of scaling dimension 3 and that its $\mathrm U(1)^3$ charges match 
those of the centered measure, given in (\ref{qqq}) with $k_1=0$. Taking into account 
that $q[\bar D\bar\Phi]=+1/2$ and 
$q^\prime[\bar D\bar\Phi]=q^{\prime\prime}[\bar D\bar\Phi]=-1$,
after some simple algebra we find that the parameters in (\ref{ansatz1}) are given by
\begin{equation}
p=-n= 1-N_0+N_1~,\quad m=1-N_0-N_1~.
\label{pmn}
\end{equation}
The instanton induced effective interactions have thus the form
\begin{equation}
{W}_{\mathrm{n.p.}} = {\mathcal C}\, \Lambda^{\beta_0}\,
\frac{\big({\bar D_{\dot\alpha}\bar\Phi\,\bar D^{\dot{\alpha}}
\bar\Phi}\big)^{p}}{{\bar\Phi}^{p}\,\Phi^{\,p+2N_0-2}}\Bigg|_{\bar\theta=0}
\label{weff2}
\end{equation}
for $p=0,1,\ldots$.

For $p=0$ (and hence for $N_1=N_0-1$) the above result reduces to the
well-known ADS superpotential for $\mathrm{SU}(N_c)$
SQCD with $N_f=N_c-1$ \cite{Affleck:1983mk}; indeed,
after using the D-flatness condition on the matter fields and
explicitly performing the integrations over all ADHM moduli in this case, 
one can prove that the overall coefficient $\mathcal C$ is non-vanishing and that
(\ref{weff2}) becomes
\begin{equation}
{W}_{\mathrm{n.p.}} =\mathcal C\,\Lambda^{2N_c+1}\,\frac{1}{\det M}
\label{wads}
\end{equation}
where $M$ is the meson superfield, in agreement with the ADS result.

For $p>0$ the above result (\ref{weff2}) reproduces 
the multi-fermion instanton induced interactions for SQCD with $N_f\geq N_c$ 
studied originally in \cite{Beasley:2004ys} in the case
$N_c=2$ and recently derived by integrating the ADHM moduli in \cite{Matsuo:2008nu} %
\footnote{See also Ref. \cite{GarciaEtxebarria:2008pi} for related considerations in the
case $N_c=1$.}. In particular, for $p=1$ and $N_c=N_f=2$, Eq. (\ref{weff2}) yields the form
\begin{equation}
{W}_{\mathrm{n.p.}}= {\mathcal C}\, \Lambda^{4}\,
\frac{{\bar D_{\dot\alpha}\bar\Phi\,\bar D^{\dot{\alpha}}
\bar\Phi}}{{\bar\Phi}\,\Phi^3}\Bigg|_{\bar\theta=0}~,
\label{BW}
\end{equation}
in accordance with the explicit result of the moduli integral 
\cite{Beasley:2004ys,Matsuo:2008nu}, which can be written as
\begin{equation}
{W}_{\mathrm{n.p.}}= \mathcal C\,\Lambda^{4}\,
\frac{\epsilon_{{f_1}f'_1}\,
\epsilon^{f_2f'_2}\,\bar D_{\dot\alpha}{\bar M}^{f_1}_{~f_2}\,
\bar D^{\dot{\alpha}}{\bar M}^{f'_1}_{~f'_2} +2\,\bar D_{\dot\alpha}\bar B
\bar D^{\dot\alpha}\bar{\tilde B}}{\big(\tr \bar M M+\bar B B+ \bar {\tilde B}\tilde B\big)^{3/2}}\Bigg|_{\bar\theta=0}~,
\label{BW2}
\end{equation}
in terms of the $\mathrm{SU}(2)$ meson and baryon fields 
(see Eq.s (\ref{meson}) and (\ref{baryon}) for $N_f=N_c=2$). 
For $p>1$ one obtains more general multi-fermion terms.
As proved in Ref. \cite{Beasley:2004ys}, these multi-fermion terms, despite
being non-holomorphic in the matter fields,
are annihilated by the anti-chiral supercharges $\overline Q_{\dot\alpha}$, and as such they are genuine F-terms even if they do not correspond to a superpotential.

\section{Non-perturbative flux-induced effective interactions}
\label{secn:fluxeffects}

\begin{figure}[htb]
 \begin{center}
  \begin{picture}(0,0)%
\includegraphics{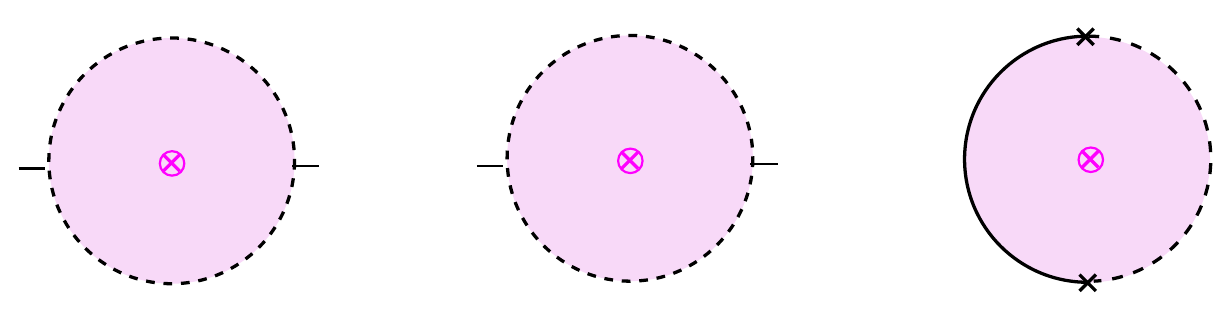}%
\end{picture}%
\setlength{\unitlength}{1381sp}%
\begingroup\makeatletter\ifx\SetFigFontNFSS\undefined%
\gdef\SetFigFontNFSS#1#2#3#4#5{%
  \reset@font\fontsize{#1}{#2pt}%
  \fontfamily{#3}\fontseries{#4}\fontshape{#5}%
  \selectfont}%
\fi\endgroup%
\begin{picture}(16636,4446)(796,-3754)
\put(11236,-1976){\makebox(0,0)[lb]{\smash{{\SetFigFontNFSS{8}{9.6}{\familydefault}{\mddefault}{\updefault}$\lda$}}}}
\put(7096,-1391){\makebox(0,0)[lb]{\smash{{\SetFigFontNFSS{8}{9.6}{\familydefault}{\mddefault}{\updefault}$\ldaup$}}}}
\put(15301,389){\makebox(0,0)[lb]{\smash{{\SetFigFontNFSS{8}{9.6}{\familydefault}{\mddefault}{\updefault}$\bar\mu$}}}}
\put(15421,-3631){\makebox(0,0)[lb]{\smash{{\SetFigFontNFSS{8}{9.6}{\familydefault}{\mddefault}{\updefault}$\mu$}}}}
\put(4951,-2011){\makebox(0,0)[lb]{\smash{{\SetFigFontNFSS{8}{9.6}{\familydefault}{\mddefault}{\updefault}$\taup$}}}}
\put(811,-1426){\makebox(0,0)[lb]{\smash{{\SetFigFontNFSS{8}{9.6}{\familydefault}{\mddefault}{\updefault}$\ta$}}}}
\put(2536,-1131){\makebox(0,0)[lb]{\smash{{\SetFigFontNFSS{8}{9.6}{\familydefault}{\mddefault}{\updefault}$G_{(3,0)}$}}}}
\put(9001,-1066){\makebox(0,0)[lb]{\smash{{\SetFigFontNFSS{8}{9.6}{\familydefault}{\mddefault}{\updefault}$G_{(0,3)}$}}}}
\put(15256,-1081){\makebox(0,0)[lb]{\smash{{\SetFigFontNFSS{8}{9.6}{\familydefault}{\mddefault}{\updefault}$G_{(3,0)}$}}}}
\end{picture}%
 \end{center}
\caption{Diagrams encoding the linear couplings of bulk fluxes of type $(0,3)$ and $(3,0)$ to the instanton moduli reported in Eq. (\ref{sflux1}).}
\label{fig:fluxed}
\end{figure}
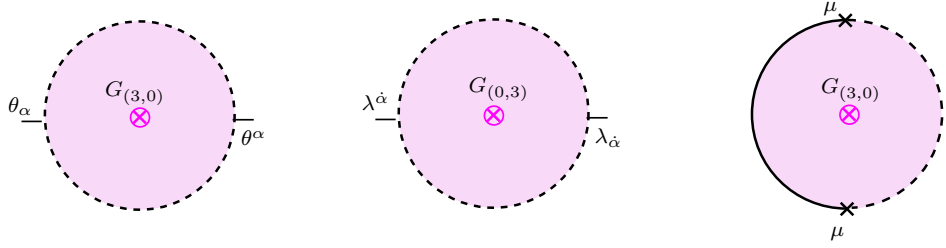

We now generalize the analysis of the previous sections and investigate the
non-perturbative effects produced in the gauge theory by adding R-R and NS-NS fluxes
in the internal space.
In particular we consider the 3-form flux
\begin{equation}
 G_3 = F - \tau \,H
\end{equation}
which is made out of the R-R 3-form $F$, the NS-NS 3-form $H$, and the axion-dilaton $\tau$. 
The general couplings of 
closed string fluxes to open string fermionic bilinears were derived in the previous Chapter
by evaluating mixed open/closed string amplitudes on disks with generalized mixed
boundary conditions like those represented in Fig. \ref{fig:fluxed}. 
\footnote{Notice that, differently from before, here the transverse space
to the D3-branes is non-compact. This implies that the normalization coefficients
to be used here are like those computed before, but without the factors of
the compactification volume $\mathcal V$.}.
Computing the mixed open/closed string diagrams of Fig. \ref{fig:fluxed} 
in the $\mathbb Z_2 \times \mathbb Z_2$ orbifold, one finds that
the flux induced interactions on the instanton moduli space
are encoded in the action 
\begin{equation}
S^{\mathrm{flux}} = {2\pi\ii}\, \left[
\frac{2G_{(3,0)}}{\sqrt{g_s}}\,\theta^{\alpha} \theta_{\alpha}
-\frac{2G_{(0,3)}}{\sqrt{g_s}}\,\frac{\pi^2{\alpha'}^2}{2}\,\lambda_{\dot\alpha}\lambda^{\dot\alpha}
\right]
+{\ii}\sqrt{g_s}\,G_{(3,0)}\,{\bar\mu}\mu~,
\label{sflux1}
\end{equation}
where we have denoted by $G_{(3,0)}$ and $G_{(0,3)}$ the $(3,0)$ and $(0,3)$ components 
of $G_3$ in the natural complex structure of the transverse space. 
These components satisfy, respectively, an imaginary self-duality and anti-self-duality 
condition and are responsible for the soft supersymmetry breaking terms 
related to the gravitino and gaugino masses, see {\it e.g.} \cite{Camara:2003ku,Camara:2004jj}.
Note that in the first and last
terms of (\ref{sflux1}) the scaling dimension of (length)$^{-1}$ carried by the $G$-flux is 
compensated by the dimensions of $\theta$, $\mu$ and $\bar\mu$, while in the second term explicit $\alpha'$ factors are needed. This is perfectly consistent with the fact that, while the $G_{(3,0)}$ flux components have a natural field theory interpretation as gaugino masses, the 
$G_{(0,3)}$ components instead have no counterpart on the gauge field theory. Thus, from the open
string point of view their presence in (\ref{sflux1}) is a genuine string effect, 
as revealed also by the explicit factors of $\alpha'$.

The action (\ref{sflux1}) can be conveniently rewritten as
\begin{equation}
S^{\mathrm{flux}} = \frac{2\pi\ii}{g_s}\, \Big[\,G\,\theta^{\alpha} \theta_{\alpha}
-\bar G\,\frac{\pi^2{\alpha'}^2}{2}\,\lambda_{\dot\alpha}\lambda^{\dot\alpha}
\Big]
+\frac{\ii}{2}\,G\,{\bar\mu}\mu~,
\label{sflux2}
\end{equation}
where we have defined
\begin{equation}
 G = 2\sqrt{g_s} \,G_{(3,0)}~,\quad
\bar G = 2\sqrt{g_s} \,G_{(0,3)}~.
\label{GbarG}
\end{equation}
Eq. (\ref{sflux2}) is the form of the flux induced moduli action which we
will use in the following to study the non-perturbative interactions generated by
fractional D-instantons in the presence of bulk fluxes.
In particular we will consider terms at the linear order in $G$ or $\bar G$ where the world-sheet derivation of the moduli action (\ref{sflux2})
are reliable.
We therefore have two possibilities depending on whether we keep $G$ or $\bar G$
different from zero, which we are going to analyze in turn.

\subsection{One-instanton effects with $G\neq 0$}
\label{subsecn:Gflux}
In this case we can set $\bar G=0$ and look for the non-perturbative interactions proportional
to $G$, assuming again that the fractional D-instanton is of type 0, {\it i.e.} that $k_0=1$,
as in Section \ref{secn:1inst}.
A class of such interactions is obtained by exploiting the $\frac{\ii}{2}G\bar\mu\mu$ term of the
flux action (\ref{sflux2}). At first order in $G$ this leads to
\begin{equation}
S_{\mathrm{n.p.}}(G) = 
\Lambda^{\beta_0}\int d^4x\,d^2\theta
\,d\,{\widehat{\mathfrak M}}
~\ee^{-S_{\mathrm{D3/D(-1)}}(\Phi,\bar\Phi)}\,\left(\frac{\ii}{2}\,G\bar\mu\,\mu\right)
\label{seffg}
\end{equation}
where $S_{\mathrm{D3/D(-1)}}(\Phi,\bar\Phi)$ is the instanton action (\ref{sk2}).
By taking the limit $\alpha'\to 0$ we
obtain non-perturbative flux-induced terms in the effective action of the form
\begin{eqnarray}
S_{\mathrm{n.p.}}(G) &=& \int d^4x\,d^2\theta \,{W}_{\mathrm{n.p.}}(G)
~,\\
{W}_{\mathrm{n.p.}}(G) &=& \Lambda^{\beta_0}
\int d\,{\widehat{\mathfrak M}}
\,~\ee^{-S^{(0)}_{\mathrm{D3/D(-1)}}(\Phi,\bar\Phi)}\,\left(\frac{\ii}{2}\,G\bar\mu\,\mu\right)
\label{weffg}
\end{eqnarray}
where $S^{(0)}_{\mathrm{D3/D(-1)}}(\Phi,\bar\Phi)$ is the moduli action in the field theory
limit given in (\ref{sk1}).
After performing the integration over all centered moduli,
in the effective field theory we expect to find an interaction of the following 
schematic form
\begin{equation}
{W}_{\mathrm{n.p.}}(G)= {\mathcal C}\, G\,\Lambda^{\beta_0}\,
{\bar\Phi}^{n}\,\Phi^{m}\,\big({\bar D_{\dot\alpha}
\bar\Phi\,\bar D^{\dot\alpha}\bar\Phi}
\big)^{p}\Big|_{\bar\theta=0}
\label{weffg2}
\end{equation}
with $\beta_0+n+m+3p=2$ in order to have an operator of mass dimension 3 (remember that
$G$ has dimensions of a mass). As before, we restrict to positive values of $p$ in order to
avoid the appearance of fermionic fields in the denominator. Requiring that the three $\mathrm U(1)$ charges of ${W}_{\mathrm{n.p.}}(G)$ match those of 
the centered instanton measure for consistency with (\ref{weffg}), and using the
information that $q(G)=-3$ and $q^\prime(G)=q^{\prime\prime}(G)=0$, it is easy to find that
the parameters in (\ref{weffg2}) are given by
\begin{equation}
 p=-n-2=2-N_0+N_1\quad\mbox{and}\quad m=-N_0-N_1~.
\label{mnpG}
\end{equation}
The resulting multi-fermion interactions are non-supersymmetric as can
be easily seen by noticing that they are non-holomorphic for $p=0$.

The case $p=1$ ({\it i.e.} $N_1=N_0-1$) is particularly interesting, 
since it corresponds to $\mathrm{SU}(N_c)$ SQCD with $N_f=N_c-1$. 
We have already recalled that in this case the
gauge instanton induces the ADS superpotential; now we see that in the presence of a $G$-flux
which softly breaks supersymmetry by giving a mass to the gaugino, the gauge instanton produces
new types of low-energy effective interactions which are of the form
\begin{equation}
 {W}_{\mathrm{n.p.}}(G)= \left.{\mathcal C}\, G\,\Lambda^{2N_c+1}\,
\frac{{\bar D_{\dot\alpha}
\bar\Phi\,\bar D^{\dot\alpha}\bar\Phi}}{\bar\Phi^3\,\Phi^{2N_c-1}}\right|_{\bar\theta=0}~.
\label{weffG1}
\end{equation}
We stress that this is a formal expression which only indicates the powers of the various
fields that appear in the result; the precise structure of ${W}_{\mathrm{n.p.}}(G)$ 
should be given in terms of the appropriate variables of the low-energy effective theory
(the meson superfields in this case) and can be obtained by explicitly performing the integral
over the instanton moduli which also dictates how color and flavor indices must be saturated. 
This task is particularly easy to do for $\mathrm{SU}(2)$ SQCD with
one flavor, and some details can be found in Appendix \ref{app:intG}. There we show that for
$N_c=2$ and $N_f=1$ the flux induced non-perturbative term (\ref{weffG1}) can be written as
\begin{equation}
{W}_{\mathrm{n.p.}}(G)= \left. \mathcal C\,G\,\Lambda^5\,\frac{\bar D^2 \bar M}{(\bar M M)^{3/2}}
\right|_{\bar\theta=0}
\label{N2G}
\end{equation}
where $M$ is the meson superfield of the effective theory. We can regard this interaction
as a low-energy non-perturbative effect of the soft supersymmetry breaking realized by the
$G$-flux in the microscopic high-energy theory.
Finally, we observe that one can alternatively exploit the
$G\theta^2$ term of the  flux action (5.3) to produce non-supersymmetric
interactions of the same type as the ones described here.

\subsection{One-instanton effects with $\bar G \neq 0$}
\label{subsecn:barGflux}
The contribution to the effective action linear in $\bar G$ in presence of a single fractional D-instanton of type 0 is given, in analogy to Eq. (\ref{seffg}), by
\begin{equation}
S_{\mathrm{n.p.}}(\bar G) = \Lambda^{\beta_0}\,\int d^4x\,d^2\theta
\,d^2\lambda\,\,d\,{\widehat{\mathfrak M}'}
~\ee^{-S_{\mathrm{D3/D(-1)}}(\Phi,\bar\Phi)}\,
\left(-\frac{2\pi\ii}{g_s}\,\frac{(\pi\alpha')^2}{2}\,\bar G \lambda_{\dot\alpha}\lambda^{\dot\alpha}\right)~.
\label{seffgbar}
\end{equation}
Here we have denoted by ${\widehat{\mathfrak M}'}$ all centered moduli but $\lambda$. 
Performing the Grassmannian integration over $d^2\lambda$, we can write
\begin{equation}
S_{\mathrm{n.p.}}(\bar G) = \int d^4x\,d^2\theta
\,\,W_{\mathrm{n.p.}}(\bar G)~,
\label{seffgbar2}
\end{equation}
where
\begin{equation}
W_{\mathrm{n.p.}}(\bar G)=(\pi\alpha')^2 \,\frac{2\pi\ii}{g_s}\,\Lambda^{\beta_0}\,\bar G\, 
\int d\,{\widehat{\mathfrak M}'} \,~\ee^{-\left.S'_{\mathrm{D3/D(-1)}}(\Phi,\bar\Phi)\right.
}
\label{weffgbar}
\end{equation}
with $S'_{\mathrm{D3/D(-1)}}(\Phi,\bar\Phi)$ being the action (\ref{sk2}) without the fermionic
ADHM constraint term since the Grassmannian integration over $\lambda$ has killed it. {From} (\ref{weffgbar})
we therefore expect to find a result of the schematic form
\begin{equation}
{W}_{\mathrm{n.p.}}(\bar G) = {\mathcal C}\,\alpha'^2\, \bar G\,\Lambda^{\beta_0}\,
{\bar\Phi}^{n}\,\Phi^{m}\,\big({\bar D_{\dot\alpha}
\bar\Phi\,\bar D^{\dot\alpha}\bar\Phi}
\big)^{p}\Big|_{\bar\theta=0} + \cdots~,
\label{weffgbar2}
\end{equation}
where the dots stand for possible higher order string corrections.
Requiring the equality of dimensions and $\mathrm{U}(1)^3$ charges between the definition
(\ref{weffgbar}) and the expression (\ref{weffgbar2}), one finds
\begin{equation}
 \label{parbar}
m= 3 -N_0 - N_1\quad,\quad n = 3 + N_0 - N_1\quad,\quad p = N_1 - N_0~.
\end{equation}

Let us focus on the simple case $p=0$ ({\it i.e.} $N_0=N_1$), which corresponds to a SQCD with $N_f=N_c$ flavors. 
In this case, in absence of fluxes, one gets only multi-fermion terms of Beasley-Witten type, 
like the ones displayed in Eq.s (\ref{BW}) and (\ref{BW2}). In the presence of a $\bar G$-flux
we have also a non-holomorphic contribution of the form
\begin{equation}
 \label{cp0}
{W}_{\mathrm{n.p.}} = \mathcal{C} \,\alpha'^2\, \bar G\,\Lambda^{2 N_c}\,
{\bar\Phi}^{3}\,\Phi^{3 - 2 N_c}\Big|_{\bar\theta=0}
\end{equation}
which can be explicitly computed by performing the integration
over  $d\,{\widehat{\mathfrak M}'}$, as shown in
Appendix \ref{app:intGbar}. Notice that again these terms are non-holomorphic and therefore manifestly
non-supersymmetric. For $N_c=2$, the result is
\begin{equation}
 \label{GbarN2}
{W}_{\mathrm{n.p.}} = \mathcal{C} \,\alpha'^2\, \bar G\,\Lambda^{4}\,
\left.\frac{\det \bar M}{\big(\tr \bar M M+\bar B B+ \bar {\tilde B}\tilde B\big)^{1/2}} \right|_{\bar\theta=0}~,
\end{equation}
where $M$ is the meson superfield and $B$ and $\widetilde B$ are the baryon superfields.

\section{Stringy instanton effects in presence of fluxes}
\label{secn:strinst}
D-instantons of type $2$ and $3$ are of different type with respect to
the D3 branes where the SQCD-like $\mathrm{SU}(N_0)\times \mathrm{SU}(N_1)$ theory
is defined, and lead thus to ``stringy'' or ``exotic'' non-perturbative effects.
In this case we have only fermionic mixed moduli $\mu^2$, $\bar\mu^2$ and
$\mu^3$, $\bar\mu^3$, while there are no $w_{\dot\alpha}$ and $\bar w_{\dot\alpha}$'s
from the NS sectors. 

Let us now derive the general form of the centered partition function
in dependence of the vacuum expectation value of the scalar $\phi$, in analogy to
what we did for the gauge instantons in Section \ref{sec:effint}.
The moduli action (\ref{cometipare}) drastically
simplifies. In particular, it is holomorphic in $\phi$
and does not contain any $\lambda$ dependence.
Therefore, unless one introduces an orientifold projection \cite{Argurio:2007qk,Argurio:2007vqa,Bianchi:2007wy}
or invokes other mechanisms \cite{Blumenhagen:2007bn,Petersson:2007sc,GarciaEtxebarria:2008pi},
the only way to get a non-zero result is to include
the flux-induced $\bar G\lambda\lambda$ term of Eq. (\ref{sflux1}) and
use it to perform the $\lambda$ integration. At the linear level in the fluxes,
the other flux interactions in (\ref{sflux2}) become then irrelevant.
Neglecting as usual numerical prefactors, we can write the centered partition function for
a stringy instanton configuration with instanton numbers $k_2$ and $k_3$ as
\begin{equation}
 \label{Wsi}
W_{\mathrm{n.p.}}(\bar G) = \alpha'^2\,\bar G\,M_s^{k_2\beta_2 +
k_3\beta_3}\,\ee^{2\pi\ii(k_2\tau_2 + k_3\tau_3)}\int d\,{\widehat{\mathfrak
M}'}\, \ee^{-S_{\mathrm{D3/D(-1)}}}~,
\end{equation}
leading to the following general Ansatz 
\begin{equation}
 \label{ansWsi}
W_{\mathrm{n.p.}}(\bar G)= \mathcal C\,\bar G\, M_s^{k_2\beta_2 +
k_3\beta_3 + n}\,\ee^{2\pi\ii(k_2\tau_2 + k_3\tau_3)}\, \phi^m~. 
\end{equation}
We have not fixed a priori the power of $M_s\sim 1/\sqrt{\alpha'}$ since
in this case the moduli integration can produce extra factors of $\alpha'$ with
respect to those appearing in Eq. (\ref{Wsi}), because of the $S_G$ part of the
moduli action (\ref{cometipare}) which appears with an explicit $(\alpha')^2$ in
front%
\footnote{In particular, the ``center of mass'' part of the $D_c$'s appears
only through the quadratic term $\sim\alpha'^2D_c D^c$ and the gaussian
integration over it produces negative powers of $\sqrt{\alpha'}$. This is different
with respect to the gauge instanton cases considered in Sections \ref{secn:1inst} and 
\ref{secn:fluxeffects}, where the $D_c$'s couple also to the bosonic moduli $w$ and
$\bar w$, leading to a completely different type of integral.}.

The equality between the $q,q'$ and $q''$ charges following from this ansatz
and those implied by the definition (\ref{Wsi}) plus the request
that the mass dimension of $W_{\mathrm{n.p.}}$ be equal to 3 impose that
\begin{equation}
\label{sicond}
 n= 2\quad,\quad m = 2(k_2 N_0 + k_3 N_1)~,
\end{equation}
and
\begin{equation}
\label{k2k3N0N1}
 (k_2 - k_3)(N_0 - N_1)  = 0~;
\end{equation}
To derive these equations we have used the fact that for our brane configuration
$\beta_2 = \beta_3 = -(N_0 + N_1)$.
The condition (\ref{k2k3N0N1}) admits the following solutions:
\begin{equation}
 \label{sisol}
\begin{aligned}
 k_2 &= k_3\quad\mbox{with}\quad N_0~\mbox{and}~N_1~\mbox{arbitrary}~,\\
 N_0 &= N_1\quad\mbox{with}\quad k_2~\mbox{and}~k_3~\mbox{arbitrary}~. 
\end{aligned}
\end{equation}
The centered partition function in stringy instanton sectors can thus be written
in the form
\begin{equation}
 \label{Wsires}
W_{\mathrm{n.p.}} (\bar G)= \mathcal C\,
\bar G\, M_s^{k_2\beta_2 + k_3\beta_3 + 2}\,\ee^{2\pi\ii(k_2\tau_2 + k_3\tau_3)}\,
\phi^{-(k_2\beta_2 + k_3\beta_3)}~.
\end{equation}

Let us now concentrate on the set-up containing a single stringy instanton
described in Fig. \ref{fk0k2}\emph{b)}, namely let us set $k_2=1$, $k_3=0$.
In this case it is easy to promote the vacuum expectation value $\phi$ 
to the full superfield $\Phi(x,\theta)$ through
diagrams such as those of Fig. \ref{fig:phimu2mu3}.
and the moduli integration can be explicitly done. 
As remarked above, 
the only way to saturate the Grassmannian integration over
$d\lambda_{\dot\alpha}$ is via their $\bar G$ interaction
and the non-perturbative contribution to the effective action of this 
``stringy'' instanton sector is
\begin{equation}
 \label{Seffsi}
S_{\mathrm{n.p.}} = 
\int d^4x\,d^2\theta\, W_{\mathrm{n.p.}}(\bar G)
\end{equation}
where the superpotential is given by
\begin{equation}
{W_{\mathrm{n.p.}}}= \mathcal C\,{\alpha'}^2\,
M_s^{-(N_0 + N_1)}\,\ee^{2\pi\ii\tau_2}\,\bar G\,
\int d\,{\widehat{\mathfrak M}'}
\,\ee^{-S_{\mathrm{D3/D(-1)}}(\Phi)}~.
\label{Weffsi}
\end{equation}
Notice that the dimensional prefactor does not combine
with the exponential of the classical action  to form the
dynamically generated scale of the gauge theory, since
$\tau_2$ is the complexified coupling of D3-branes of type 2, which are not the ones that
support the gauge theory we are considering.

The moduli $\widehat{\mathfrak M}'$ appearing in (\ref{Weffsi}) are
simply $\{D_c,\mu^2\,{\bar\mu}^2,\mu^3\,{\bar\mu}^3\}$, with $\mu^2$ and $\mu^3$
transforming in the fundamental representations of
$\mathrm{U}(N_0)$ and $\mathrm{U}(N_1)$ respectively. Thus, the moduli action to be used in (\ref{Weffsi}) simply 
reduces to
\begin{equation}
 \label{seffex}
S_{\mathrm{D3/D(-1)}}(\Phi) =
\frac{2\pi^3{\alpha'}^2}{g_s}\,D_cD^c -\frac{\ii}{2} 
\left({\bar\mu}^3\Phi\mu^2 -{\bar\mu}^2\Phi\mu^3 \right)~. 
\end{equation}
Hence, the integral in (\ref{Weffsi}) explicitly reads
\begin{equation}
 \label{modintsi}
\int d^3D\,d^{N_0}\mu^2\, d^{N_0}{\bar\mu}^2\, d^{N_1}\mu^3\, d^{N_1}{\bar\mu}^3
\ee^{-\frac{2\pi^3{\alpha'}^2}{g_s}\,D_cD^c +\frac{\ii}{2} 
\left({\bar\mu}^3\Phi\mu^2 -{\bar\mu}^2\Phi\mu^3 \right) }~.
\end{equation}
The integration over the $\mu$'s clearly vanishes unless $N_0=N_1$, in which
case we get, after performing also the gaussian integration over the $D$'s,
\begin{equation}
 \label{risintsi}
{\alpha'}^{-3} \det Q\, \det \tilde Q =
{\alpha'}^{-3} \det M~.
\end{equation}
Here we have used the form (\ref{Phi}) of $\Phi$ and in the last step we have introduced
the meson field $M=\tilde Q Q$. We have also disregarded all numerical constants and kept track
only of the powers of $\alpha'\propto M_s^{-2}$, since in all of
our treatment we have specified completely only the dimensional part of the prefactors 
in the moduli measure.

Inserting (\ref{risintsi}) into (\ref{Weffsi}) we find therefore that a single
stringy instanton in presence of an imaginary self-dual three-form flux produces for $N_0=N_1$ 
({\it i.e.} for a SQCD with $N_f=N_c$ flavors) a
\emph{holomorphic} superpotential
\begin{equation}
 \label{risweffsi}
W_{\mathrm{n.p.}}= \mathcal{C}\, M_s^{2 - 2
N_c}\ee^{2\pi\ii\tau_2}\, \bar G\, \det M~.
\end{equation}
Interestingly, the interactions generated by stringy instantons
are still holomorphic and therefore supersymmetric even in the
presence of the supersymmetry breaking flux $\bar{G}$.

\chapter{Conclusions on Part I}
\label{conclflux}

In the first part of this thesis we have computed the couplings of NS-NS and R-R fluxes to
fermionic bilinears living on general brane intersections (including
instantonic ones). The couplings have been extracted in Chapter \ref{flux} from disk amplitudes
among two open string vertex operators and one closed string vertex
representing the background fluxes. The results for the R-R and
NS-NS amplitudes are given in Eqs. (\ref{amplFfinal}) and (\ref{amplHfinal}).
At leading order in $\alpha'$ they describe fermionic mass terms for open string modes, with boundary conditions which are encoded in the magnetized reflection
matrices $R_0,{\cal R}_0$ and in the twists
$\vec\vartheta$,
at linear order in the R-R and NS-NS fluxes.

The case $\vec\vartheta=0$ corresponds to open strings starting and
ending on two parallel D-branes. The result in this case can be
written in the simple form
\begin{equation}
\mathcal A  =-\frac{2\pi\ii}{3!}\,c_F\, \Theta\Gamma^{MNP}\Theta\,
T_{MNP} 
\label{ampltot1s}
\end{equation}
where $c_F$ is a normalization factor and
\begin{equation}
T_{MNP} = \big(F\mathcal{R}_0\big)_{MNP}+\frac{3}{g_s}\,\big(
\partial B R_0\big)_{[MNP]} ~. 
\label{tmnls}
\end{equation}
This formula shows that different branes couple to different
combinations of the R-R and NS-NS fields.
For compactifications to $d=4$ in presence of 3-form internal fluxes
the explicit form of the $T$ tensors are displayed in
Tab.~\ref{Dbranes} for spacetime filling D-branes and in
Tab.~\ref{Ebranes} for instantonic branes. For spacetime filling
branes, the $T$-tensor describes the structure of soft fermionic
mass terms for a general D-brane intersection.  For Euclidean branes, they
accounts for fermionic mass terms in the instanton moduli space
action modifying the fermionic zero mode structure of the instanton.
Our results are in perfect agreement with
those of Refs.
\cite{Grana:2002tu,Marolf:2003ye,Camara:2003ku,Camara:2004jj,Martucci:2005rb,Tripathy:2005hv,Bergshoeff:2005yp}
that have been derived with pure supergravity methods and generalize
them to generic (instantonic or not) D-brane intersections. The
effects of open string magnetic fluxes can be easily incorporated
into these formulas via the reflection matrices  $R_0({\mathcal F})$ and
${\mathcal R}_0({\mathcal F})$. 
As an example, the explicit form of the $T$-tensor for Euclidean magnetized E5-branes 
has been given in \eq{te5m}.

The cases of open strings ending on D3-branes and D-instantons
have been studied in detail.  For D3-branes in flat space one obtains
\beqa
{\mathcal A}_{\mathrm D3} &=& \frac{2\pi \ii}{3!}\,c_F(\Lambda)\,{\mathrm{Tr}}
\Big[\, \Lambda^{\alpha A}
\Lambda_{\alpha}^{{\phantom \alpha}B}
\big(\overline\Sigma^{mnp}\big)_{AB}\,G_{mnp}^{\mathrm{IASD}} \Big. \nn \\
 && \Big. -
\bar\Lambda_{\dot\alpha A}\bar\Lambda^{\dot\alpha}_{{\phantom\alpha}B}
\big(\Sigma^{mnp}\big)^{AB}\,\big(G_{mnp}^{\mathrm{IASD}}\big)^*
\,\Big]
\label{massD3s}
\eeqa
with $G=F-\tau H$, $G^{\mathrm{IASD}}$ its imaginary
anti-self-dual part and $c_F(\Lambda)$ a normalization factor discussed in the text.
This formula encodes the structure of soft
symmetry breaking terms in ${\cal N}=4$ gauge theory induced by
NS-NS and R-R fluxes.

The coupling of fluxes to D-instantons is given instead by
\beqa
{\mathcal A}_{{\mathrm D}(-1)} &=& \frac{2\pi\ii}{3!}\, \Big[\,c_F(\theta)\,
\theta^{\alpha A} \theta_{\alpha}^{{\phantom\alpha} B}
\big(\overline\Sigma^{mnp}\big)_{AB}\,G_{mnp}^{\mathrm{IASD}} \Big. \nn \\
&& \Big. + c_F(\lambda)\,\lambda_{\dot\alpha A}\lambda^{\dot\alpha}_{{\phantom
\alpha}B} \big(\Sigma^{mnp}\big)^{AB}\,G_{mnp}^{\mathrm{ISD}}
\,\Big]~.
\label{massD-1s}
\eeqa

The case  $\vec\vartheta\neq 0$ describes the couplings of open strings stretching between
non-parallel stacks of D-branes. For spacetime filling D-branes the corresponding
open string excitations describe chiral matter transforming in bi-fundamental
representations of the gauge group and always contain massless chiral fermions.
The case, where open strings are
twisted by $\vartheta=\frac{1}{2}$ along the spacetime directions,
describes the charged moduli of gauge or exotic instantons. For gauge
instantons in ${\cal N}=4$ gauge theory one finds the flux induced action
\begin{equation}
{\mathcal A}_{\mathrm{D3/D(-1)}} =\frac{4\pi\ii
}{3!}\,c_F(\mu)\,{\bar\mu}^{A}\mu^{B}
\,\big(\overline\Sigma^{mnp}\big)_{AB}\,G_{mnp}^{\mathrm{IASD}}~.
\label{mumubtots}
\end{equation}
The results obtained here extend straightforwardly to less
supersymmetric theories and to exotic instantons. In particular for pure ${\cal N}=1$ SYM,
the flux couplings for both gauge and exotic instantons follow from
(\ref{massD3s},\ref{massD-1s},\ref{mumubtots}) by restricting the
spinor components to $A=B=0$. The only contributions to fermionic
mass terms come in this case from the components $G_{(3,0)}$ and
$G_{(0,3)}$ related to the gaugino and
gravitino masses. Explicitly for
$\mathcal{T}_6/(\mathbb{Z}_2\times\mathbb{Z}_2)$ we have
\begin{equation}
 \label{gravmasscon}
m_\Lambda
= 4 \,\frac{\ee^{\varphi/2}}{\mathcal{V}} \,\big|G_{(3,0)}\big| \quad,\quad
m_{3/2}
= 4\,\frac{\ee^{\varphi/2}}{\mathcal{V}} \,\big|G_{(0,3)}\big|~,
\end{equation}
and the fermionic flux couplings can be written as
\begin{eqnarray}
{\mathcal A}_{\mathrm D3} &=& -\ii\,\frac{\ee^{-\varphi/2}G_{(3,0)}}{4 \pi{\mathcal V}}
 {\mathrm{Tr}}\big[\, \Lambda^\alpha \Lambda_\alpha 
\big]+{\rm c.c.} \nonumber \\
&=& -\frac{\ii}{16\pi}\,m_\Lambda\,\ee^{-\varphi}\,
 {\mathrm{Tr}}\big[\, \Lambda^\alpha \Lambda_\alpha
\big]+{\rm c.c.}~,\label{massfs1} \\
{\mathcal A}_{{\mathrm D}(-1)} &=&
-4\pi\ii\,\frac{\ee^{-\varphi/2}{G}_{(3,0)}}{\mathcal V}\,\theta^\alpha\theta_\alpha\,
+\,\pi\ii\,(2\pi{\alpha'})^2\,\frac{\ee^{-\varphi/2} G_{(0,3)}}{2{\mathcal V}} \,\lambda_{\dot\alpha}\lambda^{\dot\alpha}~\nonumber\\
&=&-\pi\ii\,m_\Lambda\,\ee^{-\varphi}\,\theta^\alpha\theta_\alpha\,
+\frac{\pi\ii}{8}\,(2\pi{\alpha'})^2\,m_{3/2}\,\ee^{-\varphi}\, \lambda_{\dot\alpha}\lambda^{\dot\alpha}~,\label{massfs2}\\
{\mathcal A}_{\mathrm{D3/D(-1)}} &=&-\ii\, \frac{\ee^{\varphi/2}
G_{(3,0)}}{2{\mathcal V}} \,\bar\mu_u \mu^u~ \nonumber\\
&=& -\frac{\ii}{8}\,m_\Lambda
\,\bar\mu_u \mu^u~.
\label{massfs3}
\end{eqnarray}

These flux couplings modify the zero mode structure of the instanton and
allow for new low energy coupling in the D3-brane action. 

In Chapter \ref{nonp} we therefore studied the non-perturbative superpotentials induced by gauge and exotic instantons in presence of this closed string background fluxes. For sake of simplicity we focused on a $\NN = 1$ quiver gauge theory with gauge group $U(N_0) \times U(N_1)$ and matter in the bi-fundamental, engineered by  fractional D3-branes at a $\mathcal{C}^3/(\mathbb{Z}_2\times\mathbb{Z}_2)$ singularity, where the non-perturbative sector is described \emph{via} fractional D-instantons. This can be thought as a local description of a 
compactification on the toroidal orbifold $T^3/(\mathbb{Z}_2\times\mathbb{Z}_2)$; for this reason we could neglect global constraints such as the introduction of orientifold planes.

In particular we focused our attention on one-instanton sectors. 
By means of formul\ae ~(\ref{massfs2}) and (\ref{massfs3}), the \emph{flux action} could be written as 
\begin{equation}
S^{\mathrm{flux}} = {2\pi\ii}\, \left[
\frac{2G_{(3,0)}}{\sqrt{g_s}}\,\theta^{\alpha} \theta_{\alpha}
-\frac{2G_{(0,3)}}{\sqrt{g_s}}\,\frac{\pi^2{\alpha'}^2}{2}\,\lambda_{\dot\alpha}\lambda^{\dot\alpha}
\right]
+{\ii}\sqrt{g_s}\,G_{(3,0)}\,{\bar\mu}\mu~.
\end{equation}

In the case with  $G_{(3,0)}\neq 0$ and $G_{(0,3)}= 0$, for a gauge instanton one 
obtains, through symmetry requirements, the general ansatz
\begin{equation}
{W}_{\mathrm{n.p.}}(G_{(3,0)})= {\mathcal C}\, \sqrt{g_s} \,G_{(3,0)}\,\Lambda^{2N_0+p-2}\,
\frac{\big({\bar D_{\dot\alpha} 
\bar\Phi\,\bar D^{\dot\alpha}\bar\Phi}
\big)^{p}}
{{\bar\Phi}^{p+2}\,\Phi^{2N_0+p-2}}
\Big|_{\bar\theta=0}
\label{G30}
\end{equation}
with $p \geq 1$ and $N_1=N_0-2+p$. For $p=1$ (\emph{i.e.} $N_1 = N_0 -1$) it corresponds to SQCD with $N_f = N_c -1$. This ansatz is a formal expression which should be rewritten in terms of the appropriate gauge invariant combinations of low energy matter fields. By enforcing the D-flatness conditions and performing the integrations over the ADHM moduli, the superpotential can then be expressed as a function of mesons and baryons superfields. In particular, if $N_c =2$
\begin{equation}
{W}_{\mathrm{n.p.}}(G_{(3,0)})= \left. \mathcal C\,\sqrt{g_s} \,G_{(3,0)}\,\Lambda^5\,\frac{\bar D^2 \bar M}{(\bar M M)^{3/2}}
\right|_{\bar\theta=0}.
\end{equation}
If instead only the  $G_{(0,3)}$ flux is switched on
the general ansatz becomes
\begin{equation}
{W}_{\mathrm{n.p.}}(G_{(0,3)}) = {\mathcal C}\,\alpha'^2\, \sqrt{g_s} \,G_{(0,3)} \,\Lambda^{2N_0-p}\,
\frac{\big({\bar D_{\dot\alpha}
\bar\Phi\,\bar D^{\dot\alpha}\bar\Phi}
\big)^{p}}{{\bar\Phi}^{p-3}\,\Phi^{2N_0+p-3}}  \Big|_{\bar\theta=0} ~,
\label{G03}
\end{equation}
with $p \geq 0$ and $N_1=N_0+p$.
For SQCD with $N_f = N_c =2$ the result of the integration is
\begin{equation}
{W}_{\mathrm{n.p.}} (G_{(0,3)}) = \mathcal{C} \,\alpha'^2\,\sqrt{g_s} \,G_{(0,3)} \,\Lambda^{4}\,
\left.\frac{\det \bar M}{\big(\tr \bar M M+\bar B B+ \bar {\tilde B}\tilde B\big)^{1/2}} \right|_{\bar\theta=0}~.
\end{equation}

In the case of exotic instantons, the only way to saturate the Grassmannian integration over the fermionic zero modes $\lambda$ by introducing a suitable flux, is to include a $G_{(0,3)}\neq 0$ closed string background, as one can easily deduce from eq.(\ref{massfs2}). The case of SQCD with $N_0 = N_1 = N_f = N_c$ and one D-instanton in the node $2$ gives

\begin{equation}
W_{\mathrm{n.p.}}(G_{(0,3)})= \mathcal{C}\, M_s^{2 - 2
N_c}\ee^{2\pi\ii\tau_2}\, \sqrt{g_s} \,G_{(0,3)}\, \det M~.
\end{equation}

Further investigations are needed to show how such a non-perturbative term should affect the low-energy effective action.


\chapter{Effective String and Statistical Mechanics}
\label{introeff}

As already discussed in Chapter \ref{intro}, the idea that string theory may provide the effective description of confining gauge theories
in their strong-coupling regime is an old and well motivated one \cite{early,lsw,l81}; in this context, 
the string degrees of freedom describe the fluctuations of the colour flux tube. 

In the last years much effort has been devoted to test this conjecture. 
In particular, several results have been obtained in an \emph{effective string} approach, 
in which the conformal anomaly due to the fact that the theory is quantized in 
a non-critical value $d\not= 26$ of the space-time dimensionality is neglected. 
In principle, this is a very problematic simplification, since conformal
invariance is at the very heart of the quantization procedure (see the
Conclusions section for some remarks on this issue). 
However, it was observed very early \cite{Olesen:1985pv} that the coefficient 
of the conformal anomaly vanishes for large distances, \emph{i.e.} for world-sheets of large size in target space.
In fact, in recent years, thanks to  various improvements in lattice simulations
~\cite{lw01,cfghp97,fep00,chp03} the effective string picture has been tested with a 
very high precision and confidence \cite{chp03}-\cite{mt04} by considering observables such as Wilson 
loops and Polyakov loop correlators. It turns out that at large inter-quark
distances and low temperatures the effective string, and in particular the simplest model, the Nambu-Goto one, correctly describes the Monte Carlo data. 
As distances are  decreased,  clear deviations from this picture are
observed~\cite{chp04,chp05,jkm03,jkm04}. For a recent general discussion on  the effective action of confining strings, see \cite{Aharony:2009gg}.

Within the effective string framework, the standard procedure to treat the effective Nambu-Goto string \cite{df83}
in the last twenty years was
to fix a \emph{physical gauge} (see section \ref{subsec:ng} for some more details) and re-express a generic effective string model as a 2d (interacting) conformal field 
theory of $d-2$ bosons. In this set-up, the inverse of the product of the string tension $\sigma$ times the 
minimal area $\mathcal{A}$ of the world sheet spanned by the string represents the parameter of a loop expansion around the classical solution  for the inter-quark potential. The first term of this expansion yields the well known \emph{L\"uscher correction} \cite{lsw}. 
The second term was evaluated more than twenty years ago in~\cite{Dietz:1982uc}, 
with a remarkable theoretical effort, for several different classes of effective string actions. 
Higher order corrections would require very complicated calculations and have never appeared in the literature.

Recently, a set of simulations both in SU(2) and SU(3) lattice gauge
theories (LGT's) \cite{lw02,lw04,cpr04,jkm03,jkm04,ns01,lt01,m02,maj04} 
and in the 3d gauge Ising 
model \cite{cfghp97,chp03,chp04,chp05,Caselle:2005vq}, 
thanks to powerful new algorithms, estimated the inter-quark potential with
precision high enough to distinguish among different effective string actions and to observe
the contribution of higher string modes. To compare the effective string predictions with these new
data in a meaningful way it is mandatory to go beyond the perturbative expansion. 

So far this has been done only for the simplest effective string action, the Nambu-Goto one
\cite{nambu-goto}, 
and for the cylindric geometry, which physically corresponds to the expectation value of the correlator of
two Polyakov loops. In this case it has been possible to build the partition function 
corresponding to the spectrum of the  Nambu-Goto string with the appropriate boundary conditions 
derived long ago in \cite{alvarez81,Arvis:1983fp}, and to make a successfull comparison 
with the simulations \cite{chp05,Caselle:2005vq}. In~\cite{Billo:2005iv} this partition function was 
re-derived via standard covariant quantization, showing that it indeed represents the 
exact operatorial result which re-sums the loop expansion of the model in the physical gauge. 

As a further step in this direction, we want to derive the exact partition 
function of the Nambu-Goto effective string for a toroidal world-sheet geometry, and 
 compare this prediction  to a set of  high precision Monte Carlo
results \cite{nuovomc} for the corresponding observable in the 3d gauge Ising model, namely the interface expectation 
value.

Indeed, the toroidal geometry corresponds  in LGT's to the \emph{maximal 't~Hooft loop}  (see, 
for instance,~\cite{deForcrand:2005pb,Bursa:2005yv}). However a much simpler, yet physically very interesting,
observable can be associated to the same geometry if we consider a three-dimensional LGT with a discrete
abelian gauge group (like the 3d gauge Ising model). In this case the gauge model is mapped by duality into a 
three-dimensional spin model (like the 3d spin Ising model). 
In particular, the confining regime of the gauge model is
mapped into the broken symmetry phase of the spin model. Any extended gauge observable, 
like the Wilson loop or the correlator of two Polyakov loops, is mapped into a set of suitably chosen
anti-ferromagnetic bonds. The torus geometry we are interested in corresponds to the case in which the set of 
anti-ferromagnetic bonds pierces a complete slice of the lattice, \emph{i.e.} to the case in which we simply impose
anti-periodic boundary conditions in one of the lattice directions. 
In a spin model with discrete symmetry group this type of boundary conditions is known to create, in the broken symmetry phase, an \emph{interface} between two
different vacua. 

The appearance of these interfaces in statistical systems under particular conditions
has always raised much interest in various fields of research, ranging
from condensed matter to high energy physics. Recently, there have been remarkable
theoretical and computational improvements in the study of this phenomenon.

On the theoretical side, different effective models can be used to describe
the behaviour of interfaces in 3d systems 
and in particular to evaluate their free energy. 
The most popular is the {\em
capillary wave model} (CWM)\cite{BLS,rw82} which is based on the assumption
of an action proportional to the area of the surface swept by the interface
(for a review, see for instance \cite{Privman:1992zv}).
This model is tantamount to consider the interfaces as bosonic strings
embedded in three dimensions, with a Nambu-Goto
action~\cite{Goto:1971ce,Nambu:1974zg}. Although this
effective string approach neglects the conformal anomaly which appears for
target space dimensions $d\not= 26$, the effects of this
approximation appear to be subleading \cite{Olesen:1985pv} for large
worldsheets. This description has exactly the same nature of the effective
string description of certain observables, such as Polyakov and Wilson loops,
in LGT (see for instance \cite{Caselle:2005xy,Kuti:2005xg} and references
therein).

The (logarithm of the) expectation value of the slice of 
anti-ferromagnetic bonds, which we expect to be described by the Nambu-Goto string, is thus proportional to the interface free energy, an observable which has been the subject of several numerical and experimental studies in condensed matter literature. We shall briefly recall these results
at the beginning of next Chapter.

Numerical investigations of interfaces in statistical systems
(and of their analogue in Lattice Gauge Theories) have attained a great level
of accuracy and reliability. In particular, the interface free energy in the
Ising 3d model has been thoroughly studied 
by means of Monte Carlo
simulations. Indeed, spin models provide a simple realization of interfaces
since in their broken symmetry phase
an interface separating
coexisting vacua of different magnetization can be easily induced in the system 
by suitably choosing the boundary conditions.

To compute the partition function, in \cite{Billo:2006zg} we
follow the philosophy of \cite{Billo:2005iv} and resort to standard covariant quantization in the 
first order formulation of the Nambu-Goto theory, of which we briefly recall some aspects in section 
\ref{subsec:ng}. In section \ref{sec:bos_string} we integrate appropriate sectors of the bosonic string partition function over the world-sheet modular parameter. In this way we describe in an exact manner the string fluctuations around a specified target space surface, in our case representing an interface in a compact space. 
In section \ref{sec:comp_func} we show that our expression reproduces the result obtained in
\cite{Dietz:1982uc} via a perturbative expansion (up to two loops) of the NG functional integral; in fact, our expression re-sums the loop expansion.  

In section \ref{sec:mc} we compare our predictions with the data for the free energy of interfaces in the 3d Ising presented in \cite{nuovomc}. Our NG prediction agrees remarkably to the data 
for values of the area larger than (approximately) four times the inverse string tension, which is the same distance scale below which deviations from the NG model emerged in the studies of other observables cited above \cite{chp05}. 
In the range where there is agreement, we find that 
the NG result is largely dominated by the lowest level mode of the bosonic string; the free energy 
associated to this mode already accounts very well for the MC data, apart from a shift of the overall normalization. 
Though corresponding to a  single particle mode, this contribution is essentially stringy, 
since this mode pertains to a wrapped string.

This lower bound on
the world-sheet area sets the limit below which one cannot neglect
the effects of the conformal anomaly.

Then we focus on the
universal properties of periodic interfaces arising in general 2d models, presenting the results of \cite{Billo:2007fm}. Following 
the philosophy of the CWM we
make the simple assumption that the weight of the interface is
proportional to its length (which indeed corresponds to a 2d version of the  CWM)
and  treat
at the quantum level this action by means of its first-order formulation.
  In
this way we show in section \ref{sec:model} that the dominant term in the
partition function acquires an universal form, proportional to to $LmK_1(mR)$,
where $m$ is the inverse of the correlation length, $R$ is the lattice size in
the direction of the interface, $L$ the lattice size orthogonal to the interface
and $K_1$ is the modified Bessel function of the first order.

In section \ref{sec:ising} we consider an explicit statistical model, the 2d
Ising model. Upon a suitable choice of boundary conditions, this system allows
the formation of interfaces. The 2d Ising model is
under full analytic control, and we can derive directly the form of the
dominant term in the interface partition function. We find agreement with the
universal expression predicted by our 2d CWM.

In section \ref{sec:dimrid}, we consider the exact expression of the free energy
for interfaces in 3d previously obtained, and we perform a
dimensional reduction in one of the directions along the interface. We retrieve
in this way the proposed 2d expression proportional to $K_1(mR)$. This dimensional reduction is the exact analogue for the
interface boundary conditions of the dimensional reduction from Polyakov loop
correlators to spin-spin correlators studied in~\cite{cdgjm06,cgm06}.
A non trivial consistency test of this correspondence is that the relation which
links the 2d mass $m$ with the 3d string tension $\sigma$ is the same in the
case of interfaces and of Polyakov loops.

Recently, a new set of high precision Monte Carlo data for the free energy
of interfaces in the 3d Ising model became available~\cite{chp07}. These 
data include asymmetric geometries in which one of the sides of the interface
becomes much smaller than the other; in this situation one expects that the
dimensionally reduced expression can describe the data accurately. In the last
section, we test this expectation comparing these data with the predictions of
the full 3d Nambu-Goto treatment and of the 2d simple model described here.

Finally, Chapter \ref{concl1} is devoted to summary and conclusions, while we moved some technical points and useful formul\ae~in the Appendix.

\chapter{Effective String and Interfaces}
\label{eff}

The properties of interfaces in three-dimensional statistical systems have been a 
long-standing subject of research. In particular the interest of people 
working in the subject has been attracted by the so called \emph{fluid} interfaces
whose dynamics is dominated by massless excitations
(for a review see for instance \cite{GFP,p92}). For 
this class of interfaces, thanks to the presence of long range massless modes,
microscopic details such as the  lattice structure of the spin model or the
chemical composition of the components of the binary mixture 
become irrelevant and the physics  can be rather accurately 
described by field theoretic methods. 

An effective model widely used to describe a
rough interface is the {\em  capillary wave model} (CWM) \cite{BLS,rw82}. 
Actually this model (which was proposed well before the Nambu-Goto papers)
exactly coincides~\cite{p92} with the Nambu-Goto one, since 
it assumes an effective Hamiltonian
proportional to the variation of the area of the surface with respect to the
classical solution. 

A simple realization of fluid interfaces is represented by 3d spin models.
In the broken-symmetry phase at low temperature, these models admit different 
vacua which, for a suitable choice of the boundary
conditions, can  occupy macroscopic regions and are 
separated by domain walls which behave as interfaces. For
temperatures  between the \emph{roughening} and the \emph{critical} one, 
interfaces are dominated by long wavelength fluctuations (i.e.\ they exactly  behave as
{\em  fluid} interfaces); all the simulations which we shall discuss below were performed in this
 region.

In these last years the 3d Ising model has played a prominent r\^ole
among the various realizations of fluid 
interfaces, for several reasons. The universality class of the Ising model
includes many physical systems, ranging from binary 
mixtures to amphiphilic membranes. Its universality class
is also the same of the $\phi^4$ theory; this allows a QFT approach to the description of 
the interface physics~\cite{m90,pv95,hm98}.
Last but not least, the Ising model, due to 
its intrinsic simplicity, allows fast and high statistics Monte Carlo
simulations, so that very precise 
comparisons can be made between theoretical predictions and numerical results.

Following this line, during the past years some high precision tests 
of the capillary wave model were performed~\cite{cfhgpv,hp97,nuovomc}. 
A remarkable agreement was found between
the numerical results for interfaces of large enough size
and the next to leading order approximation of the CWM. 

We
shall be able to compare the numerical data of ~\cite{nuovomc} with the
exact prediction of the Nambu-Goto effective model for the interface free energy.

In next Section we will review some basic notion of the Nambu-Goto model.

\section{The Nambu-Goto model and the first order formulation}
\label{subsec:ng}

As we have already seen in Chapter \ref{intro}, the most natural model to describe fluctuating surfaces is the Nambu-Goto bosonic string
\cite{nambu-goto}, in which the action is proportional, via the string tension $\sigma$, to the
induced area of a surface embedded in a $d$-dimensional target space:
\begin{equation}
\label{ngaction}
S = \sigma \int d^2\xi \sqrt{\det g}~,
\hskip 0.6cm g_{\alpha\beta} = \frac{\partial X^i}{\partial\xi^\alpha}\frac{\partial X^j}{\partial \xi^\beta} G_{ij}~.
\end{equation}
For simplicity of notation and to not confuse the string tension $\sigma$ with the world-sheet coordinate, here we parametrize the surface by proper coordinates $\xi^\alpha$. As usual,
$X^i(\xi)$ ($i=1,\ldots,d$) describes the target space position of a point specified by $\xi$.
For us the target space metric $G_{ij}$ will always be the flat one. 

The invariance under re-parametrizations of the action
(\ref{ngaction}) can be used to fix a so-called \emph{static} gauge where the proper coordinates are identified with two of the target space coordinates, say $X^0$ and $X^1$. The quantum version of the NG theory 
can then be defined through the functional integration over the $d-2$ transverse d.o.f. $\vec X(X^0,X^1)$ of the gauge-fixed action. The partition function for a surface $\Sigma$ with prescribed boundary conditions
is given by
\begin{equation}
\label{ngpart}
\begin{aligned}
Z_{\Sigma} & =\int_{\partial\Sigma}\!\!\!\! DX^i \exp\left\{-\sigma \int_\Sigma dX^0 dX^1 \left[1 + (\partial_0\vec X)^2
 + (\partial_1\vec X)^2 + (\partial_0 \vec X \wedge \partial_1 \vec X)^2\right]^{\frac 12}\right\}
\\
& =\int_{\partial\Sigma}\!\!\!\! DX^i  \exp\left\{-\sigma \int_\Sigma dX^0 dX^1 \left[
1 +\frac 12 (\partial_0\vec X)^2 + \frac 12 (\partial_1\vec X)^2 + \mbox{interactions} \right]\right\}.
\end{aligned}
\end{equation}

Expanding the square root as in the second line above, the classical area law $\exp(-\sigma\mathcal{A})$ 
(where $\mathcal{A}$ is the area of the minimal surface $\Sigma$) is singled out. It multiplies
the quantum fluctuations of the 
fields $\vec X$, which have a series of higher order (derivative) interactions. 
The functional integration can be performed perturbatively, the loop expansion parameter being $1/(\sigma\mathcal{A})$, and it 
depends on the boundary conditions imposed on the fields $\vec X$, i.e., on the topology of the boundary $\partial\Sigma$ and hence of $\Sigma$. The cases in which $\Sigma$ is a disk, a cylinder or a torus are
the ones relevant for an effective string description of, respectively, Wilson loops, Polyakov loop correlators and interfaces in a compact target space. 

The computation was carried out up to two loops in \cite{Dietz:1982uc}, see also \cite{cfhgpv,cp96}. 
The surface $\Sigma$ is taken to be a rectangle, with the opposite sides in none, one or 
both directions being identified to get the disk, the cylinder or the torus topology.
At each loop order, the result depends in a very non-trivial way on the geometry of $\Sigma$,
namely on its area $\mathcal{A}$ and on the ratio $u$ of its two sides: it typically involves non-trivial modular forms of the latter. 
The two loop result for the case of interfaces is reported here in section \ref{sec:comp_func}.

An alternative treatment of the NG model
takes advantage of the first order formulation (the \emph{Polyakov action} of Chapter \ref{intro}), in which the action is simply 
\beq
\label{sac}
S_{Pol} = \sigma\int d\xi^0 \int_0^{2\pi} d\xi^1\,
h^{\alpha\beta}\partial_\alpha X^i \partial_\beta X^i~,
\eeq 
where $\xi^1\in [0,2\pi]$ parametrizes the spatial extension of the string and $\xi^0$ its proper time evolution. 

As it is well known, to include the process of splitting and joining of strings, i.e., to include string interactions,
one must consider world-sheets of different topology, \emph{i.e.} Riemann surfaces of different genus $g$. The genus of the world-sheet represents thus the loop order in a string loop expansion. 
For each fixed topology of the world-sheet, instead of integrating out $h$,
we can use re-parametrization and  Weyl invariance to put the metric in a reference form 
$\ee^\phi \hat h_{\alpha\beta}$ (conformal gauge fixing). For instance, on the sphere, \emph{i.e.} at genus $g=0$, we can choose $\hat h_{\alpha\beta} = \eta_{\alpha\beta}$, while on the torus, at genus $g=1$, $\hat h_{\alpha\beta}$ is constant, but still depends on a single complex parameter $\tau$, the modulus of the torus; see later for some more details.
The scale factor $\ee^\phi$ decouples at the classical
level%
\footnote{Actually, this property persists at the quantum level only if the anomaly
parametrized by the total central charge $c= d - 26$ vanishes;
Weyl invariance is otherwise broken and the mode $\phi$, as shown by Polyakov long ago \cite{Polyakov:1981rd},
has to be thought of as a field with a Liouville-type action.
However, as argued in Chapter \ref{introeff} and in the Conclusions section, we 
can, in first instance, neglect this effect for our purposes.} 
and the action takes then the form
\begin{equation}
\label{sac2}
S = \sigma\int d\xi^0 \int_0^{2\pi} d\xi^1\,
\hat h^{\alpha\beta}\partial_\alpha X^i \partial_\beta X^i
 + S_{\mathrm{gh.}}~,
\end{equation}
where $S_{\mathrm{gh.}}$ in \eq{sac} is the action for the ghost
and anti-ghost fields (traditionally called $c$ and $b$) that arise from the
Jacobian to fix the conformal gauge; we do not really need its explicit
expression here, see \cite{gsw} or \cite{polbook} for reviews. The modes of the Virasoro constraints 
$T_{\alpha\beta}=0$, which follow from the
$h^{\alpha\beta}$ equations of motion, generate the residual conformal invariance of  the model.
The ghost system corresponds to a CFT of central charge $c_{\mathrm{gh.}} = -26$.
The fields $X^i(\tau,\sigma)$, with $i=1,\ldots,d$, describe the embedding of
the string world-sheet in the target space and form the simple, well-known two-dimensional CFT of $d$
free bosons. 

\section{The partition function for the interface from bosonic strings}
\label{sec:bos_string}
In the present section we want to describe the fluctuations of an interface in a toroidal target space 
$T^d$ by means of standard closed bosonic string theory. We use the standard first order formulation discussed in section \ref{subsec:ng} and we specify the periodicity of the target space coordinates to be $x^i\sim x^i + L^i$ ($i=1,\ldots,d$). 

The partition function for the bosonic string on the target torus $T^d$ is expressed
as
\begin{equation}
\label{bos1}
\mathcal{Z}^{(d)} = \int \frac{d^2\tau}{\tau_2} \, Z^{(d)}(q,\bar q)\, Z^{\mathrm{gh}}(q,\bar q).
\end{equation}
Here $\tau = \tau_1 + \ii\tau_2$ is the modular parameter of the world-sheet, 
which is a surface of genus $g=1$, \emph{i.e.} a torus. Moreover, $Z^{(d)}(q,\bar q)$ is the
CFT partition function of the $d$ compact bosons $X^i$ defined on such a world-sheet. 
In an operatorial formulation, this reads
\begin{equation}
\label{bos2}
Z^{(d)}(q,\bar q) = \Tr\, q^{L_0 - \frac{d}{24}}\, \bar q^{\tilde L_0 - \frac{d}{24}}~,
\end{equation}
where 
\begin{equation}
\label{bos2bis}
q = \exp (2\pi\ii\tau)~,
\hskip 0.4cm
\bar q = \exp (-2\pi\ii\bar\tau)~.
\end{equation}
$L_0$ and $\tilde L_0$ (particular modes of the Virasoro constraints) are the left
and right-moving dilation generators. With 
$Z^{\mathrm{gh}}(q,\bar q)$ we denote the CFT partition function for the ghost system, defined on the same world-sheet.

The modular parameter $\tau$ is the \emph{Teichm\"uller parameter} of the world-sheet surface: as discussed
in subsection \ref{subsec:ng}, using Weyl invariance and diffeomorphisms we can choose the reference metric $\hat h_{\alpha\beta}$ to be constant and of unit determinant, 
but the complex parameter $\tau$, with $\mathrm{Im}\tau\ge 0$,  which characterizes the complex structure, \emph{i.e.} the shape of the torus, cannot be fixed.
It is thus necessary to integrate over it, as indicated in \eq{bos1}, the integration domain%
\footnote{We will discuss later the issue of discrete modular transformations acting on $\tau$.} 
being the upper half-plane. In the Polyakov approach \cite{Polyakov:1981rd},
this integration is the remnant of the functional integration over the independent world-sheet metric $h_{\alpha\beta}$ after the invariances of the model have been used as described above.
The measure used in \eq{bos1} ensures, as we will see, the modular invariance of the integrand. 

The CFT partition function for a single boson defined on a circle 
\begin{equation}
\label{bos3} X(\xi^0,\xi^1) \sim X(\xi^0,\xi^1) + L
\end{equation}
is given, with the action defined as in \eq{sac2}, by%
\footnote{With an abuse of notation, and for the sake of convenience, we will sometimes denote the Dedekind eta function $\eta(\tau)$ defined in 
\eq{defeta1} as $\eta(q)$, where $q=\exp(2\pi\ii\tau)$.} 
\begin{equation}
\label{bos4}
Z(q,\bar q) = \Tr\, q^{L_0 - \frac{d}{24}}\, \bar q^{\tilde L_0 - \frac{d}{24}}
= \sum_{n,w\in\mathbb{Z}} q^{\frac{1}{8\pi\sigma}\left(\frac{2\pi n}{L} + \sigma w L\right)^2}
\bar q^{\frac{1}{8\pi\sigma}\left(\frac{2\pi n}{L} - \sigma w L\right)^2}
\frac{1}{\eta(q)} \frac{1}{\eta(\bar q)}~.
\end{equation}
The integers $n$ and $w$ are zero-modes of the field $X$, describing respectively its discrete momentum
$p = 2\pi n/L$ and its winding around the compact target space: $X$ must be periodic in $\xi^1$, but 
the target space identification \eq{bos3} allows the possibility that
\begin{equation}
\label{bos5}
X(\xi^0,\xi^1 + 2\pi) = X(\xi^0,\xi^1) + w L~.
\end{equation}
The factors of $1/\eta(q)$ and $1/\eta(\bar q)$ result from the trace over the left and right moving 
non-zero modes, which after canonical quantization become just bosonic oscillators contributing 
to $L_0$ and $\tilde L_0$ their total occupation numbers.

The Hamiltonian trace \eq{bos4} can be re-summed \emph{\'a la Poisson}, see \eq{poisson_res}, over the integer $n$, after which it becomes
\begin{equation}
\label{bos6}
Z(q,\bar q) = \sqrt{\frac{\sigma}{2\pi}}L\sum_{m,w\in\mathbb{Z}} 
\ee^{- \frac{\sigma L^2}{2\tau_2} |m - \tau w|^2}
\frac{1}{\sqrt{\tau_2}\eta(q)\eta(\bar q)}~,
\end{equation}
an expression which is naturally obtained from the path-integral formulation. The discrete
sum over $m,w$ represents the sum over \emph{world-sheet instantons}, namely classical solutions of the field $X$ which, beside their wrapping number $w$ over the $\xi^1$ direction, are characterized also by their wrapping $m$ along the compact propagation direction:
\begin{equation}
\label{bos7}
X(\xi^0 + 2\pi\tau_2,\xi^1 + 2\pi\tau_1) = X(\xi^0,\xi^1) + m L~.
\end{equation}

The form \eq{bos6} of $Z(q,\bar q)$ makes its modular invariance manifest. In fact, the combination
$\sqrt{\tau_2}\eta(q)\eta(\bar q)$ is modular invariant, as it follows from the properties of the Dedekind eta function 
given in \eq{etaS}. Moreover, from the exponential term we infer the effect of modular transformations of the parameter $\tau$ on the wrapping integers $w,m$: they act as $\mathrm{SL}(2,\mathbb{Z})$ matrices on the vector $(m,w)$. In particular, the $S$ and $T$ generators of the modular group are represented as follows:
\begin{align}
\label{bos8}
S:&\null & \tau &\to -\frac{1}{\tau}~, & 
\begin{pmatrix}
m \\ w
\end{pmatrix}
 &\to 
\begin{pmatrix}
0 & -1\\ 1 & 0
\end{pmatrix}
\begin{pmatrix}
m \\ w
\end{pmatrix}~,
\\
T:&\null & \tau &\to \tau+1~, & 
\begin{pmatrix}
m \\ w
\end{pmatrix}
 &\to 
\begin{pmatrix}
1 & -1\\ 0 & 1
\end{pmatrix}
\begin{pmatrix}
m \\ w
\end{pmatrix}~.
\end{align}
This allows to reabsorb the effect of modular transformations by relabelling the sums over $m$ and $w$. 

The partition function for the ghost system is given by
\begin{equation}
\label{bos9}
Z^{\mathrm{gh}}(q,\bar q) = \left(\eta(q)\eta(\bar q)\right)^2~,
\end{equation}
namely it coincides with the inverse of the non zero-mode contributions of two bosons.
Notice that the string partition function \eq{bos1} can be rewritten, substituting the
above expression for $Z^{\mathrm{gh}}(q,\bar q)$, in an explicitly modular-invariant way:
\begin{equation}
\label{bos10}
\mathcal{Z}^{(d)} = \int \frac{d^2\tau}{(\tau_2)^2} \, \left(\sqrt{\tau_2}\eta(q)\eta(\bar q)\right)^2\, 
Z^{(d)}(q,\bar q)~,
\end{equation}
so that the integration over the modular parameter $\tau$ in \eq{bos10} has to be restricted to
the fundamental cell of the modular group. 
Indeed, the Poincar\'e measure $d^2\tau/(\tau_2)^2$ 
and the combination $\sqrt{\tau_2}\eta(q)\eta(\bar q)$ are modular invariant.
The bosonic partition function $Z^{(d)}(q,\bar q)$ is also modular invariant, being 
the product of $d$ expressions of the type \eq{bos4}. It depends
on integers $n^i$ and $w^i$, the discrete momentum and winding number 
for each direction. For each direction $i$, we can Poisson re-sum over the discrete momentum $n^i$, 
as in \eq{bos6}, replacing it with the topological number $m^i$. 

\begin{figure}
\begin{center}

\begin{picture}(0,0)%
\includegraphics{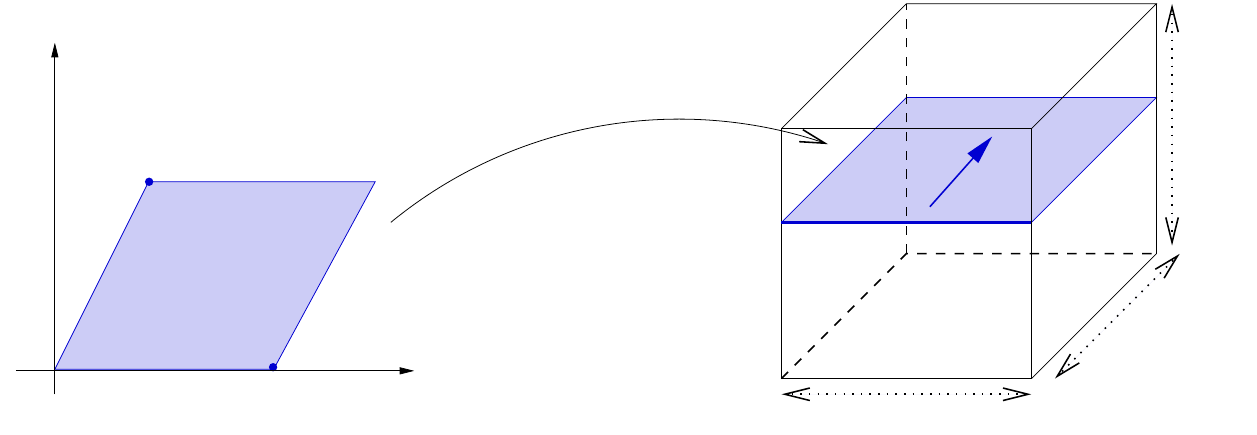}%
\end{picture}%
\setlength{\unitlength}{1973sp}%
\begingroup\makeatletter\ifx\SetFigFont\undefined%
\gdef\SetFigFont#1#2#3#4#5{%
  \reset@font\fontsize{#1}{#2pt}%
  \fontfamily{#3}\fontseries{#4}\fontshape{#5}%
  \selectfont}%
\fi\endgroup%
\begin{picture}(12048,4153)(676,-3667)
\put(12001,-586){\makebox(0,0)[lb]{\smash{{\SetFigFont{9}{10.8}{\rmdefault}{\mddefault}{\updefault}$L_3$}}}}
\put(8401,-2011){\makebox(0,0)[lb]{\smash{{\SetFigFont{9}{10.8}{\rmdefault}{\mddefault}{\updefault}$w_1=1$}}}}
\put(11101,-1411){\makebox(0,0)[lb]{\smash{{\SetFigFont{9}{10.8}{\rmdefault}{\mddefault}{\updefault}$m_2=1$}}}}
\put(9076,-3586){\makebox(0,0)[lb]{\smash{{\SetFigFont{9}{10.8}{\rmdefault}{\mddefault}{\updefault}$L_1$}}}}
\put(11401,-2836){\makebox(0,0)[lb]{\smash{{\SetFigFont{9}{10.8}{\rmdefault}{\mddefault}{\updefault}$L_2$}}}}
\put(676,-361){\makebox(0,0)[lb]{\smash{{\SetFigFont{9}{10.8}{\rmdefault}{\mddefault}{\updefault}$\xi^0$}}}}
\put(4351,-3436){\makebox(0,0)[lb]{\smash{{\SetFigFont{9}{10.8}{\rmdefault}{\mddefault}{\updefault}$\xi^1$}}}}
\put(3001,-3436){\makebox(0,0)[lb]{\smash{{\SetFigFont{9}{10.8}{\rmdefault}{\mddefault}{\updefault}$2\pi$}}}}
\put(1801,-1111){\makebox(0,0)[lb]{\smash{{\SetFigFont{9}{10.8}{\rmdefault}{\mddefault}{\updefault}$2\pi\tau$}}}}
\end{picture}%

\end{center}
\caption{\label{fig:sector}
\small The mapping of the toroidal string world-sheet, of modular parameter $\tau$, into the target space
is organized in many distinct sectors, labeled by the integers $w_i$ and $m_i$ (see the text). By 
selecting the sector with (say) $w_1=1$ and $m_2=1$ we are considering the fluctuations of an 
extended interface, which is a torus because of the target space periodicity.}
\end{figure}
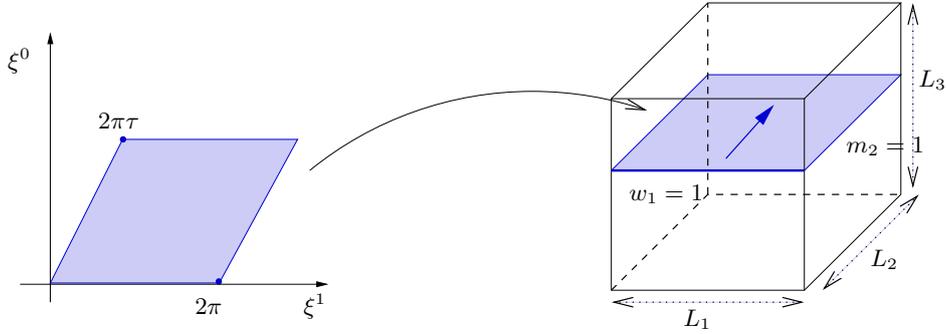

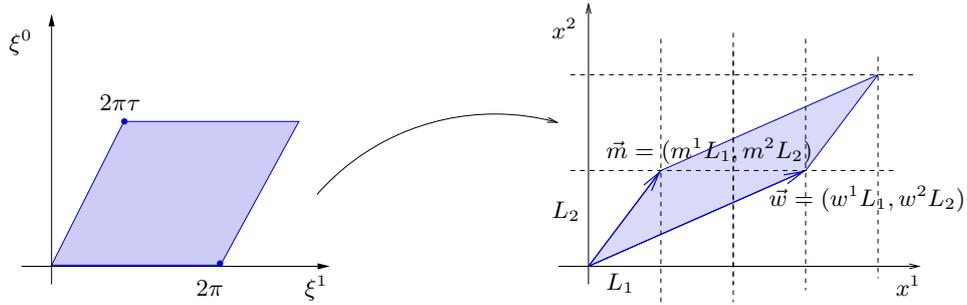
\begin{figure}
\begin{center}

\begin{picture}(0,0)%
\includegraphics{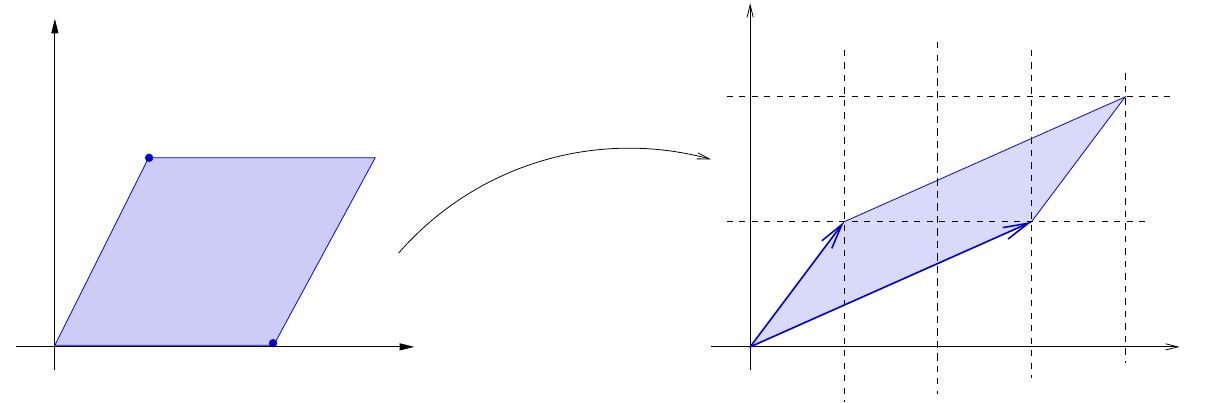}%
\end{picture}%
\setlength{\unitlength}{1973sp}%
\begingroup\makeatletter\ifx\SetFigFont\undefined%
\gdef\SetFigFont#1#2#3#4#5{%
  \reset@font\fontsize{#1}{#2pt}%
  \fontfamily{#3}\fontseries{#4}\fontshape{#5}%
  \selectfont}%
\fi\endgroup%
\begin{picture}(11717,3849)(676,-3598)
\put(676,-361){\makebox(0,0)[lb]{\smash{{\SetFigFont{9}{10.8}{\rmdefault}{\mddefault}{\updefault}$\xi^0$}}}}
\put(4351,-3436){\makebox(0,0)[lb]{\smash{{\SetFigFont{9}{10.8}{\rmdefault}{\mddefault}{\updefault}$\xi^1$}}}}
\put(3001,-3436){\makebox(0,0)[lb]{\smash{{\SetFigFont{9}{10.8}{\rmdefault}{\mddefault}{\updefault}$2\pi$}}}}
\put(1801,-1111){\makebox(0,0)[lb]{\smash{{\SetFigFont{9}{10.8}{\rmdefault}{\mddefault}{\updefault}$2\pi\tau$}}}}
\put(11701,-3436){\makebox(0,0)[lb]{\smash{{\SetFigFont{9}{10.8}{\rmdefault}{\mddefault}{\updefault}$x^1$}}}}
\put(7426,-211){\makebox(0,0)[lb]{\smash{{\SetFigFont{9}{10.8}{\rmdefault}{\mddefault}{\updefault}$x^2$}}}}
\put(8101,-3361){\makebox(0,0)[lb]{\smash{{\SetFigFont{9}{10.8}{\rmdefault}{\mddefault}{\updefault}$L_1$}}}}
\put(7426,-2461){\makebox(0,0)[lb]{\smash{{\SetFigFont{9}{10.8}{\rmdefault}{\mddefault}{\updefault}$L_2$}}}}
\put(8101,-1711){\makebox(0,0)[lb]{\smash{{\SetFigFont{9}{10.8}{\rmdefault}{\mddefault}{\updefault}$\vec m = (m^1 L_1,m^2 L_2)$}}}}
\put(10126,-2311){\makebox(0,0)[lb]{\smash{{\SetFigFont{9}{10.8}{\rmdefault}{\mddefault}{\updefault}$\vec w = (w^1 L_1,w^2 L_2)$}}}}
\end{picture}%

\caption{\label{fig:maps}
\small Consider an embedding of the world-sheet torus into the target space $T^2$ aligned along the directions $x^1,x^2$, characterized by the wrapping numbers $(m^1,w^1)$ and $(m^2,w^2)$. 
It corresponds, in the covering space of this $T^2$, to a parallelogram defined by the vectors 
$\vec w = (w^1 L_1,w^2 L_2)$ and $\vec m = (m^1 L_1,m^2 L_2)$, whose area is $\vec w\wedge \vec m = L_1 L_2(w^1 m^2 - w^2 m^1)$.}
\end{center}
\end{figure}

We want to single out the contributions to the partition function \eq{bos10}
which describe the fluctuations of an interface aligned along a  two-cycle $T^2$ inside $T^d$, 
say the one in the $x^1,x^2$ directions. The world-sheet torus parametrized by $\xi^0,\xi^1$ must be
mapped onto the target space torus by embedding functions $X^1(\xi^0,\xi^1),X^2(\xi^0,\xi^1)$ with 
non-trivial wrapping numbers $(m^1,w^1)$ and $(m^2,w^2)$. The wrapping numbers of $X^i(\xi^0,\xi^1)$, $i>2$, must instead vanish. The minimal area spanned by such a wrapped torus is given by (see Fig. \ref{fig:maps})
\begin{equation}
\label{areamw}
L_1 L_2\, 
(w^1 m^2 - m^1 w^2) =
L_1 L_2
\begin{pmatrix}
m^1 & w^1 
\end{pmatrix}
\begin{pmatrix}
0 & -1 \\
1 & 0 
\end{pmatrix}
\begin{pmatrix}
m^2 \\ w^2 
\end{pmatrix}~.
\end{equation}

\begin{figure}
\begin{center}

\begin{picture}(0,0)%
\includegraphics{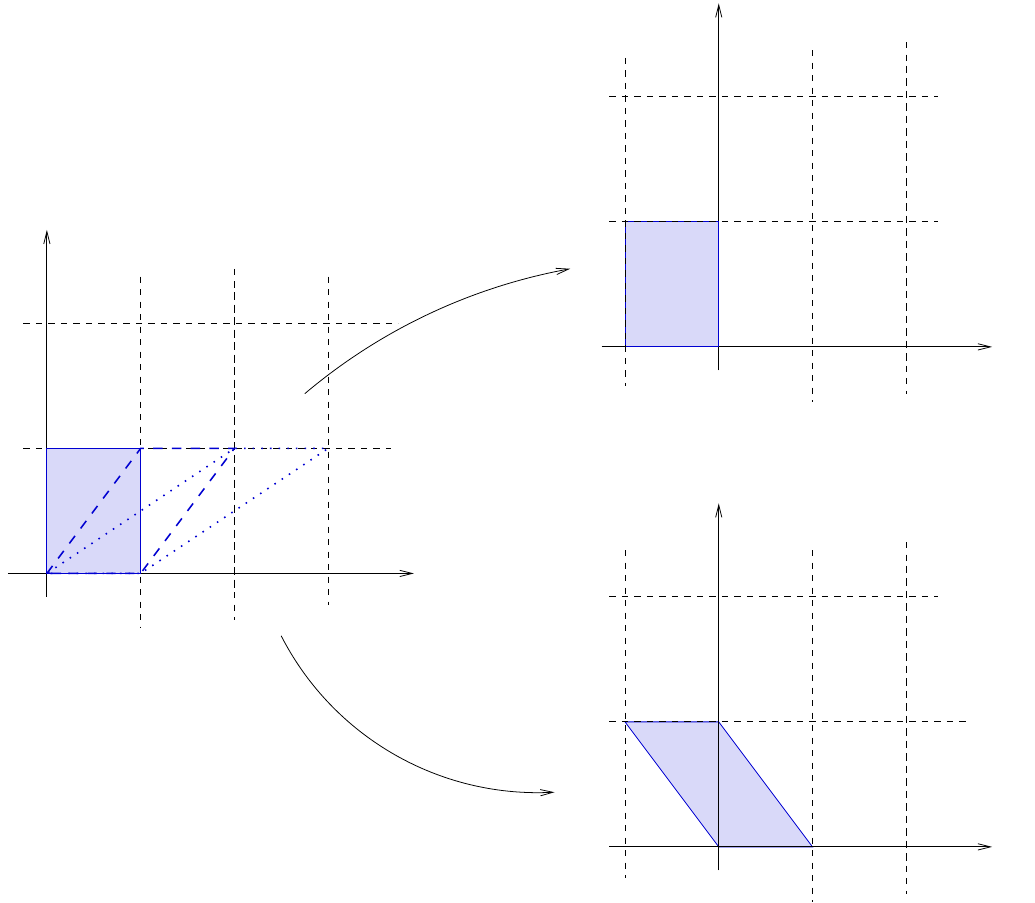}%
\end{picture}%
\setlength{\unitlength}{1973sp}%
\begingroup\makeatletter\ifx\SetFigFont\undefined%
\gdef\SetFigFont#1#2#3#4#5{%
  \reset@font\fontsize{#1}{#2pt}%
  \fontfamily{#3}\fontseries{#4}\fontshape{#5}%
  \selectfont}%
\fi\endgroup%
\begin{picture}(9917,8649)(301,-8098)
\put(7351,-736){\makebox(0,0)[lb]{\smash{{\SetFigFont{9}{10.8}{\rmdefault}{\mddefault}{\updefault}$(m^1,w^1)=(-1,0)$ }}}}
\put(7351,-1186){\makebox(0,0)[lb]{\smash{{\SetFigFont{9}{10.8}{\rmdefault}{\mddefault}{\updefault}$(m^2,w^2)=(0,1)$ }}}}
\put(301,-2086){\makebox(0,0)[lb]{\smash{{\SetFigFont{9}{10.8}{\rmdefault}{\mddefault}{\updefault}$x^2$}}}}
\put(976,-5236){\makebox(0,0)[lb]{\smash{{\SetFigFont{9}{10.8}{\rmdefault}{\mddefault}{\updefault}$L_1$}}}}
\put(301,-4336){\makebox(0,0)[lb]{\smash{{\SetFigFont{9}{10.8}{\rmdefault}{\mddefault}{\updefault}$L_2$}}}}
\put(3976,-5236){\makebox(0,0)[lb]{\smash{{\SetFigFont{9}{10.8}{\rmdefault}{\mddefault}{\updefault}$x^1$}}}}
\put(6751, 89){\makebox(0,0)[lb]{\smash{{\SetFigFont{9}{10.8}{\rmdefault}{\mddefault}{\updefault}$x^2$}}}}
\put(7426,-3061){\makebox(0,0)[lb]{\smash{{\SetFigFont{9}{10.8}{\rmdefault}{\mddefault}{\updefault}$L_1$}}}}
\put(6751,-2161){\makebox(0,0)[lb]{\smash{{\SetFigFont{9}{10.8}{\rmdefault}{\mddefault}{\updefault}$L_2$}}}}
\put(6751,-4711){\makebox(0,0)[lb]{\smash{{\SetFigFont{9}{10.8}{\rmdefault}{\mddefault}{\updefault}$x^2$}}}}
\put(7426,-7861){\makebox(0,0)[lb]{\smash{{\SetFigFont{9}{10.8}{\rmdefault}{\mddefault}{\updefault}$L_1$}}}}
\put(6751,-6961){\makebox(0,0)[lb]{\smash{{\SetFigFont{9}{10.8}{\rmdefault}{\mddefault}{\updefault}$L_2$}}}}
\put(9526,-3061){\makebox(0,0)[lb]{\smash{{\SetFigFont{9}{10.8}{\rmdefault}{\mddefault}{\updefault}$x^1$}}}}
\put(9451,-7861){\makebox(0,0)[lb]{\smash{{\SetFigFont{9}{10.8}{\rmdefault}{\mddefault}{\updefault}$x^1$}}}}
\put(4426,-2086){\makebox(0,0)[lb]{\smash{{\SetFigFont{9}{10.8}{\rmdefault}{\mddefault}{\updefault}$S$}}}}
\put(3826,-7111){\makebox(0,0)[lb]{\smash{{\SetFigFont{9}{10.8}{\rmdefault}{\mddefault}{\updefault}$T$}}}}
\put(7351,-5536){\makebox(0,0)[lb]{\smash{{\SetFigFont{9}{10.8}{\rmdefault}{\mddefault}{\updefault}$(m^1,w^1)=(-1,1)$ }}}}
\put(7351,-5986){\makebox(0,0)[lb]{\smash{{\SetFigFont{9}{10.8}{\rmdefault}{\mddefault}{\updefault}$(m^2,w^2)=(1,0)$ }}}}
\end{picture}%

\caption{\label{fig:mod_transf}
\small On the left, (some of) the coverings corresponding to the chosen sector of the partition function are depicted. Beside the one which exactly corresponds to the fundamental cell of the target space $T^2$, i.e. the case with $(m^1,w^1)=(0,1)$ and $(m^2,w^2)=(1,0)$ there are  ``slanted'' ones corresponding to generic values of $m^1$.
The generators $S$ and $T$ of the world-sheet modular group, which act by $\mathrm{SL}(2,\mathbb{Z})$ matrices as indicated in \eq{bos8}, map these coverings to different ones with the same area. For instance, on the right, we draw the $S$- and $T$-transform of the ``fundamental'' covering discussed above (the solid one in the leftmost drawing).}
\end{center}
\end{figure}
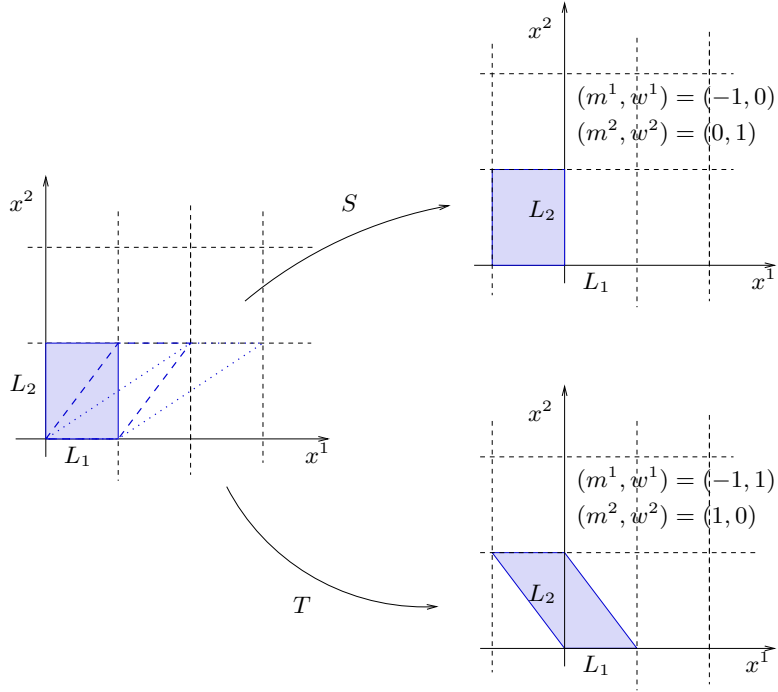

A particular sector which contributes to such a target-space configuration can be obtained 
(see Figure \ref{fig:sector})
by considering a string winding once in, say, the $x^1$ direction:
\begin{equation}
\label{bos11}
w_1 = 1~,\hskip 0.4cm
w_2 = w_3 = \ldots = w_d = 0
\end{equation}
and, upon Poisson re-summation along the directions $x^2$ to $x^d$, selecting the integers
\begin{equation}
\label{bos12}
m_2 = 1~,\hskip 0.4cm
m_3 = m_4 = \ldots = m_d = 0~.
\end{equation}
This corresponds to wrapping numbers in the $x^1,x^2$ directions given by $(m^1,w^1)=(m^1,1)$
and $(m^2,w^2) = (1,0)$ and hence, according to \eq{areamw}, to a minimal area $L_1 L_2$: such a configuration covers the two-cycle in the $x^1,x^2$ directions just once. We have not fixed the value of the wrapping number $m^1$, therefore the sector we consider contains infinite embeddings that corresponds to ``slanted'' coverings, see Fig. \ref{fig:mod_transf}.

The choice $w_i=m_i=0$ for $i=3,\ldots, d$ is preserved under the action 
\eq{bos8} of the modular group. Moreover, this $\mathrm{SL}(2,\mathbb{Z})$ action preserves
the area \eq{areamw}, as it is easy to see%
\footnote{The expression of the area \eq{areamw} contains the symplectic product of the two vectors
$(m^1,w^1)$ and $(m^2,w^2)$. The simultaneous action of $\mathrm{SL}(2,\mathbb{Z})$ on both vectors
preserves this symplectic product: $\mathrm{SL}(2)\sim \mathrm{Sp}(1)$.}. 
The orbit of the modular group which contains the configurations chosen in eq.s (\ref{bos11}-\ref{bos12}) spans thus infinite other string configurations characterized by different wrapping numbers in the directions $x^1$, $x^2$; all these configurations, however, always wrap the $T^2$ in target space just once, and correspond to equivalent descriptions of the interface; see Fig. \ref{fig:mod_transf}.

To obtain the partition function for this interface we should include all the wrapping numbers $(m^1,w^1)$ and $(m^2,w^2)$ which are in the modular orbit of the configurations in eq.s (\ref{bos11}-\ref{bos12}), but we should 
integrate the $\tau$ parameter in \eq{bos13} over the fundamental cell of the modular group only, as usual, to avoid overcounting. We can, equivalently, consider only the configurations of eq.s (\ref{bos11}-\ref{bos12}), fixing
in this way the degeneracy associated to modular transformations, and integrate $\tau$ all over the upper half-plane, which contains all the images under modular transformations of the fundamental cell.

The second choice turns out to be convenient, as the integrals become very simple.
In this way, we get the following expression for the interface partition function:
\begin{eqnarray}
\label{bos13}
\mathcal{I}^{(d)}\!  &=& \!
\int\! \frac{d^2\tau}{(\tau_2)^{\frac{d+1}{2}}} 
\left[\frac{1}{\eta(q)}\frac{1}{\eta(\bar{q})}\right]^{d-2}\!
\sum_{n_1\in\mathbb{Z}} q^{\frac{1}{8\pi\sigma}\left(\frac{2\pi n_1}{L_1} +  \sigma L_1\right)^{2}}
\bar q^{\frac{1}{8\pi\sigma}\left(\frac{2\pi n_1}{L_1} -  \sigma L_1\right)^{2}} \nn \\
&& \times~
\prod_{i=2}^d \left(\sqrt{\frac{\sigma}{2\pi}} L_i\right)
\ee^{-\frac{\sigma L_2^2}{2\tau_2}}~.
\end{eqnarray}

Expanding in series the Dedekind's $\eta$ functions:
\begin{equation}
\label{bos14}
\left[\eta(q)\right]^{2-d} = \sum_{k=0}^{\infty}c_k q^{k-\frac{d-2}{24}}~,
\end{equation}
and expressing $q$ and $\bar q$ in terms of $\tau$, we rewrite \eq{bos13} as
\begin{equation}
\label{bos15}
\begin{aligned}
\mathcal{I}^{(d)}  & = \prod_{i=2}^d \left(\sqrt{\frac{\sigma}{2\pi}}L_i\right)
\sum_{k,k'=0}^\infty \sum_{n_1\in \mathbb{Z}} c_k c_{k'}\int_{-\infty}^\infty d\tau_1
\ee^{2\pi\ii(k-k'+n_1)} \int_0^\infty \frac{d\tau_2}{(\tau_2)^{\frac{d+1}{2}}}
\\
& \times
\exp\left\{ -\tau_{2}\left[\frac{\sigma L_{1}^{2}}{2} + \frac{2\pi^{2} n_{1}^{2}}{\sigma L_{1}^{2}} + 2\pi(k+k'-\frac{d-2}{12})\right]-\frac{1}{\tau_{2}}\left[\frac{\sigma L_{2}^{2}}{2}\right]\right\}~.
\end{aligned}
\end{equation}
The integration over $\tau_2$ can be carried out in terms of modified Bessel functions
using the formula
\begin{equation}
\label{bos17}
\int_0^\infty\frac{d\tau_{2}}{\tau_{2}^{\frac{d+1}{2}}}\exp\left\{ -A^{2}\tau_{2}-\frac{B^{2}}{\tau_{2}}\right\} =2\left(\frac{A}{B}\right)^{\frac{d-1}{2}}K_{\frac{d-1}{2}}(2AB)~,
\end{equation}
with 
\begin{equation} 
\label{bos18}
A  =\sqrt{\frac{\sigma}{2}} L_{1}\, \cale~,
\hskip 0.4cm
B  = \sqrt{\frac{\sigma}{2}} L_{2}
\end{equation}
and
\begin{eqnarray}
\label{bos19}
\cale  &=&
\sqrt{1 + \frac{4\pi}{\sigma L_1^2}(k+k'-\frac{d-2}{12}) + 
\frac{4\pi^{2}\, n_{1}^{2}}{\sigma^2 L_{1}^{4}}}\nn\\
&=& \sqrt{1 + \frac{4\pi\, u}{\sigma \mathcal{A}}(k+k'-\frac{d-2}{12}) + 
\frac{4\pi^{2} \, u^2\, n_{1}^{2}}{(\sigma\mathcal{A})^2}}~.
\end{eqnarray}
In the second step, we introduced the area $\mathcal{A}$ and the modular parameter $\ii u$ of the torus swept out by the string in the target space, namely the interface:
\begin{equation}
\label{Au}
\mathcal{A} = L_1 L_2~, \hskip 0.4cm
u = \frac{L_2}{L_1}~.
\end{equation}

The integration over $\tau_1$ produces a $\delta(k -k' + n_1)$ factor, and thus,
using Eq.s \ref{bos17}--\ref{bos18}, we rewrite the interface partition function 
\eq{bos15} as
\begin{equation}
\label{boskkp}
\mathcal{I}^{(d)} = 2 \left(\frac{\sigma}{2\pi}\right)^ {\frac{d-2}{2}}\, V_T \, \sqrt{\sigma\mathcal{A}u}
\sum_{k,k'=0}^\infty  c_k c_{k'} 
\left(\frac{\cale}{u}\right)^{\frac{d-1}{2}}\, K_{\frac{d-1}{2}}\left(\sigma\mathcal{A}\cale\right)~,
\end{equation}
where $\cale$, introduced in \eq{bos19}, is now to be written replacing $k - k'$ for $n_1$.
In \eq{boskkp} we have introduced the transverse volume 
$V_T = \prod_{i=3}^d L_i$, and we have re-written the prefactor $\sqrt{\sigma}L_2$ as 
$\sqrt{\sigma\mathcal{A}u}$.

Eq. (\ref{boskkp})
gives the exact expression for the fluctuations of the interface aligned along 
the $x^1,x^2$ directions, in our hypothesis that they can be described from standard bosonic string theory.
This expression depends only on the target space geometric data, namely the transverse volume $V_T$ and
the area $\mathcal{A}$ and the shape parameter $u$ of the interface.

Considering the leading exponential behaviour of the Bessel functions only, \eq{boskkp} would reduce
to a partition function constructed by summing over closed string states, characterized by the left- and right-moving occupation numbers $k$ and $k'$ and by the momentum $n_1$ which is fixed to $k-k'$ by the level-matching condition, with the usual bosonic multiplicities $c_k$, $c_{k'}$ and with exponential weights
$\exp(-L_2 E_{k,k'})$, where the energies are given by
\begin{equation}
\label{spectrum}
E_{k,k'} = \sigma L_1 \cale = \sqrt{\sigma^2 L_1^2 + 4\pi\sigma (k+k'-\frac{d-2}{12}) + 
\frac{4\pi^{2}\, n_{1}^{2}}{L_{1}^{2}}}~.
\end{equation}
This spectrum substantially agrees with the expression proposed in \cite{Kuti:2005xg}; however, the correct expression of the partition function, \eq{boskkp}, involves Bessel functions rather than exponentials 
and it also contains extra factors of $\cale$. These modifications are crucial in reproducing correctly the loop expansion of the functional approach, as we will see in the next section.

\subsection{Comparison with the functional integral approach} 
\label{sec:comp_func}
As discussed in Section \ref{subsec:ng}, the Nambu-Goto interface partition function
can also be computed in a functional integral 
approach with a physical gauge-fixing. This leaves just the $d-2$ bosonic d.o.f. corresponding to the transverse fluctuations of the interface, but the action is not a free one. One can 
evaluate the path integral perturbatively, the loop expansion parameter being the inverse of $\sigma\mathcal{A}$ \cite{Dietz:1982uc}. The result of this computation
up to the 2-d loop order was given in
\cite{Dietz:1982uc} and reads
\begin{equation}
\label{df1}
\mathcal{I}^{(d)} \propto\sigma^{\frac{d-2}{2}} \frac{\ee^{-\sigma\mathcal{A}}}{\left[\sqrt{u}
\eta^2(\ii u)\right]^{d-2}} \left\{1 + \frac{f_1(u)}{\sigma\mathcal{A}} + \ldots\right\}~. 
\end{equation}
where the dots stand for higher loop contributions and
\begin{equation}
\label{fuis}
f_1(u) = \frac{(d-2)^2}{2} \left[\left(\frac{\pi}{6}\right)^2 u^2 E_2^2(\ii u) - \frac{\pi}{6} E_2(\ii u)\right]
+ \frac{d(d-2)}{8}~.
\end{equation}
Here $E_2$ denotes the 2nd Eisenstein series (see Appendix \ref{app:useful}). 
Actually, the constant term in \eq{fuis} is different from the one given in 
\cite{Dietz:1982uc}; in Appendix \ref{app:2loop} we show, by reconsidering the computation of the second loop 
term, that in fact the correct result is the one we quote in \eq{fuis}%
\footnote{It is interesting to observe that the missing contribution in \cite{Dietz:1982uc} is proportional to $(d-3)$ and thus disappears in three dimensions. 
This is the reason for which it was not found in the calculations reported
in~\cite{cfhgpv,cp96} which evaluated this 2 loop correction with three other types of regularization in the $d=3$ case. Moreover the difference is modular invariant and thus it could not be detected by Dietz and Filk in the tests of their calculation which they made in \cite{Dietz:1982uc}.}

Our exact expression \eq{boskkp} should reproduce the perturbative expansion of the functional integral result when asymptotically expanded for large $\sigma\mathcal{A}$. To this effect, let us re-organize \eq{boskkp} in a suitable way. 
Using the asymptotic expansion of Bessel functions for large arguments:
\begin{equation}
\label{bfexp}
K_{j}(z) \sim \sqrt{\frac{\pi}{2z}}\ee^{-z}
\left(1 + \sum_{r=1}^\infty a_r^{(j)} z^{-r}\right)~,
\end{equation}
where the coefficients $a_r^{(j)}$ are well known, we obtain
\begin{equation}
\label{Idexp1}
\mathcal{I}^{(d)} = \sqrt{2\pi} \left(\frac{\sigma}{2\pi}\right)^{\frac{d-2}{2}} V_T 
\sum_{k,k'=0}^\infty  c_k c_{k'} 
\left(\frac{\cale}{u}\right)^{\frac{d-2}{2}} \ee^{-\sigma \mathcal{A}\cale} \left(1 + \sum_{r=1}^\infty a_r^{(\frac{d-1}{2})} (\sigma\mathcal{A}\cale)^{-r}\right).
\end{equation}
We can then Taylor expand the expression \eq{bos19} of $\cale$ and write it in the form
\begin{equation}
\label{Xexp}
\cale = 1 + \sum_{t=1}^\infty d_t\left(a,b;u\right)(\sigma\mathcal{A})^{-t}~,
\end{equation}
having introduced, for notational simplicity,
\begin{equation}
\label{ab}
a = k + k' - \frac{d-2}{12}~,
\hskip 0.4cm
b = k - k'~.
\end{equation}
In particular one has
\begin{equation}
\label{d1}
d_1 = 2\pi u a = 2\pi u\left(k + k' - \frac{d-2}{12}\right)~.
\end{equation}
For the sake of readability, we leave to Appendix \ref{app:a} most of the details of the present computation, such as explicit expressions of the various expansion coefficients, describing here
only the basic steps.
 
Plugging the expansion \eq{Xexp} into \eq{Idexp1} we obtain
\begin{equation}
\label{iexp1}
\mathcal{I}^{(d)} \propto
\sigma^{\frac{d-2}{2}}\,
\frac{\ee^{-\sigma\mathcal{A}}}{u^{\frac{d-2}{2}}}\sum_{k,k'=0}^\infty  c_k c_{k'} 
\ee^{-2\pi u \left(k + k' - \frac{d-2}{12}\right)}
\left\{1 + \sum_{s=1}^\infty \frac{g_s\left(a,b;u\right)}{(\sigma\mathcal{A})^{s}}\right\}
~,
\end{equation}
where the coefficients $g_s$ can be re-constructed 
starting from the coefficients appearing in eq.s (\ref{bfexp}) and (\ref{Xexp}); 
see Appendix \ref{app:a} for more details.

We can introduce $Q = \exp(-2\pi u)$ and re-write \eq{iexp1} as follows:
\begin{equation}
\label{iexp2}
\mathcal{I}^{(d)} \propto
\sigma^{\frac{d-2}{2}}\,
\frac{\ee^{-\sigma\mathcal{A}}}{u^{\frac{d-2}{2}}} 
\lim_{\overline Q\to Q}
\sum_{k,k'=0}^\infty  c_k c_{k'}
\left\{1 + \sum_{s=1}^\infty \frac{g_s(a,b;u)}{(\sigma\mathcal{A})^{s}}\right\}
Q^{k - \frac{d-2}{24}}\overline Q^{k' - \frac{d-2}{24}}~.
\end{equation}
At this point, we can effectively replace
\begin{equation}
\label{subder}
a = k+k'- \frac{d-2}{12} \longrightarrow Q\frac{d~}{dQ} +  \overline Q\frac{d~}{d\overline Q}~,
\hskip 0.8cm
b = k - k' \longrightarrow Q\frac{d~}{dQ} -  \overline Q\frac{d~}{d\overline Q}
\end{equation}
and take into account  the definition, \eq{bos14}, of the coefficients $c_{k}$ to obtain
a form of $\mathcal{I}^{(d)}$ proportional to
\begin{equation}
\label{iexp3}
\sigma^{\frac{d-2}{2}} \,
\frac{\ee^{-\sigma\mathcal{A}}}{u^{\frac{d-2}{2}}} 
\lim_{\overline Q\to Q} \left\{1 + \sum_{s=1}^\infty 
\frac{g_s\left(Q\frac{d~}{dQ} - \overline Q\frac{d~}{d\overline Q},Q\frac{d~}{dQ} + \overline Q\frac{d~}{d\overline Q} ;u\right)}{(\sigma\mathcal{A})^{s}}\right\}
\left[\eta(Q)\eta(\overline Q)\right]^{2-d}
\end{equation}
with, in the end, $Q$ being set to $\exp(-2\pi u)$. Finally, we can rewrite the above expression as
\begin{equation}
\label{iexp4}
\mathcal{I}^{(d)} \propto
\sigma^ {\frac{d-2}{2}}\,
\frac{\ee^{-\sigma\mathcal{A}}}{\left[\sqrt{u}
\eta^2(\ii u)\right]^{d-2}} \left\{1 + \sum_{s=1}^\infty \frac{f_s(u)}{(\sigma\mathcal{A})^{s}} \right\}
\end{equation}
where the coefficients $f_s(u)$ are given by
\begin{equation}
\label{fcoeff}
\eta^{2d-4}(\ii u) \lim_{\overline Q\to Q= \ee^{-2\pi u}} 
g_s\left(Q\frac{d~}{dQ} - \overline Q\frac{d~}{d\overline Q},Q\frac{d~}{dQ} + \overline Q\frac{d~}{d\overline Q} ;u\right)
\left[\eta(Q)\eta(\overline Q)\right]^{2-d}
\end{equation}
and can be related to Eisenstein series. In Appendix \ref{app:a} we derive the expression of the 2-nd loop and 3-rd loop 
coefficients, $f_1$ and $f_2$, but it would be straightforward to push the computation to higher orders.
Our expression for $f_1$ agrees completely with the (corrected) Dietz-Filk result \eq{fuis}. Let us remark 
that our result eq.(\ref{iexp4}) is proportional (with no need of \emph{ad hoc} modifications) to $\sigma^ {\frac{d-2}{2}}$ , \emph{i.e.} to $\sqrt{\sigma}$ in three dimensions. This 
result agrees with the (completely independent) field theoretical calculations in the framework of the $\phi^4$ model of \cite{m90,pv95}  and also, as we will see in the next Section, with the Monte Carlo simulations.

\subsection{Comparison with Monte Carlo data}
\label{sec:mc}
In a very recent publication \cite{nuovomc}, precise Monte Carlo data on the free energy $F_s$ of interfaces 
in the 3d Ising model were presented. The most accurate sets of data were obtained for square lattices
(\emph{i.e.} in the case $u=1$), for values of the Ising coupling given by 
$\beta= 0.226102$ (Table 4 of \cite{nuovomc}; we shall refer to this data set as the set n. 1) and 
$\beta=0.236025$ (Table 2 of \cite{nuovomc}, set 2). 
The data set 1 contains 21 points, obtained at lattice sizes $L_1=L_2\equiv L$ 
ranging from $L=10$ to $L=30$; the 9 points in data set 2 are%
\footnote{We consider here only the data in Table 2 of \cite{nuovomc} which, for a given value of $L$, are 
obtained with $L_0$, the size of the lattice in the transverse direction, being equal to $3L$, to ensure
 uniformity with the data set 1, which is obtained with this choice.}
for $L=6$ to $L=14$. 

Previous work regarding other observables \cite{chp05} 
has made it clear that, in appropriate regimes, the 3d Ising model can be successfully described by an 
effective string theory. For the two values of $\beta$ which we study here, very precise estimates for the 
string tension exist~\cite{Caselle:2004jq} (see Table 1 of 
\cite{nuovomc}). For set 1 one has $\sigma = 0.0105241$; for set 2,  $\sigma=0.044023$. This 
entails that the points in data set 1 correspond to values of $\sqrt{\sigma\mathcal{A}}$ ranging from $1.02$ 
to $3.07$, while the points in data set 2 correspond to values from $1.26$ to $2.94$.

\begin{table}
\begin{center}
\begin{tabular}{cccccc}
\hline
\hline 
& & \multicolumn{2}{c}{$N=100$} & \multicolumn{2}{c}{$N=0$} \\
\cline{3-6}
$L_{\mathrm{min}}$ & $(\sqrt{\sigma\mathcal{A}})_{\mathrm{min}}$ & $\mathcal{N}$ & $\chi^2/(\mathrm{d.o.f})$ 
& $\mathcal{N}$ & $\chi^2/(\mathrm{d.o.f})$ 
\\ 
\hline
\multicolumn{6}{c}{Data set 1}
\\
\hline
19 & 1.949 & 0.91957(18) & 4.22 & 0.91413(18) & 1.60\\ 
20 & 2.051 & 0.91891(22) & 1.84 & 0.91377(22) & 0.88\\ 
21 & 2.154 & 0.91836(27) & 0.63 & 0.91344(27) & 0.47\\ 
22 & 2.257 & 0.91829(33) & 0.70 & 0.91354(33) & 0.50\\ 
23 & 2.359 & 0.91797(45) & 0.63 & 0.91339(45) & 0.53\\ 
24 & 2.462 & 0.91762(57) & 0.57 & 0.91316(57) & 0.55\\ 
25 & 2.565 & 0.91715(75) & 0.50 & 0.91279(75) & 0.55\\
\hline
\multicolumn{6}{c}{Data set 2}
\\
\hline 
9 & 1.888 & 0.91052(21) & 7.22 & 0.90466(21) & 2.22 \\
10 & 2.098 & 0.90924(33) & 2.71 & 0.90413(33) & 1.69\\
11 & 2.308 & 0.90820(51) & 1.12 & 0.90349(51) & 1.33\\
\hline 
\hline
\end{tabular}
\end{center}
\caption{\small The fit of the NG free energy \eq{Fc0} with normalization $\mathcal{N}$ to the two data set of Ref. \cite{nuovomc}, performed using only the points in Table 4 and 2 of \cite{nuovomc} corresponding 
to lattice sizes $L\geq L_{\mathrm{min}}$, \emph{i.e.} those with $\sqrt{\sigma\mathcal{A}}\geq (\sqrt{\sigma\mathcal{A}})_{\mathrm{min}}$.
The reduced $\chi^2$ becomes of order unity for $(\sqrt{\sigma\mathcal{A}})_{\mathrm{min}}\gtrsim 2$. 
The fit is performed by truncating the sum over the oscillator levels at $N=100$ or at $N=0$, i.e. keeping 
only the 0-mode contributions (last two columns).}
\label{tab:fits}
\end{table}

Using the information above, it is possible to compare the MC values of the free energy $F_s$ in data set 1
and 2 to the free energy $F$ corresponding to our partition function
\eq{boskkp} (in $d=3$, and for $u=1$), where we factor out the transverse volume factor $V_T$ and allow for
an overall normalization $\ee^{-\mathcal{N}}$:
\begin{equation}
\label{Fc0}
F = -\log \left(\frac{\mathcal{I}^{(3)}}{V_T}\right) + \mathcal{N}~.
\end{equation}
The constant $\mathcal{N}$ will be the only free parameter to be fitted to the data.
For numerical evaluation of the free energy \eq{Fc0}, we
re-arrange the sums appearing in \eq{boskkp} using the property 
\begin{equation}
\label{bos20}
\sum_{k,k'=0}^\infty c_k c_{k'}\,f((k-k')^2,k+k') 
=
\sum_{m=0}^\infty \sum_{k=0}^m c_k c_{m-k} f\left((2k-m)^2,m\right)~,
\end{equation}
valid for any function $f$ of the specified arguments, and write \eq{boskkp} as
\begin{equation}
\label{bos21}
\mathcal{I}^{(d)} = 2
\left(\frac{\sigma}{2\pi}\right)^{\frac{d-2}{2}}\, V_T \, \sqrt{\sigma\mathcal{A}u}
\sum_{m=0}^\infty \sum_{k=0}^m c_k c_{m-k} 
\left(\frac{\cale}{u}\right)^{\frac{d-1}{2}}\, K_{\frac{d-1}{2}}\left(\sigma\mathcal{A}\cale\right)~,
\end{equation}
where $\cale$ is now $\cale((2k-m)^2,m)$, 
since we transformed the summation indices as in \eq{bos20}. The integer $m$ represents the total (left-  
plus right-moving) oscillator number; the contributions to the partition function from states of a given $m$
are suppressed by exponentials of the type $\exp(-2\pi u\, m)$, as it can easily be seen from \eq{bos21} 
upon expanding the argument of the Bessel function using eq.s (\ref{Xexp},\ref{d1}).
We can approximate the exact expression of $F$ by truncating the sum over $m$ at a certain $N$, \emph{i.e.}
neglecting the contributions of string states with total occupation number larger that $N$. 

\begin{figure}
\begin{center}

\begin{picture}(0,0)%
\includegraphics{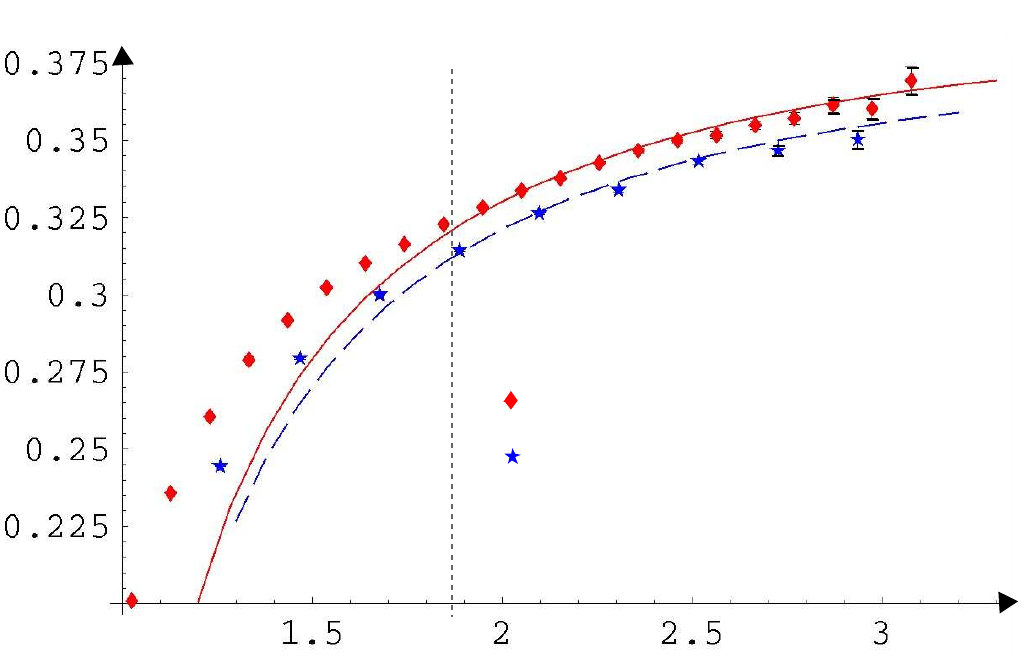}%
\end{picture}%
\setlength{\unitlength}{1184sp}%
\begin{picture}(18227,11603)(810,-10997)
\put(1217,22){\makebox(0,0)[lb]{\small $F_s - \sigma\mathcal{A} + \frac 12 \log \sigma$}}
\put(17776,-10711){\makebox(0,0)[lb]{\small $\sqrt{\sigma\mathcal{A}}$}}
\put(9276,-7991){\makebox(0,0)[lb]{\small $\beta=0.236025, \sigma= 0.044023$ (data set 2)}}
\put(9266,-6971){\makebox(0,0)[lb]{\small $\beta=0.226102, \sigma= 0.0105241$ (data set 1)}}
\end{picture}%

\end{center}
\caption{\label{fig:fit}
\small Two sets of Monte Carlo data for the interface free energy data provided in \cite{nuovomc} are compared to our theoretical predictions following from eq.s (\ref{Fc0},\ref{bos21}),
represented by the solid line for the data set 1 and by the dashed line for data set 2. 
The only free parameter is the additive constant $\mathcal{N}$, corresponding to an overall normalization of 
the NG partition function, fitted to the data using the points to the right of the vertical dashed line
(see the text for more details). 
The error bars in the MC data are visible only for the rightmost points, but they were kept into
account in the fit. 
The sum over the level $m$ in \eq{bos21} was truncated at $N=100$.}
\end{figure}
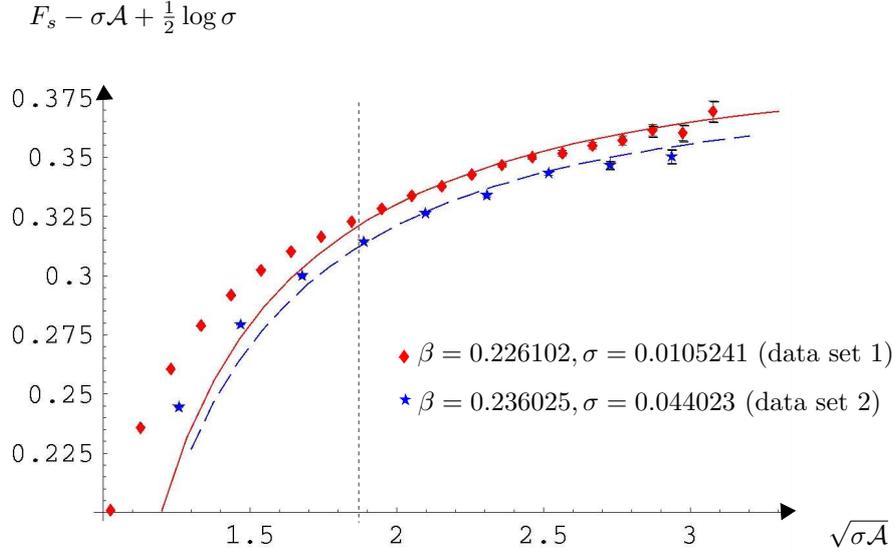

Having fixed $u=1$ and having specified $N$, the free energy $F$ of \eq{Fc0} is just a function of $\sigma\mathcal{A}$, which we can fit to the MC points
in the two data sets referred to above.

The results of this comparison are summarized in Table \ref{tab:fits}. 
For each data set, we consider only the data corresponding to lattice sizes $L\geq L_{\mathrm{min}}$,
\emph{i.e.} to values of $\sqrt{\sigma\mathcal{A}}\geq (\sqrt{\sigma\mathcal{A}})_{\mathrm{min}}$, 
and fix the value of the normalization constant $\mathcal{N}$ so as to minimize the $\chi^2$; we repeat
the analysis for various values of $L_{\mathrm{min}}$. Also, we consider the case where we
truncate the sum over the oscillator number at $N=100$ and the case $N=0$ where we keep only the zero-mode 
contributions. 

For both data sets 1 and 2, the reduced $\chi^2$ becomes of order one when
we consider in the fit only values of $\sqrt{\sigma\mathcal{A}}\gtrsim 2$. This is therefore the region 
in which our theoretical expression, derived from the Nambu-Goto model, describes well the data. 
This pattern is fully consistent with previous analysis of other observables in the 3d Ising models. In particular, the case of Polyakov loop correlators was  discussed in
\cite{chp05}, where two different
behaviours were observed.  In the compact direction corresponding to the ``temperature", a good agreement with  the Nambu-Goto predictions was found
exactly in the same range of distances as in the present case. On the contrary,
in the direction with Dirichlet boundary conditions a clear disagreement 
was observed up to very large distances. This is probably due to some type of \emph{boundary effect}
associated to the Dirichlet boundary conditions. It is exactly the absence of this type of effects which makes the case that we study, periodic in both directions and hence free of boundary corrections, particularly well suited to study the effective string contributions, disentangled from other
spurious effects.

Considering the values of the $\chi^2$, we can take as our best estimates for the normalization constant 
$\mathcal{N}$ the value obtained at $L_{\mathrm{min}}=21$ for the data set 1, namely $\mathcal{N}=0.9184(3)$, 
and the value at $L_{\mathrm{min}}=11$ for data set 2, i.e. $\mathcal{N}=0.9082(5)$. It is
interesting to observe that truncating to $N=0$ the mode expansion gives
comparable values for the $\chi^2$ but leads to a sizeable change (with respect
to the statistical errors) of $\mathcal{N}$ which becomes
$\mathcal{N}=0.9134(3)$ and $\mathcal{N}=0.9035(5)$ for the data sets 1 and 2 respectively.
This difference is due to the fact that the truncation of the sum at $N=0$ replaces the area-independent 
1-loop determinant $\eta^{4-2d}(\ii u)$ appearing in the exact partition function \eq{iexp4} with 
its approximation $\exp(\pi u (d-2)/6)$. This can be seen by keeping the $m=0,k=0$ term only in \eq{bos21},
looking at the exponential behaviour $\exp(-\sigma\mathcal{A}\cale)$ and expanding $\cale$
to the first order, see Eq.s (\ref{Xexp}-\ref{d1}). In our situation, where $d=3$ and $u=1$, the 1-loop determinant (which is essentially reproduced summing the contributions up to $N=100$) contributes 
to the free energy a constant term $2 \log\left[\eta(\ii)\right] = - 0.527344...$ ; in the $N=0$ truncation, this constant becomes
$-\pi/6 = -0.523599$. The difference between these two values is $0.00374536...$, and  accounts for most of the difference between the estimates of the overall constant $\mathcal{N}$ at $N=100$ and $N=0$ given above.

It is instructive to compare our estimates for the overall constant with those obtained in \cite{nuovomc} by truncating the perturbative expansion in $1/\sigma \mathcal{A}$ to the second
order. Keeping into account the difference in the definition of the normalizations, which 
simply amounts to the contribution $2 \log\left[\eta^2(\ii)\right]$ of the one-loop determinant, we
would obtain, using the notations of~\cite{nuovomc},
$c_0+\frac12\log{\sigma}=0.3910(3)$ for the data set 1 and $N=100$, 
which should be compared with the value $c_0+\frac12\log{\sigma}=0.388(5)$. The estimates are compatible
within the errors. Since the fit we performed depends on one free parameter only, our estimate for
$c_0+\frac12\log{\sigma}$ turns out to be more precise than the one of
\cite{nuovomc}. As discussed in~\cite{nuovomc} the result we obtain is of the
order of magnitude but somehow larger than the prediction obtained in
~\cite{hp97,m90} in the framework of the $\phi^4$ theory:
$c_0+\frac12\log{\sigma}\sim0.29$.

To make the above discussion visually appreciable, in Figure \ref{fig:fit} we plot our theoretical curve 
against the MC points, having fixed the value of the 
normalization $\mathcal{N}$ according to the fit at $L_{\mathrm{min}}=21$ for the data set 1 and at 
$L_{\mathrm{min}}=11$ for data set 2 and having truncated the sum over the oscillators at $N=100$ in both cases. 
In fact, to draw a readable plot, it is convenient to subtract from both the Monte Carlo data and the 
theoretical prediction for the free energy the dominant scaling term $\sigma\mathcal{A}$
and a constant%
\footnote{This term is related to the natural appearance of a multiplicative factor
$\sigma^{\frac{d-2}{2}}$ in the effective string partition function, see \eq{bos21}. 
Furthermore we prefer to exhibit the same data plots appearing in \cite{nuovomc}.} 
term $-\frac 12 \log\sigma$. 

In Table 9 of ref. \cite{nuovomc} some data regarding rectangular lattices, i.e., 
with $u>1$ were also presented. Such data represent an important test of the form of our expression \ref{Fc0}. We present the 
results of this test in Table \ref{tab:rect}; they are quite encouraging, since we basically get 
agreement with the data within the
error bars, at least when including the contributions of higher oscillator numbers (up to $N=100$). 
Notice that in this test we use the value of the normalization constant $\mathcal{N}$ already determined 
by fitting it to the data for squared lattices of data set 2, see Table \ref{tab:fits}, since the data
for asymmetric lattices in ref. \cite{nuovomc} were obtained at exactly the same value of the coupling constant $\beta$ used for the data set 2. This makes the observed agreement even more remarkable, since it involved  no adjustable free parameter.

\begin{table}[tb]
\begin{center}
\begin{tabular}{cccclll}
\hline
\hline 
$L_1$ & $L_2$ & $\sqrt{\sigma\mathcal{A}}$ & $u$ &  \multicolumn{1}{c}{$F_s$} & \multicolumn{1}{c}{diff $(N=100)$} 
& \multicolumn{1}{c}{diff $(N=0)$} \\
\hline 
$10$ & $12$ & $2.29843$ & $6/5$  & $\phantom{1}7.1670(6) $ &  $\phantom{-}0.0016$ & $0.0049$\\ 
$10$ & $15$ & $2.56972$ & $3/2$  & $\phantom{1}8.4449(12)$ &  $          -0.0004$ & $0.0042$\\ 
$10$ & $18$ & $2.81498$ & $9/5$  & $\phantom{1}9.6976(17)$ &  $          -0.0009$ & $0.0037$\\ 
$10$ & $20$ & $2.96725$ & $2$    & $          10.5235(25)$ &  $          -0.0012$ & $0.0035$\\ 
$10$ & $22$ & $3.11208$ & $11/5$ & $          11.3466(36)$ &  $\phantom{-}0.0017$ & $0.0064$\\ 
\hline
\end{tabular}
\end{center}
\caption{\small Comparison between the data presented in \cite{nuovomc} for rectangular lattices and our 
predictions.
The data were extracted at $\beta=0.236025$, corresponding to $\sigma=0.044023$, \emph{i.e.} in the same situation as for the square lattices considered in data set 2 considered above. 
In the last columns we give the MC values  of the 
free energy $F_s$, according to Table 9 of \cite{nuovomc}, and the difference between these values and the 
values obtained from our expression \eq{Fc0}, with the sum over the oscillator number truncated at $N=100$ 
or at $N=0$. 
The normalization constant $\mathcal{N}$ was already determined by
the fit to the data set 2 presented in Table \ref{tab:fits}; we chose here the value obtained setting 
$L_{\mathrm{min}}=11$.}
\label{tab:rect}
\end{table}

One remarkable feature of the comparisons presented here in Tables \ref{tab:fits} and \ref{tab:rect} is 
that keeping just the zero modes of the string, \emph{i.e.} setting $N=0$,
seems to yield a very good agreement with the data, in particular for symmetric interfaces. 
The contribution of non-zero modes modifies the overall normalization coming from the one-loop determinant, 
as discussed above, but it is much less important for higher loops, so that the shape of the free energy 
as a function of $\sigma\mathcal{A}$ is little changed. 

Keeping only the zero-modes, the quantity $\cale$ of \eq{bos19} which appears in the partition function
reduces effectively to
\begin{equation}
\label{Xzm}
\widehat \cale = \sqrt{1 - \frac{\pi u (d-2)}{3\sigma\mathcal{A}}}
\end{equation}
and the partition function \eq{bos21} simply becomes 
\begin{equation}
\label{boszero}
\widehat{\mathcal{I}}^{(d)} = 2
\left(\frac{\sigma}{2\pi}\right)^{\frac{d-2}{2}}\, V_T \, \sqrt{\sigma\mathcal{A}u} 
\left(\frac{\widehat \cale}{u}\right)^{\frac{d-1}{2}}\, K_{\frac{d-1}{2}}\left(\sigma\mathcal{A} \widehat \cale \right)~.
\end{equation}

\section{A simple model for interfaces in 2d}
\label{sec:model}
We want now consider interfaces in 2d and describe them by means of a simple model.

\begin{figure}
\begin{center}
\begin{picture}(0,0)%
\includegraphics{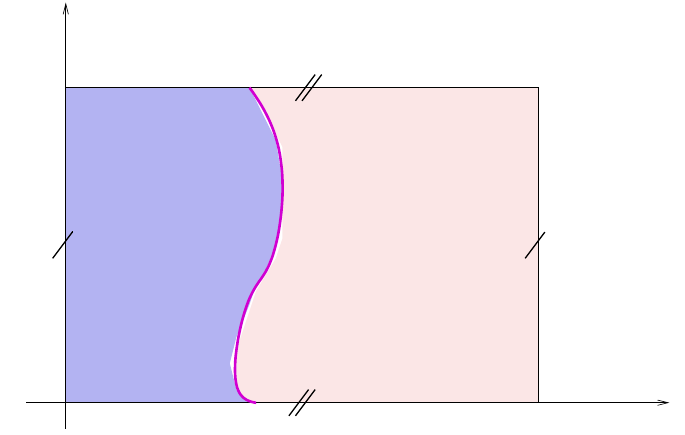}%
\end{picture}%
\setlength{\unitlength}{1658sp}%
\begingroup\makeatletter\ifx\SetFigFont\undefined%
\gdef\SetFigFont#1#2#3#4#5{%
  \reset@font\fontsize{#1}{#2pt}%
  \fontfamily{#3}\fontseries{#4}\fontshape{#5}%
  \selectfont}%
\fi\endgroup%
\begin{picture}(7811,4945)(451,-4919)
\put(6151,-4861){\makebox(0,0)[lb]{\smash{{\SetFigFont{5}{6.0}{\familydefault}{\mddefault}{\updefault}$L_1\equiv L$}}}}
\put(7801,-4861){\makebox(0,0)[lb]{\smash{{\SetFigFont{5}{6.0}{\familydefault}{\mddefault}{\updefault}$x_1$}}}}
\put(676,-136){\makebox(0,0)[lb]{\smash{{\SetFigFont{5}{6.0}{\familydefault}{\mddefault}{\updefault}$x_1$}}}}
\put(451,-961){\makebox(0,0)[lb]{\smash{{\SetFigFont{5}{6.0}{\familydefault}{\mddefault}{\updefault}$L_2=R$}}}}
\put(3751,-2986){\makebox(0,0)[lb]{\smash{{\SetFigFont{5}{6.0}{\familydefault}{\mddefault}{\updefault}{$\Gamma$}%
}}}}
\end{picture}%
\end{center}
 \caption{An interface (the line $\Gamma$) separating two phases on a two-dimensional torus.}
\label{fig:int}
\end{figure}
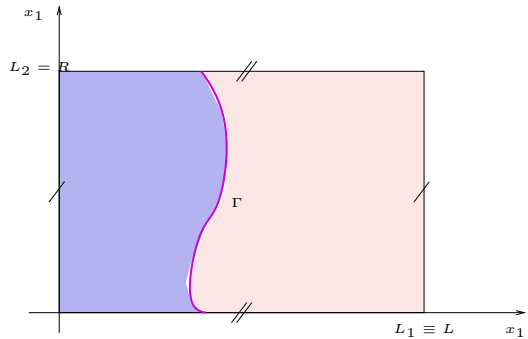

Consider indeed a physical system, defined on a two-dimensional space with periodic
boundary conditions in both directions, in which two different phases, separated
by a one-dimensional interface, may coexist (see Fig. \ref{fig:int}).

We make the assumption that the effective action is just given by
the length of the interface $\Gamma$:
\begin{equation}
 \label{sel}
S[x] = m \int_\Gamma ds = m \int_0^1 d\tau \sqrt{(\dot x})^2~,
\end{equation}
where $\tau$ parametrizes the curve $\Gamma$, the dots denotes $\tau$
derivatives,  $(\dot x)^2 \equiv \dot x_i \dot x^i$ and $m$ is a constant%
\footnote{We can view $m$ as the ``mass'' of
a relativistic particle in an Euclidean target space (the torus)
whose world-line is the interface.}
with the dimension of an inverse length.
This is just the analogue of the conjecture which, in three dimensions, leads
to the CWM \cite{BLS,rw82}.

The corresponding partition function is given by the functional integral
\begin{equation}
 \label{pfx}
Z = \int Dx\, \ee^{-S[x]}~,
\end{equation}
and it should correspond to the interface free energy, in analogy to what
happens in the stringy description
of 3d interfaces, as previously seen.

The action \eq{sel}, as well known, admits a first-order reformulation:
\begin{equation}
 \label{sel1}
S[x,e] = \frac{m}{2}\int_0^1 d\tau \left(\frac{(\dot x)ì2}{e} + e\right)~,
 \end{equation}
where $e(\tau)$ is the einbein%
\footnote{Namely one has $ds^2 = e^2(\tau)d\tau^2$},
so that the partition function can be written as
\begin{equation}
 \label{pfxe}
Z = \int De\, Dx\, \ee^{-S[x,e]}~.
\end{equation}

The action \eq{sel1} is invariant under reparametrizations of $\Gamma$:
\begin{equation}
 \label{rep}
\begin{aligned}
\delta\tau & = \epsilon~,\\
\delta e & = \dot\epsilon e + \epsilon \dot e~.
\end{aligned}
\end{equation}
We can gauge-fix this invariance%
\footnote{Notice that in the first-order formulation
we have to take into account the constraints which follow from the variation of
$S$ w.r.t. the einbein
$e$, which (in the chosen gauge) are simply $(\dot x)^2 = \lambda$.}
 by choosing $e(\tau) = \lambda = \mathrm{const}$, so that
the action becomes
\begin{equation}
 \label{acxl}
S[x,e] \to \frac{m\lambda}{2} + \frac{m}{2\lambda}\int_0^1 d\tau (\dot x)^2~.
\end{equation}
The constant $\lambda$ represents the length of the path
and is a \emph{Teichm\"uller parameter}: it cannot be changed using the
reparametrizations
in \eq{rep}. As such, it must be integrated over in the partition function,
which now
reads
\begin{equation}
 \label{pfxl}
Z = \int_0^\infty d\lambda\, \ee^{-\frac{m\lambda}{2}}
\,Z_x\,
Z_{\mathrm{gh}}~.
\end{equation}
Here
\begin{equation}
 \label{Zx}
Z_x = \int Dx\, \exp\left\{-\frac{m}{2\lambda} \int_0^1 d\tau (\dot x)^2\right\}
\end{equation}
and we took into account the Faddeev-Popov ghosts $b,c$ for the chosen
gauge-fixing of the invariances in \eq{rep}.
For them we have, by standard treatment,
\begin{equation}
 \label{zgh}
Z_{\mathrm{gh}} = \int Db\, Dc \, \exp\left\{-\lambda\int_0^1 d\tau\,
b\partial_\tau c\right\}~.
\end{equation}
With the chosen periodic boundary conditions on $\tau$, it turns out 
that
\begin{equation}
 \label{Zghl}
Z_{\mathrm{gh}}= \frac{1}{\lambda}~.
\end{equation}

The partition function $Z_x$ is simply that of a non-relativistic particle of
mass
$\mu = m/\lambda$, whose Hamiltonian is
\begin{equation}
 \label{Hx}
H = \frac{p^2}{2\mu}~.
\end{equation}
Since the particle moves on a two-dimensional torus of sides $L_1,L_2$, the
components of
the momentum are quantized:
\begin{equation}
 \label{qm}
p^i = \frac{2\pi n^i}{L_i}~,
\end{equation}
for $i=1,2$, with $n^i\in\mathbb{Z}$. The partition function of this system is
\begin{equation}
 \label{pfn}
\sum_{\{n_i\}} \exp\left\{-\frac{4\pi^2}{2\mu}\left(\frac{n_1^2}{L_1^2} +
\frac{n_2^2}{L_2^2} \right)\right\} =
\frac{L_1 L_2\,\mu}{2\pi}
\sum_{\{m_i\}} \exp\left\{-\frac{\mu^2}{2}\left(L_1^2 n_1^2 +
L_2^2 n_2^2 \right)\right\}~.
\end{equation}
In the second step above we performed a Poisson re-summation, which reorganizes
the sum in contributions
of sectors describing quantum fluctuations around classical solutions of
wrapping numbers $m_i$.

To describe an interface such as the one depicted in Fig. \ref{fig:int}, we set
\begin{equation}
 \label{setm}
m_1 = 0~,~~~~ m_2 = 1~,
\end{equation}
getting therefore (if we remember that $\mu = m/\lambda$)
\begin{equation}
 \label{Zxms}
Z_x = L_1 L_2 \frac{m}{2\pi\lambda} \exp\left\{-\frac{m}{2\lambda}
L_2^2\right\}~.
\end{equation}

Inserting \eq{Zghl} and \eq{Zxms} into the expression \eq{pfxl} of the total
partition function
we obtain finally%
\footnote{We use the integral
\begin{equation}
 \label{K1int}
\int_0^\infty \frac{d\lambda}{\lambda^\alpha}\ee^{-A^2\lambda -
\frac{B^2}{\lambda}} = 2 \left(\frac{A}{B}\right)^{\alpha-1} K_{\alpha-1}(2AB)~.
\end{equation}
}
\begin{equation}
 \label{Zxfe}
Z = L_1 L_2 \frac{m}{2\pi}\int_0^\infty
\frac{d\lambda}{\lambda^2}\exp\left\{-\frac{m}{2}\lambda -
\frac{m L_2^2}{2}\frac{1}{\lambda}\right\} \\
= \frac{L_1 m}{\pi} K_1(m L_2)~,
\end{equation}
where $K_1$ is the modified Bessel function of first order.

In the next section, we will consider a concrete physical system where the
interface free energy can be computed directly, namely the 2d Ising
model, and we will find that our proposed general expression \eq{Zxfe} does
indeed capture its dominant behaviour.

\section{Interfaces in the 2d Ising model}
\label{sec:ising}
Interfaces can be realized in the context of spin models,
since under peculiar boundary conditions their broken symmetry phase
allows separated but coexisting vacua of different magnetization.
Being interested in interfaces in 2d systems, it is natural to consider
the 2d Ising model, where an explicit analytical treatment is viable.

To evaluate the interface free energy of the 2d Ising model in a simple way, we
can take advantage of the fact that the continuum limit of the Ising model
is described by the Quantum Field Theory  of a free fermionic field of mass $m$.
This framework allows the calculation of the partition function with any
type of boundary conditions. Being interested in estimating the
interface free energy, we will impose periodic boundary conditions
in one direction and antiperiodic ones in the other direction. Explicit
expressions of various partition functions can be found in~\cite{km90} and
(using the $\zeta$ function regularization) in~\cite{is87}\footnote{Notice
however that in~\cite{is87} there is a sign mistake which was
later corrected in~\cite{km90}.}. It's worth pointing out that the solution
proposed in~\cite{km90} is nothing else than the continuum limit of Kaufmann
solution~\cite{k49}. Such continuum limit was performed by Ferdinand and Fischer
in~\cite{ff69} at the critical point and the result of~\cite{km90} is
essentially the extension of~\cite{ff69} outside the critical point.

Using the notations of \cite{km90}, eq.s (91)-(105), the partition function of a
fermion on a rectangular torus of sizes $L_1$ and $L_2$, which for
notational simplicity we shall henceforth denote as $L$ and $R$, is given
by the so-called \emph{massive fermion determinant}:
\beq
\label{fermdet}
D_{\alpha,\beta}(m|L,R)=
\ee^{-\delta_\beta \pi L c_\beta(r)/6R} \prod_{n\in \mathbb{Z}+\beta}
(1-\delta_\alpha e^{-L\epsilon_n(r)/R})~.
\eeq
Here  $\alpha,\beta$ can take the values $0, 1/2$ and label the
boundary conditions in the $L$ and $R$ directions respectively:
the value $0$ corresponds to periodic b.c.s, $1/2$ to antiperiodic ones.
Moreover, $\delta_\alpha=e^{2\pi i \alpha}$ and
\beq
\frac{\epsilon_n(r)}{R}=\sqrt{m^2+(\frac{2\pi n}{R})^2}~.
\eeq
The adimensional variable $r\equiv mR$ sets the scale of the theory.

Depending on the boundary conditions, the coefficient $c_{\beta}(r)$ can
assume the following values:
\beq
c_{\frac12}(r)=\frac{6r}{\pi^2}\sum_{k=1}^\infty \frac{(-1)^{k-1}}{k}K_1(kr)
\eeq
or
\beq
c_{0}(r)=\frac{6r}{\pi^2}\sum_{k=1}^\infty \frac{1}{k}K_1(kr)~.
\eeq

The partition function of the Ising model with periodic boundary conditions in
both directions can be written as follows:
\beq
\label{Zp}
Z_{\mathrm{Ising}}(m|L,R)=
\frac12 \left( D_{\frac12,\frac12}+D_{0,\frac12}+D_{\frac12,0}-D_{0,0
} \right)~.
\eeq
If the $L$ direction is antiperiodic, the partition function becomes
\beq
\label{Zap}
Z_{\mathrm{Ising,ap}}(m|L,R)=
\frac12 \left( D_{\frac12,\frac12}+D_{0,\frac12}
-D_ { \frac12 , 0 } -D_ { 0 ,0}\right)~.
\eeq
This is the situation which generates an odd number of interfaces along the $R$
direction. The interface free energy will be given by the
following expression:
\beq
\label{Fising}
\ee^{-F_{\mathrm{interface}}} \equiv Z_{\mathrm{interface}}
= \frac{Z_{\mathrm{Ising,ap}}}{Z_{\mathrm{Ising}}}~.
\eeq

Notice, as a side remark, that taking the limit%
\footnote{In this regime, one must keep in mind that the parameter $\tau$ of
\cite{ff69} corresponds to $r/2$.}
$r\to 0$, one moves to the critical point and all the expressions written above 
flow toward the conformal invariant ones and agree with the standard CFT results
 and with those reported in the  pioneering work of
 Ferdinand and Fisher \cite{ff69}.

We are now interested in the large $mL$ and $mR$ limit of eq.(\ref{Fising}).
The large $mL$ limit allows us to neglect the infinite product in
eq.(\ref{fermdet}) while in the large $mR$ we approximate
\beq
c_{\frac12}(r)\sim c_0(r)\sim \frac{6r}{\pi^2}K_1(r)~.
\eeq
The  partition functions in eq.(\ref{Zp}) and (\ref{Zap}) simplify and we
end up with
\beq
\begin{aligned}
Z_{\mathrm{Ising}} & \sim 1~,
\\
Z_{\mathrm{Ising,ap}} & \sim \frac{\pi L}{6R} \left( c_{\frac12}(r) + c_{0}(r)
\right)~,
\end{aligned}
\eeq
from which we find
\beq
\label{Zisint}
Z_{\mathrm{interface}}\sim \frac{\pi L}{6R} ( c_{\frac12}(r) + c_{0}(r) )
\sim\frac{2 L r}{\pi R} K_1(r) =\frac{2 L m}{\pi } K_1(mR)~.
\eeq
By comparing this result to \eq{Zxfe}, we see that the behaviour of the
interface free energy in the 2d Ising model for large scales is perfectly
captured by our simple effective model.

\section{Dimensional reduction of the NG effective description of 
interfaces in 3d}
\label{sec:dimrid}
In this section we want to discuss the behaviour of the Nambu-Goto effective
string description of
fluctuating interfaces when dimensional reduction occurs along
one of the two lattice sizes which define the interface.

As we discussed above, interfaces can be realized in spin
models. In this realization, the physical ideas underlying the
dimensional reduction process are perhaps best discussed. Focusing on the 3d
Ising model, let us recall that it is mapped by duality into the 3d Ising gauge
model. Under this mapping, the interface free energy becomes a close
relative of the Wilson loop expectation value of the 3d gauge Ising model, the
only difference being in the boundary conditions of the two observable: fixed in
the Wilson loop case and periodic in the interface case.

In lattice gauge theories, \emph{dimensional reduction} has a very important physical
meaning. The size of the lattice direction which is reduced is interpreted as
the inverse temperature of the system and the point in which dimensional
reduction occurs coincides with the \emph{deconfinement phase transition}. The
behaviour of the model in the vicinity of the deconfinement transition is well
described, according to the Svetitsky-Yaffe conjecture \cite{sy82}, by the 2d Ising
spin model. More generally, for other gauge groups with continuous
deconfinement phase transitions, it is described by the 2d spin model with
global symmetry group the center of the original gauge group.

As a consequence of this observation, we expect that the interface free
energy of the 3d Ising model
 should smoothly map into the 2d one when the dimensional reduction scale
is approached from above, while it should vanish below it, since in the dual
model this is the deconfined phase. Any reliable effective description of the
free energy should therefore be compatible with these two requirements. This is
a rather non trivial test for the effective string models which are expected to
effectively describe both the Wilson and Polyakov loop expectation value and the
interface free energy.

We start from the expression for the interface
free energy in 3d obtained in Section \ref{sec:bos_string}  and consider its behaviour
under dimensional reduction; we
find that it does indeed smoothly reduce to the 2d expression given here in
\eq{Zxfe}. This represents a test of our effective approach to the interface
free energy both in three and in two dimensions.

In the next section, we will refine this test by comparing both the 3d and the
2d theoretical expressions to Monte Carlo data for asymmetric interfaces: the
more asymmetric are the interfaces, the more we expect the prediction of our 3d
and 2d expression to become compatible and reliable.

Let us consider a three-dimensional torus of sides $L_{1,2,3}$, and assume that
the interface lies in the plane orthogonal to $L_1$.
Our starting point is the expression for the interface partition function in
the Nambu-Goto approximation, Eq. (\ref{boskkp}), where we set $d=3$:
\begin{equation}
\label{I3}
\mathcal{I}^{(3)} = 2 \left(\frac{\sigma}{2\pi}\right)^ {\frac{1}{2}}\, L_1 \,
\sqrt{\sigma\mathcal{A}u}
\sum_{k,k'=0}^\infty  c_k c_{k'}
\left(\frac{\cale}{u}\right)  K_{1}\left(\sigma\mathcal{A}\cale\right)~,
\end{equation}
where now
\begin{equation}
\mathcal{A}=L_3 L_2~,~u=\frac{L_2}{L_3}
\end{equation}
and remember that
\begin{equation}
\cale  =
\sqrt{1 + \frac{4\pi\, u}{\sigma \mathcal{A}}(k+k'-\frac{1}{12}) +
\frac{4\pi^{2} \, u^2\, (k-k')^{2}}{(\sigma\mathcal{A})^2}}
\end{equation}
is a function of the occupation numbers $k$ and $k'$.

The dimensional reduction of \eq{I3} can be most easily discussed by writing the
interface free energy as a function of $L_3$. 
Let us introduce again the \emph{energy levels} 
\beq
E_{k,k'} \equiv\sigma L_3 \cale.
\eeq 
With this choice the argument of the Bessel functions in
eq.(\ref{I3}) becomes $L_2 E_{k,k'}$ and we have:
\begin{equation}
\label{spectrum1}
E_{k,k'} = \sqrt{\sigma^2 L_3^2 + 4\pi\sigma (k+k'-\frac{1}{12}) +
\frac{4\pi^{2}\, (k-k')^{2}}{L_{3}^{2}}}~.
\end{equation}
From this expression it is evident that, as the ratio $L_2/L_3$ increases, the
separation between consecutive levels in the spectrum becomes larger and larger
and, in particular, the gap between the lowest state $k=k'=0$ and the second one
increases.

In the dimensional reduction limit $L_3\ll L_2$ we can thus truncate the sum
in eq.(\ref{I3}) to the first term. In this limit, the argument of the
Bessel function in eq.(\ref{I3})
becomes large,  allowing us to use the asymptotic expansion
\begin{equation}
\label{bessel}
K_j(z) \sim \sqrt{\frac{\pi}{2z}}e^{-z}\left\{1+O\left(\frac{1}{z}\right)\right\},
\end{equation}
from which we see that higher states are exponentially suppressed with respect
to the $k=k'=0$ one. Setting $k=k'=0$, we get
\begin{equation}
\label{I3_00}
\mathcal{I}^{(3)}_{(0,0)} = 2 \left(\frac{\sigma}{2\pi}\right)^ {\frac{1}{2}}\,
L_1 \, \sqrt{\sigma\mathcal{A}u}
\left(\frac{\cale}{u}\right)  K_{1}\left(\sigma\mathcal{A}\cale\right)~,
\end{equation}
where now
\begin{equation}
\label{cale00}
\cale  =
\sqrt{1 - \frac{\pi}{3\sigma \mathcal{A}}u}=\sqrt{1 - \frac{\pi}{3\sigma L_3^{2}}}~.
\end{equation}
Equation (\ref{I3_00}) can therefore be rewritten as:
\begin{equation}
\mathcal{I}^{(3)}_ {(0,0)} =  \left(\frac{2}{\pi}\right)^ {\frac{1}{2}}\, L_1 \, \sigma L_3
 \sqrt{1 - \frac{\pi}{3\sigma L_3^{2}}} K_{1}\left(\sigma L_3 L_2\sqrt{1 - \frac{\pi}{3\sigma
 L_3^{2}}}\right)~.
\end{equation}
If we now  define
\begin{equation}
\label{meff}
m_{\mathrm{eff}} = \sigma L_3 \sqrt{1 - \frac{\pi}{3\sigma L_3^{2}}}
\end{equation}
we find
\begin{equation}
\mathcal{I}^{(3)}_ {(0,0)} =  \left(\frac{2}{\pi}\right)^ {\frac{1}{2}}\, L_1
\, m_{\mathrm{eff}} K_{1}\left(m_{\mathrm{eff}} L_2\right)~,
\end{equation}
which is in perfect agreement with the partition function given by (\ref{Zxfe})
a part from the normalization constant.

It is clear from \eq{cale00} that the above discussion is consistent
only for
\begin{equation}
L_3~\geq~\sqrt{\frac{\pi}{3\sigma}}~.
\end{equation}
In the string language, this bound represents the tachyonic singularity of the
bosonic string: for $L_3~\leq~\sqrt{\frac{\pi}{3\sigma}}$ the lowest state has a
negative energy. In the 3d Ising model framework, the critical value
\begin{equation}
\label{lc}
 L_{3,\mathrm{c}}= \sqrt{\frac{\pi}{3\sigma}}=1.023...
\times\frac{1}{\sqrt{\sigma}}
\end{equation}
at which the mass of the lowest state vanishes can be considered as the
effective string prediction for the dimensional reduction scale which we were
looking for. We thus see that this scale is naturally embedded in the NG formulation of
the interface
free energy. The same argument in the dual case of the Polyakov loop correlator
leads to the identification of this scale with deconfinement temperature
\cite{olesen}.
The value obtained in this way for the dimensional reduction
scale (or, equivalently, for the deconfinement temperature in the dual model) is
in rather good agreement with the Monte Carlo estimates. For the 3d gauge Ising
model the disagreement is only of about $20\%$, as the Monte Carlo estimate
turns out to be $\sqrt{\sigma}L_c=0.8186(16)$~\cite{ch96}  instead of 1.023...)
and it further decreases if one looks at $SU(N)$ pure Yang Mills theories.

It is important to stress that the mapping between 3d and 2d observables which
emerges from this comparison is exactly the same which is found considering the
finite temperature behaviour of the Polyakov loop correlators in the vicinity 
of the deconfinement transition in the 
3d gauge Ising model. In particular, the
relation between the mass scale of the 2d model and the string tension (in the
present case the interface tension) of the 3d one reported in eq. (\ref{meff})
is the same which one finds in the Polyakov loop case. This is a
rather non trivial consistency check of the whole dimensional reduction picture.

As a last remark let us stress that the smooth flow of the 3d Nambu-Goto result
towards the 2d ones under dimensional reduction has various important
implications on our understanding of the Nambu-Goto effective action. Recall
that the treatment leading to eq. (\ref{I3}) neglected the conformal anomaly,
\emph{i.e.} equivalently, did not take into account the \emph{Liouville field} which does not
decouple in the Polyakov formulation of the string when $d\not=26$. Most
probably
the smooth flow toward the 2d result occurs exactly because we neglected the
Liouville field contribution. It is thus interesting to compare simultaneously
the Nambu-Goto expression (without Liouville field), the 2d result presented
here and the output of the Montecarlo simulations for the free energy of
interfaces on lattices with a value of $L_3$ small enough to make the 2d
approximation a reliable one. We shall devote the next section to this
comparison.

\section{Comparison with the numerical data}
\label{sec:num}
Following the above discussion we compare in this section 
the expressions for the 2d interface free energy given 
in \eq{Zxfe} and for the 3d free energy \eq{I3}, with a set of high-precision Monte Carlo data
\cite{chp07} obtained in a 3d Ising model in which interfaces are created by
imposing  antiperiodic boundary conditions in one of the three lattice
directions. 

In particular we isolated two sets of data. The first one is reported in Tab.
\ref{tab16} and displays the values of the free energies obtained in the 3d
Ising model at $\beta = 0.223101$. The interface string tension for this value
of $\beta$ is $\sigma = 0.0026083$ and the dimensional reduction scale is
exactly $16$ lattice spacings~\cite{ch96}. The data  correspond to values of the
$L_3$ size ranging from a minimum value of $3/2$ the dimensional reduction scale
up to $3$ times.

In the table we report, from left to right, the lattice sizes $L_{1,2,3}$ , the
numerical estimate $F_{\mathrm{num}}$ for the free energy (with its statistical
error), the theoretical estimate $F$ according to \eq{I3}, the estimate
$F_{(0,0)}$ obtained in the full dimensional reduction limit by truncating
\eq{I3} to $k=k'=0$, which corresponds to our 2d formula \eq{Zxfe},
and the estimate $F_{1\mathrm{st}}$ in which also the first excited states,
$k+k'=1$, are kept into account. Finally, as a comparison, in the last two
columns we report the one loop and two loop perturbative approximations to the
whole NG action discussed, for instance, in \cite{df83,cfghpv94}.

In Tab. \ref{tab8} we report the same set of data evaluated at $\beta =
0.226102$, corresponding to $\sigma = 0.0105254$ and to a dimensional reduction
scale of $8$ lattice spacings~\cite{ch96}.

\begin{table}
\begin{small}
\begin{center}
\begin{tabular}{cccccccccc}
\hline
$L_3$ & $L_2$ & $L_1$ & $F_{\mathrm{num}}$ & $F$ & $F_{(0,0)}$ &
$F_{1\mathrm{st}}$ & $F_{\mathrm{1-loop}}$ & $F_{\mathrm{2-loop}}$ \\
\hline
24 & 64 & 96 & 6.8855(20) & 6.7495 & 6.7495 & 6.7495 & 6.9974 & 6.8347\\
28 & 64 & 96 & 7.6929(21) & 7.6537 & 7.6537 & 7.6537 & 7.7875 & 7.6821\\
32 & 64 & 96 & 8.4626(20) & 8.4518 & 8.4518 & 8.4518 & 8.5380 & 8.4632\\
36 & 64 & 96 & 9.1996(21) & 9.2012 & 9.2012 & 9.2012 & 9.2632 & 9.2062\\
40 & 64 & 96 & 9.9227(23) & 9.9231 & 9.9232 & 9.9231 & 9.9713 & 9.9253\\
44 & 64 & 96 & 10.6203(23) & 10.6278 & 10.6280 & 10.6278 & 10.6674 & 10.6288\\
48 & 64 & 96 & 11.3138(25) & 11.3209 & 11.3214 & 11.3209 & 11.3548 & 11.3213\\
\hline
\end{tabular}
\end{center}
\end{small}
\caption{Numerical results of the free energy in the 3d
Ising model at $\beta = 0.223101$ are compared to various theoretical
estimates. See the main text for detailed explanations.}
\label{tab16}
\end{table}

\begin{table}
\begin{small}
\begin{center}
\begin{tabular}{cccccccccc}
\hline
$L_3$ & $L_2$ & $L_1$ & $F_{\mathrm{num}}$ & $F$ & $F_{(0,0)}$ &
$F_{1\mathrm{st}}$ & $F_{\mathrm{1-loop}}$ & $F_{\mathrm{2-loop}}$ \\
\hline
24 & 64 & 96 & 18.4131(26) & 18.4121 & 18.4121 & 18.4121 & 18.4555 & 18.4152\\
28 & 64 & 96 & 21.2414(27) & 21.2450 & 21.2450 &  21.245 & 21.2724 & 21.2463\\
32 & 64 & 96 & 24.0310(27) & 24.0306 & 24.0306 & 24.0306 & 24.0497 & 24.0312\\
36 & 64 & 96 & 26.7859(28) & 26.7873 & 26.7873 &  26.7873& 26.8017 & 26.7875\\
40 & 64 & 96 & 29.5271(30) & 29.5250 & 29.5251 & 29.5250 & 29.5365 & 29.5251\\
44 & 64 & 96 & 32.2449(31) & 32.2498 & 32.2500 & 32.2498 & 32.2594 & 32.2498\\
48 & 64 & 96 & 34.9623(33) & 34.9653 & 34.9657 & 34.9653 & 34.9736 & 34.9653\\
\hline
\end{tabular}
\end{center}
\end{small}
\caption{Numerical results of the free energy in the 3d
Ising model at $\beta = 0.226102$ are compared to various theoretical
estimates. See the main text for detailed explanations.}
\label{tab8}
\end{table}

Looking at these tables one can see that the 2d expression $F_{(0,0)}$ always
gives a very good approximation of the whole NG result $F$: the difference
between the two is always smaller than the statistical error on the numerical
data. One can also notice that the 2d approximation is more reliable when the
dimensional reduction scale $\sqrt{\pi/(3\sigma)}$ is smaller, \emph{i.e.} in Tab.
\ref{tab8}, and that within each data set it becomes slightly worse as $L_3$
increases. Looking at the seventh column we see that the deviation from the full
NG result is completely due to the first excited state.

It is indeed easy to evaluate the gap $\Delta$ between the lowest state and the
first excitation which appears at the exponent of the asymptotic expansion of
the Bessel function in \eq{I3}. In the large $\sigma L_3^2$ limit, the first
excited states correspond to $k=1$, $k'=0$ or $k=0$, $k'=1$ and
$\Delta$ takes the value $2\pi L_2/L_3$. When  $\pi/3 \leqslant \sigma
L_3^2 \leqslant \pi$ the first excited state is instead the one with $k=1$,
$k'=1$ and the variation of the gap is
\beq
\sqrt{\frac{8}{3}}\pi\frac{L_2}{L_3} \geqslant \Delta
\geqslant \pi\frac{L_2}{L_3}~.
\eeq
We see that the dominating effect is due to the asymmetry ratio $L_2/L_3$
(which in our data ranges from $8/3$ to $4/3$) and that a subdominant r\^ole is
played by the $\sigma L_3^2$ combination, in complete agreement with the results
reported in Table \ref{tab16} and \ref{tab8}.

Given this agreement between 2d and whole NG results it is very interesting to
compare them both to the Montecarlo data. The agreement is rather good for large
values of $F$, see in particular the results reported in Tab. \ref{tab8}. As $F$
decreases, however, the NG predictions compare less favorably to the data, see
Table \ref{tab16}. As a matter of fact, in this regime the Monte Carlo data
seem to better agree with the two loop perturbative expansion than with the
whole NG result (and its 2d approximation). This fact has already been discussed
in \cite{chp07} and is most probably due to the fact that the two loop result is
actually consistent also at the quantum level, in the sense that at this order
the contribution due to the Liouville field vanishes
\cite{ps91,Luscher:2004ib,d04,hdm06}. 

As expected, these problems  disappear
if one works exactly in two dimensions, and indeed the direct calculation of
the interface partition function in the 2d Ising model reported in sec.
\ref{sec:ising} shows that the expression \eq{Zxfe} obtained in section
\ref{sec:model} should be valid to all orders.

\chapter{Conclusions on Part II}
\label{concl1}

In the second part of this thesis, we studied the free energy of interfaces in a compact $d$-dimensional target space $T^d$, modeling them with a Nambu-Goto string effective action; this basically corresponds, for the $d=3$ case,
to use the capillary wave model. Performing standard covariant quantization and explicitly integrating over the modular parameter of the world-sheet, we were able to derive an exact expression for the sector of the NG partition function that describes the interface fluctuations.\\
Focusing on the three-dimensional case, we compared the exact predictions of the NG model with the free energy of interfaces obtained in \cite{nuovomc} through very precise Monte Carlo simulations in the Ising model. 
The Monte Carlo data are very well accounted for when we consider lattices of sufficiently large sizes, 
typically with $L\geq 2/\sqrt{\sigma}$; here $\sigma$ is the string tension, which can be associated with great accuracy to the value of the Ising coupling \cite{Caselle:2004jq}.\\
Our analysis, which allows to compare the data with the exact theoretical prediction,
confirms and strengthens the evidence that the NG model is a very reliable effective model for such 
lattice sizes. Previously,
comparisons were made to the two-loop expansion of the NG model \cite{Dietz:1982uc}, which already gives a very good approximation for square interfaces; as pointed out in \cite{nuovomc}, however, and as confirmed here, higher order corrections are more important for asymmetric lattices. Indeed we found
extremely good agreement of the exact expression (which re-sums all the loop corrections) for the few data with $u\not=1$ presented in \cite{nuovomc}.\\

For smaller lattice sizes, rather large deviations from the NG predictions are observed in the data. This
pattern is in perfect agreement with what has been observed in previous studies of the NG predictions for other observables, such as the Polyakov loop correlator \cite{chp05}, and is in fact to be expected for deep
theoretical motivations. \\
Indeed, we have studied the Nambu-Goto bosonic string in $d$ dimensions 
paying no attention to the fact that 
in a consistent quantum treatment, when $d\not= 26$ further degrees of freedom have to be taken into account, beside the $d-2$ transverse oscillations. In the Polyakov first order formulation \cite{Polyakov:1981rd}, 
the scale $\ee^\phi$ of the world-sheet metric $h$ does not decouple, and gets in fact a Liouville-type action; in the Virasoro treatment, longitudinal modes of the string enter the game \cite{Goddard:1972iy,Brower:1972wj} (the two effects are  
not disconnected, see \cite{Marnelius:1986wp}).
For this reason these quantum bosonic string models have not been considered in the past 
\cite{Polchinski:1992vg,Polchinski:1991ax} 
as viable dual descriptions of situations, such as the confining regime of gauge theories or the fluid interfaces, where only transverse fluctuations are expected on physical grounds.\\
In a modern perspective, the extra degree of freedom corresponding to the Liouville mode could play somehow the r\^ole of the renormalization scale \cite{Polyakov:1997tj,Polyakov:1998ju}, in the spirit of (a non-conformal version of) AdS/CFT duality \cite{Maldacena:1997re}. 
Some effort to take into account the contributions to the partition function
of the extra d.o.f. arising in $d=26$ at the quantum level, and to see whether they can account for the deviations of data, at small lattice sizes, from the prediction obtained via the na\"ive treatment of the NG model, has been done in \cite{Dalley:2005fw}, focusing on the relevance of the longitudinal modes for the description of large $N$ gauge theories.\\
In \cite{Polchinski:1991ax} Polchinski and Strominger put forward a different proposal to circumvent the 
mismatch of d.o.f. between the bosonic string model and the transverse oscillations. An effective string was built, with an action analogous to the Polyakov one, where the independent scale 
$\ee^\phi$ of the metric gets replaced by the induced metric. This model is certainly very interesting, and it is receiving nowadays a renewed attention \cite{Drummond:2004yp,Kuti:2005xg}. It is noticeable that
the excitation spectra obtained within this model agree with the na\"ive NG ones up to two loops in the 
$1/(\sigma\mathcal{A})$ expansion; the prediction of this model would therefore nicely agree with NG in the region where the latter is valid. Higher corrections, whose computation does not appear to be simple,
could in principle better explain the data for smaller lattices.\\
Then we addressed the description of some universal properties of
interfaces in two-dimensional physical systems. Using a simple model in which
one assumes an action proportional to the length of the interface, in analogy
with the 3d capillary wave model, we proposed a general expression for the
interface free energy. This expression agrees with an exact calculation in the
case of the 2d Ising model. We also shew that our 2d result represents the
dimensional reduction of the interface 3d free energy obtained using the
capillary wave model, which is tantamount to the Nambu-Goto effective string
description, when the effects of the conformal anomaly are neglected. Finally,
we compared the 3d and 2d theoretical predictions to Monte Carlo numerical
estimates for the interface free energy in the case of the 3d Ising model.
In the range of values that we studied the difference between the 3d and 2d
estimates is always smaller than the statistical error of the numerical data.
Moreover we were able to show that this difference is completely due to the
first excited state of the Nambu-Goto action.\\
In conclusion, there is a renewed interest in the problem of identifying the correct QCD string dual and of testing the various proposals, see for instance
\cite{Juge:2004xr,Drummond:2004yp,Brower:2005pb,Brower:2005cd,Kuti:2005xg,Dalley:2005fw}. As already said, 
our contribution corroborates
the ``experimental'' result that the na\"ive treatment of the NG model provides a very accurate description of fluctuating surfaces for sufficiently large sizes, and that, for smaller lattices, clear deviations appear  which should be accounted for
by the correct dual model (at least down to some lower scale where the string description itself breaks down). 
On the theoretical side, our derivation of the exact partition function, and not just of its perturbative expansion, can offer some insight for a similar treatment of consistent models including also the extra d.o.f. arising in $d\not=26$, or of the Polchinski-Strominger string.

\appendix

 \chapter{Remarks on flux compactifications}
 \label{appendix}

\section{Notations and conventions}
\label{app:conventions}

We use the following notations for space-time indices in the real basis:
\begin{itemize}
\item
$d=10$ vector indices: $M,N, ... \in \{0, ..., 9\}$;
\item
$d=4$ vector indices: $\mu,\nu, ... \in \{0, ..., 3\}$;
\item
$d=6$ vector indices: $m,n, ... \in \{4, ..., 9\}$.
\end{itemize}
The corresponding complex indices are denoted by  $I, J=1, ..., 5$ with $I=i=1,2,3$ referring
to the coordinates of the six-dimensional space and $I=4,5$ referring to the four-dimensional
space-time directions. Even with Euclidean signatures we use the real index 0.

\paragraph{$\Gamma$-matrices in ten dimensions}
In a $d=10$ Euclidean space the $\Gamma$ matrices which satisfy
$\acomm{\Gamma^M}{\Gamma^N} = 2\delta^{MN}$
can be given the following explicit representation in terms of the Pauli matrices $\tau^c$:
\begin{equation}
\label{gm10}
\begin{aligned}
\Gamma^0 & = \tau^1 \otimes ~\mathbf{1} \otimes ~\mathbf{1} \otimes ~\mathbf{1} \otimes ~\mathbf{1}
\\
\Gamma^1 & = \tau^2 \otimes ~\mathbf{1} \otimes ~\mathbf{1} ~\otimes \mathbf{1}~ \otimes \mathbf{1}
\\
\Gamma^2 & = \tau^3 \otimes \tau^1 \otimes ~\mathbf{1} \otimes ~\mathbf{1} \otimes ~\mathbf{1}
\\
\Gamma^3 & = \tau^3 \otimes \tau^2  \otimes ~\mathbf{1} \otimes ~\mathbf{1} \otimes ~\mathbf{1}
\\
& ~~~~~~~~~~~~\vdots
\\
\Gamma^{8} & =  \tau^3 \otimes \tau^3 \otimes \tau^3 \otimes \tau^3  \otimes \tau^1
\\
\Gamma^{9} & = \tau^3 \otimes \tau^3 \otimes \tau^3 \otimes \tau^3  \otimes \tau^2
\end{aligned}
\end{equation}
The charge conjugation matrix $C$ satisfies
\begin{equation}
\label{Cdef}
C  \Gamma^M C^{-1} = - (\Gamma^M)^t\quad\mbox{with}\quad
C^t = - C~,
\end{equation}
and in the above representation is
\begin{equation}
\label{ccm2n}
C = \tau^2 \otimes \tau^1 \otimes \tau^2 \otimes \tau^1 \otimes \tau^2 ~.
\end{equation}
The charge conjugation matrix is used to raise and lower the 32-dimensional spinor indices
(${\mathcal{\widehat A}}, {\mathcal {\widehat B}},\ldots$) of the
$\Gamma$-matrices according to
\begin{equation}
(\Gamma^{M})^{{\mathcal{\widehat A}}{\mathcal {\widehat B}}} \,\equiv\,
(\Gamma^{M})^{\mathcal{\widehat A}}_{~\,\,\mathcal{\widehat C}} \,\,
(C^{-1})^{{\mathcal {\widehat C}}{\mathcal {\widehat B}}}\quad\mbox{and}
\quad
(\Gamma^{M})_{{\mathcal {\widehat A}}{\mathcal {\widehat B}}} \,\equiv\,
(C)_{{\mathcal {\widehat A}}{\mathcal {\widehat C}}} \,\, (\Gamma^{M})^{\mathcal{
\widehat C}}_{~\,\,\mathcal {\widehat B}} ~~.
\end{equation}
The chirality matrix is defined by
\begin{equation}
\label{chir2n}
 \Gamma_{(11)}^{\mathrm E}= -\ii \, \Gamma^0 \Gamma^1 \ldots \Gamma^{9}
=\tau^3 \otimes \tau^3 \otimes \tau^3 \otimes \tau^3 \otimes \tau^3 ~.
\end{equation}

The above expressions are useful to obtain the factorization of the $d=10$ matrices
when the ten-dimensional space is splitted into $4+6$. In fact, by writing
\begin{equation}
\Gamma^\mu =\gamma^\mu \otimes \mathbf{1}~~,~~
\Gamma^m =\gamma_{(5)} \otimes \ \gamma^m~~,~~
\Gamma_{(11)}^{\mathrm E} =\gamma_{(5)} \otimes \gamma_{(7)}~~,~~
C =C_4\otimes C_6~,
\label{split}
\end{equation}
we can read off the explicit representation of the Dirac matrices $\gamma^\mu$ and
$\gamma^m$ for $d=4$ and $d=6$, respectively, of the corresponding chirality matrices
$\gamma_{(5)}$ and $\gamma_{(7)}$, and of the charge conjugation matrices
$C_4$ and $C_6$.

\paragraph{$\Gamma$-matrices in four dimensions} The $d=4$ matrices
$\gamma^\mu$ which can be read from Eqs. (\ref{gm10}) and (\ref{split}) are
\begin{equation}
\label{gm4}
\gamma^0 = \tau^1 \otimes \mathbf{1}~~,~~
\gamma^1 = \tau^2 \otimes \mathbf{1}~~,~~
\gamma^2 = \tau^3 \otimes \tau^1~~,~~
\gamma^3 = \tau^3 \otimes \tau^2~,
\end{equation}
while the chirality and charge conjugation matrices are
\begin{equation}
\label{gamma5}
\gamma_{(5)} = -\, \gamma^0 \gamma^1 \gamma^2 \gamma^3= \tau^3\otimes\tau^3
\quad\mbox{and}\quad
C_4=\tau^2\otimes\tau^1~.
\end{equation}
In this tensor product basis the four spinor indices are ordered as
\begin{center}
\begin{tabular}{c|c}
$2 \,\vec{\epsilon}$ & chirality \\
\hline
$(++)$ & $+$ \\
$(-+)$ & $-$ \\
$(+-)$ & $-$ \\
$(--)$ & $+$ \\
\end{tabular}
\end{center}
However, it is often useful to rearrange them in order to have first the two chiral indices
$\alpha \in \{(++),(--)\}$ and then the two anti-chiral ones $\dot\alpha\in\{(-+),(+-)\}$, in
such a way that the chirality matrix takes the more conventional form
$\gamma_{(5)} =\mathbf{1}\otimes \tau^3$.
With such a rearrangement the above Euclidean Dirac matrices $\gamma^\mu$ become
\begin{equation}
\label{gamma4def}
\gamma^\mu =
\begin{pmatrix}
0 & \sigma^\mu \\
\overline \sigma^\mu & 0
\end{pmatrix}
\end{equation}
with $\sigma^\mu = \big( \mathbf{1},\, -\ii \tau^3,\, \ii \tau^2,\, -\ii \tau^1\big)$
and $\overline\sigma^\mu  =  \big( \mathbf{1},\, \ii \tau^3,\, -\ii \tau^2,\, \ii \tau^1\big)$.
The matrices $\big(\sigma^\mu\big)_{\alpha\dot\beta}$ and
$\big(\overline\sigma^\mu\big)^{\dot\alpha\beta}$ act on spinors of definite chirality $\psi_{\alpha}$ and $\psi^{\dot\alpha}$ as
\begin{equation}
\big(\sigma^\mu\big)_{\alpha\dot\beta} \,\psi^{\dot\beta}\quad\mbox{and}\quad
\big(\overline\sigma^\mu\big)^{\dot\alpha\beta} \,\psi_{\beta}~.
\end{equation}
After the rearrangement of indices, the charge conjugation matrix becomes
\begin{equation}
C_4= \tau^2 \otimes \tau^3~;
\label{c44}
\end{equation}
thus it is block diagonal with $\big(C_4\big)^{\alpha\beta}=- \ii \,\epsilon^{\alpha\beta}$ and
$\big(C_4\big)_{\dot\alpha\dot\beta}= \ii \, \epsilon_{\dot\alpha\dot\beta}$.

\paragraph{$\Gamma$-matrices in six dimensions}
The $d=6$ matrices
$\gamma^m$ which can be read from Eqs. (\ref{gm10}) and (\ref{split}) are
\begin{equation}
\begin{aligned}
&\gamma^4 = \tau^1 \otimes \mathbf{1} \otimes \mathbf{1} ~~,~~~
\gamma^5 = \tau^2 \otimes \mathbf{1} \otimes \mathbf{1}  ~~,~~~~
\gamma^6 = \tau^3 \otimes \tau^1 \otimes \mathbf{1}\\
&\gamma^7 = \tau^3 \otimes \tau^2 \otimes \mathbf{1} ~~,~~
\gamma^8 = \tau^3 \otimes \tau^3 \otimes \tau^1 ~~,~~
\gamma^9 = \tau^3 \otimes \tau^3 \otimes \tau^2
\end{aligned}
\label{gm2}
\end{equation}
while the corresponding chirality and charge conjugation matrices are
\begin{equation}
\label{gamma7}
\gamma_{(7)} = \ii\, \gamma^4 \gamma^5 \ldots \gamma^9\,=\,
\tau^3 \otimes \tau^3 \otimes \tau^3\quad\mbox{and}\quad
C_6= \tau^2 \otimes \tau^1 \otimes \tau^2~.
\end{equation}
In this case the eight spinor indices are ordered according to
\begin{equation}
\begin{tabular}{c|c}
\label{stati}
$2\vec{\epsilon}$ & chirality \\
\hline
$(+++)$ & $ + $ \\
$(-++)$ & $ - $ \\
$(+-+)$ & $ - $ \\
$(--+)$  & $ + $ \\
$(++-)$  & $ - $ \\
$(-+-)$  & $ + $ \\
$(+--)$  & $ + $ \\
$(---)$  & $ - $
\end{tabular}
\end{equation}
but again they can be rearranged in such a way to put first the chiral ones and then the anti-chiral
ones, and have the chirality matrix in the standard
form $\gamma_{(7)}= \mathbf{1} \otimes \mathbf{1} \otimes \tau^3$. In this basis the
matrices $\gamma^m\,C_6^{-1}$ may be written in the block diagonal form
\begin{equation}
\label{gamma6cdef}
\gamma^m \, C_6^{-1}=
\begin{pmatrix}
\Sigma^m & 0 \\
 0 & \overline\Sigma^{\,m}
\end{pmatrix}
\end{equation}
where $(\Sigma^m)^{AB}$ and $(\overline\Sigma^{\,m})_{AB}$ are $4\times 4$ anti-symmetric matrices.

If we order the four chiral indices as $\{(+++),(+--),(-+-),(--+)\}$ and the four anti-chiral indices
as $\{(---),(-++),(+-+),(++-)\}$ (see also \eq{spintransf} below), we have
\begin{equation}
\begin{aligned}
\Sigma^m &= \big(\eta^3, -\ii\overline\eta^3,\eta^2,
 -\ii\overline\eta^2,\eta^1,\ii\overline\eta^1\big)~,
\\
\overline\Sigma^{\,m} &= \big(\eta^3,\ii\overline\eta^3,-\eta^2,
-\ii\overline\eta^2,\eta^1,-\ii\overline\eta^1\big)~,
\end{aligned}
\label{Sigma}
\end{equation}
where $\eta^c$ and $\overline\eta^c$ are, respectively, the self-dual and anti-self-dual 't Hooft
symbols. Proceeding in a similar way for the antisimmetrized product of three matrices, we find
\begin{equation}
\label{g36cdef}
\gamma^{mnp} \, C_6^{-1}=
\begin{pmatrix}
\Sigma^{mnp} & 0 \\
0 &  \overline\Sigma^{\,mnp}
\end{pmatrix}~,
\end{equation}
where $(\Sigma^{mnp})^{AB}$ and $(\overline\Sigma^{\,mnp})_{AB}$ the $4\times 4$ symmetric matrices
that appear in Section \ref{sec:effects}. Using the properties of the chirality and charge
conjugation matrices, it is easy to show the following imaginary self-duality properties
\begin{equation}
*_6\Sigma^{mnp} = -\ii\,\Sigma^{mnp}\quad,\quad
*_6\overline\Sigma^{mnp} = +\ii\,\overline\Sigma^{mnp}~.
\label{sigmaSD1}
\end{equation}

\paragraph{Useful formulas}
The previous formulas allow us to obtain the explicit expressions for the fermion bilinears which have been discussed in Sections \ref{sec:CFT} and \ref{sec:effects}.
In this respect we point out that in writing a fermion bilinear, like the one appearing for instance
in Eq.(\ref{ampltot2}), we always understand the inverse charge conjugation matrix $C^{-1}$.
The precise expression of the bilinear is then
\begin{equation}
\label{th}
\Theta\Gamma^{mnp}\Theta \equiv
\Theta_{\cal A}\, (\Gamma^{mnp}C^{-1})^{{\cal A}{\cal B}} \,\Theta_{\cal B}~.
\end{equation}
Using the 4+6 decomposition discussed above, we obtain
\begin{equation}
\Gamma^{mnp}C^{-1} = \big(\gamma_{(5)} C_4^{-1} \big)\otimes \big(\gamma^{mnp}C_6^{-1}\big)
=
\begin{pmatrix}
\tau^2 &  0 \\
0 & \tau^2
 \end{pmatrix}
\otimes
\begin{pmatrix}
\Sigma^{mnp} & 0 \\
0 &  \overline\Sigma^{\,mnp}
\end{pmatrix}~,
\end{equation}
so that Eq. (\ref{th}) can be rewritten as
\begin{equation}
\label{th1}
\Theta_{\cal A}\, (\Gamma^{mnp}C^{-1})^{{\cal A}{\cal B}} \,\Theta_{\cal B} =
- \ii\,\Theta^{\alpha  A}\,\epsilon_{\alpha \beta} \Theta^{\beta B}
(\overline\Sigma^{\,mnp})_{AB} \,- \ii \,
\Theta_{\dot \alpha  A}\,\epsilon^{\dot \alpha \dot \beta} \Theta_{\dot \beta B}
(\Sigma^{mnp})^{AB}
\end{equation}
which coincides with Eq. (\ref{decomp}).

Finally, we observe that the natural ordering (\ref{stati}) of the spinor indices
for the tensor product representation (\ref{gm2}) is particularly convenient if one
uses the complex basis in the internal six-dimensional space. Indeed, computing the holomorphic
and anti-holomorphic products $\gamma^{123}$ and $\gamma^{\bar 1 \bar 2 \bar 3}$ and combining
them with the charge conjugation matrix we find
\begin{equation}
\begin{aligned}
\gamma^{123} \,C_6^{-1} &= -\,\begin{pmatrix}
0\phantom 0&  0 \\
0\phantom 0 & 1
 \end{pmatrix}
\otimes
\begin{pmatrix}
0\phantom 0&  0 \\
0\phantom 0 & 1
 \end{pmatrix}
\otimes\begin{pmatrix}
0\phantom 0&  0 \\
0\phantom 0 & 1
 \end{pmatrix}
\\
\gamma^{\bar 1\bar 2 \bar 3} \,C_6^{-1} &= +\,\begin{pmatrix}
1\phantom 0&  0\\
0\phantom 0 & 0
 \end{pmatrix}
\otimes
\begin{pmatrix}
1\phantom 0&  0 \\
0\phantom 0 & 0
 \end{pmatrix}
\otimes\begin{pmatrix}
1\phantom 0&  0 \\
0\phantom 0& 0
 \end{pmatrix}
\end{aligned}
\end{equation}
from which we immediately see that $\Sigma^{123}=\overline\Sigma^{\,\bar 1\bar 2\bar3}=0$ and that
the only non-vanishing entries of the matrices $\Sigma^{\bar 1\bar 2 \bar 3}$ and $\overline\Sigma^{\,123}$ are, respectively, the upper most left and the lower most right, that is
\begin{equation}
\big(\Sigma^{\bar 1\bar 2 \bar 3})^{+++,+++}=1\quad\mbox{and}\quad
\big(\overline\Sigma^{\,123}\big)_{---,---}=-1~.
\label{sigma00}
\end{equation}

\section{The orbifold $\mathcal{T}_6/(\mathbb{Z}_2\times\mathbb{Z}_2)$}
\label{subapp:T6orb} The orbifold group
$\mathbb{Z}_2\times\mathbb{Z}_2$ acting on the orthonormal complex
coordinates $z^i$ of $\mathcal{T}_6$ as in Table \ref{tablez2z2} is
a discrete subgroup of $\mathrm{SO}(6)$ that contains 4 elements
$h_I$ ($I=0,1,2,3$), with $h^0\equiv e$ being the identity element,
and
\begin{equation}
\label{frac2bis} h^1 = \ee^{\ii\pi (J_3 - J_2)}~,~~~~ h^2 =
\ee^{\ii\pi (J_1 - J_3)}~,~~~~ h^3\equiv h_1 h_2 = \ee^{\ii\pi (J_1
- J_2)}
\end{equation}
where $J_{1,2,3}$ are the generators of rotations in the 4-5, 6-7
and 8-9 planes respectively. We may summarize the transformation
properties  for the conformal fields $\partial Z^i$ and $\Psi^i$
($i=1,2,3$) in the Neveu-Schwarz sector by means of the following
table:
\begin{equation}
\label{ZPsirep}
\begin{tabular}{c|c}
conf. field & irrep \\
\hline
$\Big.\partial Z^i$, $\Psi^i$ & $R_i$
\end{tabular}~,
\end{equation}
where $\{R_A\}=\{R_0,R_i\}$ are the irreducible representations of $\mathbb{Z}_2\times
\mathbb{Z}_2$, identified by writing the character table $\ch_A^I = \tr_{R_A}\big(h^I\big)$
of the group
\begin{equation}
\label{frac3}
\begin{tabular}{c|cccc}
 & $e$ & $h_1$ & $h_2$ & $h_3$ \\
\hline
$R_0$ & 1 & ~1 & ~1 & ~1 \\
$R_1$ & 1 & ~1 & $-1$ & $-1$ \\
$R_2$ & 1 & $-1$ & ~1 & $-1$ \\
$R_3$ & 1 & $-1$ & $-1$ & ~1
\end{tabular}
\end{equation}
The Clebsch-Gordan series for these representations is simply given by
\begin{equation}
\label{frac4}
R_0\otimes R_A = R_A~~~~,~~~~
R_i\otimes R_j = \delta_{ij}  R_0 + |\epsilon_{ijk}| R_k~~,
\end{equation}
and is crucial in determining the open string spectrum.

Recall that through the bosonization procedure \cite{Friedan:1985ge}
the chiral spin fields $S^A \sim \ee^{\ii
\vec{\epsilon}^{\,A}\cdot\vec\varphi}$ of $\mathrm{SO}(6)$
and the anti-chiral ones $S_{A} \sim \ee^{\ii
\vec{\epsilon}_{\,A}\cdot\vec\varphi}$ are associated respectively to the
$\mathrm{SO}(6)$ spinor weights
$\vec{\epsilon}^{\,A}= \frac 12(\pm,\pm,\pm)$ with the product of signs being positive,
and $\vec{\epsilon}_{\,A}= \frac 12(\pm,\pm,\pm)$ with the product of signs being negative.
Using this information, we easily deduce from (\ref{frac2bis}) the
transformation properties of the various spin fields, which are
summarized in the following table
\begin{equation}
\label{spintransf}
\begin{tabular}{c|c|c}
irrep $R_A$ & $S^A$ &  $S_A$  \\
\hline
$R_0$ &  $S^0 \equiv S^{+++}$
& $ \Big.S_{0} \equiv S_{---}$
\\
$R_1$ & $S^1 \equiv S^{+--}$
& $\Big.S_{1} \equiv S_{-++}$
\\
$R_2$ & $S^2\equiv S^{-+-}$
& $\Big. S_{2}\equiv S_{+-+}$
\\
$R_3$ &  $S^3\equiv S^{--+}$
& $\Big.S_{3}\equiv S_{++-}$
\end{tabular}
\end{equation}
In other words, we can order the internal spinor indices so that
$S^A$ and $S_{A}$ transform in the irrep $R_A$.

Closed strings on the orbifold have different sectors. The untwisted
sector simply contains  the closed string states defined on the
covering space $\mathcal{T}_6$ which are invariant under the
orbifold action. The twisted sectors are in correspondence with the
16 fixed planes ($a=1,\ldots,16$) of the action of a nontrivial
element $h^i$. The vertex operators in a twisted sector contain
left- and right-moving twist fields $\Delta^i_a(z)$ and
$\tilde\Delta^i_a(z)$, and must be invariant under the orbifold. If,
as explained in the main text, we assume that the orbifold group
does not act on the twist fields, it is not difficult to write down
all the massless vertices.

Let us also notice that the orbifold projection leaves only two bulk supercharges, whose Weyl components in the $-1/2$ picture are
\begin{equation}
 \label{sclm}
Q_{\alpha} = \oint \frac{dz}{2\pi\ii}\, S_\alpha S_0 \,\ee^{-\frac{\phi}{2}}(z)~,~~~
Q^{\dot\alpha} = \oint \frac{dz}{2\pi\ii} \,S^{\dot\alpha} S^{0}\, \ee^{-\frac{\phi}{2}}(z)
\end{equation}
for the left-moving ones, and
\begin{equation}
 \label{scrm}
{\tilde Q}_{\alpha} = \oint \frac{d\bar z}{2\pi\ii}\, {\tilde S}_\alpha {\tilde S}_0
\,\ee^{-\frac{\tilde\phi}{2}}(z)~,~~~
{\tilde Q}^{\dot\alpha} = \oint \frac{d\bar z}{2\pi\ii} \,{\tilde S}^{\dot\alpha} {\tilde S}^{0} \,\ee^{-\frac{\tilde\phi}{2}}(z)
\end{equation}
for the right-moving ones.

\section{Soft supersymmetry breaking on fractional D9 branes}
\label{subsec:fD9}
In the orbifold $\mathcal T_6/{\mathbb Z_2}\times {\mathbb Z_2}$, we can realize an $\mathcal{N}=1$ $d=4$ gauge theory using fractional D9 branes that completely wrap the internal compact space.
Such brane configuration preserves a different  $\mathcal{N}=1$ supersymmetry
with respect to the fractional D3 branes considered in Section \ref{sec:n1int}, and thus
the moduli fields organize into chiral multiplets with respect to this new supersymmetry.
The bulk Lagrangian (which is in fact the same since we have not changed the compactification manifold) can be rewritten in terms of these multiplets via a different
K\"ahler potential and superpotential. Again, this allows to relate the flux-induced gaugino mass term to the value of the auxiliary component of the gauge kinetic function.

For the reasons already explained in the case of D3 branes, we are interested mostly
in the untwisted couplings, which can be deduced by reducing to four
dimensions the usual DBI-WZ action of a D9 brane
on $\mathcal{T}_6/(\mathbb{Z}_2\times\mathbb{Z}_2)$ given by
\begin{equation}
 \label{D9BI}
-T_9 \int_{D3}\int_{\mathcal{T}_6/(\mathbb{Z}_2\times\mathbb{Z}_2)}
  \ee^{-\varphi}\sqrt{-{\rm det}(G_{(10)} + \mathcal{F} )} + T_9
\int_{D3}\int_{\mathcal{T}_6/(\mathbb{Z}_2\times\mathbb{Z}_2)}
\sum_{n=0}^5 C_{2n} \,\ee^{  \mathcal{F} }
\end{equation}
where $T_9 = (4\pi)^{-1}
(2\pi\alpha')^{-2}(2\pi\sqrt{\alpha'})^{-6}$. From this expression
it follows that the quadratic part in the gauge fields $F$, after
promoting the latter to the non-abelian case and switching to the
Einstein frame, is
\begin{equation}
 \label{YMD9}
-\frac{\ee^{\varphi/2}\mathcal{V}}{32\pi}\int_{D_3} d^4x \,{\mathrm{Tr}} \big(F_{\mu\nu} F^{\mu\nu}
\big) + \frac{\widetilde{C}}{32\pi}\int_{D_3} d^4x\,
{\mathrm{Tr}} \big(F_{\mu\nu}{}^*F^{\mu\nu}\big)~,
\end{equation}
where $\widetilde C$ is%
\footnote{This pseudoscalar modulus is in fact related, through the duality between $F_7$ and $F_3$, to
the two-form $C_2$ with indices in the space-time directions. Notice also that
\begin{equation*}
 \label{intg6}
\int_{\mathcal{T}_6/(\mathbb{Z}_2\times\mathbb{Z}_2)}\!\!\! d^6y \sqrt{-g_{(6)}} =
\frac 14\, (2\pi\sqrt{\alpha'})^6\, \ee^{3\varphi/2}\mathcal{V}
\end{equation*}
since it corresponds to the internal volume in the string frame, which is related by the factor of $\ee^{3\phi/2}$
to the Einstein frame volume $\mathcal{V}$. This explains the prefactor in \eq{YMD9}.}
\begin{equation}
 \label{defct}
\int_{\mathcal{T}_6/(\mathbb{Z}_2\times\mathbb{Z}_2)} C_6 =
\frac 14 (2\pi\sqrt{\alpha'})^{-6}\, \tilde{C}~.
\end{equation}
Comparing with \eq{biwzf}, we see that the untwisted part of the gauge kinetic function $f_A^{(9)}$
for any type $A$ of fractional D9 branes reads
\begin{equation}
 \label{gkf9}
f_A^{(9)}=\frac{s}{4}\quad\mbox{with}\quad
s = \tilde{C} + \ii \ee^{\varphi/2}\mathcal{V}~.
\end{equation}
In a similar way one can consider D5 branes wrapped on untwisted
cycles $e^i$, which preserve the same supersymmetry of the D9
branes. The way to combine the  untwisted moduli into chiral
multiplets is again suggested by the gauge kinetic functions
$f^{(5)}_i$. Extracting the quadratic terms in the gauge fields from
the wrapped D5-brane DBI-WZ action, it is straightforward to obtain
\begin{equation}
 \label{f5i}
f^{(5)}_i = \frac 14 r^i \quad\mbox{with}\quad r^i \equiv c^i + \ii
\ee^{-\varphi/2}\,v^i~.
\end{equation}
Notice that $c^i$ and $v^i$  are invariant under the O9 orientifold
projection appropriate to our situation.

By supersymmetry, the complex scalars  represents the lowest
component of a chiral superfield, and the supersymmetric
$\mathcal{N}=1$ Lagrangian (\ref{gkf}), contains the coupling of the
gaugino to the auxiliary components of this multiplet. For D9-branes
\begin{equation}
\label{gaugaux9}
-\frac{\ii}{32\pi} F^s\,\Tr \big(\Lambda^\alpha \Lambda_\alpha\big) + \cc
\end{equation}
This has to be compared to the gaugino mass term which follows from the D9 flux coupling
indicated in Table \ref{Dbranes}. To do so we have to adapt the steps
used to arrive at \eq{massD3} in the D3 case,
since now the normalization $c_F$ contains the topological factor $\mathcal C_{(10)}$
suitable for D9 disk amplitude, namely \cite{Billo:2002hm}
\begin{equation}
\mathcal C_{(10)} = \frac{\mathcal C_{(4)}}{(2\pi\sqrt{\alpha'})^6}~.
\label{c10}
\end{equation}
On the other hand to obtain the four dimensional couplings, we have to
dimensionally reduce on $\mathcal{T}_6/(\mathbb{Z}_2\times\mathbb{Z}_2)$, gaining a factor of
$ (2\pi\sqrt{\alpha'})^6\,\ee^{3\varphi/2}\,\mathcal{V}$, to obtain the 4d coupling. In the end, we find the term
\begin{equation}
 \label{gc9}
-\frac{\ii}{4\pi}\,\ee^{\varphi/2}\,\mathcal{V}\,(2\pi\alpha')^{-\frac12}\,\mathcal{N}_F\, F_{(3,0)}
=-\frac{\ii}{4\pi}\,\ee^{\varphi}\, F_{(3,0)}
\end{equation}
where in the second step we used the normalization of the flux vertex already fixed in \eq{NFvalue}
so that by comparison with \eq{gaugaux9} the auxiliary field must be given by
\begin{equation}
\label{Fs}
F^s = 8 \,\ee^{\varphi}\, F_{(3,0)}~.
\end{equation}

Analogously to what we did in the D3-brane case, we restrict to the
slice of moduli space spanned by $s$ and by the overall scale
\begin{equation}
 \label{runico}
r \equiv r^1 = r^2 = r^3~;
\end{equation}
this scale is related to the volume by $(\im r)^3 =
\ee^{-3\varphi/2}\,\mathcal{V}$, as it follows from \eq{vol}. In
this slice of the moduli space the bulk theory can be rewritten in
the standard $\mathcal{N}=1$ form employing the chiral fields
$s(\theta)$ and $r(\theta)$. The K\"ahler potential reads
\begin{equation}
 \label{Kahlersr}
K = - \log(\im s) - 3 \log (\im r)~,
\end{equation}
and it coincides with \eq{KP} re-expressed in the new set of variables. The superpotential is
given by
\begin{equation}
 \label{suppotrs}
W = \frac{1}{\kappa_{10}^2}\int F\wedge \Omega~,
\end{equation}
where, with respect to \eq{WGO},  the O9 projection eliminates the NS-NS flux.
It is straightforward to check that
\begin{equation}
 \label{FWs}
{\overline F}^{\,\bar s} = -\ii \kappa_4^2\,\ee^{K/2} K^{\bar{s} s} D_{s} W
= 8 \,\ee^{\varphi}\, F_{(0,3)}~,
\end{equation}
in agreement with \eq{Fs}.

\section{Vertex operators for gauge fields and instanton moduli}
\label{app:vert}

In this section we list the vertex operators of the various fields and moduli
of our model, including their normalizations which we express 
in terms of the unit of length $(2\pi\alpha')^{\frac{1}{2}}$. 
In our conventions the vertex operators are always dimensionless and, in general, we
assign canonical dimensions to their polarizations, namely dimensions of (length)$^{-1}$ 
to bosonic fields and dimensions of (length)$^{-{3}/{2}}$ to fermionic ones. However,
in the instanton sector, some of the ADHM moduli acquire different dimensions as indicated
in Tab. \ref{dimensions}. The vertex operators we list in the following are written using the
open string fields appropriate for the D3/D$(-1)$ system, namely the space-time bosonic and fermionic string coordinates $X^\mu$ and $\psi^\mu$, the transverse bosonic and fermionic string coordinates in the complex basis $Z^I$ and $\Psi^I$, the space-time spin-fields $S_\alpha$ and $S^{\dot\alpha}$, 
and the internal ones $S_A$ and $S^A$. Moreover we denote by $\phi$ the bosonic field of
the superghost system. For more details we refer to \cite{Billo:2002hm,Billo:2007py}.

\paragraph{Vertices for gauge fields}
The gauge fields originate from D3/D3 strings. They include a gauge vector $A_\mu$ and a gaugino
$\Lambda^\alpha$ with its conjugate $\bar\Lambda_{\dot\alpha}$ transforming in the adjoint representation, plus bi-fundamental matter fields.
The corresponding vertices at momentum $p$ are
\begin{equation}
\label{gaugevert}
\begin{aligned}
V_A & = (\pi \alpha')^{\frac{1}{2}} \left(A_\mu\right)^{u_A}_{~v_A}\, 
\psi^\mu \ee^{-\phi}\,\ee^{\ii p\cdot X}~,\\
V_{\Lambda}& =  (2 \pi \alpha')^{\frac{3}{4}} 
\left(\Lambda^\alpha\right)^{u_A}_{~v_A}\, S_\alpha S_0 \ee^{-\phi/2}\,\ee^{\ii p\cdot X}~,\\
V_{\bar\Lambda} & = (2 \pi \alpha')^{\frac{3}{4}} \left(\bar\Lambda_{\dot\alpha}\right)^{u_A}_{~v_A}\, S^{\dot\alpha} S^0 \ee^{-\phi/2}\,\ee^{\ii p\cdot X}~,
\end{aligned}
\end{equation}
for the adjoint fields, and
\begin{equation}
 \label{mattervert}
\begin{aligned} 
V_{\phi^I}& =  (\pi \alpha')^{\frac{1}{2}} \left(\phi^I\right)^{u_A}_{~v_{A\otimes I}}\, 
{\bar\Psi}_I \,\ee^{-\phi}\,\ee^{\ii p\cdot X}~,\\
V_{ {\bar\phi}_I}& = (\pi \alpha')^{\frac{1}{2}}  \left({\bar\phi}_I\right)^{u_{A\otimes I}}_{~~\,v_A}\,
\Psi^I \,\ee^{-\phi}\,\ee^{\ii p\cdot X}~,\\
V_{\psi^I}& =  (2 \pi \alpha')^{\frac{3}{4}} \left(\psi^{\alpha I}\right)^{u_A}_{~v_{A\otimes I}}\,
S_\alpha S_I \,\ee^{-\phi/2}\,\ee^{\ii p\cdot X}~,\\
V_{{\bar\psi}_I} & = (2 \pi \alpha')^{\frac{3}{4}} \left({\bar\psi}_{\dot\alpha I}\right)^{u_{A\otimes
 I}}_{~~\,v_A}\, S^{\dot\alpha} S^I \,\ee^{-\phi/2}\,\ee^{\ii p\cdot X}~,\\
V_{F^I}& = {\pi \alpha'} \left(F^I\right)^{u_A}_{~v_{A\otimes I}}\, 
\epsilon_{IJK}\,{\Psi}^J{\Psi}^K \,\ee^{\ii p\cdot X}~,\\
V_{{\bar F}_I}& = {\pi \alpha'} \left({\bar F}_I\right)^{u_A}_{~v_{A\otimes I}}\, 
\epsilon^{IJK}\,{\bar\Psi}_J{\bar\Psi}_K \,\ee^{\ii p\cdot X}~,
\end{aligned}
\end{equation}
for the bi-fundamental matter fields. The internal spin fields appearing in the above formulas are
\begin{equation}
 \begin{aligned}
S^0&=S^{+++}~,~S^1=S^{+--}~,~S^2=S^{-+-}~,~S^3=S^{--+}~,\\
S_0&=S_{---}~,~S_1=S_{-++}~,~S_2=S_{+-+}~,~S_3=S_{++-}~,
 \end{aligned}
\label{spinfields}
\end{equation}
where the $\pm$'s denote the signs of the spinorial weights of $\mathrm{SO}(6)$. Notice that
$S^A$ and $S_A$ trasform in the  representation $R_A$ of the orbifold group.

\paragraph{Vertices for instanton moduli}
Strings with at least one end-point on a D$(-1)$ brane give rise to moduli rather 
than dynamical fields because either they do not have longitudinal Neumann directions at all
or they have mixed boundary conditions. The vertex operators for
D$(-1)$/D$(-1)$ moduli with alike end-points are
\begin{equation}
 \label{D-1vert}
\begin{aligned}
 V_a& = \sqrt{2} g_0 (\pi \alpha')^{\frac{1}{2}}  \left(a_\mu\right)^{i_A}_{~j_A}\,
\psi^\mu\,\ee^{-\phi}~,\\
 V_D& = 2\pi \alpha' \left(D_c\right)^{i_A}_{~j_A}\bar\eta^c_{\mu\nu}\, \psi^\mu\psi^\nu~,\\
 V_M& = \frac{g_0}{\sqrt{2}} (2 \pi \alpha')^{\frac{3}{4}} 
\left(M_\alpha\right)^{i_A}_{~j_A}\, S_\alpha S_0 \ee^{-\phi/2}~,\\
 V_\lambda& = (2 \pi \alpha')^{\frac{3}{4}} \left(\lambda_{\dot\alpha}\right)^{i_A}_{~j_A}\,
S^{\dot\alpha} S^0 \,\ee^{-\phi/2}~,
\end{aligned}
\end{equation} 
where D-instanton gauge coupling $g_0$ is expressed in terms of $\alpha'$ and $g_s$ as in (\ref{g0}).
If the end-points are different we instead have
\begin{equation}
 \label{D-1Ivert}
\begin{aligned}
 V_{\chi^I}& = (\pi \alpha')^{\frac{1}{2}} \left(\chi^I\right)^{i_A}_{~j_{A\otimes I}}\,
 {\bar\Psi}_I \,\ee^{-\phi}~,\\
 V_{ {\bar\chi}_I}& = (\pi \alpha')^{\frac{1}{2}} \left({\bar\chi}_I\right)^{i_{A\otimes I}}_{~\,j_A}\,
\Psi^I \,\ee^{-\phi}~,\\
 V_{M^I}& =  \frac{g_0}{\sqrt{2}} (2 \pi \alpha')^{\frac{3}{4}}
\left(M^{\alpha I}\right)^{i_A}_{~j_{A\otimes I}}\, S_\alpha S_I \,\ee^{-\phi/2}~,\\
 V_{\lambda_I}& = (2 \pi \alpha')^{\frac{3}{4}} \left(\lambda_{\dot\alpha I}\right)^{i_{A\otimes 
I}}_{~\,j_A}\, S^{\dot\alpha} S^I\,\ee^{-\phi/2}~.
\end{aligned}
\end{equation}
The moduli of the charged sector arise from open strings with D3/D$(-1)$ or D$(-1)$/D3 boundary conditions. The vertex operators for those which in the field theory limit describe gauge instantons
are
\begin{equation}
\label{mixgaugevert}
\begin{aligned}
V_w& = \frac{g_0}{\sqrt{2}} (2 \pi \alpha')^{\frac{1}{2}}  \left(w_{\dot\alpha}\right)^{u_A}_{~j_A}\, 
\Delta \,S^{\dot\alpha} \,\ee^{-\phi}~,\\
V_{\bar w}& = \frac{g_0}{\sqrt{2}} (2 \pi \alpha')^{\frac{1}{2}}  \left({\bar w}_{\dot\alpha}\right)^{i_A}_{~u_A}\, \bar\Delta \,S^{\dot\alpha} \,\ee^{-\phi}~,\\
V_\mu& = \frac{g_0}{\sqrt{2}} (2 \pi \alpha')^{\frac{3}{4}} 
\left(\mu\right)^{u_A}_{~j_A}\, \Delta \,S_0 \,\ee^{-\phi/2}~,\\
V_{\bar \mu}& = \frac{g_0}{\sqrt{2}} (2 \pi \alpha')^{\frac{3}{4}} \left(\mu\right)^{i_A}_{~u_A}\, {\bar\Delta} \,S_0 \,\ee^{-\phi/2}~,
\end{aligned}
\end{equation}
while those related to stringy instantons are
\begin{equation}
\label{mixstringvert}
\begin{aligned}
V_{\mu^I} & = \frac{g_0}{\sqrt{2}} (2 \pi \alpha')^{\frac{3}{4}} 
\left(\mu^I\right)^{u_A}_{~i_{A\otimes I}}\, \Delta\, S_I \,\ee^{-\phi/2}~,\\
V_{{\bar\mu}_I}& = \frac{g_0}{\sqrt{2}} (2 \pi \alpha')^{\frac{3}{4}} \left({\bar \mu}_I\right)^{i_{A\otimes I}}_{~u_A}\, \bar\Delta\,S_I \,\ee^{-\phi/2}~,
\end{aligned}
\end{equation}
where $\Delta$ and $\bar{\Delta}$ are the bosonic twist and anti-twist operators respectively and encode the change of boundary condition from Neumann to Dirichlet and \emph{vice-versa}.

\section{Derivation of the non-perturbative flux effects for $N_f=N_c-1$}
\label{app:intG}

As discussed in Section \ref{subsecn:Gflux}, the non-perturbative interactions induced by the flux
$G$ are obtained at linear order by computing the integral (see Eq. (\ref{weffg}))
\begin{equation}
{W}_{\mathrm{n.p.}}(G) = \Lambda^{\beta_0}
\int d\,{\widehat{\mathfrak M}}
\,~\ee^{-S^{(0)}_{\mathrm{D3/D(-1)}}(\Phi,\bar\Phi)}\,\left(\frac{\ii}{2}\,G\bar\mu\,\mu\right)
\label{weffg1} 
\end{equation}
where the action on the moduli space is given in Eq. (\ref{sk1}). As we have seen we expect a non-vanishing contribution to this integral when $N_f=N_c-1$, whose schematic form is given in Eq. (\ref{weffG1}).

Here we consider in detail the case of the $\mathrm{SU}(2)$ theory, {\it i.e.} $N_c=2$ and $N_f=1$.
Denoting by $Q^u$ and $\widetilde Q_u$ (with $u=1,2$) the fundamental
and anti-fundamental blocks of the matter superfield $\Phi$ (see Eq. (\ref{Phi})), and by
$\mu'$ and $\bar\mu'$ the non-trivial components of $\mu^1$ and $\bar\mu^1$ respectively,
the moduli action (\ref{sk1}) in this case becomes
\begin{equation}
\begin{aligned}
S^{(0)}_{\mathrm{D3/D(-1)}} =&
~\ii\,D_{c}\big(\bar{w}_{\dot\alpha u}
(\tau^c)^{\dot\alpha}_{~\dot\beta} w^{\dot\beta u}\big)
+\ii \,\lambda_{\dot\alpha} \big(\bar{\mu}_u\,w^{\dot{\alpha} u}
+\bar{w}^{\dot{\alpha}}_{~u}\,\mu^u\big)\\
&~+\left[\frac12\,\bar{w}_{\dot{\alpha}u}\big(Q^u \bar Q_v+\bar{\widetilde Q}^{u}\widetilde Q_v\big)
w^{\dot{\alpha}v}
+\frac{\ii}{2}\,\bar\mu'\,
\bar Q_u\,\mu^u - \frac{\ii}{2}\,\bar\mu_u\,\bar{\widetilde Q}^{u} \,\mu'\right.\\
&\left.\phantom{\frac12}+\ii\,\bar w_{\dot \alpha u}\big(\bar D^{\dot\alpha}\bar{\widetilde Q}^{u}\big)\mu'
-\ii\,\bar\mu'\big(\bar D_{\dot\alpha}\bar Q_u\big) w^{\dot\alpha u}\right]_{\bar\theta=0}
\end{aligned}
\label{smod30}
\end{equation}
where we have explicitly indicated also the two-valued color indices.

The integration over $d^2\lambda$
allows to soak up the two fermionic zero-modes $\mu$ and $\bar \mu$ left after the $G$-flux insertion.
After some elementary algebra, one finds 
a contribution simply proportional to
\begin{equation}
\bar{w}_{\dot{\alpha}u}{w}^{\dot{\alpha}u}~.
\label{contr1}
\end{equation}
To saturate the Grassmannian integrals over $d\mu'$ and $d\bar\mu'$ the only option is to bring down the terms
containing $\bar D_{\dot\alpha}\bar{\widetilde Q}$ and $\bar D_{\dot\alpha}\bar Q$ from the moduli action, thus obtaining a contribution proportional to
\begin{equation}
\left.\big( \bar D^{\dot\alpha}\bar{\widetilde Q}^{u}
\bar D^{\dot\beta}\bar Q_v \big)\right|_{\bar\theta=0}\,
\bar{w}_{\dot{\alpha}u}{w}_{\dot{\beta}}^{~v}~.
\label{contr2}
\end{equation}
Thus, after integrating over all fermionic instanton moduli, we are left with
\begin{eqnarray}
{W}_{\mathrm{n.p.}}(G) &=& {\mathcal C}\,G\,\Lambda^{5}\,
\big( \bar D^{\dot\alpha}\bar{\widetilde Q}^{u}
\bar D^{\dot\beta}\bar Q_v \big)
\int \!\!d^4w\,d^4\bar w\,d^3D \nn \\
&& \cdot~\ee^{-\ii D_c \bar w \tau^c w -\frac{1}{2}\bar w(Q\bar Q+\bar{\widetilde Q}\widetilde Q)w}
~\bar{w}_{\dot{\alpha}u}{w}_{\dot{\beta}}^{~v}\,
\bar{w}_{\dot{\gamma}r}{w}^{\dot{\gamma}r}
\label{weffg20}
\end{eqnarray}
where we have clumped all numerical constants in the normalization factor $\mathcal C$ and understood that we must set $\bar\theta=0$ in the right hand side.
The bosonic integral (\ref{weffg20}) has been evaluated in Ref. \cite{Matsuo:2008nu} (see
in particular Eq. (5.7) of the published version) and the result is
\begin{equation}
\int \!\!d^4w\,d^4\bar w\,d^3D
\,~\ee^{-\ii D_c \bar w \tau^c w -\bar w A w}
~\bar{w}_{\dot{\alpha}u}{w}_{\dot{\beta}}^{~v}\,
\bar{w}_{\dot{\gamma}r}{w}^{\dot{\gamma}r} = 
\frac{2\epsilon_{\dot\alpha\dot\beta}\,\delta_u^{~v}}{\big(\mathrm{tr}\, A\big)^3}
\label{int20}
\end{equation}
where $A$ is the $2\times 2$ hermitian matrix $A=\frac{1}{2}\big(Q\bar Q+\bar{\widetilde Q}
\widetilde Q\big)$. Exploiting the D-flatness condition (\ref{D-flat1}) for $\xi=0$, it is easy to prove that
\begin{equation}
\mathrm{tr}\, A = \big(\bar M M\big)^{1/2}
\label{AM}
\end{equation}
where $M=\widetilde Q_u Q^u$ is the meson superfield. Using these results in (\ref{weffg20}), after some
simple manipulations and absorbing all numerical factors by redefining the overall coefficient $\mathcal C$, we finally obtain
\begin{equation}
{W}_{\mathrm{n.p.}}(G)= \left. G\,\Lambda^5\,\frac{\bar D^2 \bar M}{(\bar M M)^{3/2}}
\right|_{\bar\theta=0}~.
\label{N2G1}
\end{equation}
As mentioned in the main text, this explicit result is in full agreement with the general expression
(\ref{weffG1}) obtained using dimensional analysis and $\mathrm{U}(1)^3$ charge conservation.

\section{Derivation of the non-perturbative flux effects for $N_f=N_c$}
\label{app:intGbar}
Let us now consider the effective interaction induced by a $\bar G$ background flux which, according to Eq. (\ref{weffgbar}), is given by
\begin{equation}
W_{\mathrm{n.p.}}(\bar G) = (\pi\alpha')^2\,\frac{2\pi\ii}{g_s}\,\Lambda^{\beta_0}\,\bar G\, 
\int d\,{\widehat{\mathfrak M}'}
\,~\ee^{-\left.S_{\mathrm{D3/D(-1)}}(\Phi,\bar\Phi)\right.}~,
\label{weffgbar22}
\end{equation}
where in the moduli action it is understood that we have to set $\bar\theta=0$.

We focus on the case $N_0=N_1$, corresponding to SQCD with $N_c=N_f$ flavors. 
The moduli action to be used in (\ref{weffgbar22}) reads explicitly
\begin{eqnarray}
\left.S_{\mathrm{D3/D(-1)}}\right|_{\bar\theta=0} & = & \ii\,D_{c}\big(\bar{w}_{\dot\alpha u}
(\tau^c)^{\dot\alpha}_{~\dot\beta} w^{\dot\beta u}\big)
+\left[\frac{1}{2}\,\bar{w}_{\dot\alpha u}\big(Q^u_{~f} {\bar Q}^f_{~v} + 
{\bar{\widetilde Q}}^u_{~f} {\widetilde Q}^f_{~v}
\big) w^{\dot\alpha v}\right.
\nn\\
&& \!\!\left.+\frac{\ii}{2}\,\bar\mu^1_f\,
{\bar Q}^f_{~u}\,\mu^u - \frac{\ii}{2}\,{\bar\mu}_u\, {\bar{\widetilde Q}}^u_{~f}\,\mu^{1f}\right. \nn\\
&&\!\!\left.
+\ii\,\bar w_{\dot \alpha u}\big(\bar D^{\dot\alpha}{\bar{\widetilde Q}}\big)^u_{~f}\mu^{1f}
-\ii\,\bar\mu^1_f\big(\bar D_{\dot\alpha}\bar Q\big)^f_{~u} w^{\dot\alpha u}\right]_{\bar\theta=0}
\label{actbu} 
\end{eqnarray}
where $u$ and $f$ are fundamental color and flavor indices respectively.
With respect to Eq. (\ref{sk2}) we
have neglected the quadratic term in $D_c$'s since it appears with an explicit $\alpha'^2$ in front, and it leads to effects of higher order in $\alpha'$ with respect to the ones we will compute in the following.

Since we have chosen $N_f=N_c$, we have the same number of $\mu^u$, ${\bar\mu}_u$ and $\mu^{1f}$, ${\bar\mu}^1_f$
and the integral over these fermionic moduli yields simply
(up to numerical constants which we disregard)
\begin{equation}
 \label{rismm1}
\det {{\bar Q}}\, \det {\bar{\widetilde Q}} = \det \bar M~,
\end{equation}
where $M^f_{~g} = {\widetilde Q}^f_{~u} Q^u_{~g}$ is the meson superfield matrix. 
The above expression has to be evaluated at $\bar\theta=0$, so that only the scalar components appear.

For $N_c=N_f=2$ the integral over the bosonic variables has exactly the form considered in Eq. (5.6) of Ref. \cite{Matsuo:2008nu}:
\begin{equation}
 \label{intw}
 \int d^3D\, d^{4}\bar w\,  d^{4}w\,\ee^{-\ii D_c \bar w \tau^c w -\bar w A w} = \frac{1}{\tr A}~,
\end{equation}
where the $2\times 2$ matrix $A$ is given by
\begin{equation}
 \label{Ais}
A^u_{~v} = \frac{1}{2}\big(Q^u_{~f} {\bar Q}^f_{~v} + {\bar {\widetilde Q}}^u_{~f} {\widetilde Q}^f_{~v}\big)~.
\end{equation}
Using the D-flatness condition (\ref{D-flat1}), it is easy to see that the trace of $A$ can be re-expressed in terms of the low-energy degrees of freedom represented by the meson and baryon superfields $M$, $B$ and $\tilde B$
as follows:
\begin{equation}
 \label{tramb}
\tr A = \big(\tr \bar M M + \bar B B+ \bar {\widetilde B}\widetilde B\big)^{\frac 12}~.
\end{equation}
Inserting into Eq. (\ref{weffgbar22}) the result (\ref{rismm1}) of the fermionic integration and the bosonic integral
(\ref{intw}) we finally get
\begin{equation}
 \label{GbarN2bis}
{W}_{\mathrm{n.p.}} = \mathcal{C} \,\alpha'^2\, \bar G\,\Lambda^{4}\,
\left.\frac{\det \bar M}{\big(\tr \bar M M+\bar B B+ \bar {\widetilde B}\widetilde B\big)^{1/2}} \right|_{\bar\theta=0}
\end{equation}
as reported in (\ref{GbarN2}).

\chapter{Remarks on effective strings}

\section{Loop expansion of the exact result}
\label{app:a}
As discussed in section \ref{sec:bos_string} eq.(\ref{boskkp}),
the exact expression for the interface partition function
\begin{equation}
\label{eq:I(d)}
\mathcal{I}^{(d)}=2\left(\frac{\sigma}{2\pi}\right)^{\frac{d-2}{2}}V_{T}\sqrt{\sigma
\mathcal{A} u}
\sum_{m=0}^{\infty}\sum_{k=0}^{m}c_{k}c_{m-k}\left(\frac{\cale}{u}\right)^{\frac{d-1}{2}}K_{\frac{d-1}{2}}(\sigma
\mathcal{A} \cale)
\end{equation}
has to be expanded for large $\sigma\mathcal{A}$, in order to compare it with the result of the functional integral approach. Therefore we have to use the asymptotic expression eq. (\ref{bfexp}) of the modified Bessel functions  for large argument which reads, making explicit the first coefficients
$a_{1}^{(j)}$ and $a_{2}^{(j)}$, 
\begin{equation*}
K_{j}(z)\sim\sqrt{\frac{\pi}{2z}}e^{-z}\left\{
1+\frac{4j^{2}-1}{8z}+\frac{(4j^{2}-1)(4j^{2}-9)}{2!(8z)^{2}}+...\right\}~.
\end{equation*}
Inserting the expansion of Bessel functions in eq. (\ref{eq:I(d)}) gives \eq{Idexp1} (which we repeat here for commodity):
\begin{equation}
\label{expansionI}
\mathcal{I}^{(d)}=\sqrt{2\pi}\left(\frac{\sigma}{2\pi}\right)^{\frac{d-2}{2}}V_{T}\sum_{k,k'=0}^{\infty}c_{k}c_{k'}
\left(\frac{\cale}{u}\right)^{\frac{d-2}{2}}e^{-\sigma
\mathcal{A}\cale}\left(1+\sum_{r=1}^{\infty}a_{r}^{(\frac{d-1}{2})}(\sigma\mathcal{A}\cale)^{-r}\right)~.
\end{equation}
The expression of $\cale$ was given in \eq{bos19}, and can be rewritten as
\begin{equation*}
\cale(a,b)=\sqrt{1 + \frac{4\pi\, u}{\sigma \mathcal{A}}a +
\frac{4\pi^{2} \, u^2\,}{(\sigma\mathcal{A})^2}b^{2}}
\end{equation*}
in terms of $a$ and $b$ defined as in \eq{ab}:
\begin{equation}
a = k + k' - \frac{d-2}{12}~, \hskip 0.4cm b = k - k'~.
\end{equation}
We expand $\cale$ as:
\begin{equation*}
\cale(a,b)\sim 1 + \frac{u}{\sigma \mathcal{A}}2\pi a +
\frac{u^{2}}{(\sigma\mathcal{A})^{2}}2\pi^{2}\left(b^{2}-a^{2}\right)+\frac{u^{3}}{(\sigma\mathcal{A})^{3}}4\pi^{3}a\left(a^{2}-b^{2}\right)+...
\end{equation*}
and insert this expansion into eq. (\ref{expansionI}), obtaining
\begin{eqnarray*}
\mathcal{I}^{(d)}&\propto& \sigma^{\frac{d-2}{2}}\frac{e^{-\sigma
\mathcal{A}}}{u^{\frac{d-2}{2}}}\sum_{k,k'=0}^{\infty} c_{k}
c_{k'} e^{-2\pi u \left(k+k'-\frac{d-2}{12}\right)}
 \\
&\times&  \left\{1+(d-2)\left[\frac{\pi
a u}{\sigma \mathcal{A}}+\frac{u^2}{(\sigma \mathcal{A})^{2}}
\left((d-6)\frac{\pi^{2}}{2}a^{2}+\pi^{2}b^{2}\right)\right]+\mathcal{O}\left(\frac{1}{(\sigma\mathcal{A})^{4}}\right)\right\}
\\
&\times&  \left\{1+\left(a^{2}-b^{2}\right)\left[\frac{2\pi^{2} u^{2}}{\sigma\mathcal{A}}
- \frac{4\pi^{3} u^3 a -
2\pi^{4} u^4 \left(a^{2}-b^{2}\right)}{(\sigma\mathcal{A})^{2}}\right]  +
\mathcal{O}\left(\frac{1}{(\sigma\mathcal{A})^{3}}\right)\right\} \\
&\times& \left\{ 1+\frac{1}{\sigma \mathcal{A}}\frac{d(d-2)}{8} -
\frac{\frac{d(d-2)(d^{2}-2d-8)}{128} - \frac{\pi u\,d(d-2)a}{4}}{(\sigma\mathcal{A})^{2}} 
+\mathcal{O}\left(\frac{1}{(\sigma\mathcal{A})^{3}}\right)\right\}.
\end{eqnarray*}

The result up to the third order is thus of the form of \eq{iexp1}: 
\begin{equation}
\label{expansion} \mathcal{I}^{(d)}\propto
\sigma^{\frac{d-2}{2}}\frac{e^{-\sigma
\mathcal{A}}}{u^{\frac{d-2}{2}}}\sum_{k,k'=0}^{\infty} c_{k}
c_{k'} e^{-2\pi u
\left(k+k'-\frac{d-2}{12}\right)}\left\{1+\frac{g_{1}}{\sigma\mathcal{A}}+\frac{g_{2}}{(\sigma\mathcal{A})^{2}}\right\},
\end{equation}
with $g_{1}$ and $g_{2}$ explicitly given by
\begin{eqnarray}
\label{g1}
g_{1}&=& \frac{d(d-2)}{8}+
u(d-2)\pi a-u^{2}2\pi^{2}\left(b^{2}-a^{2}\right)~,\\
\label{g2}
g_{2}&=& \left[\frac{d(d-2)(d^{2}-2d-8)}{128}
\right] +
u\frac{d(d-2)(d-4)\pi}{8}a\nonumber\\
&+& u^{2}\frac{(d-2)(d-4)\pi^{2}}{4}\left(3a^{2}-b^{2}\right) + u^{3} 2\pi^{3}(d-4)a
\left(a^{2}-b^{2}\right) \nn \\
&+& u^{4}2\pi^{4}\left(b^{2}-a^{2}\right)^{2}.
\end{eqnarray}
Introducing the new variable $Q=\exp(-2\pi u)$ we can write eq.
(\ref{expansion}) in the form
\begin{equation}
\mathcal{I}^{(d)}\propto \sigma^{\frac{d-2}{2}}\frac{e^{-\sigma
\mathcal{A}}}{u^{\frac{d-2}{2}}}\lim_{\overline{Q}\rightarrow
Q}\left\{1+\frac{g_{1}}{\sigma\mathcal{A}}+\frac{g_{2}}{(\sigma\mathcal{A})^{2}}\right\}[\eta(Q)\eta(\overline{Q})]^{2-d}~,
\end{equation}
where we have replaced $a$ and $b$ with the derivatives on $Q$ and $\overline{Q}$ (see eq. (\ref{subder})):
\begin{eqnarray}
\label{ainQ}
a&\rightarrow& Q\frac{d}{dQ}+\overline{Q}\frac{d}{d\overline{Q}}~,\\
b&\rightarrow& Q\frac{d}{dQ}-\overline{Q}\frac{d}{d\overline{Q}}~.
\end{eqnarray}
As we can see from the expansion coefficients (\ref{g1}) and (\ref{g2}), the only terms involving powers of $a$ and $b$ higher than 1 are
\begin{eqnarray*}
b^{2}-a^{2}&=&-4Q\frac{d}{dQ}\overline{Q}\frac{d}{d\overline{Q}}\\
3a^{2}-b^{2}&=& 2\left(Q\frac{d}{dQ}\right)^{2}+2\left(\overline{Q}\frac{d}{d\overline{Q}}\right)^{2}+8Q\frac{d}{dQ}\overline{Q}\frac{d}{d\overline{Q}}\\
\left(b^{2}-a^{2}\right)^{2}&=& 16\left(Q\frac{d}{dQ}\right)^{2}\left(\overline{Q}\frac{d}{d\overline{Q}}\right)^{2}
\end{eqnarray*}
which have to be applied to $\eta^{2-d}(Q)\eta^{2-d}(\overline{Q})$.
Applying the expressions \eq{der1eta} and \eq{der2eta} of the first logarithmic derivatives of the eta-function
given in Appendix \ref{app:useful}, we have
\begin{eqnarray*}
Q\frac{d}{dQ}\eta^{2-d}(Q)&=&(2-d) \eta^{2-d}(Q)
\frac{E_{2}(Q)}{24}~,\\
\left(Q\frac{d}{dQ}\right)^{2}\eta^{2-d}(Q)&=&(2-d)\eta^{2-d}(Q)\frac{(4-d)E_{2}^{2}(Q)-2E_{4}(Q)}{576}~.
\end{eqnarray*} 
It is easy to check that, after the application of the derivatives, a factor of $\eta^{4-2d}(\ii u)$ 
will always appear in front of the various terms and
the total result up to the third order can be written in the following form:
\begin{equation*}
\mathcal{I}^{(d)}\propto\sigma^{\frac{d-2}{2}}\frac{\ee^{-\sigma
\mathcal{A}}}{u^{\frac{d-2}{2}}}\frac{1}{\eta^{2d-4}(\ii u)}
\left\{1+\frac{f_{1}}{\sigma\mathcal{A}}+\frac{f_{2}}{(\sigma\mathcal{A})^{2}}\right\}~.
\end{equation*}
First of all we consider the $f_{1}$ term derived from the $g_{1}$ one. As one can see, the only derivative involved in the computation is $Q\frac{d}{dQ}$: this implies that our final formula will include $E_{2}$ functions but not $E_{4}$. We can divide the calculation in
function of the powers of $u$. The term independent from it is the first one appearing in eq. (\ref{g1}):
\begin{equation*}
f_{1,0}=\frac{d(d-2)}{8}.
\end{equation*}
The next power of $u$, after the substitution (\ref{ainQ}), gives, acting on the $\eta$ functions,
\begin{equation*}
f_{1,1}= - 2\pi u
(d-2)^{2}\frac{E_{2}(\ii u)}{24}.
\end{equation*}
The last term is instead:
\begin{equation*}
f_{1,2}= 2\pi^{2} u^2 (d-2)^{2}\frac{E_{2}^{2}(\ii u)}{144}.
\end{equation*}
The final result is:
\begin{equation}
\label{f1}
f_{1} =
\left\{\frac{(d-2)^{2}}{2}\left[\left(\frac{\pi}{6}\right)^{2}u^{2}E_{2}^{2}(\ii u)-
\frac{\pi}{6}uE_{2}(\ii u) \right]+\frac{d(d-2)}{8}\right\}.
\end{equation}
Proceeding in the same way to evaluate $f_{2}$ we find:
\begin{eqnarray}
f_{2}&=&
\left\{u^{4} \frac{\pi^{4}(d-2)^{2}}{18} \left[\frac{(4-d)E_{2}^{2}-2E_{4}}{24}\right]^{2}
+u^{3} \frac{\pi^{3}}{72}(d-4)(d-2)^{2} \right. 
\nonumber\\
&\times& \left. \left[\frac{E_{2}^{3}}{12}(4-d)- \frac{E_{2}E_{4}}{6}\right]
+u^{2}\frac{\pi^{2}(d-2)^{2}(d-4)}{4}\left[\frac{E_{4}}{72}+\frac{(3d-8)E_{2}^{2}}{144}\right]
\right.
\nonumber\\
&-& \left. u\frac{\pi}{96}d(d-2)^{2}(d-4)E_{2}+\frac{1}{128}d(d-2)(d^{2}-2d-8) \right\}
\end{eqnarray}
(all modular forms above have to be evaluated at $\ii u$).

\section{The two loop contribution: functional integral computation}
\label{app:2loop}
In this appendix we check the calculation of the two-loop terms of
the free energy made by Dietz and Filk with the functional
integral method. Starting from the partition function written in
the physical gauge and expanding it in powers of $\frac{1}{\sigma
\mathcal{A}}$, one can evaluate it on a rectangular domain $B$ of sizes $(R,T)$.
Defining $\mathbf{G}$ as the
inverse Laplacian of the theory, the second order terms are given, in the notation of \cite{Dietz:1982uc},
by:
\begin{eqnarray}
\label{eq:Z_{2}}
{Z}_{\Gamma,B}^{(2)} &=& \Bigl[\lambda_{1} D (D-1)(\langle1\rangle-\langle2\rangle)+\lambda_{2} D (\langle3\rangle+\langle4\rangle+ 2 \langle1\rangle - 4 \langle2\rangle) \nonumber \\
&+& \frac{\lambda_{2}}{3} D (D-1)
(\langle3\rangle+\langle4\rangle+ 6 \langle1\rangle)\Bigr]~,
\end{eqnarray}
where $D$ is $d-2$ in our notation, $\Gamma$ speficies the topology of the boundary $B$; we are 
interested in the case where the topology is $S_1\times S_1$. 
$\lambda_{1}$ and $\lambda_{2}$ are parameters which depend on the string model under consideration 
and in particular they are given by
\begin{equation}
\lambda_{1}=1,
\hskip 0.4cm
\lambda_{2}=-\frac{1}{4}
\end{equation}
in the Nambu-Goto case. Moreover, the following definitions are used:
\begin{eqnarray*}
\langle1\rangle &=& \frac{1}{\sigma}\int_{0}^{R} dz \int_{0}^{T}
dt \frac{\partial^{2}\mathbf{G}}{\partial z \partial
z'}|_{z=z',t=t'} \frac{\partial^{2}\mathbf{G}}{\partial t \partial
t'}|_{z=z',t=t'}~,\\
\langle2\rangle &=& \frac{1}{\sigma}\int_{0}^{R} dz \int_{0}^{T}
dt \left(\frac{\partial^{2}\mathbf{G}}{\partial z
\partial
t'}\right)^{2}|_{z=z',t=t'}~, \\
\langle3\rangle &=& \frac{3}{\sigma}\int_{0}^{R} dz \int_{0}^{T}
dt \left(\frac{\partial^{2}\mathbf{G}}{\partial z \partial
z'}\right)^{2}|_{z=z',t=t'}~, \\
\langle4\rangle &=& \frac{3}{\sigma}\int_{0}^{R} dz \int_{0}^{T}
dt \left(\frac{\partial^{2}\mathbf{G}}{\partial t \partial
t'}\right)^{2}|_{z=z',t=t'}~.
\end{eqnarray*}
Notice that the formul\ae ~used by Dietz and Filk in eq. (3.4) have to be multiplied by $-\frac{1}{2}$ to get 
the right normalization of their final result eq. (3.7). 

We shall deal with the divergent terms within the $\zeta$-function regularization scheme, following Dietz and Filk.
In the Nambu-Goto model with  $S_{1}\times S_{1}$ boundary
conditions on the rectangular domain, the Green function is:
\begin{equation*}
\mathbf{G} = \frac{1}{4\pi^{2}RT} \sum_{\atopnew{m,n=-\infty}{(m,n)\neq (0,0)}}^{+\infty} \frac{\exp{\left\{2\pi \ii
\frac{m}{T}(t-t')\right\}}\exp{\left\{2\pi \ii
\frac{n}{R}(z-z')\right\}}}
{\frac{n^2}{R^{2}}+\frac{m^{2}}{T^{2}}}~.
\end{equation*}
Let us now evaluate the various terms of
${Z}_{S_1\times S_1,(R,T)}^{(2)}$. To begin with, we find
\begin{eqnarray}
\label{1prov}
\langle1\rangle &=& \frac{1}{\sigma R T}  \sum_{\atopnew{m,n,p,q=-\infty}{\atopnew{(m,n)\neq (0,0)}{(p,q)\neq (0,0)}}}^{+\infty}  \frac{\frac{n^{2}}{R^{2}}\frac{q^{2}}{T^{2}}} {\left( \frac{n^{2}}{R^{2}}+\frac{m^{2}}{T^{2}}\right)\left( \frac{p^{2}}{R^{2}}+\frac{q^{2}}{T^{2}}\right)} \nonumber\\
&=&\frac{1}{\sigma R T} \left[ 16 \sum_{m,n,p,q=1}^{+\infty}
\frac{\frac{n^{2}}{R^{2}}\frac{q^{2}}{T^{2}}} {\left(
\frac{n^{2}}{R^{2}}+\frac{m^{2}}{T^{2}}\right)\left(
\frac{p^{2}}{R^{2}}+\frac{q^{2}}{T^{2}}\right)} + (m=0,n \neq
0,p,q) \right. \nonumber\\
&+& \left.  (m,n,p=0,q \neq 0)-(m=0,n\neq 0,p=0,q\neq 0) \right]~,
\end{eqnarray}
where with $(m=0,n \neq 0,p,q)$ we indicate the sum over any $n$
different from zero, any $p$ and any $q$ when $m$ has been fixed
at zero. The last term in r.h.s. of eq. (\ref{1prov}) is to avoid
the double counting and the first one can be rewritten using the
equality
\begin{equation}
\label{equality}
\sum_{m,n=1}^{\infty}\frac{\frac{n^{2}}{R^{2}}}{\frac{n^{2}}{R^{2}}+\frac{m^{2}}{T^{2}}}=E_{2}\left(
\ii\frac{R}{T}\right) \frac{\pi}{24} \frac{R}{T}~.
\end{equation}
The result is
\begin{eqnarray*}
\langle1\rangle &=& \frac{1}{\sigma R T} \left[ \left(
\frac{\pi}{6}\right)^{2} E_{2}\left(\ii\frac{R}{T}\right)
E_{2}\left(\ii\frac{T}{R}\right) +(m=0,n \neq 0,p,q)  \right. \\
&+& \left. (m,n,p=0,q\neq 0) - (m=0,n \neq 0,p=0, q\neq 0)\right]~.
\end{eqnarray*}

The last three terms can be evaluated using (\ref{equality}) and the $\zeta$ function regularization and they sum up to
\begin{equation*}
-\frac{\pi}{6}\frac{R}{T}E_{2}\left(\ii\frac{R}{T}\right)
-\frac{\pi}{6}\frac{T}{R}E_{2}\left(\ii\frac{T}{R}\right)+1~,
\end{equation*}
which vanishes using the modular property (\ref{eq:modular E_{2}}) for the
Eisenstein series $E_{2}$.
We get thus
\begin{equation*}
\langle1\rangle=\frac{1}{\sigma R T} H~,
\end{equation*}
where we have defined for notational convenience
\begin{equation}
\label{H}
H=\left( \frac{\pi}{6}\right)^{2} E_{2}\left(
\ii\frac{T}{R}\right) E_{2}\left(\ii\frac{R}{T}\right)~.
\end{equation}

The  contribution  $\langle2\rangle$ vanishes because the terms in the sum are odd:
\begin{equation*}
\langle2\rangle=\frac{1}{\sigma RT}
\sum_{\atopnew{m,n,p,q=-\infty}{\atopnew{(m,n)\neq(0,0)}{(p,q)\neq(0,0)}}}^{+\infty}\frac{mnpq}{\left(\frac{n^{2}}{R^{2}}+\frac{m^{2}}{T^{2}}\right)\left(\frac{p^{2}}{R^{2}}+\frac{q^{2}}{T^{2}}\right)}=0~.
\end{equation*}

To compute the expression $(\langle3\rangle+\langle4\rangle+6\langle1\rangle)$ which appears in (\ref{eq:Z_{2}}), one can rewrite it as a perfect square:
\begin{eqnarray*}
(\langle3\rangle+\langle4\rangle+6\langle1\rangle)&=&\frac{3}{\sigma}\int_{0}^{R}dz\int_{0}^{T}dt\left(\frac{\partial^{2}\mathbf G}{\partial z\partial z'}+\frac{\partial^{2}\mathbf G}{\partial t \partial t'}\right)^{2}|_{z=z',t=t'}\\
&=&\frac{3}{\sigma RT}\left[\sum_{\atopnew{m,n=-\infty}{(m,n)\neq(0,
0)}}^{+\infty}\frac{\frac{n^{2}}{R^{2}}+\frac{m^{2}}{T^{2}}}{\frac{n^{2}}{R^{2}}+\frac{m^{2}}{T^{2}}}\right]^{2}
= \frac{3}{\sigma RT}~.
\end{eqnarray*}

Collecting the previous results we obtain the following system of
equations:
\begin{eqnarray*}
\langle1\rangle &=& \frac{H}{\sigma RT}~, \\
\langle2\rangle &=& 0~, \\
\langle3\rangle + \langle4\rangle + 6\langle1\rangle &=& \frac{3}{\sigma RT}
\end{eqnarray*}
which gives, for the last term in equation (\ref{eq:Z_{2}}):
\begin{equation*}
\langle3\rangle+\langle4\rangle+2\langle1\rangle -4\langle2\rangle
= \frac{3-4H}{\sigma RT}~.
\end{equation*}

The total second order correction is therefore
\begin{eqnarray}
{Z}_{S_1\times S_1,B} &=& \frac{1}{\sigma R T} \left\{ D (D-1) H-\frac{1}{4} D (3-4H) -\frac{3}{12} D (D-1)\right\} \nonumber\\
&=& \frac{1}{\sigma R T} \left\{ D^{2} H -\frac{1}{4} D
(D+2)\right\}~.
\end{eqnarray}
This corresponds to our result%
\footnote{Dietz and Filk found: $ {Z}_{S_1\times S_1,(R,T)}^{(2)}
= \frac{1}{\sigma R T} \left\{D^{2} H -\frac{1}{4} [ D
(4D-1)]\right\}$} eq. (\ref{f1}) where $D=d-2$, up to an overall factor of $-\frac{1}{2}$.

\section{Useful formul\ae}
\label{app:useful}
In this appendix we collect some useful formul\ae.

\paragraph{Dedekind $\eta$ function}
The Dedekind eta function
is defined, in terms of the quantity $q=\exp\{2\pi \ii \tau\}$, by
\begin{equation} 
\label{defeta1}
\eta(\tau)=q^{\frac{1}{24}} \prod_{n=1}^{\infty}
\left( 1-q^{n} \right)~.
\end{equation}
One can expand it in $q$-series:
\begin{equation}
\label{defeta2}
[\eta(\tau)]^{-1}=\sum_{k=0}^{\infty} p_{k}\, q^{k-\frac{1}{24}}~,
\end{equation}
where $p_{k}$'s are the number of
partitions of $k$. In the text, we often switch between the notation $\eta(\tau)$ and $\eta(q)$ for the function defined in eq.s (\ref{defeta1},\ref{defeta2}), according to the convenience.

Under the modular transformations $T$ and $S$, the
Dedekind eta function transforms in the following way:
\begin{eqnarray}
\label{etaT}
\eta(\tau+1)&=& \ee^{\frac{\ii\pi}{12}} \eta(\tau)~, \\
\label{etaS}
\eta(-\frac{1}{\tau})&=& (\ee^{-\ii\frac{\pi}{2}}\tau)^{\frac{1}{2}} \eta(\tau)~.
\end{eqnarray}
\paragraph{Eisenstein series}
The second Eisenstein function is defined by
\begin{equation*}
E_{2}(\tau)=1-24\sum_{n=1}^{\infty}\sigma_{1}(n)q^{n}=1-24\sum_{k=1}^{\infty}\frac{kq^{k}}{1-q^{k}}~,
\end{equation*}
where $\sigma_{1}(n)$ denotes the sum of the positive divisors of
$n$.
An useful property is
\begin{equation}
q\frac{d}{dq}E_{2}(\tau)=\frac{E_{2}^{2}(\tau)-E_{4}(\tau)}{12}~.
\end{equation}
where the fourth Eisenstein series is defined, in terms of the sum
of the cubes of the positive divisors of $n$, $\sigma_{3}(n)$, as
\begin{equation*}
E_{4}(\tau)=1+240\sum_{n=1}^{+\infty}\sigma_{3}\left(n\right)q^{n}~.
\end{equation*}
The modular properties of the Eisenstein functions are:
\begin{eqnarray}
\label{eq:modular E_{2}}
E_{2}(\tau) & = &\left(\frac{1}{\tau}\right)^{2}E_{2}\left(-\frac{1}{\tau}\right)+\frac{6}{\pi}
\frac{\ii}{\tau}~,
\\
E_{2k}(\tau)& = & (-1)^{k}\left(\frac{1}{\tau}\right)^{2k}E_{2k}\left(-\frac{1}{\tau}\right),k\geq 2~.
\end{eqnarray}
Multiple logarithmic derivatives of the eta function are related to Eisenstein series, and in particular we have:
\begin{eqnarray}
\label{der1eta}
q\frac{d}{dq}\eta^{\alpha}(\tau)&=&\alpha \eta^{\alpha}(\tau)
\frac{E_{2}(\tau)}{24}~,\\
\label{der2eta}
\left(q\frac{d}{dq}\right)^{2}\eta^{\alpha}(\tau)&=&\alpha\eta^{\alpha}\frac{(\alpha+2)E_{2}(\tau)^{2}-2E_{4}(\tau)}{576}~.
\end{eqnarray}
\paragraph{Poisson resummation formula}
In section \ref{sec:bos_string} the following resummation formula plays a key r\^ole:
\begin{equation}
\label{poisson_res}
\sum_{n=-\infty}^{+\infty}\exp\{-\pi
an^{2}+2\pi
ibn\}=a^{-\frac{1}{2}}\sum_{m=-\infty}^{+\infty}\exp\left\{
-\frac{\pi(m-b)^{2}}{a}\right\}
\end{equation}

\providecommand{\href}[2]{#2}\begingroup\raggedright\endgroup

\cleardoublepage

 \chapter*{Acknowledgments}
 \addcontentsline{toc}{chapter}{Acknowledgments}
  \noindent 
First of all, I am very grateful to my supervisor Marco Bill\`{o}: without his invaluable help this work would not have been possible. Despite my countless questions about Physics, Mathematics and more practical organizational issues he was always available to offer his help with precise answers and patient suggestions. \\
My gratitude also goes to Marialuisa Frau and Alberto Lerda, for many illuminating discussions and their advice, to Michele Caselle, who opened me the doors of Statistical Mechanics, and to all members of the Theoretical Physics Department who contributed to my scientific education and to make these PhD years serene.\\
I wish to thank the people at \'{E}cole Normale in Paris for their wonderful hospitality that made me feel at home. In particular, I am grateful to Costas Bachas for having given me the opportunity to work at ENS and for his kind support, to Raphael Benichou, Davide Cassani, Giuseppe Policastro, Alberto Rosso, Raoul Santachiara, Waldemar Schulgin, Jan Troost. You have made my stay in Paris an unforgettable period. The help of Carlo Angelantonj at the beginning of my stay there was very precious.
I remember with pleasure also my friends at Maison de l'Italie, my family during the parisian months.\\
Many thanks to Rugiada, simply for her continuous friendship, Giuseppe for being the best office colleague one can desire, Diego for amusing lunches (due also to his bottle of sparkling water), Gabriele for the morning coffees and the innumerable times he helped me in my fight against computers, 
Stefano Cremonesi and Francesco Benini for the tourism done together, and to all friends who shared with me these years. Thanks to Daniele Musso for having pointed out misprints in this thesis.\\
Last but not least, I would like to especially thank my parents for all the time they spent and sacrificed to allow me to dedicate myself to this work without worrying about any additional issue. Without their support to my choices and their encouragements, everything would have been more difficult to be carried on.
\\~\\
This work was partially supported by the INFN section of Torino, the European Superstring Network MRTN-CT-2004-512194, the European Commission FP6 Programme under contract MRTN-CT-2004-005104.

\end{document}